\documentclass[iop,apj,numberedappendix,appendixfloats]{emulateapj}
\usepackage{apjfonts}
\usepackage{natbib}
\usepackage{amssymb}
\usepackage{enumitem}
\usepackage{comment}
\usepackage{multirow}
\usepackage[title]{appendix}
\usepackage{gensymb}
\newcommand{\OVIdblt}{{\rm O}\kern 0.1em{\sc vi}~$\lambda\lambda 1031, 1037$} 
\newcommand{\MgIIdblt}{{\rm Mg}\kern 0.1em{\sc ii}~$\lambda\lambda 2796, 2803$}
\newcommand{\OVI}{\hbox{{\rm O}\kern 0.1em{\sc vi}}}
\newcommand{\OVII}{\hbox{{\rm O}\kern 0.1em{\sc vii}}}
\newcommand{\OVIII}{\hbox{{\rm O}\kern 0.1em{\sc viii}}}
\newcommand{\MgII}{\hbox{{\rm Mg}\kern 0.1em{\sc ii}}}
\newcommand{\CIV}{\hbox{{\rm C}\kern 0.1em{\sc iv}}}
\newcommand{\HI}{\hbox{{\rm H}\kern 0.1em{\sc i}}}
\newcommand{\Hy}{\hbox{{\rm H}}}
\newcommand{\PVdblt}{{\rm P}\kern 0.1em{\sc v}~$\lambda\lambda 1117, 1128$}
\newcommand{\CaIIdblt}{{\rm Ca}\kern 0.1em{\sc ii}~$\lambda\lambda 3934, 3969$}
\newcommand{\AlIIIdblt}{{\rm Al}\kern 0.1em{\sc iv}~$\lambda\lambda 1855, 1863$}
\newcommand{\CIVdblt}{{\rm C}\kern 0.1em{\sc iv}~$\lambda\lambda 1548, 1550$}
\newcommand{\NVdblt}{{\rm N}\kern 0.1em{\sc v}~$\lambda\lambda 1238, 1242$}  
\newcommand{\SVIdblt}{{\rm S}\kern 0.1em{\sc vi}~$\lambda\lambda 933, 944$} 
\newcommand{\SiIIdblt}{{\rm Si}\kern 0.1em{\sc ii}~$\lambda\lambda 1190, 1193$} 
\newcommand{\SiIVdblt}{{\rm Si}\kern 0.1em{\sc iv}~$\lambda\lambda 1393, 1402$} 
\newcommand{\PV}{\hbox{{\rm P}\kern 0.1em{\sc v}}}
\newcommand{\AlI}{\hbox{{\rm Al}\kern 0.1em{\sc i}}}
\newcommand{\AlII}{\hbox{{\rm Al}\kern 0.1em{\sc ii}}}
\newcommand{\AlIII}{{\hbox{\rm Al}\kern 0.1em{\sc iii}}}
\newcommand{\CaII}{\hbox{{\rm Ca}\kern 0.1em{\sc ii}}}
\newcommand{\CII}{\hbox{{\rm C}\kern 0.1em{\sc ii}}}
\newcommand{\CIIe}{\hbox{{\rm C$^{\ast}$}\kern 0.1em{\sc ii}}}
\newcommand{\CIII}{\hbox{{\rm C}\kern 0.1em{\sc iii}}}
\newcommand{\CV}{\hbox{{\rm C}\kern 0.1em{\sc v}}}
\newcommand{\HII}{\hbox{{\rm H}\kern 0.1em{\sc ii}}}
\newcommand{\Lya}{\hbox{{\rm Ly}\kern 0.1em$\alpha$}}
\newcommand{\Lyb}{\hbox{{\rm Ly}\kern 0.1em$\beta$}}
\newcommand{\Lyg}{\hbox{{\rm Ly}\kern 0.1em$\gamma$}}
\newcommand{\Lyd}{\hbox{{\rm Ly}\kern 0.1em$\delta$}}
\newcommand{\Lye}{\hbox{{\rm Ly}\kern 0.1em$\epsilon$}}
\newcommand{\Lyphi}{\hbox{{\rm Ly}\kern 0.1em$\phi$}}
\newcommand{\Lyfive}{\hbox{{\rm Ly}\kern 0.1em$5$}}
\newcommand{\Lysix}{\hbox{{\rm Ly}\kern 0.1em$6$}}
\newcommand{\Lyseven}{\hbox{{\rm Ly}\kern 0.1em$7$}}
\newcommand{\Lyeight}{\hbox{{\rm Ly}\kern 0.1em$8$}}
\newcommand{\Lynine}{\hbox{{\rm Ly}\kern 0.1em$9$}}
\newcommand{\Lyten}{\hbox{{\rm Ly}\kern 0.1em$10$}}
\newcommand{\Lyeleven}{\hbox{{\rm Ly}\kern 0.1em$11$}}
\newcommand{\HeI}{\hbox{{\rm He}\kern 0.1em{\sc i}}}
\newcommand{\HeII}{\hbox{{\rm He}\kern 0.1em{\sc ii}}}
\newcommand{\FeI}{\hbox{{\rm Fe}\kern 0.1em{\sc i}}}
\newcommand{\FeII}{\hbox{{\rm Fe}\kern 0.1em{\sc ii}}}
\newcommand{\FeIII}{\hbox{{\rm Fe}\kern 0.1em{\sc iii}}}
\newcommand{\MnII}{\hbox{{\rm Mn}\kern 0.1em{\sc ii}}}
\newcommand{\MgI}{\hbox{{\rm Mg}\kern 0.1em{\sc i}}}
\newcommand{\MgIII}{\hbox{{\rm Mg}\kern 0.1em{\sc iii}}}
\newcommand{\NI}{\hbox{{\rm N}\kern 0.1em{\sc i}}}
\newcommand{\NII}{\hbox{{\rm N}\kern 0.1em{\sc ii}}}
\newcommand{\NIII}{\hbox{{\rm N}\kern 0.1em{\sc iii}}}
\newcommand{\NV}{\hbox{{\rm N}\kern 0.1em{\sc v}}}
\newcommand{\OI}{\hbox{{\rm O}\kern 0.1em{\sc i}}}
\newcommand{\OII}{\hbox{[{\rm O}\kern 0.1em{\sc ii}]}}
\newcommand{\OIV}{\hbox{{\rm O}\kern 0.1em{\sc iv}]}}
\newcommand{\SI}{{\rm S}\kern 0.1em{\sc i}}
\newcommand{\SIV}{{\rm S}\kern 0.1em{\sc iv}}
\newcommand{\SVI}{{\rm S}\kern 0.1em{\sc vi}}
\newcommand{\SiI}{\hbox{{\rm Si}\kern 0.1em{\sc i}}}
\newcommand{\SiII}{\hbox{{\rm Si}\kern 0.1em{\sc ii}}}
\newcommand{\SiIII}{\hbox{{\rm Si}\kern 0.1em{\sc iii}}}
\newcommand{\SiIV}{\hbox{{\rm Si}\kern 0.1em{\sc iv}}}
\newcommand{\SII}{\hbox{{\rm S}\kern 0.1em{\sc ii}}}
\newcommand{\SIII}{\hbox{{\rm S}\kern 0.1em{\sc iii}}}
\newcommand{\NaI}{\hbox{{\rm Na}\kern 0.1em{\sc i}}}
\newcommand{\TiII}{\hbox{{\rm Ti}\kern 0.1em{\sc ii}}}
\newcommand{\cms}{\hbox{cm$^{-2}$}}
\newcommand{\kms}{\hbox{km~s$^{-1}$}}
\newcommand{\etal}{et~al.}
\newcommand{\NHI}{\log N_{\hbox{\tiny \HI}}}
\newcommand{\nh}{\log n_{\hbox{\tiny \Hy}}}

\shorttitle{CGM Metallicity and Galaxy Orientation}
\shortauthors{\sc Pointon {\etal}}
\slugcomment{Received 2019 January 10; revised 2019 July 29; accepted 2019 July 31}

\begin{document}

\title{Relationship between the Metallicity of the Circumgalactic Medium and Galaxy Orientation}
\author{
Stephanie K. Pointon$^{1,2}$
}
\author{Glenn G. Kacprzak$^{1,2}$
}
\author{Nikole M. Nielsen$^{1}$
}
\author{Sowgat Muzahid$^{3}$
}
\author{Michael T. Murphy$^{1}$
}
\author{Christopher W. Churchill$^{4}$
}
\author{Jane C. Charlton$^{5}$
}

\affil{$^1$ Centre for Astrophysics and Supercomputing, Swinburne
  University of Technology, Hawthorn, Victoria 3122, Australia;
  spointon@swin.edu.au
}
\affil{$^2$ ARC Centre of Excellence for All Sky Astrophysics in 3 Dimensions (ASTRO 3D)
}
\affil{$^3$ Leiden Observatory, University of Leiden, PO Box 9513, NL-2300 RA Leiden, The Netherlands
}
\affil{	$^4$ Department of Astronomy, New Mexico State University, Las Cruces, NM 88003, USA
}
\affil{$^5$ Department of Astronomy and Astrophysics, The Pennsylvania State University, State College, PA 16801, USA
}

\begin{abstract}
We investigate the geometric distribution of gas metallicities in the
circumgalactic medium (CGM) around $47$, $z<0.7$ galaxies from the
``Multiphase Galaxy Halos'' Survey. Using a combination of quasar
spectra from \textit{HST}/COS and from Keck/HIRES or VLT/UVES we
measure column densities of, or determine limits on, CGM absorption
lines. We then use a Monte-Carlo Markov chain approach with Cloudy to
estimate the metallicity of cool (T$\sim$$10^4$K) CGM gas. We also use \textit{HST} images to determine host galaxy
inclination and quasar--galaxy azimuthal angles.  Our sample spans a {\HI} column density range of $13.8~{\cms}<{\NHI}<19.9~{\cms}$. We find (1) while the metallicity
distribution appears bimodal, a Hartigan dip test cannot rule out a
unimodal distribution ($0.4\sigma$). (2) CGM metallicities are
independent of halo mass, spanning three orders of magnitude at fixed
halo mass. (3) The CGM metallicity does not depend on the galaxy
azimuthal and inclination angles regardless of {\HI} column density,
impact parameter and galaxy color. (4) Ionization parameter does not
depend on azimuthal angle. We suggest that the partial Lyman limit metallicity
bimodality is
not driven by a spatial azimuthal bimodality. Our results are
consistent with simulations where the CGM is complex and outflowing,
accreting, and recycled gas are well-homogenized at $z<0.7$. The
presence of low metallicity gas at all orientations suggests that cold
streams of accreting filaments are not necessarily aligned with the
galaxy plane at low redshifts or intergalactic transfer may
dominate. Finally, our results support simulations showing
that strong metal absorption can mask the presence of low metallicity
gas in integrated line-of-sight CGM metallicities.
\end{abstract}


\section{Introduction}
\label{sec:intro}
The circumgalactic medium (CGM) of a galaxy is a vast multi-phase gaseous halo extending out to distances of $200$ kpc \citep[e.g.,][]{kacprzak08, kacprzak11morph, chen10b, steidel10, tumlinson11, rudie12, burchett13, magiicat1, magiicat2, werk13, johnson15}. It contains roughly half of the baryonic mass of the galaxy \citep{thom11,tumlinson11,werk14}, located at the interface between the intergalactic medium (IGM) and the interstellar medium (ISM), and thus plays an important role in regulating gas flows in and out of galaxies. Understanding the physical processes and properties of the CGM is necessary in order to correctly understand and model the evolution of galaxies.

The CGM can regulate many aspects of the formation of stars within a galaxy. ``Closed box'' galaxy models, which assume that galaxies do not accrete or expel gas, cannot explain the continued star-formation rate seen in galaxies today \citep[e.g.][]{lilly13}. Instead, gas flows from the IGM as well as recycled gas from galactic outflows are required to maintain the star-formation rate of star-forming galaxies \citep{springel03, keres05, oppenheimer08, oppenheimer10, dekel09, dave11a, dave11b, dave12, stewart11, kacprzak13, ford14}. Simulations have found that galaxies accrete gas from the CGM through both hot and cold accretion \citep[e.g.][]{birnboim07, keres09, dekel09}. In cold-mode accretion, near pristine gas from the IGM spirals onto the galaxy though filaments \citep[e.g.][]{danovich12,danovich15,shen13,stewart11,stewart13,stewart17, fumagalli11, oppenheimer12, vandevoort+schaye12,kacprzak16}. The filaments were predicted to be co-planar with the major axis of the galaxy \citep{danovich15}. Cold-mode accretion is typical for $z>1$ galaxies with masses less than $M_h < 10^{12}$~$M_{\odot}$ \citep{birnboim07,keres09, dekel09, vandevoort12}, while hot-mode accretion tends to dominate after $z=1$ \citep{vandevoort11}. Metal-enriched outflows, driven by supernovae are then ejected perpendicular to the galaxy disk \citep[e.g.][]{brook11, vandevoort12, shen13}. Over time, the metal--rich outflows recycle through the CGM and mix with the accreting metal--poor IGM \citep[e.g.][]{rubin12, zheng17, oppenheimer10, angles16}. 

The distribution of metals in the CGM has been investigated using quasar absorption line detections of the {\MgII} doublet with respect to the azimuthal angle, defined as the angle between the background quasar sight--line and the galaxy projected major axis. It was determined that there is an azimuthal angle dependence on the covering fraction of {\MgII} absorption systems whereby absorption tends to be found near the major ($\Phi=0{\degree}$) and minor axes ($\Phi=90{\degree}$) of galaxies \citep{bouche12,kacprzak12}. The absorption is generally stronger along the minor axis where metal--enriched outflows are expected  \citep{bordoloi11,bordoloi14a,kacprzak12,lan14,lan18}. Furthermore, the bimodality and absorption strength of {\MgII} absorbers are driven by star-forming galaxies where high metallicity winds are expected \citep{bordoloi11, bordoloi14a,kacprzak12}. Similar bimodal azimuthal dependencies have been found for the highly ionized CGM using {\OVI} absorption \citep{kacprzak15}. 

While both the low- ({\MgII}) and high- ({\OVI}) ionization studies have shown an azimuthal bimodality, where gas tends to reside both along the major and minor axes of galaxies, this does not demonstrate that the bimodality is a result of gas flows in or out of galaxies. However, metallicity could be used to identify metal--rich outflows and metal--poor accretion. 

The CGM metallicity of individual metal-line selected galaxy--absorber pairs has been investigated in an attempt to identify accretion and outflows. In an effort to identify cold-mode accretion, studies have found metal--poor absorption systems with a metallicity range between $-2 < [\text{X}/\text{H}] < -1$ \citep[e.g.][]{tripp05,cooksey08, kacprzak10b, kacprzak14, ribaudo11, churchill12, bouche13, bouche16, crighton13, crighton15}. Some studies found that these metal--poor absorption systems were consistent with accreting filaments along the major axis \citep{crighton13,bouche13}. Similarly, high metallicity systems ($[\text{X}/\text{H}]>-0.7$) have also been found near galaxies and assumed to be outflows or recycled gas \citep[e.g.][]{chen05, lehner09, peroux11, peroux16, crighton15, muzahid15, muzahid16}. Therefore, studies of individual metal-line selected galaxy--absorber pairs have found tentative evidence for accretion and outflows in the CGM. However, the orientation of the galaxy disk was not always known and thus it was difficult to determine whether the gas is truly inflowing or outflowing. Furthermore, selecting absorption systems using metal-lines may bias observations towards metal--enriched gas, potentially limiting detections of accretion.

\citet{lehner13, lehner18} and \citet{ wotta16, wotta18} presented large, unbiased metallicity studies of the cool CGM at low redshift $(z<1)$ where over $100$ absorbers were selected using only the presence of {\HI} serendipitously discovered in the quasar spectra. They found that the metallicity was bimodal for partial Lyman limit systems (pLLS, $16.2~{\cms} < {\NHI} < 17.2~{\cms}$) with two distinct peaks of low ($[\text{Si}/\text{H}] \sim -1.7$) and high ($[\text{Si}/\text{H}] \sim -0.4$) metallicity gas. They suggested that the low metallicity gas was due to cold accretion onto galaxies from the IGM and the metal--rich gas traced outflows containing processed gas. 

\citet{prochaska17} presented the CGM metallicities of $32$, $z\sim0.2$ galaxies from the COS-Halos survey. They found a median CGM metallicity of $[Z/\text{H}] \sim -0.51$ with a $95\%$ confidence interval spanning $-1.71 < [Z/\text{H}] < 0.76$ for a {\HI} column density range of $14.7~{\cms} < {\NHI} < 19.9~{\cms}$. Interestingly, the metallicity distribution was consistent with a unimodal distribution, which overlapped with the high metallicity peak from \citet{lehner13, lehner18} and \citet{wotta16, wotta18}. \citet{prochaska17} suggested that the previously detected low metallicity peak may arise from lower mass galaxies. This suggestion was supported by \citet{johnson17} who found dwarf galaxies have fewer and weaker detections of metal absorption. However, \citet{prochaska17} and \citet{berg18} did not find any evidence that CGM metallicity is dependent on stellar mass for $\geq L_{\ast}$ galaxies. The limitation of these CGM metallicity studies is the lack of associated galaxy orientation data. If outflows and accretion have a preferred spatial relationship with respect to the galaxy plane, then imaging and identifying associated galaxies is needed to investigate how metallicity relates to the orientation of the galaxy, and hence the locations of outflows and inflows. 

Combining metallicity studies of the CGM with galaxy data, \citet{peroux16} presented the galaxy ISM to CGM gas metallicity difference as a function of azimuthal angle for a sample of 9 galaxies. Interestingly, the authors found large scatter in the CGM metallicity along the major axis of the galaxy, while they found only low metallicity along the minor axis. Thus, they suggest that accretion and outflows are complex where accreting CGM may be contaminated by recycled outflows, leading to higher metallicities along the major axis. Lower metallicity absorption systems along the minor axis, where outflows are expected may be explained by gas which is ejected from the galaxy before it can form stars. 

Following on from \citet{peroux16}, we investigate the relationship between CGM metallicity and the spatial distribution of gas around a larger sample of 47 galaxy-absorber pairs. We test the simple model that low metallicity gas is accreted from the IGM along the projected galaxy major axis while high metallicity gas is expelled along the projected minor axis. That is, we test the hypothesis that metallicity is a powerful probe of baryon cycle processes around isolated galaxies at $z<0.7$.

This paper is organized as follows: In Section \ref{sec:observations} we describe our sample of galaxy--absorber pairs, detailing our method for obtaining CGM metallicities and galaxy properties such as redshift, halo mass, inclination, and the azimuthal angle. We present the results of our analysis with absorption properties in Section \ref{sec:results}. In Section \ref{sec:discussion} we discuss the relationship between the metallicity of the CGM and other properties of the galaxy--absorber pairs. In Section \ref{sec:conclusions} we summarize our results and provide our concluding remarks. We use a standard $\Lambda$CDM cosmology with $H_o = 70$~{\kms} Mpc$^{-1}$, {$\Omega_M = 0.3$} and {$\Omega_\Lambda = 0.7$}.

\section{Observations}\label{sec:observations}
\begin{deluxetable*}{lcllcccc}
	\tablecolumns{8}
	\tablewidth{0pt}
	\setlength{\tabcolsep}{0.06in}
	\tablecaption{Quasar Observations \label{tab:obsqso}}
	\tablehead{
		\colhead{(1)}           	&
        \colhead{(2)}     &
		\colhead{(3)}        &
		\colhead{(4)}       &
		\colhead{(5)}	&
		\colhead{(6)}	 	&			
		\colhead{(7)} 	&
        \colhead{(8)}\\
		\colhead{J-Name}           	&
        \colhead{$z_{\rm qso}$}     &
		\colhead{RA (J2000)}        &
		\colhead{DEC (J2000)}       &
		\colhead{COS Gratings}	&
		\colhead{COS PID(s)}	 	&			
		\colhead{Optical Spectrograph} 	&
        \colhead{Optical PID(s)}}
	\startdata
	J$0125$	&	$1.074$	&	$01$:$25$:$28.84$	&	$-00$:$05$:$55.93$	&	G160M	        &	13398	&	UVES	&	075.A-0841(A)	\\
J$0351$	&	$0.616$	&	$03$:$51$:$28.54$	&	$-14$:$29$:$08.71$ 	&	G130M, G160M	&	13398	&	UVES	&	076.A-0860(A)	\\
J$0407$	&	$0.572$	&	$04$:$07$:$48.43$	&	$-12$:$11$:$36.66$	&	G130M, G160M	&	11541	&	HIRES	&	G01H, U68H	\\
J$0456$	&	$0.533$	&	$04$:$56$:$08.92$	&	$-21$:$59$:$09.40$	&	G160M&	12466,12252,13398	&	UVES	&	076.A-0463(A)	\\
J$0853$	&	$0.514$	&	$08$:$53$:$34.25$	&	$+43$:$49$:$02.33$	&	G130M, G160M	&	13398	&	$\cdots$	&	$\cdots$	\\
J$0914$	&	$0.735$	&	$09$:$14$:$40.38$	&	$+28$:$23$:$30.62$	&	G130M, G160M	&	11598	&	HIRES	&	U059Hb	\\
J$0943$	&	$0.564$	&	$09$:$43$:$31.61$	&	$+05$:$31$:$31.49$	&	G130M, G160M	&	11598	&	HIRES	&	U066Hb	\\
J$0950$	&	$0.589$	&	$09$:$50$:$00.73$	&	$+48$:$31$:$29.38$	&	G130M, G160M	&	11598	&	HIRES	&	U059Hb	\\
J$1004$	&	$0.327$	&	$10$:$04$:$02.61$	&	$+28$:$55$:$35.39$	&	G130M, G160M	&	12038	&	$\cdots$	&	$\cdots$	\\
J$1009$	&	$0.456$	&	$10$:$09$:$02.06$	&	$+07$:$13$:$43.87$	&	G130M, G160M	&	11598	&	HIRES	&	U066Hb	\\
J$1041$	&	$1.270$	&	$10$:$41$:$17.16$	&	$+06$:$10$:$16.92$	&	G160M	        &	12252	&	HIRES	&	C17H	\\
J$1119$	&	$0.176$	&	$11$:$19$:$08.67$	&	$+21$:$19$:$18.01$	&	G130M, G160M	&	12038	&	HIRES	&	U152Hb	\\
J$1133$	&	$0.524$	&	$11$:$33$:$27.78$	&	$+03$:$27$:$19.17$	&	G130M, G160M	&	11598	&	HIRES	&	U059Hb	\\
J$1139$	&	$0.556$	&	$11$:$39$:$10.70$	&	$-13$:$50$:$43.63$	&	G130M	        &	12275	&	$\cdots$	&	$\cdots$	\\
J$1219$ &	$0.331$	&	$12$:$19$:$20.93$	&	$+06$:$38$:$38.52$	&	G130M, G160M	&	12025	&	$\cdots$	&	$\cdots$	\\
J$1233$	&	$0.470$	&	$12$:$33$:$04.05$	&	$-00$:$31$:$34.20$	&	G130M, G160M	&	11598	&	HIRES	&	U059Hb	\\
J$1241$	&	$0.583$	&	$12$:$41$:$54.02$	&	$+57$:$21$:$07.38$	&	G130M, G160M	&	11598	&	HIRES	&	U059Hb	\\
J$1244$	&	$1.273$	&	$12$:$44$:$10.82$	&	$+17$:$21$:$04.52$	&	G160M	        &	12466	&	HIRES	&	\tablenotemark{a}	\\
J$1301$	&	$0.477$	&	$13$:$01$:$12.93$	&	$+59$:$02$:$06.75$	&	G130M, G160M	&	11541	&	$\cdots$	&	$\cdots$	\\
J$1319$	&	$1.014$	&	$13$:$19$:$56.23$	&	$+27$:$28$:$08.22$	&	G160M	        &	11667	&	HIRES	&	U074	\\
J$1322$	&	$0.374$	&	$13$:$22$:$22.68$	&	$+46$:$45$:$35.22$	&	G130M, G160M	&	11598	&	HIRES	&	U066Hb	\\
J$1342$	&	$0.326$	&	$13$:$42$:$51.60$	&	$-00$:$53$:$45.31$	&	G130M, G160M	&	11598	&	HIRES	&	U059Hb	\\
J$1357$	&	$0.720$	&	$13$:$57$:$04.43$	&	$+19$:$19$:$07.37$	&	G160M	        &	13398	&	UVES	&	076.A-0860(A)	\\
J$1547$	&	$0.264$	&	$15$:$47$:$43.53$	&	$+20$:$52$:$16.61$	&	G130M, G160M	&	13398	&	$\cdots$	&	$\cdots$	\\
J$1555$	&	$0.714$	&	$15$:$55$:$04.40$	&	$+36$:$28$:$48.04$	&	G130M, G160M	&	11598	&	HIRES	&	U059Hb	\\
J$1704$	&	$0.371$	&	$17$:$04$:$41.37$	&	$+60$:$44$:$30.50$	&	STIS/E140M	    &	8015	&	HIRES	&	G400H, U019Hb	\\
J$2131$ &	$0.501$	&	$21$:$31$:$35.26$	&	$-12$:$07$:$04.79$ 	&	G160M	        &	13398	&	HIRES	&	C54H, U51H, C99H	\\
J$2137$	&	$0.200$	&	$21$:$37$:$45.17$	&	$-14$:$32$:$55.81$	&	G130M, G160M	&	13398	&	$\cdots$	&	$\cdots$	\\
J$2253$	&	$0.859$	&	$22$:$53$:$57.74$	&	$+16$:$08$:$53.56$	&	G130M, G160M	&	13398	&	UVES	&	075.A-0841(A)	\\[-10pt]

	\enddata
	\tablenotetext{a}{Spectra from \citet{churchill01}.}
\end{deluxetable*}
%\begin{turnpage}
  \begin{deluxetable*}{llcllrcrrrccrr}
      \tablecolumns{14}
      %\tablewidth{0pt}
      %\setlength{\tabcolsep}{0.06in}
      \tablecaption{Galaxy Observations and Properties \label{tab:obsgal}}
      \tablehead{
          \colhead{(1)}     &
          \colhead{(2)}     &
          \colhead{(3)}     &
          \colhead{(4)}     &
          \colhead{(5)}     &
          \colhead{(6)}	    &
          \colhead{(7)}	    &
          \colhead{(8)}     &
          \colhead{(9)} 	&
          \colhead{(10)} 	&
          \colhead{(11)} 	&
          \colhead{(12)} 	&
          \colhead{(13)} 	&
          \colhead{(14)}    \\
          \colhead{J-Name}                  &
          \colhead{$z_{\rm gal}$}           &
          \colhead{Ref.\tablenotemark{a}}   &
          \colhead{RA }                     &
          \colhead{DEC}                     &
          \colhead{D}	                    &
          \colhead{$B-K$}	 	            &
          \colhead{$\Phi$} 	                &
          \colhead{$i$} 	                &
          \colhead{$\log M_{h}/$} 	        &
          \colhead{\textit{HST}} 	        &
          \colhead{\textit{HST}} 	        &
          \colhead{Exp.} 	                &
          \colhead{\textit{HST}}            \\
          \colhead{}                &
          \colhead{}                &
          \colhead{}                &
          \colhead{(J2000)}         &
          \colhead{(J2000)}         &
          \colhead{(kpc)}	        &
          \colhead{}	 	        &
          \colhead{$({\degree})$}   &
          \colhead{$({\degree})$} 	&
          \colhead{$M_{\odot}$} 	&
          \colhead{Filter} 	        &
          \colhead{Camera} 	        &
          \colhead{(s)} 	        &
          \colhead{PID}
          }
      \startdata
      J0125	&	$	0.398525	$	&	1	&	$	01$:$25$:$27.671	$	&	$	-00$:$05$:$31.39	$	&	$	163.0	\pm	{	0.1	}	$	&	$	1.80	$	&	$	73.4	_{	-4.7	}^{+	4.6	}	$	&	$	63.2	_{	-2.6	}^{+	1.7	}	$	&	$	12.51	_{-	0.15	}^{+	0.16	}	$	&	F702W	&	WFPC2	&	700	&	6619	\\[2pt]
J0351	&	$	0.2617	    $	&	2	&	$	03$:$51$:$28.933	$	&	$	-14$:$29$:$54.31	$	&	$	188.6	\pm	{	0.3	}	$	&	$	2.30	$	&	$	64.9	_{	-15.8	}^{+	21.1}	$	&	$	83.0	_{	-3.0	}^{+	2.0	}	$	&	$	11.56	_{-	0.21	}^{+	0.44	}	$	&	F702W	&	WFPC2	&	800	&	5949	\\[2pt]
J0351	&	$	0.356992	$	&	1	&	$	03$:$51$:$27.892	$	&	$	-14$:$28$:$57.88	$	&	$	72.3	\pm	{	0.4	}	$	&	$	0.30	$	&	$	4.9	    _{	-4.9	}^{+	33.0}	$ 	&	$	28.5	_{	-12.5	}^{+	19.8	}	$	&	$	12.00	_{-	0.19	}^{+	0.29	}	$	&	F702W	&	WFPC2	&	800	&	5949	\\[2pt]
J0407	&	$	0.1534	    $	&	3	&	$	04$:$07$:$43.930	$	&	$	-12$:$12$:$08.49	$	&	$	195.9	\pm	{	0.1	}	$	&	$	\cdots	$	&	$	26.3	_{	-1.0	}^{+	0.9	}	$	&	$	49.5	_{	-0.7	}^{+	0.5	}	$	&	$	11.94	_{-	0.20	}^{+	0.31	}	$	&	F702W	&	WFPC2	&	800	&	5949	\\[2pt]
J0407	&	$	0.3422	    $	&	3	&	$	04$:$07$:$48.481	$	&	$	-12$:$12$:$11.13	$	&	$	172.0	\pm	{	0.1	}	$	&	$	\cdots	$	&	$	48.1	_{	-0.9	}^{+	1.0	}	$	&	$	85.0	_{	-0.4	}^{+	0.1	}	$	&	$	11.62	_{-	0.21	}^{+	0.42	}	$	&	F702W	&	WFPC2	&	800	&	5949	\\[2pt]
J0407	&	$	0.495164	$	&	4	&	$	04$:$07$:$49.020	$	&	$	-12$:$11$:$20.76	$	&	$	107.6	\pm	{	0.4	}	$	&	$	\cdots	$	&	$	21.0	_{	-3.7	}^{+	5.3	}	$	&	$	67.2	_{	-7.5	}^{+	7.6	}	$	&	$	11.41	_{-	0.21	}^{+	0.45	}	$	&	F702W	&	WFPC2	&	800	&	5949	\\[2pt]
J0456	&	$	0.2784	    $	&	2	&	$	04$:$56$:$09.660	$	&	$	-21$:$59$:$03.930	$	&	$	50.7	\pm	{	0.5	}	$	&	$	0.46	$	&	$	78.4	_{	-2.1	}^{+	2.1	}	$	&	$	71.2	_{	-2.6	}^{+	2.6	}	$	&	$	11.44	_{-	0.21	}^{+	0.50	}	$	&	F702W	&	WFPC2	&	600	&	5098	\\[2pt]
J0456	&	$	0.381511	$	&	1	&	$	04$:$56$:$08.820	$	&	$	-21$:$59$:$27.400	$	&	$	103.4	\pm	{	0.3	}	$	&	$	1.78	$	&	$	63.8	_{	-2.7	}^{+	4.3	}	$	&	$	57.1	_{	-2.4	}^{+	19.9	}	$	&	$	12.00	_{-	0.19	}^{+	0.29	}	$	&	F702W	&	WFPC2	&	600	&	5098	\\[2pt]
J0456	&	$	0.4828	    $	&	5	&	$	04$:$56$:$08.913	$	&	$	-21$:$59$:$29.000	$	&	$	108.0	\pm	{	0.5	}	$	&	$	1.66	$	&	$	85.2	_{	-3.7	}^{+	3.7	}	$	&	$	42.1	_{	-3.1	}^{+	3.1	}	$	&	$	12.28	_{-	0.15	}^{+	0.19	}	$	&	F702W	&	WFPC2	&	600	&	5098	\\[2pt]
J0853	&	$	0.1635	    $	&	2	&	$	08$:$53$:$33.384	$	&	$	43$:$49$:$03.97	    $	&	$	26.2	\pm	{	0.1	}	$	&	$	1.80	$	&	$	56.0	_{	-0.8	}^{+	0.8	}	$	&	$	70.1	_{	-0.8	}^{+	1.4	}	$	&	$	11.89	_{-	0.20	}^{+	0.33	}	$	&	F702W	&	WFPC2	&	800	&	5949	\\[2pt]
J0853	&	$	0.2766	    $	&	2	&	$	08$:$53$:$36.881	$	&	$	43$:$49$:$33.32	    $	&	$	179.4	\pm	{	0.2	}	$	&	$	0.63	$	&	$	36.7	_{	-15.3	}^{+	14.9}	$	&	$	32.8	_{	-6.7	}^{+	5.7	}	$	&	$	11.59	_{-	0.21	}^{+	0.43	}	$	&	F702W	&	WFPC2	&	800	&	5949	\\[2pt]
J0853	&	$	0.4402	    $	&	2	&	$	08$:$53$:$35.160	$	&	$	43$:$48$:$59.81	    $	&	$	58.1	\pm	{	0.4	}	$	&	$	1.80	$	&	$	23.0	_{	-7.6	}^{+	6.5	}	$	&	$	73.3	_{	-3.0	}^{+	3.8	}	$	&	$	11.95	_{-	0.18	}^{+	0.27	}	$	&	F702W	&	WFPC2	&	800	&	5949	\\[2pt]
J0914	&	$	0.244312	$	&	1	&	$	09$:$14$:$41.759	$	&	$	+28$:$23$:$51.18	$	&	$	105.9	\pm	{	0.1	}	$	&	$	1.02	$	&	$	18.2	_{	-1.0	}^{+	1.1	}	$	&	$	39.0	_{	-0.2	}^{+	0.4	}	$	&	$	11.88	_{-	0.20	}^{+	0.33	}	$	&	F814W	&	ACS	&	1200	&	13024	\\[2pt]
J0943	&	$	0.1431	    $	&	6	&	$	09$:$43$:$29.210	$	&	$	+05$:$30$:$41.75	$	&	$	154.2	\pm	{	0.1	}	$	&	$	2.46	$	&	$	77.7	_{	-0.1	}^{+	0.1	}	$	&	$	75.5	_{	-0.1	}^{+	0.1	}	$	&	$	12.16	_{-	0.18	}^{+	0.25	}	$	&	F814W	&	ACS	&	1200	&	13024	\\[2pt]
J0943	&	$	0.2284	    $  	&	6	&	$	09$:$43$:$33.789	$	&	$	+05$:$31$:$22.26	$	&	$	123.3	\pm	{	0.1	}	$	&	$	1.93	$	&	$	30.4	_{	-0.4	}^{+	0.3	}	$	&	$	52.3	_{	-0.3	}^{+	0.3	}	$	&	$	12.20	_{-	0.17	}^{+	0.23	}	$	&	F814W	&	ACS	&	1200	&	13024	\\[2pt]
J0943	&	$	0.353052	$	&	1	&	$	09$:$43$:$30.671	$	&	$	+05$:$31$:$18.08	$	&	$	96.5	\pm	{	0.3	}	$	&	$	0.96	$	&	$	8.2	    _{	-5.0	}^{+	3.0	}	$	&	$	44.4	_{	-1.2	}^{+	1.1	}	$	&	$	11.66	_{-	0.21	}^{+	0.41	}	$	&	F814W	&	ACS	&	1200	&	13024	\\[2pt]
J0950	&	$	0.211866	$	&	1	&	$	09$:$50$:$00.863	$	&	$	+48$:$31$:$02.59	$	&	$	93.6	\pm	{	0.2	}	$	&	$	2.39	$	&	$	16.6	_{	-0.1	}^{+	0.1	}	$	&	$	47.7	_{	-0.1	}^{+	0.1	}	$	&	$	12.37	_{-	0.16	}^{+	0.18	}	$	&	F814W	&	ACS	&	1200	&	13024	\\[2pt]
J1004	&	$	0.1380	    $	&	2	&	$	10$:$04$:$02.353	$	&	$	+28$:$55$:$12.50	$	&	$	56.7	\pm	{	0.2	}	$	&	$	0.81	$	&	$	12.4	_{	-2.9	}^{+	2.4	}	$	&	$	79.1	_{	-2.1	}^{+	2.2	}	$	&	$	10.87	_{-	0.22	}^{+	0.63	}	$	&	F702W	&	WFPC2	&	800	&	5949	\\[2pt]
J1009	&	$	0.227855	$	&	1	&	$	10$:$09$:$01.579	$	&	$	+07$:$13$:$28.00	$	&	$	64.0	\pm	{	0.8	}	$	&	$	1.39	$	&	$	89.6	_{	-1.3	}^{+	0.4	}	$	&	$	66.3	_{	-0.9	}^{+	0.6	}	$	&	$	11.76	_{-	0.21	}^{+	0.37	}	$	&	F625W	&	WFC3	&	2256	&	11598	\\[2pt]
J1041	&	$	0.3153	    $	&	7	&	$	10$:$41$:$16.858	$	&	$	+06$:$10$:$06.13	$	&	$	54.0	\pm	{	0.5	}	$	&	$	2.20	$	&	$	77.3	_{	-1.2	}^{+	1.2	}	$	&	$	72.6	_{	-1.3	}^{+	1.3	}	$	&	$	11.57	_{-	0.22	}^{+	0.43	}	$	&	F702W	&	WFPC2	&	1300	&	5984	\\[2pt]
J1041	&	$	0.442173	$	&	1	&	$	10$:$41$:$17.801	$	&	$	+06$:$10$:$18.97	$	&	$	56.2	\pm	{	0.3	}	$	&	$	2.81	$	&	$	4.3	    _{	-1.0	}^{+	0.9	}	$	&	$	49.8	_{	-5.2	}^{+	7.4	}	$	&	$	11.99	_{-	0.18	}^{+	0.26	}	$	&	F702W	&	WFPC2	&	1300	&	5984	\\[2pt]
J1119	&	$	0.1383	    $	&	8	&	$	11$:$19$:$06.675	$	&	$	+21$:$18$:$29.56	$	&	$	138.0	\pm	{	0.2	}	$	&	$	2.21	$	&	$	34.4	_{	-0.4	}^{+	0.4	}	$	&	$	26.4	_{	-0.4	}^{+	0.8	}	$	&	$	12.24	_{-	0.17	}^{+	0.21	}	$	&	F606W	&	WFPC2	&	2200	&	5849	\\[2pt]
J1133	&	$	0.154599	$	&	4	&	$	11$:$33$:$28.218	$	&	$	+03$:$26$:$59.00	$	&	$	55.6	\pm	{	0.1	}	$	&	$	1.07	$	&	$	56.1	_{	-1.3	}^{+	1.7	}	$	&	$	23.5	_{	-0.2	}^{+	0.4	}	$	&	$	11.64	_{-	0.21	}^{+	0.41	}	$	&	F814W	&	ACS	&	1200	&	13024	\\[2pt]
J1139	&	$	0.1755	    $	&	2	&	$	11$:$39$:$10.536	$	&	$	-13$:$49$:$48.59	$	&	$	163.0	\pm	{	0.5	}	$	&	$	\cdots	$	&	$	21.4	_{	-10.7	}^{+	10.7}	$	&	$	85.0	_{	-0.2	}^{+	0.2	}	$	&	$	11.16	_{-	0.21	}^{+	0.58	}	$	&	F702W	&	ACS	&	700	&	6619	\\[2pt]
J1139	&	$	0.204194	$	&	1	&	$	11$:$39$:$11.520	$	&	$	-13$:$51$:$08.69	$	&	$	93.2	\pm	{	0.3	}	$	&	$	2.30	$	&	$	5.8	    _{	-0.5	}^{+	0.4	}	$	&	$	83.4	_{	-0.5	}^{+	0.4	}	$	&	$	11.69	_{-	0.21	}^{+	0.40	}	$	&	F702W	&	ACS	&	700	&	6619	\\[2pt]
J1139	&	$	0.212259	$	&	1	&	$	11$:$39$:$09.533	$	&	$	-13$:$51$:$31.46	$	&	$	174.8	\pm	{	0.1	}	$	&	$	2.10	$	&	$	80.4	_{	-0.5	}^{+	0.4	}	$	&	$	85.0	_{	-0.6	}^{+	5.0	}	$	&	$	11.73	_{-	0.21	}^{+	0.39	}	$	&	F702W	&	ACS	&	700	&	6619	\\[2pt]
J1139	&	$	0.219724	$	&	4	&	$	11$:$39$:$08.330	$	&	$	-13$:$50$:$45.64	$	&	$	122.0	\pm	{	0.2	}	$	&	$	2.10	$	&	$	44.9	_{	-8.1	}^{+	8.9	}	$	&	$	85.0	_{	-8.5	}^{+	5.0	}	$	&	$	11.04	_{-	0.21	}^{+	0.60	}	$	&	F702W	&	ACS	&	700	&	6619	\\[2pt]
J1139	&	$	0.319255	$	&	1	&	$	11$:$39$:$09.801	$	&	$	-13$:$50$:$53.08	$	&	$	73.3	\pm	{	0.4	}	$	&	$	1.60	$	&	$	39.1	_{	-1.7	}^{+	1.9	}	$	&	$	83.4	_{	-1.1	}^{+	1.4	}	$	&	$	11.86	_{-	0.20	}^{+	0.34	}	$	&	F702W	&	ACS	&	700	&	6619	\\[2pt]
J1219	&	$	0.1241	    $	&	8	&	$	12$:$19$:$23.469	$	&	$	+06$:$38$:$19.84	$	&	$	93.4	\pm	{	5.3	}	$	&	$	1.20	$	&	$	67.2	_{	-67.2	}^{+	22.8}	$	&	$	22.0	_{	-21.8	}^{+	18.7	}	$	&	$	11.87	_{-	0.20	}^{+	0.34	}	$	&	F702W	&	WFPC2	&	600	&	5143	\\[2pt]
J1233	&	$	0.318757	$	&	4	&	$	12$:$33$:$04.084	$	&	$	-00$:$31$:$40.20	$	&	$	88.9	\pm	{	0.2	}	$	&	$	1.15	$	&	$	17.0	_{	-2.3	}^{+	2.0	}	$	&	$	38.7	_{	-1.8	}^{+	1.6	}	$	&	$	11.91	_{-	0.20	}^{+	0.32	}	$	&	F814W	&	ACS	&	1200	&	13024	\\[2pt]
J1241	&	$	0.205267	$	&	1	&	$	12$:$41$:$53.731	$	&	$	+57$:$21$:$00.94	$	&	$	21.1	\pm	{	0.1	}	$	&	$	1.19	$	&	$	77.6	_{	-0.4	}^{+	0.3	}	$	&	$	56.4	_{	-0.5	}^{+	0.3	}	$	&	$	11.64	_{-	0.21	}^{+	0.41	}	$	&	F814W	&	ACS	&	1200	&	13024	\\[2pt]
J1241	&	$	0.217905	$	&	4	&	$	12$:$41$:$52.410	$	&	$	+57$:$20$:$43.28	$	&	$	94.6	\pm	{	0.2	}	$	&	$	1.29	$	&	$	63.0	_{	-2.1	}^{+	1.8	}	$~\tablenotemark{b}	&	$	17.4	_{	-1.6	}^{+	1.4	}	$	&	$	11.62	_{-	0.21	}^{+	0.42	}	$	&	F814W	&	ACS	&	1200	&	13024	\\[2pt]
J1244	&	$	0.5504	    $	&	2	&	$	12$:$44$:$11.045	$	&	$	+17$:$21$:$05.05	$	&	$	21.2	\pm	{	0.3	}	$	&	$	1.34	$	&	$	20.1	_{	-19.1	}^{+	16.7}	$	&	$	31.7	_{	-4.8	}^{+	16.2	}	$	&	$	11.82	_{-	0.19	}^{+	0.31	}	$	&	F702W	&	WFPC2	&	1300	&	6557	\\[2pt]
J1301	&	$	0.1967	    $	&	2	&	$	13$:$01$:$20.123	$	&	$	+59$:$01$:$35.72	$	&	$	135.5	\pm	{	0.1	}	$	&	$	1.60	$	&	$	39.7	_{	-2.2	}^{+	2.8	}	$	&	$	80.7	_{	-3.2	}^{+	4.3	}	$	&	$	11.36	_{-	0.21	}^{+	0.53	}	$	&	F702W	&	WFPC2	&	700	&	6619	\\[2pt]
J1319	&	$	0.6610	    $	&	9	&	$	13$:$19$:$55.773	$	&	$	+27$:$27$:$54.84	$	&	$	103.9	\pm	{	0.5	}	$	&	$	1.45	$	&	$	86.6	_{	-1.2	}^{+	1.5	}	$	&	$	65.8	_{	-1.2	}^{+	1.2	}	$	&	$	12.15	_{-	0.15	}^{+	0.19	}	$	&	F702W	&	WFPC2	&	1300	&	5984	\\[2pt]
J1322	&	$	0.214431	$	&	1	&	$	13$:$22$:$22.470	$	&	$	+46$:$45$:$45.98	$	&	$	38.6	\pm	{	0.2	}	$	&	$	1.73	$	&	$	13.9	_{	-0.2	}^{+	0.2	}	$	&	$	57.9	_{	-0.2	}^{+	0.1	}	$	&	$	12.13	_{-	0.18	}^{+	0.25	}	$	&	F814W	&	ACS	&	1200	&	13024	\\[2pt]
J1342	&	$	0.0708	    $	&	6	&	$	13$:$42$:$50.002	$	&	$	-00$:$53$:$28.88	$	&	$	39.4	\pm	{	0.5	}	$	&	$	\cdots	$	&	$	13.9	_{	-0.2	}^{+	0.2	}	$	&	$	57.7	_{	-0.3	}^{+	0.3	}	$	&	$	11.36	_{-	0.21	}^{+	0.53	}	$	&	F814W	&	ACS	&	1200	&	13024	\\[2pt]
J1342	&	$	0.2013	    $	&	6	&	$	13$:$42$:$52.235	$	&	$	-00$:$53$:$43.10	$	&	$	31.8	\pm	{	0.2	}	$	&	$	2.12	$	&	$	44.5	_{	-0.3	}^{+	0.1	}	$	&	$	71.6	_{	-0.2	}^{+	0.3	}	$	&	$	11.66	_{-	0.21	}^{+	0.41	}	$	&	F814W	&	ACS	&	1200	&	13024	\\[2pt]
J1342	&	$	0.227042	$	&	1	&	$	13$:$42$:$51.866	$	&	$	-00$:$53$:$54.07	$	&	$	35.3	\pm	{	0.2	}	$	&	$	1.34	$	&	$	13.2	_{	-0.4	}^{+	0.5	}	$~\tablenotemark{b}	&	$	0.1	    _{	-0.1	}^{+	0.6	}	$	&	$	12.40	_{-	0.16	}^{+	0.17	}	$	&	F814W	&	ACS	&	1200	&	13024	\\[2pt]
J1357	&	$	0.4295	    $	&	2	&	$	13$:$57$:$03.290	$	&	$	+19$:$18$:$44.41	$	&	$	157.9	\pm	{	1.5	}	$	&	$	1.69	$	&	$	8.7	    _{	-1.4	}^{+	1.6	}	$	&	$	85.0	_{	-1.7	}^{+	5.0	}	$	&	$	11.49	_{-	0.21	}^{+	0.43	}	$	&	F702W	&	WFPC2	&	800	&	5949	\\[2pt]
J1357	&	$	0.4592	    $	&	2	&	$	13$:$57$:$04.539	$	&	$	+19$:$19$:$15.15	$	&	$	45.5	\pm	{	0.7	}	$	&	$	1.40	$	&	$	64.2	_{	-13.8	}^{+	13.6}	$	&	$	24.7	_{	-6.5	}^{+	5.7	}	$	&	$	11.72	_{-	0.20	}^{+	0.34	}	$	&	F702W	&	WFPC2	&	800	&	5949	\\[2pt]
J1547	&	$	0.0949	    $	&	2	&	$	15$:$47$:$45.561	$	&	$	+20$:$51$:$41.37	$	&	$	79.8	\pm	{	0.5	}	$	&	$	1.00	$	&	$	54.7	_{	-2.4	}^{+	2.0	}	$	&	$	80.9	_{	-2.0	}^{+	1.8	}	$	&	$	10.77	_{-	0.22	}^{+	0.63	}	$	&	F702W	&	WFPC2	&	1100	&	5099	\\[2pt]
J1555	&	$	0.189201	$	&	1	&	$	15$:$55$:$05.295	$	&	$	+36$:$28$:$48.46	$	&	$	33.4	\pm	{	0.1	}	$	&	$	1.20	$	&	$	47.0	_{	-0.8	}^{+	0.3	}	$	&	$	51.8	_{	-0.7	}^{+	0.7	}	$	&	$	12.07	_{-	0.18	}^{+	0.27	}	$	&	F814W	&	ACS	&	1200	&	13024	\\[2pt]
J1704	&	$	0.0921	    $	&	2	&	$	17$:$04$:$34.330	$	&	$	+60$:$44$:$47.59	$	&	$	93.6	\pm	{	0.5	}	$	&	$	\cdots	$	&	$	53.1	_{	-0.6	}^{+	0.6	}	$	&	$	72.0	_{	-0.5	}^{+	0.5	}	$	&	$	11.48	_{-	0.21	}^{+	0.48	}	$	&	F702W	&	WFPC2	&	600	&	5949	\\[2pt]
J2131	&	$	0.430200	$	&	1	&	$	21$:$31$:$35.635	$	&	$	-12$:$06$:$58.56	$	&	$	48.4	\pm	{	0.2	}	$	&	$	2.06	$	&	$	14.9	_{	-4.9	}^{+	6.0	}	$	&	$	48.3	_{	-3.7	}^{+	3.5	}	$	&	$	12.04	_{-	0.18	}^{+	0.25	}	$	&	F702W	&	WFPC2	&	600	&	5143	\\[2pt]
J2137	&	$	0.0752	    $	&	2	&	$	21$:$37$:$45.083	$	&	$	-14$:$32$:$06.27	$	&	$	70.9	\pm	{	0.7	}	$	&	$	\cdots	$	&	$	73.2	_{	-0.5	}^{+	1.0	}	$	&	$	71.0	_{	-1.0	}^{+	0.9	}	$	&	$	11.40	_{-	0.21	}^{+	0.52	}	$	&	F702W	&	WFPC2	&	1400	&	5343	\\[2pt]
J2253	&	$	0.352787	$	&	1	&	$	22$:$54$:$00.417	$	&	$	+16$:$09$:$06.82	$	&	$	203.2	\pm	{	0.5	}	$	&	$	1.30	$	&	$	88.7	_{	-4.8	}^{+	1.3	}	$	&	$	36.7	_{	-4.6	}^{+	6.9	}	$	&	$	11.93	_{-	0.20	}^{+	0.32	}	$	&	F702W	&	WFPC2	&	700	&	6619	\\[-7pt]

      \enddata
      \tablenotetext{a}{Galaxy redshift reference: $(1)$ \citet{kacprzak19a}, $(2)$ \citet{chen01b}, $(3)$ \citet{johnson13}, $(4)$ this work, $(5)$ \citet{kacprzak10a}, $(6)$ \citet{werk12}, $(7)$ \citet{lanzetta95}, $(8)$ \citet{prochaska11}, $(9)$ \citet{kacprzak12b}.}
      \tablenotetext{b}{We note that the uncertainties in the azimuthal angles of face-on galaxies ($i<20${\degree}) are statistical errors derived from modelling in GIM2D. It is possible that the systematic errors are larger.}
  \end{deluxetable*}
%\end{turnpage}
In order to study the distribution of CGM metallicities, we use the ``Multiphase Galaxy Halos'' Survey, which is comprised of our \textit{Hubble Space Telescope (HST)} program (PID 13398) \citep{kacprzak15,kacprzak19a, muzahid15, muzahid16,nielsenovi,ng19} as well as data taken from literature \citep{yuan02, danforth10, meiring11, churchill12, tilton12, tilton13, schull12, fox13, mathes14}. All 29 quasars fields have \textit{HST} imaging and UV spectra from the Cosmic Origins Spectrograph (COS) instrument on the \textit{HST}. In addition, 22 quasars have optical spectra from Keck/HIRES or VLT/UVES.

Our sample comprises 47 galaxies with spectroscopic redshifts between $0.07 < z < 0.66$ ($\langle z \rangle = 0.27$), which have an impact parameter range of $21$~kpc~$< D < 203$~kpc from a background quasar.  We require the galaxies in this sample to be isolated by selecting those which have no neighbors within $100$~kpc and have a line-of-sight velocity separation of more than $500$~{\kms} from the nearest galaxy. The galaxies in this sample were selected to be isolated to minimize the possibility of the CGM structure being disturbed by mergers. The halo mass range of the galaxies in our sample is $10.8 < \log M_{h}/M_{\odot} < 12.5$, ($\langle \log M_{h}/M_{\odot} \rangle = 11.8$). This range represents that of typical $L_{\ast}$ galaxies.

\subsection{UV Quasar Spectra}\label{sec:cos}
The UV spectra in the ``Multiphase Galaxy Halos'' Survey are taken from the COS instrument. The spectra have a medium resolving power of $R \approx 20,000$ and cover a range of ions including the {\HI} Lyman series, {\CII}, {\CIII}, {\CIV}, {\NII}, {\NIII}, {\NV}, {\OI}, {\OVI}, {\SiII}, {\SiIII} and {\SiIV}. Details of the \textit{HST}/COS observations are shown in Table \ref{tab:obsqso}. The \textit{HST}/COS spectra were reduced using the CALCOS pipeline \citep{massa13}. The signal-to-noise ratio was improved by co-adding all spectra \citep{danforth10}\footnote{\url{http://casa.colorado.edu/~danforth/science/cos/costools.html}} and binning by three pixels. Continuum normalization was done by fitting low-order polynomials to the spectra while excluding regions with lines. The UV spectrum for J$1704$ was obtained using the E$140$M grating of the Space Telescope Imaging Spectrograph (STIS) on the \textit{HST} with a spectral resolving power of $R=45,800$.

\subsection{Optical Quasar Spectra}
We use the optical spectra to complement the UV spectra because ionic transitions including {\MgI}, {\MgII}, {\FeII}, {\MnII} and {\CaII} are especially useful in providing metallicity constraints for absorption systems at redshifts of $z_{abs} \gtrsim 0.2$. We have optical spectra from Keck/HIRES or VLT/UVES for 34 absorption systems with a resolving power of $R\approx 40,000$. The project IDs and instruments for the optical spectra are shown in Table \ref{tab:obsqso}. The HIRES spectra were reduced using either the Mauna Kea Echelle Extraction (MAKEE) package or IRAF. The UVES spectra were reduced using the European Southern Observatory (ESO) pipeline \citep{dekker00} and the UVES Post-Pipeline Echelle Reduction (UVES POPLER) code \citep{murphy16, murphy18}. 

\subsection{Galaxy Imaging}
Each of the galaxy--absorber pairs in the ``Multiphase Galaxy Halos'' Survey have high-resolution images from either \textit{HST/}WFPC2 (F702W or F606W filters), \textit{HST/}WFC3 (F625W, F390W or F702W filters) or \textit{HST/}ACS (F814W filter) to determine the morphology of the galaxies. The  details of the cameras and filters used for each of the quasar fields along with their exposure times and PIDs are shown in Table \ref{tab:obsgal}. 

Reduction of the \textit{HST}/WFPC2 images was done using the WFPC2 Associations Science Products Pipelines (WASPP) \citep[see][for more details]{kacprzak11morph}. The DrizzlePac software was used to reduce the WFC3 and ACS images \citep{gonzaga12} where cosmic rays were removed using the multidrizzle process or by using lacosmic \citep{vandokkum01}. 

The Source Extractor package \citep[SExtractor;][]{bertin96} was used to calculate the galaxy photometry. For WFPC2, the Vega $m_{HST}$ magnitudes were calculated by \citet{kacprzak11morph} and were then converted to AB $B$-band absolute magnitudes \citep{magiicat2}. The magnitudes from the ACS and WFC3 filters were calculated in AB. We obtained $B$- and $K$- band magnitudes and luminosities for each galaxy, as well as their $B-K$ color \citep{magiicat2,nielsenovi}.  Using galaxy magnitudes, we calculated the halo masses (dark + baryonic matter) using the halo abundance matching method described in \citet{magiicat3}. The galaxy properties are detailed in Table \ref{tab:obsgal}. 

Morphologies and orientations of the galaxies were determined using GIM2D \citep[Galaxy IMage 2D;][]{simard02} to fit a two-component (disk+bulge) model. The disk component is fit with an exponential profile while the bulge component is fit with a S\'ersic profile. The S\'ersic parameter varied between $0.2 \leq n \leq 4.0$. The details of the models are described in \citet{kacprzak15}. The azimuthal angle is then defined as the angle between the semi-major axis of the galaxy and the quasar sight--line where $\Phi = 0{\degree}$ indicates that the quasar lies along the projected major axis of the galaxy, while $\Phi = 90{\degree}$ is where the quasar is located along the projected minor axis. Galaxy inclination angles are defined such that $i=0{\degree}$ represents face-on galaxies while $i=90{\degree}$ indicates edge-on galaxies.

\subsection{Galaxy Spectra}
The Keck Echelle Spectrograph and Imager (ESI) \citep{sheinis02} was used to obtain spectra of 27 galaxies. The method of reduction used is presented in \citet{kacprzak19a}. The slits used in the observations were $20^{\prime\prime}$ by $1^{\prime\prime}$ and the data were binned by two, which resulted in pixel sizes of $0.27^{\prime\prime} - 0.34^{\prime\prime}$ in the spatial direction and a spectral resolution of $22~\kms$. The wavelength range of the ESI spectra is $4000$~{\AA} to $10,000$~{\AA}, which cover a range of emission and absorption lines. The ESI data were reduced using IRAF and then vacuum and heliocentric corrections were applied. Galaxy redshifts were then calculated to be the velocity centroid of the emission lines and are shown in column (2) of Table \ref{tab:obsgal} and are labeled `1' or `4' in column (3). The remaining galaxy redshifts were taken from the literature indicated in Table \ref{tab:obsqso}.

\subsection{Spectral Analysis}\label{sec:spec_an}
The absorption systems were modeled with Voigt profiles using the VPFIT software \citep{carswell14}. To account for the non-Gaussian line spread function (LSF) of the COS spectrograph, we use the LSF from \citet{kriss11}. The LSF for each absorption profile was calculated and convolved with the model profile in the fitting process. This method of calculating the LSF also takes into account the life-time position of COS at the time of observation for each spectrum. The velocity resolution of the HIRES and UVES spectrographs is $\sim 6.6$~{\kms} and we assumed a Gaussian LSF.

We searched for 40 different ionic transitions in each spectrum within $\pm400$~{\kms} of the associated galaxy redshift. We required that the {\HI} absorption was measurable and that any absorption from other ions were associated with the {\HI}. Additionally, since we often cover multiple transitions of the same ion, such as the {\MgII} doublet and the {\SiII} quintet, we expect to observe similar structure in the transitions. These checks help to rule out coincidental absorption features which may be due to gas at other redshifts. 

We initially fit the typically unsaturated low ionisation states such as {\MgII} and {\SiII}. While fitting the absorption profiles, we could encounter a number of scenarios. In the first scenario, the absorption system is uncontaminated by other absorption features. This could be determined by consistency in the shape the absorption profile compared to others from a similar ionization state. To optimize the chi-squared value of each fit, we attempted to use the minimum number of Voigt profile components possible while maintaining reasonable Doppler parameters for each component. 

In the second scenario, one or more transitions are blended with gas at other redshifts or with ions at similar rest-frame wavelengths (e.g. the {\CII} $\lambda_o = 1036.34$~{\AA} and {\OVI} $\lambda_o = 1037.62$~{\AA} lines). Where possible, additional components were added to the fit to account for the blend in the absorption profile. We then follow the same process used for unblended absorption profiles. For example, if there is unblended absorption in the Ly$\alpha$ and Ly$\gamma$ lines while Ly$\beta$ has a dominant blend, we would only use the former transitions to calculate the fit to the data. As a check, we overlay the fit onto the blended transition to ensure that it is consistent with the data.

It was also quite common that many of the {\HI} Lyman series lines were saturated, providing only lower limits on the {\HI} column density. An accurate measurement of the {\HI} column density was possible where a part of the series was unsaturated. Where we only detected saturated {\HI} absorption across the entire series available we had two options:

(1) A basic one or two component fit was applied to the absorption profile. Then, applying the curve-of-growth relationship between the column density and Doppler parameter gave us a lower limit on the {\HI} column density. Due to the absence of damping wings in the absorption profile, the upper limit on the {\HI} column density is then ${\NHI} < 19.0$~{\cms}, above which we detect sub-DLAs, which are notable for the presence of wings.\footnote{We follow the definition in \citet{lehner18,wotta18} for the classification of {\HI} absorbers. The {\HI} column density ranges for pLLSs are $16.2~{\cms} < {\NHI} < 17.2~{\cms}$, LLS have $17.2~{\cms} < {\NHI} < 19.0~{\cms}$, sub--DLAs have $19.0~{\cms} < {\NHI} < 20.3~{\cms}$ and DLAs have ${\NHI} > 20.3~{\cms}$.} In some cases, we instead apply an upper limit of ${\NHI} < 17.2$~{\cms}, classifying the system as a pLLS. This occurred where the {\HI} absorption profile was unlikely to be saturated at the Lyman limit.

(2) Saturated {\HI} absorption was modeled using the fits to cool gas tracers such as {\MgII} and {\SiII} as a template. In this case, we would calculate a fit to the {\MgII} or {\SiII} absorption profiles following the method described for unsaturated systems. We then assume that {\HI} has the same kinematic stricture as {\MgII} or {\SiII}, where the Voigt profile component redshifts are fixed between ions. We further assume that the ions are at the same temperature, such that thermal broadening dominates. This required that the ratio between {\MgII} or {\SiII} and {\HI} for a given component is the square root of the ion mass ratio. The column densities of the {\HI} components were permitted to vary. We note that method (2) assumes that all of the {\HI} in the absorption profile is associated with the low temperature gas traced by {\MgII} and {\SiII}. However, this method of fitting the absorption profile is consistent with the assumption of a single--phase ionization model to calculate the CGM metallicity which also associates all {\HI} gas with low ionization gas.

The {\HI} column density obtained from  VPFIT for both methods (1) and (2) was then assumed to be a lower limit. A comparison of methods (1) and (2) found that they produced consistent column densities. 

In most cases, $3\sigma$ upper limits on the column densities were calculated for ions where no absorption profile was measurable. To calculate the limits we assume a single cloud with a Doppler parameter $b\sim8~\kms$. We chose this Doppler parameter as it is the average width of a {\SiII} transition in our survey. We found that adopting a larger Doppler parameter did not significantly change the calculated metallicities, indicating that the ionization approach described in Section \ref{sec:ion} is not overly sensitive to the column density limits. Where an ion is significantly blended we fit a Voigt profile to the absorption profile and use the column density as a conservative upper limit.

\begin{figure*}
	\centering
	\includegraphics[width=\linewidth]{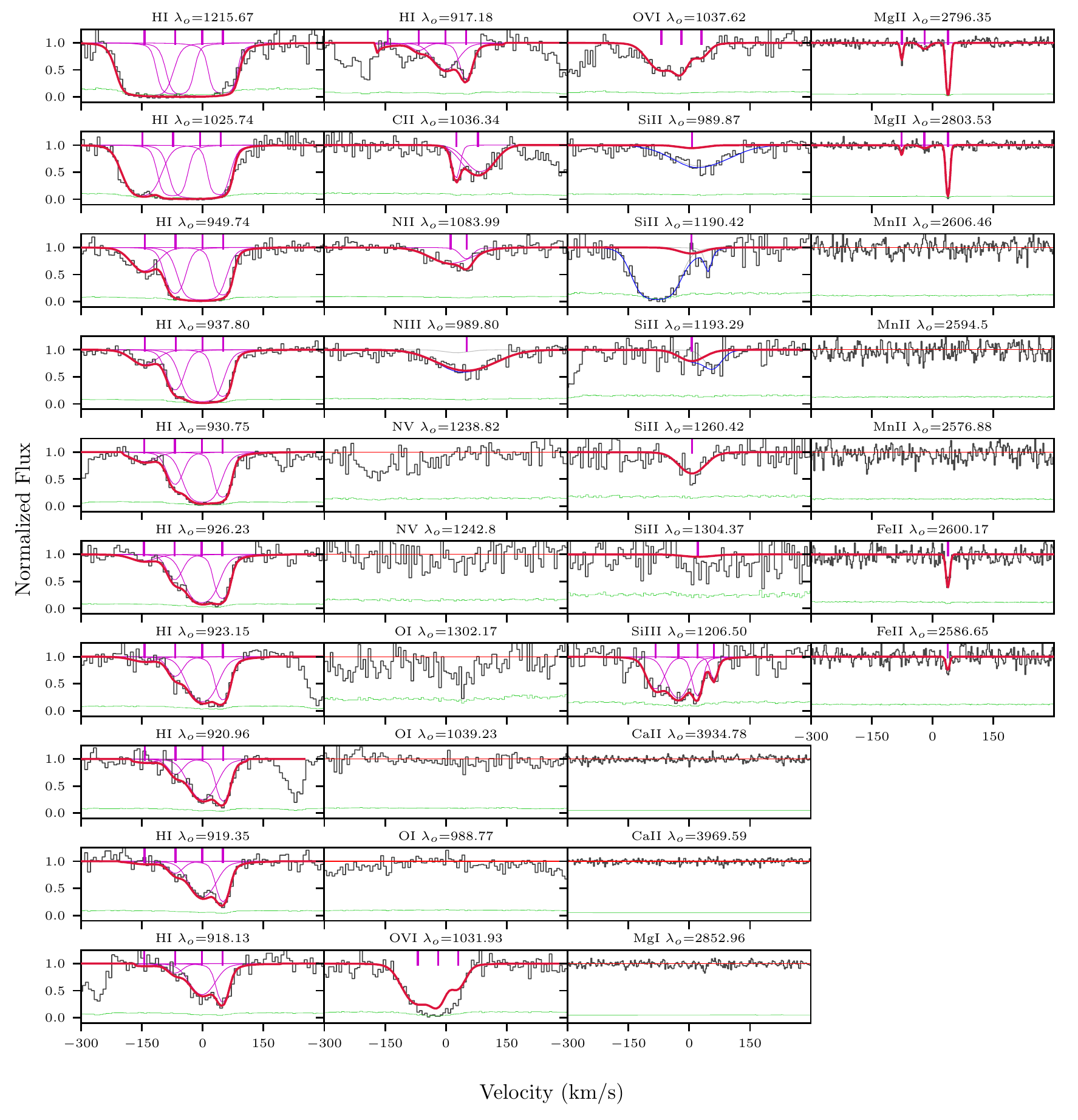}
	\caption{The fits for the system J$0351$, $z_{gal} = 0.356992$. The data for each ion (labeled above each panel) is shown in black. The fit to the absorption profile  is shown in red while the components are shown in pink. The centers of each Voigt profile used to fit the absorption profile are marked with a pink tick. The total profile fit, incorporating all ion components and blended components, is shown in blue. The fits to each of the blended components are shown in gray. The error spectrum is shown in green. The zero-point of the velocity is defined by the associated galaxy's redshift. For ions where we calculate limits, we show the continuum level as a red line. The {\SiII} $989$~{\AA} and {\NIII} $989$~{\AA} absorption profiles are blended together, while {\SiII} $1190$~{\AA} and {\SiII} $1193$~{\AA} lines are blended with unknown lines. However, the shape and depth of the {\SiII} absorption is constrained by the {\SiII} 1260~{\AA} and 1304~{\AA} lines. The total {\OVI} fits from \citet{nielsenovi} are shown here for completeness, although they are not used in the ionization modelling. Plots for the rest of the sample are shown in Appendix \ref{app:appA}.}
	\label{fig:Q0122_0.2119}
\end{figure*}

From this analysis, we were able to obtain column densities for the {\HI} Lyman series, {\CII}, {\CIII}, {\CIV}, {\NII}, {\NIII}, {\NV}, {\OI}, {\OVI}, {\SiII}, {\SiIII}, {\SiIV}, {\CaII}, {\MgI}, {\MgII} and {\FeII}. In Figure \ref{fig:Q0122_0.2119}, we present the results of the Voigt profile fitting for the galaxy--absorber pair J$0351$, $z_{gal} = 0.356992$. The black line represents the data, while the red line shows the fit to the absorption profiles for the ionic transition labeled above the plot. The pink lines indicate the individual components used in the fit while the pink ticks indicate the central position of each component. Where additional components were added to the fit to de-blend the absorption profile, the total fit is shown in blue with each additional component represented by a grey line. From the fitting process, we extract the column density of each ion and list them in Table \ref{tab:Q0122_0.2119} for this galaxy--absorber pair. The plots and tabulated data for all the other 46 systems are shown in Appendix \ref{app:appA}. We note that the {\OVI} column density measurements are presented in \citet{kacprzak15} and the total fits from \citet{nielsenovi} are shown in this work for completeness but are not used in the ionization modeling. 

\subsection{Ionization Modelling} \label{sec:ion}
The CGM metallicity for each galaxy--absorber pair is then obtained by calculating a likelihood function using the measured column densities and a grid of ionization properties generated by the ionization modeling suite Cloudy \citep{ferland13}. Cloudy predicts the column densities of the ions at each grid point given a particular combination of {\HI} column density, ${\NHI}$, hydrogen density, $n_{\text{H}}$ and metallicity, [Si/H]. Typical grids cover a range $-5.0~\text{cm}^{-3} < \log n_{\text{H}} < -1.0~\text{cm}^{-3}$, $13.0~{\cms} < {\NHI} < 20.0~\text{cm}^{-2}$ and $-4.0 < [\text{Si}/\text{H}] < 1.5$. We model a uniform layer of gas that is irradiated by background UV radiation, by assuming a single-phase model with no dust and solar abundance ratios \citep{crighton13,crighton15}. 

Previous studies have investigated the difference between the ionizing backgrounds from Haardt and Madau 2005, as implemented in Cloudy, and \citet{haardt12} (hereafter HM05 and HM12, respectively) \citep{howk09,werk14,wotta16,wotta18,chen17, zahedy18}. Most recently, \citet{wotta18} found a mean difference between the metallicities derived from the two ionizing backgrounds of $\text{[Z/H]}_{\text{HM}12} - \text{[Z/H]}_{\text{HM}05} = +0.37 \pm 0.19 $ for the entire ${\NHI}$ column density range in their sample. They also recomputed the metallicities from COS-HALOS \citep{prochaska17} using the HM05 ionizing background and compared them to the HM12 metallicities. The mean difference between the metallicities from COS-HALOS calculated from the two ionizing backgrounds was found to be $\text{[Z/H]}_{\text{HM}12} - \text{[Z/H]}_{\text{HM}05} = +0.26 \pm 0.19$. The smaller difference in metallicities between the two ionizing backgrounds for the COS-Halos sample, compared to \citet{wotta18} is attributed to the difference {\HI} column density ranges probed by the samples. The harder spectrum of ionizing photons from the HM12 background is due to a lower escape fraction of radiation from galaxies compared to the HM05 background, which leads to higher metallicity estimates. 

We investigate the difference between the HM05 and HM12 metallicity measurements in Appendix \ref{app:appB}. We find no significant difference between the metallicities calculated using the HM05 ionizing background compared to the HM12 ionizing background, although most of the data does tend to reside above the 1--1 line. Therefore, for consistency with the {\HI} absorption-selected surveys by \citet{lehner13, lehner18} and \citet{wotta16, wotta18}, where a bimodal metallicity distribution was observed, we use the HM05 ionizing background. The shape of the ionization background, which impacts the metallicity and ionization parameter values \citep{fechner11}, is evolved with redshift.

We used the Markov Chain Monte Carlo (MCMC) technique described by \citet{crighton13} to find the most likely range of metallicities and ionization parameters associated with the measured column densities. The likelihood function takes into account upper and lower limits of column densities, which are treated as one-sided Gaussians. In general, any priors on the grid variables applied to the likelihood analysis are boundaries of the Cloudy ionization grids. In such cases, the priors on the metallicity and gas density are flat. The priors placed on the {\HI} column density of galaxy-absorber pairs are Gaussian where we measured a column density and its associated uncertainty. When only upper and lower limits of the {\HI} column density were found, we applied them as bounds on a flat prior. The column densities used in the MCMC analysis are shown in Table~\ref{tab:Q0122_0.2119} and Table~\ref{tab:ionisation_param}. The {\OVI} measurements are shown in the table for completeness. However, as a single--phase ionization model is assumed, we do not include the {\OVI} column densities in the MCMC analysis. For each MCMC analysis we initialize 100 walkers with a burn-in stage of 200 steps. We then run the MCMC walkers for another 200 steps to calculate the final distributions.

\begin{deluxetable}{crc}
	\tablecolumns{3}
	\tablewidth{0.95\linewidth}
	\setlength{\tabcolsep}{0.06in}
	\tablecaption{J$0351$, $z_{gal} = 0.356992$ Measured Column Densities \label{tab:Q0122_0.2119}}
	\tablehead{
		\colhead{Ion}          &
        \colhead{$\log N~({\cms})$}    &
		\colhead{$\log N$ Error~({\cms})}}
	\startdata
	{\HI}   & $16.86$   &$0.03$\\
{\CII}  & $14.45$   &$0.03$\\
{\NII}  & $14.23$   &$0.04$\\
{\NIII} & $14.40$   &$0.03$\\
{\NV}   & $<13.34$  &$\cdots$\\
{\OI}   & $<13.69$  &$\cdots$\\
{\SiII} & $13.01$   &$0.08$\\
{\SiIII}& $13.68$   &$0.15$\\
{\CaII} & $<11.31$  &$\cdots$\\
{\MgI}  & $<11.02$  &$\cdots$\\
{\MgII} & $13.09$   &$0.02$\\
{\MnII} & $<12.22$  &$\cdots$\\
{\FeII} & $12.78$   &$0.05$\\[-5pt]

	\enddata
\end{deluxetable}

\begin{figure}
	\centering
	\includegraphics[width=\linewidth]{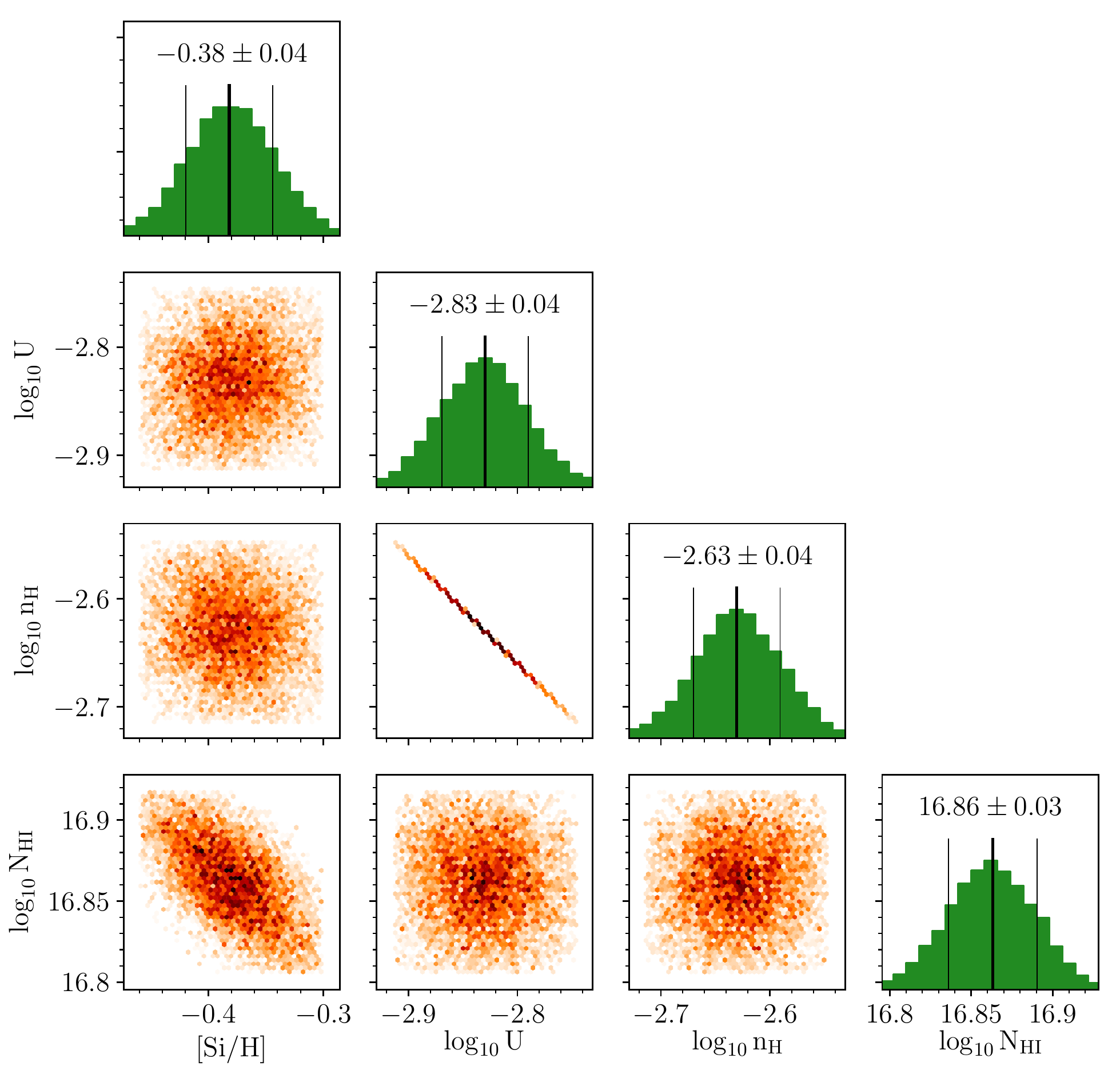}
	\caption{The posterior distribution profiles from the MCMC analysis of the Cloudy grids for J$0351$, $z_{gal} = 0.356992$ are shown as the orange hexbin plots where the model parameters on the x-axis are plotted as a function of the model parameters on the y-axis. The model parameters shown are $[\text{Si}/\text{H}]$, $\log U$, ${\nh}$ and ${\NHI}$. On the end of each row, the distributions of each of those parameters are shown in green where the average and width of the $68\%$ confidence intervals are shown above and indicated by the black lines. Plots for the rest of the sample are shown in Appendix \ref{app:appA}.}
	\label{fig:Q0122_0.2119_par}
\end{figure}

In Figure \ref{fig:Q0122_0.2119_par} the MCMC posterior distributions and histograms are plotted for J$0351$, $z_{gal} = 0.356992$. From left to right, the columns are plotted as a function of the metallicity [Si/H], ionization parameter $\log U$, hydrogen number density ${\nh}$ and {\HI} column density ${\NHI}$. The green histograms on the end of each row are the distributions of the final positions of the walkers in the MCMC analysis for each corresponding parameter. The $68\%$ confidence interval and the average of these two values for each parameter are labeled at the top of the histogram panels, and indicated by the black lines. The remaining plots show the posterior distributions of the walkers at the end of the MCMC analysis for the parameter listed on the y-axis as a function of the parameter on the x-axis. Darker colors indicate a higher probability. Posterior distribution plots for the remaining systems are plotted in Appendix \ref{app:appA}. The inferred model parameters, shown as the maximum likelihood values with $68\%$ confidence interval uncertainties, for the full sample are tabulated in Table \ref{tab:ionisation_param}. Upper or lower limits are calculated such that we are $95\%$ confident that the true value should be located below or above, respectively.

\begin{deluxetable*}{clrrrrrrrrrrr}
	\tablecolumns{13}
	\tablewidth{0pt}
	\setlength{\tabcolsep}{0.06in}
	\tablecaption{MCMC Output \label{tab:ionisation_param}}
	\tablehead{
		\colhead{}           	&
        \colhead{}     &
        \colhead{Meas.}     &
        \multicolumn{4}{c}{HM05}    &
        \colhead{}                  &
        \multicolumn{4}{c}{HM12}    \\
        \cline{4-7} \cline{9-12}\\[-5pt]
        \colhead{J-name}            &
        \colhead{$z_{\rm gal}$}    &
        \colhead{${\NHI}$\tablenotemark{a}}&
		\colhead{[Si/H]\tablenotemark{b}}               &
		\colhead{${\NHI}$\tablenotemark{b}}    &
		\colhead{${\nh}$\tablenotemark{b}}	&
		\colhead{$\log U$\tablenotemark{b}}               &
		\colhead{}                  &
		\colhead{[Si/H]\tablenotemark{b}}               &
		\colhead{${\NHI}$\tablenotemark{b}}    &
		\colhead{${\nh}$\tablenotemark{b}}	&
		\colhead{$\log U$\tablenotemark{b}}}
	\startdata
	J0125	&	$	0.398525	$	&	$	[18.85, 19.00]			$	&	$		-1.56	^{+	0.03	}_{-	0.03	}	$	&	$		18.85	^{+	0.04	}_{-	0.01	}	$	&	$		-3.018	^{+	0.003	}_{-	0.001	}	$	&	$		-2.41	^{+	0.01	}_{-	0.01	}	$	&	&	$		-1.60	^{+	0.03	}_{-	0.04	}	$	&	$		18.90	^{+	0.08	}_{-	0.01	}	$	&	$		-3.228	^{+	0.005	}_{-	0.001	}	$	&	$		-3.21	^{+	0.01	}_{-	0.01	}	$	\\	[2pt]
J0351	&	$	0.2617	$	&	$	14.51	\pm	0.13	$	&	$	<	0.82						$	&	$		14.51	^{+	0.09	}_{-	0.15	}	$	&	$	<	-2.001						$	&	$	<	-2.91						$	&	&	$	<	1.15						$	&	$		14.51	^{+	0.11	}_{-	0.14	}	$	&	$	<	-2.005						$	&	$	<	-2.85						$	\\	[2pt]
J0351	&	$	0.356992	$	&	$	16.86	\pm	0.03	$	&	$		-0.38	^{+	0.04	}_{-	0.04	}	$	&	$		16.86	^{+	0.03	}_{-	0.03	}	$	&	$		-2.628	^{+	0.039	}_{-	0.039	}	$	&	$		-2.91	^{+	0.12	}_{-	0.04	}	$	&	&	$		-0.38	^{+	0.04	}_{-	0.04	}	$	&	$		16.86	^{+	0.33	}_{-	0.27	}	$	&	$		-2.629	^{+	0.048	}_{-	0.034	}	$	&	$		-3.34	^{+	0.12	}_{-	0.04	}	$	\\	[2pt]
J0407	&	$	0.1534	$	&	$	13.79	\pm	0.01	$	&	$	<	0.45						$	&	$		13.79	^{+	0.01	}_{-	0.01	}	$	&	$	<	-2.001						$	&	$	<	-2.14						$	&	&	$	<	0.51						$	&	$		13.79	^{+	0.01	}_{-	0.01	}	$	&	$	<	-2.001						$	&	$	<	-2.17						$	\\	[2pt]
J0407	&	$	0.3422	$	&	$	13.78	\pm	0.01	$	&	$	<	-0.04						$	&	$		13.79	^{+	0.02	}_{-	0.02	}	$	&	$	<	-2.002						$	&	$	<	-2.11						$	&	&	$	<	0.48						$	&	$		13.77	^{+	0.04	}_{-	0.01	}	$	&	$	<	-2.012						$	&	$	<	-2.38						$	\\	[2pt]
J0407	&	$	0.495164	$	&	$	14.34	\pm	0.56	$	&	$		-1.10	^{+	0.49	}_{-	0.55	}	$	&	$		14.35	^{+	0.35	}_{-	0.35	}	$	&	$		-3.790	^{+	0.754	}_{-	0.384	}	$	&	$		-2.68	^{+	1.53	}_{-	0.39	}	$	&	&	$		-0.23	^{+	0.42	}_{-	0.70	}	$	&	$		14.34	^{+	0.39	}_{-	0.32	}	$	&	$		-3.947	^{+	0.659	}_{-	0.520	}	$	&	$		-2.91	^{+	1.69	}_{-	0.50	}	$	\\	[2pt]
J0456	&	$	0.2784	$	&	$	[15.06, 19.00]			$	&	$	<	-1.40						$	&	$		15.71	^{+	1.55	}_{-	0.73	}	$	&	$	<	-2.000						$	&	$	<	-2.68						$	&	&	$	<	-1.08						$	&	$		15.13	^{+	2.11	}_{-	0.14	}	$	&	$	<	-2.015						$	&	$	<	-2.94						$	\\	[2pt]
J0456	&	$	0.381511	$	&	$	15.10	\pm	0.39	$	&	$		-0.06	^{+	0.03	}_{-	1.01	}	$	&	$		15.13	^{+	0.38	}_{-	0.35	}	$	&	$		-3.238	^{+	0.459	}_{-	0.381	}	$	&	$		-3.11	^{+	1.29	}_{-	0.46	}	$	&	&	$		0.19	^{+	0.01	}_{-	0.98	}	$	&	$		15.12	^{+	0.38	}_{-	0.30	}	$	&	$		-3.389	^{+	0.447	}_{-	0.286	}	$	&	$		-3.29	^{+	1.11	}_{-	0.38	}	$	\\	[2pt]
J0456	&	$	0.4828	$	&	$	[16.53, 19.00]			$	&	$		-1.32	^{+	0.15	}_{-	0.15	}	$	&	$		17.65	^{+	0.18	}_{-	0.17	}	$	&	$		-2.381	^{+	0.053	}_{-	0.056	}	$	&	$		-3.06	^{+	0.17	}_{-	0.06	}	$	&	&	$		-1.14	^{+	0.11	}_{-	0.14	}	$	&	$		17.63	^{+	0.16	}_{-	0.10	}	$	&	$		-2.597	^{+	0.053	}_{-	0.066	}	$	&	$		-3.23	^{+	0.16	}_{-	0.04	}	$	\\	[2pt]
J0853	&	$	0.1635	$	&	$	19.93	\pm	0.01	$	&	$		-1.70	^{+	0.06	}_{-	0.05	}	$	&	$		19.93	^{+	0.01	}_{-	0.01	}	$	&	$		-2.631	^{+	0.052	}_{-	0.045	}	$	&	$		-3.17	^{+	0.14	}_{-	0.04	}	$	&	&	$		-1.69	^{+	0.07	}_{-	0.05	}	$	&	$		19.93	^{+	0.01	}_{-	0.01	}	$	&	$		-3.024	^{+	0.067	}_{-	0.053	}	$	&	$		-3.28	^{+	0.19	}_{-	0.07	}	$	\\	[2pt]
J0853	&	$	0.2766	$	&	$	14.15	\pm	0.03	$	&	$		-0.30	^{+	0.04	}_{-	0.92	}	$	&	$		14.15	^{+	0.04	}_{-	0.02	}	$	&	$		-3.225	^{+	0.357	}_{-	0.830	}	$	&	$		-2.67	^{+	1.18	}_{-	0.01	}	$	&	&	$		-0.11	^{+	0.37	}_{-	0.17	}	$	&	$		14.15	^{+	0.03	}_{-	0.03	}	$	&	$		-4.211	^{+	0.950	}_{-	0.202	}	$	&	$		-2.77	^{+	0.76	}_{-	0.02	}	$	\\	[2pt]
J0853	&	$	0.4402	$	&	$	17.30	\pm	0.20	$	&	$	<	-1.19						$	&	$		17.23	^{+	0.21	}_{-	0.19	}	$	&	$	<	-2.000						$	&	$	<	-3.21						$	&	&	$		-1.58	^{+	0.19	}_{-	0.27	}	$	&	$		17.32	^{+	0.34	}_{-	0.76	}	$	&	$		-2.440	^{+	0.055	}_{-	0.052	}	$	&	$		-3.44	^{+	0.15	}_{-	0.05	}	$	\\	[2pt]
J0914	&	$	0.244312	$	&	$	15.55	\pm	0.03	$	&	$		-0.78	^{+	0.09	}_{-	0.10	}	$	&	$		15.55	^{+	0.04	}_{-	0.03	}	$	&	$		-3.436	^{+	0.037	}_{-	0.206	}	$	&	$		-2.29	^{+	0.33	}_{-	0.09	}	$	&	&	$		-0.03	^{+	0.06	}_{-	0.31	}	$	&	$		15.45	^{+	0.14	}_{-	0.12	}	$	&	$		-3.508	^{+	0.137	}_{-	0.063	}	$	&	$		-2.36	^{+	0.33	}_{-	0.13	}	$	\\	[2pt]
J0943	&	$	0.1431	$	&	$	[15.45, 17.00]			$	&	$	<	-1.14						$	&	$	<	16.88						$	&	$	<	-2.000						$	&	$	<	-2.56						$	&	&	$	<	-0.92						$	&	$	<	16.93						$	&	$	<	-2.004						$	&	$	<	-2.76						$	\\	[2pt]
J0943	&	$	0.2284	$	&	$	16.03	\pm	0.67	$	&	$		-1.33	^{+	0.66	}_{-	0.71	}	$	&	$		16.04	^{+	0.66	}_{-	0.48	}	$	&	$		-3.242	^{+	0.151	}_{-	0.344	}	$	&	$		-2.69	^{+	0.66	}_{-	0.16	}	$	&	&	$		-0.79	^{+	0.60	}_{-	0.66	}	$	&	$		16.02	^{+	0.51	}_{-	0.51	}	$	&	$		-3.385	^{+	0.159	}_{-	0.266	}	$	&	$		-2.95	^{+	0.59	}_{-	0.17	}	$	\\	[2pt]
J0943	&	$	0.353052	$	&	$	16.46	\pm	0.03	$	&	$	<	-1.69						$	&	$		16.38	^{+	0.11	}_{-	0.01	}	$	&	$	<	-2.155						$	&	$	<	-1.79						$	&	&	$		-0.88	^{+	0.05	}_{-	0.06	}	$	&	$		16.46	^{+	0.03	}_{-	0.03	}	$	&	$		-2.577	^{+	0.059	}_{-	0.043	}	$	&	$		-3.43	^{+	0.15	}_{-	0.05	}	$	\\	[2pt]
J0950	&	$	0.211866	$	&	$	[16.28, 19.00]			$	&	$		-1.48	^{+	0.04	}_{-	0.02	}	$	&	$		19.00	^{+	0.01	}_{-	0.09	}	$	&	$		-3.140	^{+	0.003	}_{-	0.001	}	$	&	$	<	-2.51						$	&	&	$		-1.35	^{+	0.03	}_{-	0.02	}	$	&	$		18.78	^{+	2.56	}_{-	2.40	}	$	&	$		-3.374	^{+	0.005	}_{-	0.001	}	$	&	$		-2.74	^{+	0.01	}_{-	0.01	}	$	\\	[2pt]
J1004	&	$	0.1380	$	&	$	14.91	\pm	0.14	$	&	$		-0.23	^{+	0.05	}_{-	0.07	}	$	&	$		15.08	^{+	0.01	}_{-	0.06	}	$	&	$		-3.291	^{+	0.031	}_{-	0.039	}	$	&	$		-2.52	^{+	0.10	}_{-	0.03	}	$	&	&	$		-0.06	^{+	0.08	}_{-	0.07	}	$	&	$		15.18	^{+	0.37	}_{-	0.53	}	$	&	$		-3.484	^{+	0.031	}_{-	0.041	}	$	&	$		-2.80	^{+	0.11	}_{-	0.04	}	$	\\	[2pt]
J1009	&	$	0.227855	$	&	$	[17.51, 19.00]			$	&	$		-2.00	^{+	0.07	}_{-	0.04	}	$	&	$		18.26	^{+	0.10	}_{-	0.13	}	$	&	$		-3.131	^{+	0.028	}_{-	0.001	}	$	&	$		-2.56	^{+	0.07	}_{-	0.01	}	$	&	&	$		-1.77	^{+	0.15	}_{-	0.03	}	$	&	$		18.22	^{+	0.04	}_{-	0.22	}	$	&	$		-3.371	^{+	0.090	}_{-	0.004	}	$	&	$		-2.93	^{+	0.22	}_{-	0.01	}	$	\\	[2pt]
J1041	&	$	0.3153	$	&	$	[14.43, 17.00]			$	&	$		-0.42	^{+	0.05	}_{-	0.05	}	$	&	$		16.12	^{+	0.05	}_{-	0.06	}	$	&	$		-2.005	^{+	0.005	}_{-	0.071	}	$	&	$		-3.51	^{+	0.08	}_{-	0.00	}	$	&	&	$		-1.02	^{+	0.24	}_{-	0.01	}	$	&	$		17.00	^{+	2.18	}_{-	2.40	}	$	&	$		-2.209	^{+	0.062	}_{-	0.061	}	$	&	$		-3.91	^{+	0.23	}_{-	0.11	}	$	\\	[2pt]
J1041	&	$	0.442173	$	&	$	[16.77, 19.00]			$	&	$		-1.77	^{+	0.04	}_{-	0.03	}	$	&	$		18.91	^{+	0.04	}_{-	0.12	}	$	&	$		-2.988	^{+	0.053	}_{-	0.003	}	$	&	$		-2.51	^{+	0.13	}_{-	0.01	}	$	&	&	$		-1.60	^{+	0.05	}_{-	0.03	}	$	&	$		18.69	^{+	0.06	}_{-	0.08	}	$	&	$		-3.196	^{+	0.021	}_{-	0.001	}	$	&	$		-2.64	^{+	0.05	}_{-	0.01	}	$	\\	[2pt]
J1119	&	$	0.1383	$	&	$	15.64	\pm	0.32	$	&	$		-0.23	^{+	0.09	}_{-	0.10	}	$	&	$		15.82	^{+	0.08	}_{-	0.11	}	$	&	$		-2.536	^{+	0.046	}_{-	0.080	}	$	&	$		-3.31	^{+	0.18	}_{-	0.05	}	$	&	&	$		0.02	^{+	0.10	}_{-	0.06	}	$	&	$		15.78	^{+	0.71	}_{-	0.92	}	$	&	$		-2.974	^{+	0.069	}_{-	0.054	}	$	&	$		-3.37	^{+	0.17	}_{-	0.05	}	$	\\	[2pt]
J1133	&	$	0.154599	$	&	$	[15.82, 17.00]			$	&	$	<	-1.98						$	&	$		16.11	^{+	0.42	}_{-	0.29	}	$	&	$	<	-2.001						$	&	$	<	-2.69						$	&	&	$		-2.87	^{+	0.47	}_{-	1.07	}	$	&	$		16.02	^{+	0.49	}_{-	0.20	}	$	&	$	<	-2.023						$	&	$	<	-2.71						$	\\	[2pt]
J1139	&	$	0.1755	$	&	$	14.15	\pm	0.05	$	&	$	<	0.69						$	&	$		14.15	^{+	0.04	}_{-	0.05	}	$	&	$	<	-2.001						$	&	$	<	-2.34						$	&	&	$	<	0.65						$	&	$		14.15	^{+	0.05	}_{-	0.05	}	$	&	$	<	-2.074						$	&	$	<	-2.49						$	\\	[2pt]
J1139	&	$	0.204194	$	&	$	[16.04, 17.00]			$	&	$		-0.35	^{+	0.03	}_{-	0.07	}	$	&	$		16.04	^{+	0.04	}_{-	0.01	}	$	&	$		-3.040	^{+	0.056	}_{-	0.077	}	$	&	$		-2.74	^{+	0.20	}_{-	0.06	}	$	&	&	$		-0.07	^{+	0.04	}_{-	0.08	}	$	&	$		16.04	^{+	0.59	}_{-	0.55	}	$	&	$		-3.362	^{+	0.062	}_{-	0.058	}	$	&	$		-2.87	^{+	0.18	}_{-	0.06	}	$	\\	[2pt]
J1139	&	$	0.212259	$	&	$	15.33	\pm	0.04	$	&	$	<	0.60						$	&	$		15.33	^{+	0.03	}_{-	0.05	}	$	&	$	<	-2.053						$	&	$	<	-2.46						$	&	&	$	<	0.56						$	&	$		15.33	^{+	0.04	}_{-	0.04	}	$	&	$	<	-2.019						$	&	$	<	-2.41						$	\\	[2pt]
J1139	&	$	0.219724	$	&	$	14.20	\pm	0.07	$	&	$	<	0.63						$	&	$		14.30	^{+	0.01	}_{-	0.28	}	$	&	$	<	-2.001						$	&	$	<	-2.42						$	&	&	$	<	0.62						$	&	$		14.21	^{+	0.07	}_{-	0.21	}	$	&	$	<	-2.005						$	&	$	<	-2.38						$	\\	[2pt]
J1139	&	$	0.319255	$	&	$	16.19	\pm	0.03	$	&	$		-2.59	^{+	0.58	}_{-	0.04	}	$	&	$		16.19	^{+	0.03	}_{-	0.03	}	$	&	$		-3.626	^{+	0.497	}_{-	0.077	}	$	&	$		-2.81	^{+	0.99	}_{-	0.42	}	$	&	&	$		-1.91	^{+	0.20	}_{-	0.13	}	$	&	$		16.19	^{+	0.14	}_{-	0.20	}	$	&	$		-3.992	^{+	0.542	}_{-	0.015	}	$	&	$		-3.29	^{+	1.36	}_{-	0.80	}	$	\\	[2pt]
J1219	&	$	0.1241	$	&	$	15.25	\pm	0.03	$	&	$		-0.72	^{+	0.20	}_{-	1.39	}	$	&	$		15.25	^{+	0.95	}_{-	0.01	}	$	&	$		-3.437	^{+	0.031	}_{-	0.154	}	$	&	$		-2.48	^{+	0.30	}_{-	0.11	}	$	&	&	$		-0.40	^{+	0.36	}_{-	1.58	}	$	&	$		15.51	^{+	1.39	}_{-	0.28	}	$	&	$		-3.434	^{+	0.033	}_{-	0.111	}	$	&	$		-2.91	^{+	0.20	}_{-	0.06	}	$	\\	[2pt]
J1233	&	$	0.318757	$	&	$	15.72	\pm	0.02	$	&	$		-1.14	^{+	0.13	}_{-	0.09	}	$	&	$		15.72	^{+	0.02	}_{-	0.02	}	$	&	$		-3.445	^{+	0.167	}_{-	0.159	}	$	&	$		-2.39	^{+	0.49	}_{-	0.16	}	$	&	&	$		-0.54	^{+	0.16	}_{-	0.08	}	$	&	$		15.72	^{+	0.02	}_{-	0.02	}	$	&	$		-3.535	^{+	0.215	}_{-	0.112	}	$	&	$		-2.75	^{+	0.46	}_{-	0.13	}	$	\\	[2pt]
J1241	&	$	0.205267	$	&	$	[16.63, 19.00]			$	&	$		-0.32	^{+	0.05	}_{-	0.03	}	$	&	$		17.43	^{+	0.02	}_{-	0.03	}	$	&	$		-3.593	^{+	0.011	}_{-	0.012	}	$	&	$		-2.54	^{+	0.03	}_{-	0.01	}	$	&	&	$		-0.28	^{+	0.03	}_{-	0.04	}	$	&	$		17.43	^{+	0.02	}_{-	0.03	}	$	&	$		-3.601	^{+	0.016	}_{-	0.008	}	$	&	$		-2.54	^{+	0.04	}_{-	0.01	}	$	\\	[2pt]
J1241	&	$	0.217905	$	&	$	15.59	\pm	0.12	$	&	$		-0.57	^{+	0.16	}_{-	0.09	}	$	&	$		15.72	^{+	0.09	}_{-	0.11	}	$	&	$		-3.879	^{+	0.069	}_{-	0.063	}	$	&	$		-2.37	^{+	0.22	}_{-	0.09	}	$	&	&	$		-0.39	^{+	0.18	}_{-	0.18	}	$	&	$		15.47	^{+	0.12	}_{-	0.12	}	$	&	$		-3.520	^{+	0.105	}_{-	0.123	}	$	&	$		-2.82	^{+	0.36	}_{-	0.13	}	$	\\	[2pt]
J1244	&	$	0.5504	$	&	$	[17.00, 19.00]			$	&	$		-1.20	^{+	0.07	}_{-	0.03	}	$	&	$		18.96	^{+	0.04	}_{-	0.21	}	$	&	$		-2.801	^{+	0.075	}_{-	0.130	}	$	&	$		-2.99	^{+	0.27	}_{-	0.07	}	$	&	&	$		-1.20	^{+	0.08	}_{-	0.03	}	$	&	$		18.95	^{+	0.05	}_{-	0.18	}	$	&	$		-2.826	^{+	0.071	}_{-	0.116	}	$	&	$		-2.95	^{+	0.24	}_{-	0.05	}	$	\\	[2pt]
J1301	&	$	0.1967	$	&	$	13.86	\pm	0.01	$	&	$	<	0.56						$	&	$	<	16.01						$	&	$	<	-2.002						$	&	$	<	-2.50						$	&	&	$	<	1.43						$	&	$	<	15.40						$	&	$	<	-2.002						$	&	$	<	-2.80						$	\\	[2pt]
J1319	&	$	0.6610	$	&	$	18.30	\pm	0.30	$	&	$		-2.18	^{+	0.03	}_{-	0.04	}	$	&	$		18.60	^{+	0.01	}_{-	0.05	}	$	&	$		-2.016	^{+	0.016	}_{-	0.053	}	$	&	$		-3.16	^{+	0.07	}_{-	0.00	}	$	&	&	$		-1.96	^{+	0.11	}_{-	0.04	}	$	&	$		18.61	^{+	0.12	}_{-	0.17	}	$	&	$		-2.303	^{+	0.109	}_{-	0.032	}	$	&	$		-3.46	^{+	0.26	}_{-	0.12	}	$	\\	[2pt]
J1322	&	$	0.214431	$	&	$	[16.97, 19.00]			$	&	$		-1.90	^{+	0.04	}_{-	0.03	}	$	&	$		19.00	^{+	0.01	}_{-	0.12	}	$	&	$		-2.827	^{+	0.055	}_{-	0.104	}	$	&	$		-2.96	^{+	0.26	}_{-	0.09	}	$	&	&	$		-1.64	^{+	0.07	}_{-	0.06	}	$	&	$		18.77	^{+	0.22	}_{-	0.11	}	$	&	$		-3.072	^{+	0.093	}_{-	0.104	}	$	&	$		-3.17	^{+	0.25	}_{-	0.05	}	$	\\	[2pt]
J1342	&	$	0.0708	$	&	$	14.61	\pm	0.47	$	&	$		-0.02	^{+	0.57	}_{-	0.33	}	$	&	$		15.33	^{+	0.26	}_{-	0.69	}	$	&	$		-3.812	^{+	0.133	}_{-	0.174	}	$	&	$		-2.21	^{+	0.34	}_{-	0.03	}	$	&	&	$		0.38	^{+	0.33	}_{-	0.85	}	$	&	$		15.35	^{+	0.35	}_{-	0.90	}	$	&	$		-4.015	^{+	0.316	}_{-	0.006	}	$	&	$		-2.64	^{+	0.31	}_{-	0.01	}	$	\\	[2pt]
J1342	&	$	0.2013	$	&	$	14.22	\pm	0.03	$	&	$	<	-0.12						$	&	$		14.30	^{+	0.03	}_{-	0.15	}	$	&	$	<	-2.000						$	&	$	<	-2.15						$	&	&	$	<	0.16						$	&	$		14.15	^{+	0.11	}_{-	0.01	}	$	&	$	<	-2.016						$	&	$	<	-2.16						$	\\	[2pt]
J1342	&	$	0.227042	$	&	$	18.83	\pm	0.05	$	&	$		-0.36	^{+	0.04	}_{-	0.05	}	$	&	$		18.88	^{+	0.06	}_{-	0.04	}	$	&	$		-2.500	^{+	0.030	}_{-	0.037	}	$	&	$		-3.19	^{+	0.10	}_{-	0.03	}	$	&	&	$		-0.28	^{+	0.04	}_{-	0.05	}	$	&	$		18.87	^{+	0.33	}_{-	0.42	}	$	&	$		-2.703	^{+	0.038	}_{-	0.044	}	$	&	$		-3.46	^{+	0.12	}_{-	0.04	}	$	\\	[2pt]
J1357	&	$	0.4295	$	&	$	14.25	\pm	0.05	$	&	$	<	0.32						$	&	$		14.26	^{+	0.04	}_{-	0.16	}	$	&	$	<	-2.001						$	&	$	<	-2.27						$	&	&	$	<	0.42						$	&	$		14.33	^{+	0.03	}_{-	0.23	}	$	&	$	<	-2.007						$	&	$	<	-1.94						$	\\	[2pt]
J1357	&	$	0.4592	$	&	$	[16.87, 19.00]			$	&	$		-1.38	^{+	0.03	}_{-	0.02	}	$	&	$		18.60	^{+	0.03	}_{-	0.04	}	$	&	$		-2.991	^{+	0.012	}_{-	0.001	}	$	&	$		-2.41	^{+	0.04	}_{-	0.01	}	$	&	&	$		-1.18	^{+	0.04	}_{-	0.02	}	$	&	$		18.50	^{+	0.03	}_{-	0.03	}	$	&	$		-3.204	^{+	0.008	}_{-	0.001	}	$	&	$		-2.57	^{+	0.02	}_{-	0.01	}	$	\\	[2pt]
J1547	&	$	0.0949	$	&	$	13.75	\pm	0.03	$	&	$	<	0.79						$	&	$		13.74	^{+	0.06	}_{-	0.06	}	$	&	$	<	-2.000						$	&	$	<	-2.08						$	&	&	$	<	0.80						$	&	$		13.68	^{+	0.16	}_{-	0.04	}	$	&	$	<	-2.007						$	&	$	<	-2.42						$	\\	[2pt]
J1555	&	$	0.189201	$	&	$	[16.37, 19.00]			$	&	$		-1.43	^{+	0.71	}_{-	0.04	}	$	&	$		18.04	^{+	0.01	}_{-	0.90	}	$	&	$		-3.176	^{+	0.268	}_{-	0.055	}	$	&	$		-2.82	^{+	0.26	}_{-	0.05	}	$	&	&	$		-1.20	^{+	0.30	}_{-	0.05	}	$	&	$		18.08	^{+	0.02	}_{-	0.35	}	$	&	$		-3.371	^{+	0.130	}_{-	0.043	}	$	&	$		-3.01	^{+	0.28	}_{-	0.11	}	$	\\	[2pt]
J1704	&	$	0.0921	$	&	$	14.27	\pm	0.02	$	&	$	<	0.62						$	&	$		14.25	^{+	0.03	}_{-	0.04	}	$	&	$	<	-2.000						$	&	$	<	-2.70						$	&	&	$	<	0.71						$	&	$		14.27	^{+	0.08	}_{-	0.01	}	$	&	$	<	-2.010						$	&	$	<	-2.52						$	\\	[2pt]
J2131	&	$	0.430200	$	&	$	19.88	\pm	0.10	$	&	$		-1.96	^{+	0.03	}_{-	0.03	}	$	&	$		19.78	^{+	0.01	}_{-	0.01	}	$	&	$		-2.594	^{+	0.043	}_{-	0.031	}	$	&	$		-2.86	^{+	0.11	}_{-	0.03	}	$	&	&	$		-1.85	^{+	0.02	}_{-	0.03	}	$	&	$		19.78	^{+	0.28	}_{-	0.28	}	$	&	$		-2.874	^{+	0.051	}_{-	0.053	}	$	&	$		-3.00	^{+	0.13	}_{-	0.03	}	$	\\	[2pt]
J2137	&	$	0.0752	$	&	$	13.96	\pm	0.02	$	&	$	<	0.78						$	&	$		13.91	^{+	0.07	}_{-	0.01	}	$	&	$	<	-2.006						$	&	$	<	-2.23						$	&	&	$	<	0.81						$	&	$		14.01	^{+	0.03	}_{-	0.11	}	$	&	$	<	-2.007						$	&	$	<	-2.56						$	\\	[2pt]
J2253	&	$	0.352787	$	&	$	14.53	\pm	0.05	$	&	$	<	-0.22						$	&	$		14.56	^{+	0.02	}_{-	0.19	}	$	&	$	<	-2.012						$	&	$	<	-1.88						$	&	&	$	<	0.35						$	&	$		14.57	^{+	0.01	}_{-	0.20	}	$	&	$	<	-2.033						$	&	$	<	-2.19						$	\\	[-5pt]
	\enddata
	\tablenotetext{a}{The {\HI} column density calculated from the Voigt profile models of the absorption, which were then used to constrain the cloudy models. A range of {\HI} values indicated where we have used a flat prior for the MCMC analysis of the ionization models.}
	\tablenotetext{b}{The maximum likelihood value with the $68\%$ uncertainties from the MCMC analysis. For upper limits, we take the $95\%$ upper uncertainty.}
\end{deluxetable*}

\section{Results}
\label{sec:results}

Here, we present the metallicity analysis of the ``Multiphase Galaxy Halos'' survey. We first present the distribution of {\HI} column density and investigate its relationship with the CGM metallicity, galaxy impact parameter and galaxy azimuthal angle. We then explore the distribution of CGM metallicities and probe the relationship with other intrinsic properties of the sample including {\HI} column densities and galaxy impact parameters, halo masses and redshifts. The relationship between the orientation of the galaxies and the CGM metallicity is then investigated.

\subsection{${\NHI}$}
The {\HI} column density of each system was calculated using the Voigt profile fitting process described in Section \ref{sec:observations}. The hydrogen column density ranges from $13.8\ \cms < {\NHI} < 19.9\ \cms$ with an average of $\langle {\NHI} \rangle = 15.8\  \cms $. 

\begin{figure*}
	\centering
	\includegraphics[width=\linewidth]{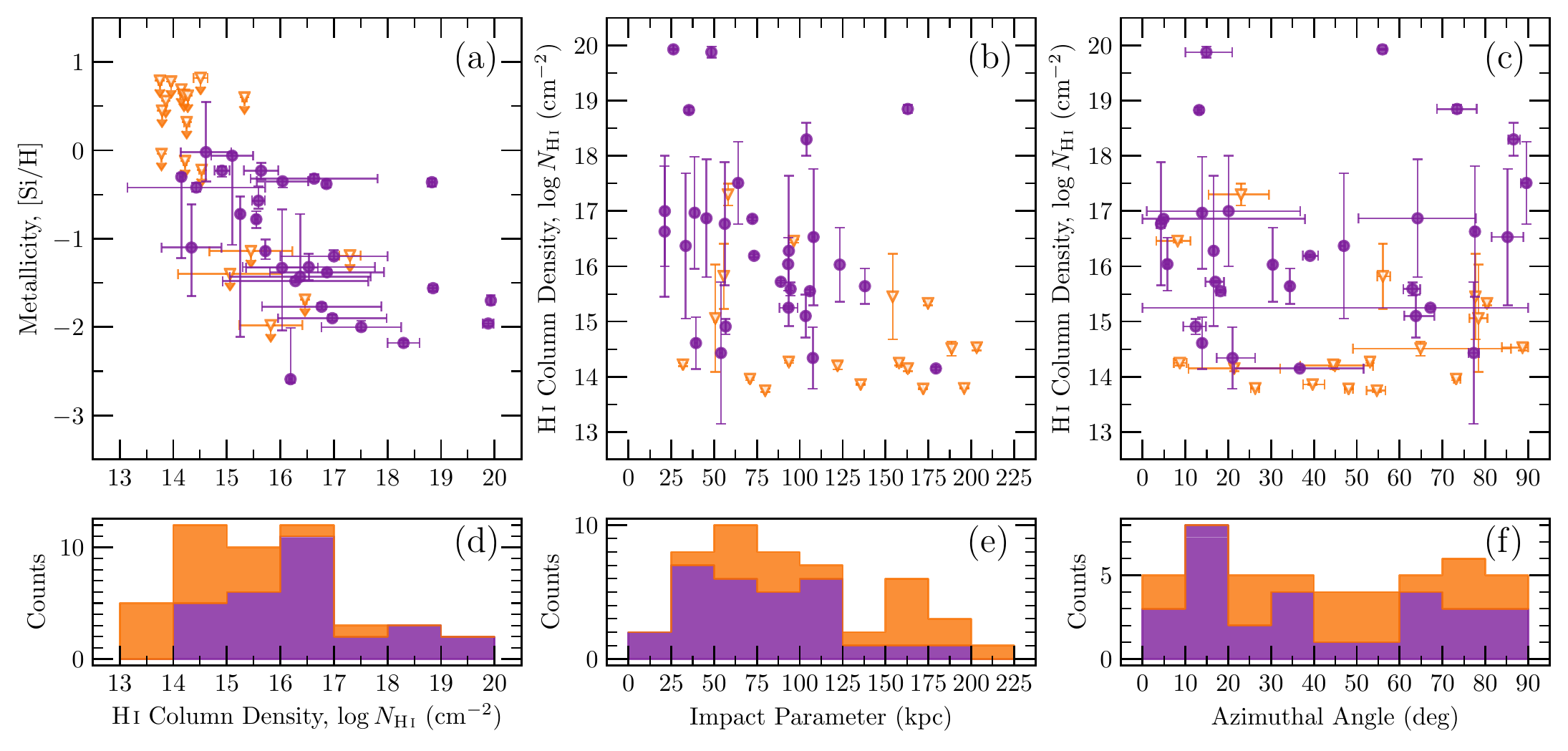}
	\caption{The metallicity, [Si/H], is shown as a function of {\HI} column densities, ${\NHI}$, in (a). The {\HI} column densities, ${\NHI}$, are shown as a function of (b) impact parameter and (c) azimuthal angle. Purple circles represent systems which have constrained metallicity values while orange triangles indicate where there are only upper limits on the metallicity. In the bottom row of panels, the distribution of (d) {\HI} column density, (e) impact parameter and (f) azimuthal angle are shown. The purple histogram shows the systems which have [Si/H] measurements, while the orange histogram shows systems with [Si/H] upper limits.}
	\label{fig:HI_prop}
\end{figure*}

The relationship between the hydrogen column density and the CGM metallicity, galaxy impact parameter and galaxy azimuthal angle are shown in the top panels of Figure \ref{fig:HI_prop}. Purple points indicate systems where we were able to constrain the metallicity, while orange points indicate those where we were only able to calculate metallicity upper limits. The bottom panels show the distributions of ${\NHI}$ (d), impact parameter (e) and azimuthal angle (f). Similarly, the purple and orange histograms correspond to systems where we have metallicity measurements or upper limits, respectively. 

In Figure \ref{fig:HI_prop}(a), we show the metallicity as a function of the hydrogen column density. The distribution of data appears to show a slight anti--correlation with about 1~dex of scatter. The majority of the metallicity upper limits are found for ${\NHI} < 17.0$~cm$^{-2}$, which would be due to the difficulty in detecting metals in lower {\HI} column density systems with the signal--to--noise ratios in our sample. To test for a correlation between {\HI} column density and metallicity, we perform a Kendall-{$\tau$} rank correlation test on the sample, taking into account the metallicity upper limits. We do not detect a significant ($0.1\sigma$) anti-correlation between the metallicity and {\HI} column density. This is consistent with \citet{zahedy18}, who performed a similar analysis for {\HI} associated with luminous red galaxies. In contrast, \citet{prochaska17} reported a significant ($>4 \sigma$) anti-correlation between the {\HI} column density and metallicity for $L_{\ast}$ galaxies. These inconsistent results may be due to a different selection of the ionizing background since \citet{zahedy18} and our work uses HM05 while \citet{prochaska17} uses HM12. Interestingly, when we perform a Kendall-{$\tau$} rank correlation test between the {\HI} column density measurements and the metallicities derived using the HM12 ionizing background, we do find a significant anti-correlation ($3.3\sigma$). The harder HM12 background seems to produce an anti-correlation between {\HI} column density and CGM metallicity \citep[see Appendix \ref{app:appB},][]{chen17,wotta18,zahedy18}.

In Figure \ref{fig:HI_prop}(d) we show the distribution of ${\HI}$ column density, which has a range of $13.8~{\cms}< {\NHI}< 19.9~{\cms}$. The majority of the {\HI} detections are outside of the pLLS range of $16.2~{\cms} < {\NHI} < 17.2~{\cms}$ from \citet{lehner13, lehner18} and \citet{wotta16, wotta18}. The spread of {\HI} column densities is similar to the range observed in COS-Halos \citep[$ 14.7~{\cms} < {\NHI}< 19.9~{\cms}$;][]{prochaska17}. %We note that there is an overlap of 10 galaxy-absorber pairs between our sample and COS-Halos.

In Figure \ref{fig:HI_prop}(b), ${\NHI}$ is plotted as a function of impact parameter. A Kendall-{$\tau$} rank correlation test, which accounts for ${\NHI}$ upper limits, indicates that there is a significant ($3.4\sigma$) anti-correlation. This is consistent with other studies \citep[e.g.,][]{lanzetta95, tripp98, chen01a, rao11, borthakur15, curran16, prochaska17} who also find that the {\HI} column density decreases with increasing impact parameter. We also show the distribution of impact parameters in Figure \ref{fig:HI_prop}(e). The majority of the systems with metallicity measurements are located within $125$~kpc of the quasar sight-line, while many of the metallicity upper limits are located at larger impact parameters. Absorbers at higher impact parameters have lower {\HI} column density, and thus we are less likely to measure metal lines with our current spectra to determine the metal content, as shown in Figure \ref{fig:HI_prop}.

\begin{figure*}
	\centering
	\includegraphics[width=\linewidth]{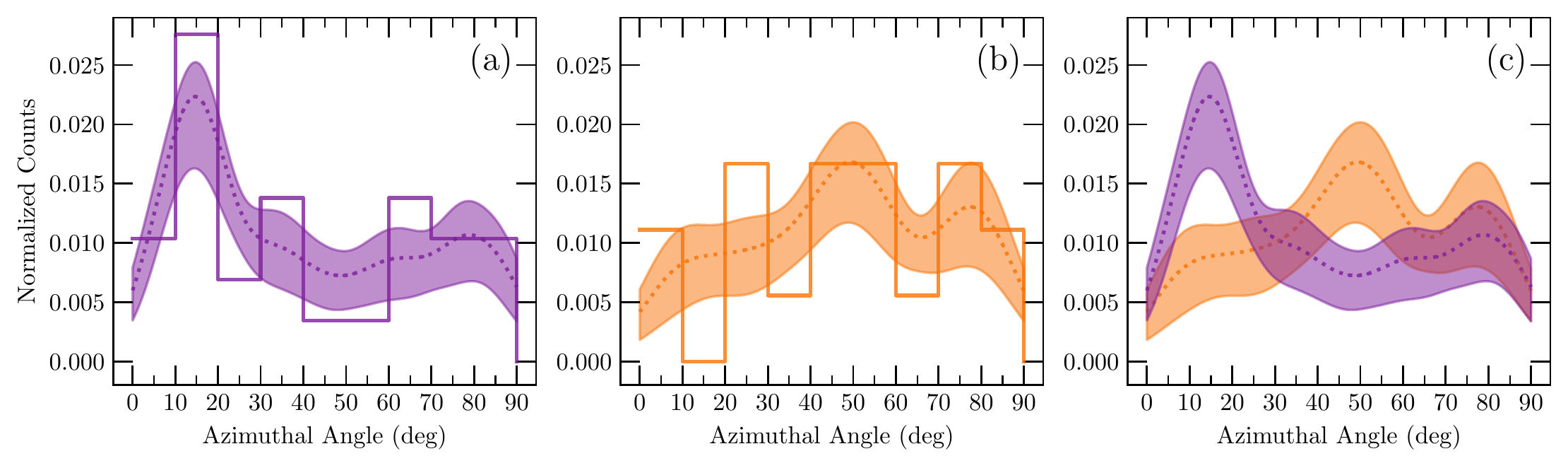}
	\caption{The azimuthal angle histograms for metallicity measurements (purple; (a)) and metallicity upper limits (orange; (b)). The dashed line represents the smoothed PDF of the data. The shaded regions indicate the $1\sigma$ errors calculated from 1000 bootstrap realizations. Each galaxy azimuthal angle was described by a Gaussian with mean and standard deviation defined by the azimuthal angle and its uncertainty, respectively. The PDF is then the area normalized sum of the individual Gaussians. The function was then smoothed using a convolution with a $15{\degree}$ FWHM Gaussian. Panel (c) compares the smoothed PDFs for azimuthal angles for metallicity measurements and upper limits.}
	\label{fig:AA_dist}
\end{figure*}

In Figure \ref{fig:HI_prop}(c), we show ${\NHI}$ as a function of the galaxy azimuthal angle. There is no relationship between the {\HI} column densities and the galaxy azimuthal angles, consistent with results from \citet{borthakur15}. Figure \ref{fig:HI_prop}(f) shows that all systems (purple+orange) are evenly distributed across all azimuthal angles. However, the distribution of absorbers with metallicity measurements (purple) appear to be clustered towards low and high azimuthal angles. \citet{bouche12} and \citet{kacprzak12} found that {\MgII} absorbers were more likely to be detected along the major and minor axes. If it is assumed that metallicity measurements (purple) are equivalent to the detection of metal absorption, such as {\MgII}, this result suggests that the azimuthal angle distribution may be bimodal.

Applying direct binning to the azimuthal angle distribution, as shown in Figure \ref{fig:HI_prop}, does not take into account the affect of uncertainties which are sometimes wider than the bin width \citep{kacprzak12}. 
Therefore, to test the presence of a bimodality in the azimuthal angle we divide the sample into a metallicity measurement subsample and a metallicity upper limit subsample. 
The azimuthal angle histograms of the metallicity measurements (purple) and metallicity upper limits (orange) are shown as solid lines in Figure \ref{fig:AA_dist}(a) and (b), respectively. 
We further create a probability density function (PDF) for each galaxy azimuthal angle using a Gaussian where the mean and standard deviation are defined by the azimuthal angle and its uncertainty, respectively. The PDF for the subsample is then calculated by summing the individual Gaussians and area normalizing. This function is then smoothed by a convolution with a Gaussian with a full width at half maximum (FWHM) of $15{\degree}$ and is shown as the dashed line. To calculate the uncertainty, we bootstrap the data values of both subsample with replacement 1000 times and calculate the PDF each time. The shaded regions then show the $1\sigma$ error range. This method results in galaxies with more precise azimuthal angles to have a higher weight in the distribution. 

In Figure \ref{fig:AA_dist}(a), we find that there is a higher probability of observing absorbers with metallicity measurements along the major axis, where the peak at $15{\degree}$ is consistent with {\MgII} \citep{kacprzak12} and {\OVI} \citep{kacprzak15b} absorbers. Assuming the toy model of the CGM is correct, we would expect to observed another peak at high azimuthal angles, along the minor axis. However, the smoothed PDF and the histogram of metallicity measurements indicate that there is not a significant number of absorbers at high azimuthal angles. Furthermore, a Hartigan's dip test did not find significant ($0.4\sigma$) evidence for a bimodal azimuthal angle distribution distribution \citep{hartigan85}.

In Figure \ref{fig:AA_dist}(b), we show the histogram and smoothed PDF of the absorption systems with metallicity upper limits. The azimuthal distribution of metallicity upper limits is reasonably uniform, although there may be a peak in metallicity upper limits at $\sim 50{\degree}$. In Figure \ref{fig:AA_dist}(c), we compare the smoothed PDFs of the metallicity measurements and upper limits. It is clear that metallicity measurements dominate over upper limits along the major axis.  These results do suggest that metals are more likely to be detected along the major axis. 

\begin{figure}
	\centering
	\includegraphics[width=0.48\textwidth]{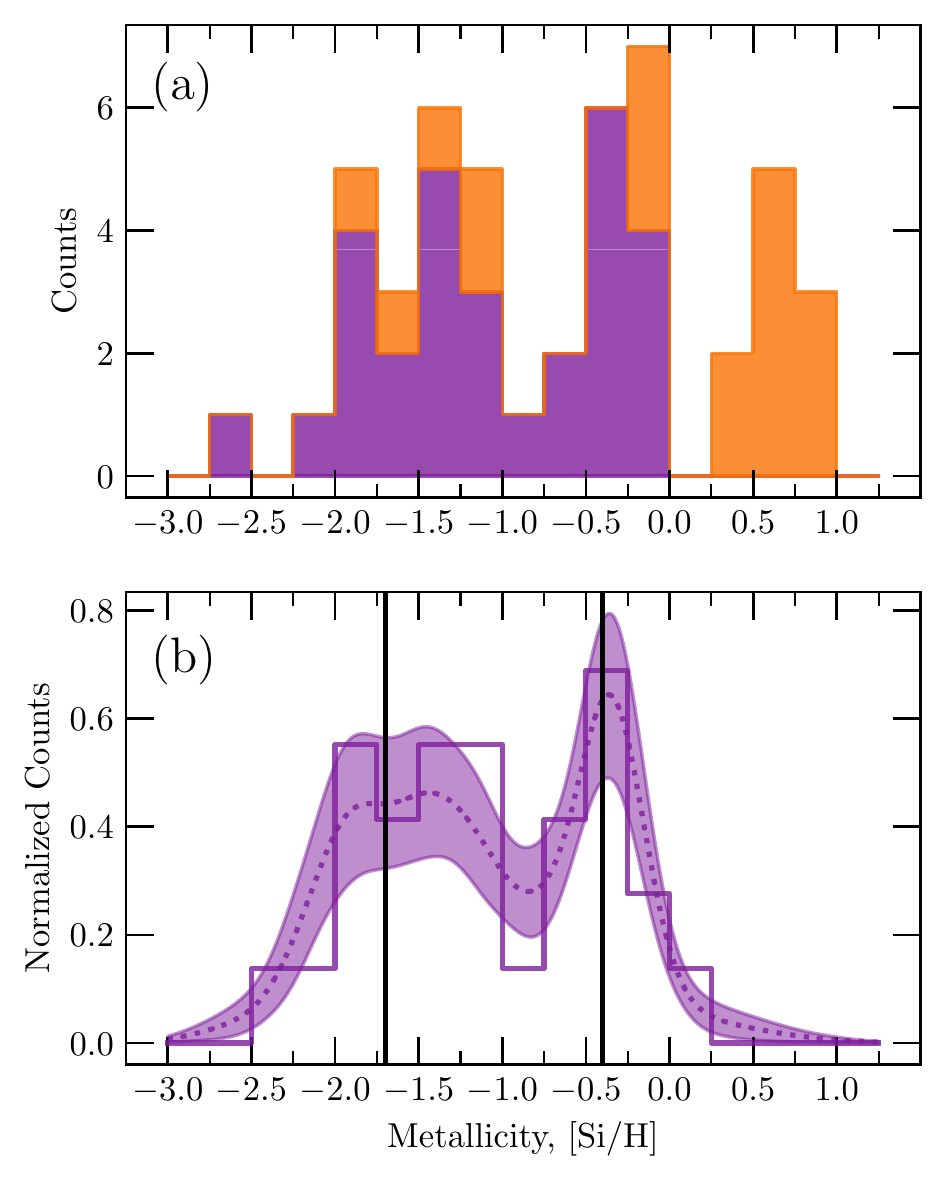}
	\caption{The distribution of metallicity for the total sample is presented in panel (a). The purple histogram represents the systems which have [Si/H] measurements, while the orange histogram shows systems with [Si/H] upper limits. In panel (b), the histogram of metallicity measurements is indicated by the solid purple line. The dashed line represents the PDF of the metallicity data, formed by making a normalized sum of Gaussians where each data point is represented by an individual Gaussian, smoothed by convolution with a Gaussian with a FWHM of $0.4$. The shaded region represents the $1\sigma$ uncertainty calculated from $1000$ bootstrap realizations. The vertical black lines are the locations of the peaks in the pLLS distribution found by \citet{wotta18}.}
	\label{fig:Z_dist}
\end{figure}

\begin{figure}
	\centering
	\includegraphics[width=0.48\textwidth]{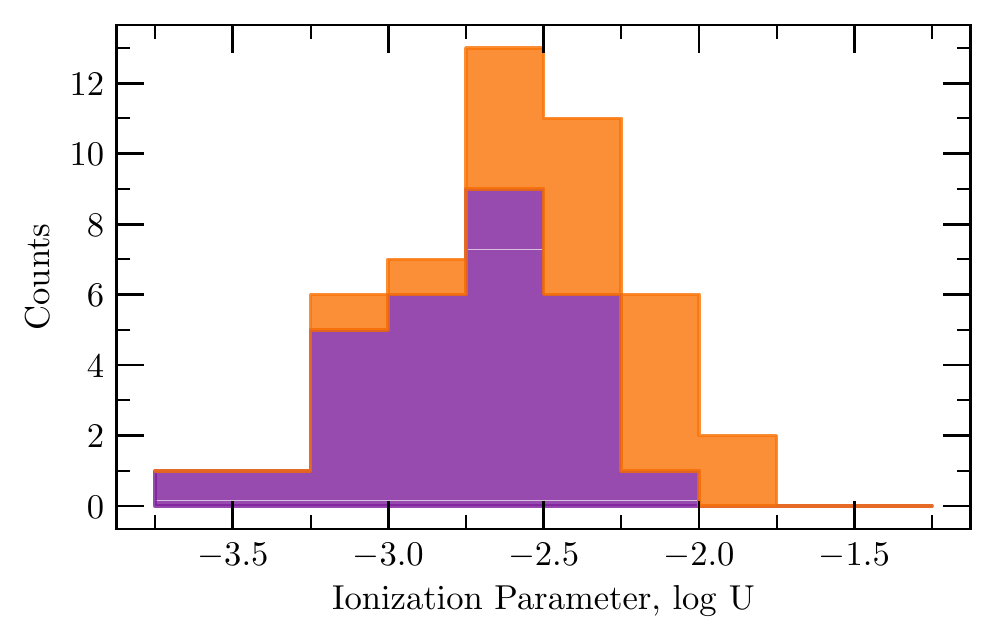}
	\caption{The ionization parameter distribution. The purple histogram shows the systems which have [Si/H] measurements, while the orange histogram shows absorbers with [Si/H] upper limits.}
	\label{fig:U_dist}
\end{figure}

\subsection{Metallicity Distributions}

The metallicity distribution, presented in Figure \ref{fig:Z_dist}(a), ranges from $-2.6 < [\text{Si}/\text{H}] < 0.8$ with a median metallicity of $[\text{Si}/\text{H}] = -1.3$. The purple and orange histograms represent metallicity measurements and upper limits, respectively.

In Figure \ref{fig:Z_dist}(b), we show the histogram of the metallicity measurements as the solid purple line. In order to fold in the uncertainty information and reduce the effect of binning, we also show the smoothed PDF of metallicity measurements as the purple dashed line. We applied the same method which was used to create the azimuthal angle smoothed PDF. Each metallicity measurement was assumed to be represented by a Gaussian, with a width described by the $68\%$ confidence interval reported by the MCMC analysis. The mean of the Gaussian was taken to be the midpoint of the confidence interval (shown on the posterior distribution plots in Figure \ref{fig:Q0122_0.2119_par} and in Appendix \ref{app:appA}). The PDF was formed by area normalizing the sum of the individual Gaussians. We then smoothed the PDF using a convolution with a Gaussian with a FWHM of $0.4$. The shaded regions represent the $1\sigma$ errors calculated using $1000$ bootstrap realizations, where the metallicity sample was randomly drawn with replacement and the smoothed PDF was calculated for each bootstrap step. 

The distribution of metallicity measurements appears bimodal. Therefore, we test for the presence of a bimodal distribution of metallicity measurements using a Hartigan dip test and find that a unimodal distribution cannot be ruled out ($0.4\sigma$). We note that this results is dependent on the sample size.

Interestingly, the bimodal peaks in pLLS from {\HI} absorption selected surveys are at [X/H]$\sim -1.7$ and $-0.4$ \citep{wotta18}, shown as black lines on Figure \ref{fig:Z_dist}(b). These values are located near to peaks in our metallicity distribution. However, we note that the bimodal distribution peaks from \citet{wotta18} were calculated using only pLLS ($16.2~{\cms} < {\NHI} < 17.2~{\cms}$), while the full sample was used in our calculations. 

%We test for the presence of a bimodal distribution of metallicity measurements, excluding the metallicity upper limits, by comparing the fit of uniform, unimodal and bimodal models. The unimodal fit a single normal distribution with a mean of $[\text{Si}/\text{H}] = -1.2$ and standard deviation of $\sigma = 0.8$. The bimodal distribution had peaks at $[\text{Si}/\text{H}] = -1.5$ and $[\text{Si}/\text{H}] = -0.4$. Random samples were then generated from a uniform distribution bound by the measured metallicity range, the unimodal model and the bimodal model. The model sample sizes were matched to that of the measured metallicity sample. We then performed a Kolmogorov-Smirnov test, comparing the distributions of each model sample to the measured metallicity sample. We cannot rule out the possibility of a uniform ($1.6\sigma$), unimodal ($0.04\sigma$) or bimodal ($0.9\sigma$) model from representing the data. 

We probe a wider metallicity range than \citet{prochaska17} who found a median metallicity of $[\text{Si}/\text{H}] = -0.51$ with a range of $-1.9 < [\text{Si}/\text{H}] < 1.00$ using the HM12 ionizing background. However, our metallicities calculated from the HM12 ionizing background span $-2.87 < [\text{Si}/\text{H}] < 1.43$. The differences between the metallicity ranges from our study and \citet{prochaska17} may then be attributed to a wider {\HI} column density range and redshift range.

Similarly, \citet{zahedy18} found a metallicity median of $[\text{M}/\text{H}] = -0.7$ with a range of $-2.47 < [\text{M}/\text{H}] < 0.75$ for luminous red galaxies. Unlike our study, \citet{zahedy18} investigates galaxies of higher mass, as well as reporting component-by-component metallicities in contrast to the integrated line-of-sight values presented in this work.

We also show the ionization parameter distribution in Figure \ref{fig:U_dist}. The ionization parameters found using the MCMC analysis have a range of $-3.51 < \log U < -1.79$ with a mean of $\langle \log U \rangle = -2.54$, compared to the COS-Halos sample which had a range of $-3.8 < \log U < -1.6$ and a mean of $\langle \log U \rangle = -2.8$ \citep{werk14}. The median width between the upper and lower $68$ percentiles for our sample is $\log U = 0.65$~dex. All systems with measured metallicities have ionization parameters of $\log U < -2.0$. The distribution appears to peak at $\log U \sim -2.75$, unlike the distributions found by \citet{lehner13} and \citet{wotta16} which peaked at $\log U\sim -3$ for a narrower {\HI} column density range of ($16~{\cms} < {\NHI} < 17.7~{\cms}$). They applied a Gaussian prior for $\log U$, while we use a flat prior, which could account for the slight difference.

\begin{figure*}[t]
	\centering
	\includegraphics[width=\linewidth]{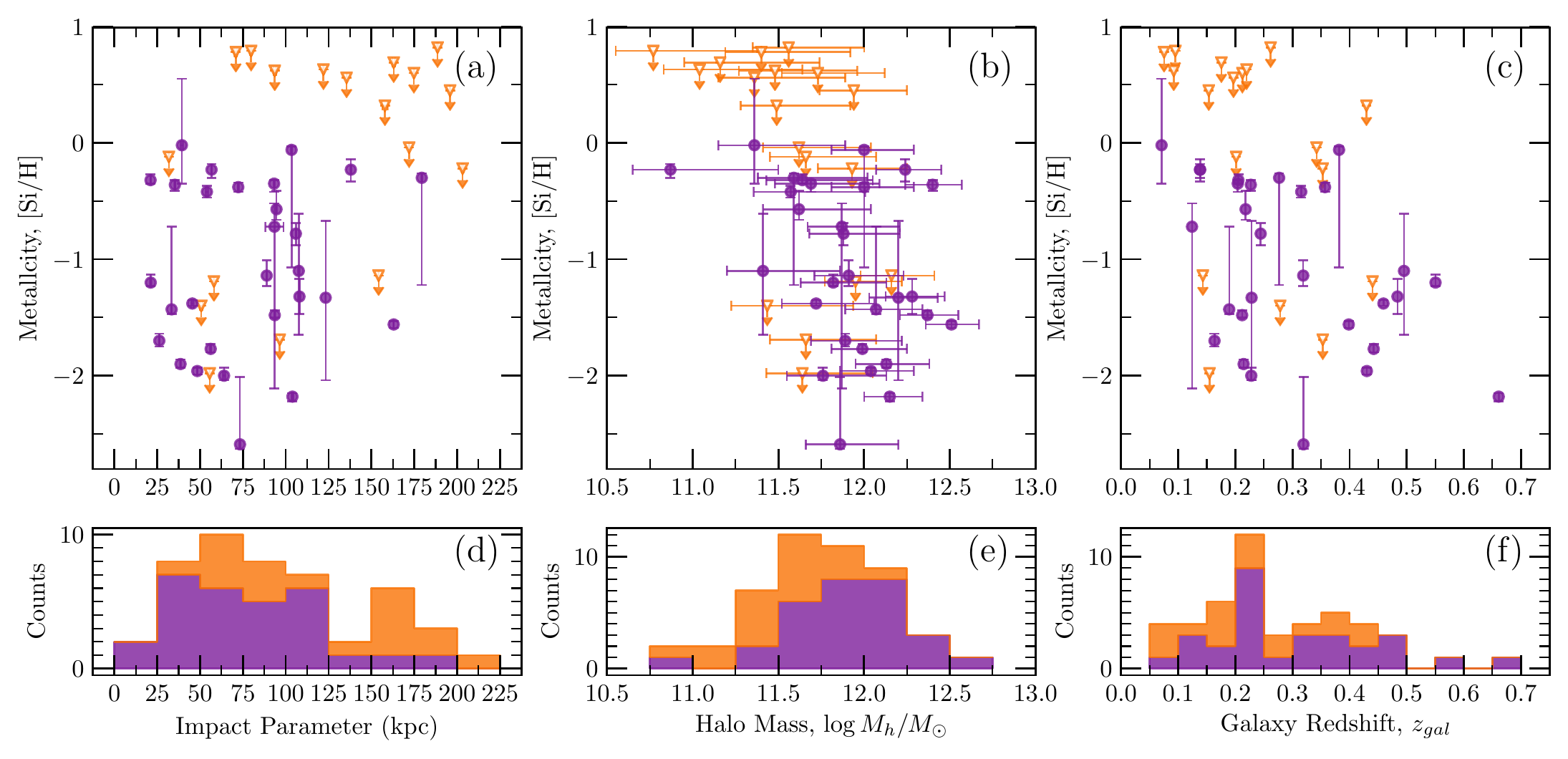}
	\caption{The relationships between the metallicity distribution and (a) impact parameter, (b) halo mass and (c) galaxy redshift. Purple circles represent systems which have constrained metallicity values while orange circles indicate upper limits on metallicity. Histograms of the (d) impact parameter, (e) halo mass and (f) galaxy redshift distributions are shown where the purple histogram represents metallicity measurements and the orange represents metallicity upper limits.}
	\label{fig:met_oth}
\end{figure*}

We also investigate the relationship between the metallicity and galaxy properties including the impact parameter, halo mass and redshift. We show the metallicity as a function of the impact parameter in Figure \ref{fig:met_oth}(a) and the distribution of impact parameters in Figure \ref{fig:met_oth}(d). The purple points and histogram correspond to metallicity measurements while the orange points and histogram correspond to the metallicity upper limits. Most of the detections are within $125$~kpc. We do not see a metallicity gradient, indicating that there is a full range of metallicities at all impact parameters. The proportion of systems which have metallicity upper limits increases at larger impact parameters, which is a result of these absorbers having low ${\NHI}$ and no detectable metals. This is consistent with the lack of metals at large impact parameters \citep[e.g.][]{chen10b, magiicat2}. Interestingly, it appears that for $D<75$~kpc, the metallicity distribution is bimodal, while at higher impact parameters, the distribution converges to mid-range metallicity values. 

In Figure \ref{fig:met_oth}(b) we show the metallicity as a function of halo mass, while in Figure \ref{fig:met_oth}(e), we show the halo mass distribution. The galaxies in our sample have a narrow halo mass range which is representative of $L_{\ast}$ galaxies. Over the narrow mass range of $10.77~M_{\odot} < \log M_{h}/M_{\odot} < 12.51~M_{\odot}$, we find that galaxies contain a full range of CGM metallicities, which could be expected for halos that have active star-formation driven outflows, gas accretion and gas recycling. The scatter in the metallicities shows that the CGM is quite complex. Interestingly, we find low metallicity CGM gas for high mass halos, ($\log M_h/M_{\odot} > 12~M_{\odot}$), for which cold--mode accretion is unlikely to occur \citep{fumagalli11, vandevoort11, vandevoort12, vandevoort+schaye12, faucher15, hafen17, halfen18}.

Finally, we investigate the influence of redshift on the metallicity of the CGM. In Figure \ref{fig:met_oth}(c) the metallicity is plotted as a function of redshift, while the redshift distribution is presented in \ref{fig:met_oth}(f). The relationship appears to be relatively flat. We performed a Kendall-{$\tau$} rank correlation test on the metallicity measurements and upper limits and find that there is no significant ($0.3\sigma$) anti-correlation between the metallicity and galaxy redshift. The range of {\HI} column densities where {\SiII} absorbers are measured is $14.6~{\cms} < {\NHI} < 19.9~{\cms}$ with an average of $\langle {\NHI} = 17.1~{\cms} \rangle$ We find that a majority of metallicity upper limits are found for low redshifts ($z_{gal}<0.3$) and low {\HI} column densities (${\NHI} <16.5$~cm$^{-2}$). 

\subsection{Metallicity and Orientation}
\begin{figure*}[t]
	\centering
	\includegraphics[width=\linewidth]{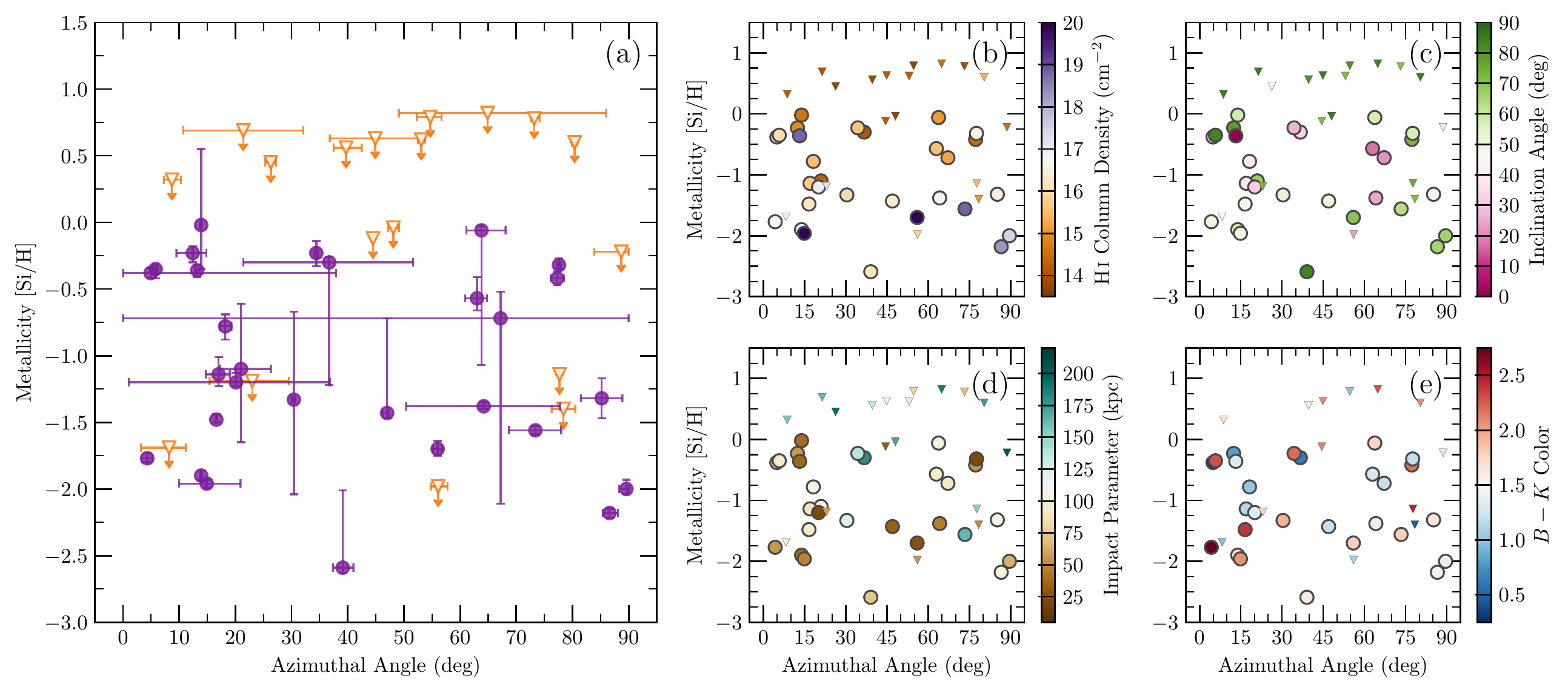}
	\caption{The relationship between metallicity and the azimuthal angle of the associated galaxies. Note that the major and minor axes correspond to azimuthal angles of $0{\degree}$ and $90{\degree}$, respectively. Similarly face-on and edge-on galaxies have inclination angles of $0{\degree}$ and $90{\degree}$, respectively. In panel (a) metallicity measurements are purple filled circles while the upper limits are orange open triangles. In the rest of the panels, the points are colored by the (b) {\HI} column density, (c) inclination angle, (d) impact parameter and (e) $B-K$ color while the metallicity measurements are circles and the upper limits are triangles.}  
	\label{fig:met_az}
\end{figure*}

Models of the CGM suggest that gas accretion along cosmic filaments should occur along the major axis and be low metallicity since it is expected that the gas has not yet been influenced by star formation \citep{fumagalli11, vandevoort+schaye12, shen13}. Similarly, outflowing gas located perpendicular to the galaxy plane is expected to be more metal--enriched since it is being ejected from the host galaxy by winds \citep{brook11,vandevoort+schaye12, peeples18}. In this section, we explore how the CGM metallicities behave as a function of the orientation of the quasar sightline with respect to the galaxy.

\subsubsection{Metallicity and Azimuthal Angle}
To probe the spatial distribution of metallicity in the CGM, we plot it as a function of azimuthal angle in Figure \ref{fig:met_az}(a) where metallicity measurements are shown as purple circles and upper limits are shown as orange triangles. There are two clusters of metallicity measurements showing the bimodal distribution of metallicity for azimuthal angle. The major and minor axes both appear to have absorption systems with metallicities which span from $-2.0 < \text{[Si/H]} < -0.1$ with similar distributions. To test this, we separated the sample into two bins using an azimuthal angle cut of $\Phi = 45{\degree}$ and performed a Kolmogorov-Smirnov test on the metallicity measurements to compare the two samples. We find no significant ($0.5\sigma$) difference between the major and minor axis metallicity distributions. The lack of a trend between the metallicity and azimuthal angle, combined with the results from the Kolmogorov-Smirnov test indicates that a wide range of metallicities exist for gas at all azimuthal angles, and that no significant difference exists between the metallicity of gas along the minor axis compared to the major axis. This is contrary to the simplistic model of CGM structure. 

The presence of scatter in the metallicity distribution as a function of azimuthal angle could due to a dependence on other galaxy or gas properties. To investigate this, we explore the relationship between metallicity and azimuthal angle while considering the {\HI} column density, inclination angle, impact parameter and $B-K$ galaxy color in Figures \ref{fig:met_az}(b)-(e).

In Figure \ref{fig:met_az}(b), we show the relationship between metallicity and azimuthal angle where the points are colored by {\HI} column density. The full range of {\HI} column densities reside along both the major and minor axes. However, no clear population of absorbers selected by {\HI} column density has a trend in the relationship between metallicity and azimuthal angle. Even for pLLS, where we expect the metallicity bimodality to be the strongest \citep{wotta18}, we do not see any dependence on azimuthal angle. There is a large scatter in the metallicity for both major and minor axes, which prevents a general conclusion about accretion occurring along the major axis and outflows along the minor axis. Instead, this suggests that the CGM is well mixed at all azimuthal angles, for all {\HI} column densities.

We also consider the effects of the inclination angle on the relationship between metallicity and the azimuthal angle. Assuming the simple model of the CGM, outflows and inflows would be more distinguishable if the galaxy was edge-on since the cross--section of the gas flows on the sky are minimized and do not overlap. As the inclination of the galaxy becomes more face-on, the cross--sections of the gas flows increase and multiple structures along the line--of--sight are present, making it more difficult to determine if the quasar sight-line probes the major or minor axis \citep{churchill15, peeples18, kacprzak19a}. Therefore, considering the inclination angle of a galaxy is important for understanding whether a trend is present between metallicity and azimuthal angle. In Figure \ref{fig:met_az}(c) we plot the metallicity as a function of azimuthal angle with points colored by galaxy inclination angle. The sample contains 23 edge-on galaxies, $i > 60{\degree}$, with both high and low azimuthal angles. However, edge-on galaxies with quasar sight--lines along the major and minor axes probe gas that spans the full metallicity range. While we may detect cold gas flows for individual galaxies, it is clear that the CGM is more complex than suggested by simple models. It is further interesting to note that, while the metallicity limits span all azimuthal angles, they tend to be found more often for edge-on galaxies.

We also investigate the effect of impact parameter on the relationship between metallicity and azimuthal angle. Previous studies found that the equivalent width of {\MgII} absorbers was strongest along the minor axis for impact parameters less than $50$~kpc \citep{bordoloi11} and $100$~kpc \citep{lan18}, indicating that the azimuthal distribution of the CGM metallicity could be more bimodal at smaller radii. Therefore, absorption systems with lower impact parameters might have an orientation-dependent metallicity structure. In Figure \ref{fig:met_az}(d), we show the metallicity as a function of azimuthal angle with the points colored by impact parameter. Low impact parameter systems tend to exist along the major and minor axes as expected for our simple model. However, there is no clear indication of a relationship between the metallicity and azimuthal angle for low impact parameters.

\citet{kacprzak12} also found a bimodality in the {\MgII} absorber azimuthal angle distribution, which was driven by blue star-forming galaxies. This result suggests that the simple model may be most valid in blue, star-forming galaxies. In Figure \ref{fig:met_az}(e) we show the metallicity as a function of azimuthal angle colored by $B-K$ color. While blue star-forming galaxies ($B-K<1.5$) have quasar sight--lines probing all azimuthal angles, the projected major and minor axes exhibit similar metallicity distributions. We even find unexpected low metallicity systems along the minor axis where we expect outflows to be dominated by metal--enrichment. Thus, we do not find that the blue galaxies have a relationship between the CGM metallicity and azimuthal angle of the galaxy. 

\begin{figure*}[t]
	\centering
	\includegraphics[width=\linewidth]{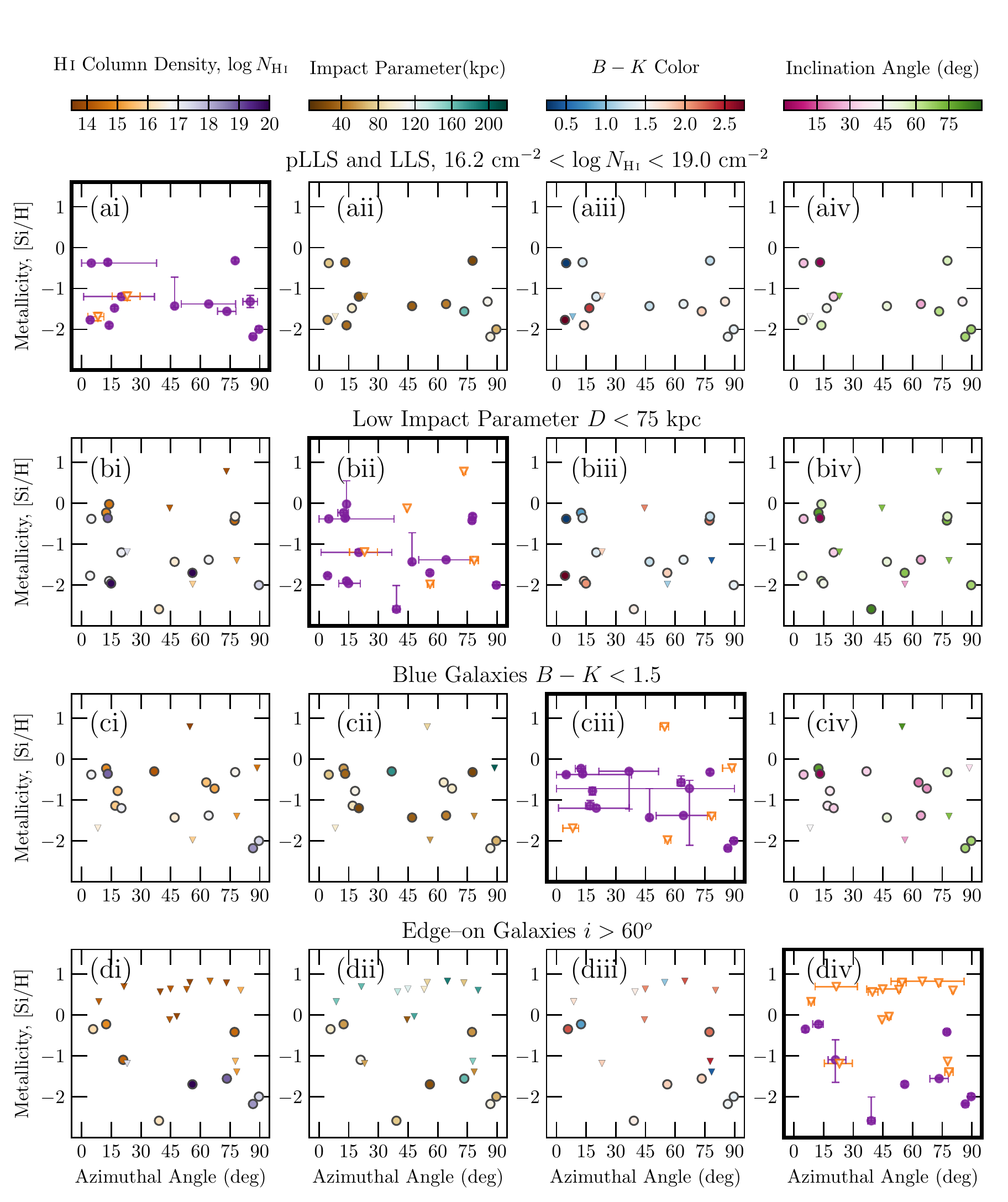}
	\caption{The relationship between metallicity and azimuthal angle is shown for different cuts of {\HI} column density, impact parameter, galaxy color, and inclination. Note that the major and minor axes correspond to azimuthal angles of $0{\degree}$ and $90{\degree}$, respectively. Similarly face-on and edge-on galaxies have inclination angles of $0{\degree}$ and $90{\degree}$, respectively. The top row shows the subset of data made up of pLLS and LLS. The second row shows the sub--sample for small impact parameters, $D<75$~kpc. The third row is for the subset of blue, star--forming galaxies with $B-K<1.5$. The bottom row shows edge-on galaxies with $i>60^{\degree}$. The plots along the diagonal with the bold frame have points plotted such that metallicity measurements are purple circles while metallicity upper limits are orange triangles. The remaining panels are colored by {\HI} column density (first column), impact parameter (second column), galaxy color (third column) and inclination angle (fourth column). Note that for even the most optimized sub--samples, we do not see any trends between the metallicity and azimuthal angle.}
	\label{fig:subsamp}
\end{figure*}

\begin{figure}[t]
	\centering
	\includegraphics[width=\linewidth]{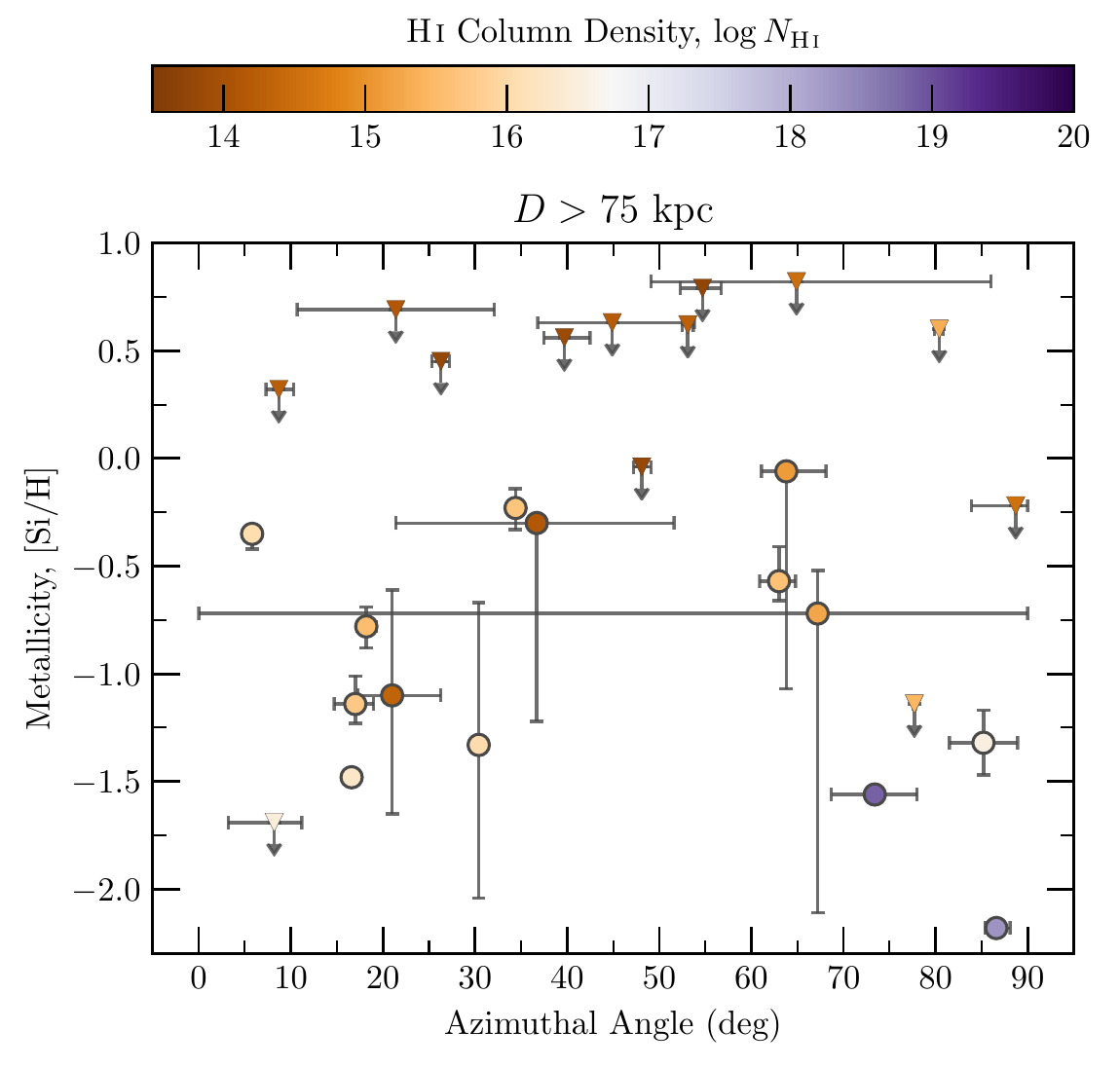}
	\caption{The relationship between metallicity and azimuthal angle is shown for high impact parameter ($D\geq 75$~kpc) galaxy-absorber pairs. Circles indicate metallicity measurements while triangles represent metallicity upper limits. The color of the points indicates the {\HI} column density of the system. }
	\label{fig:highD}
\end{figure}

To further investigate the impact of {\HI} column density, inclination angle, impact parameter and $B-K$ galaxy color on the relationship between metallicity and azimuthal angle, we isolate sub--samples of the absorber and galaxy properties described above, where a relationship is most likely to be observed. 

The metallicity bimodality was only found for pLLS and LLS ($16.2~{\cms} < {\NHI} < 19.0~{\cms}$) absorption systems by \citet{lehner13, wotta16, wotta18}, which is therefore the {\HI} column density range where a metallicity--azimuthal angle relationship would be expected. Also, edge--on galaxies ($i>60{\degree}$) should have a metallicity--azimuthal angle relationship due to the cross--section of inflows and outflows becoming minimized and ceasing to overlap. Additionally, simulations have found that lower impact parameter ($D<50-100$~kpc) absorption systems have low metallicity accretion along the major axis and metals present in the outflows \citep{danovich15}, while observations have found that the equivalent width of {\MgII} absorbers is strongest along the minor axis \citep{bordoloi11, lan18}. Finally, \citet{bordoloi11}, \citet{kacprzak12} and \citet{lan18} found that the bimodality in the azimuthal angle distribution of {\MgII} absorbers was driven by blue star--forming galaxies ($B-K < 1.5$). These four conditions then represent the optimal conditions for which we expect to observe a bimodal azimuthal angle distribution of CGM metallicities. 

In Figure \ref{fig:subsamp}, we investigate these sub--samples, with the sub--sample of pLLS and LLS metallicities as a function of azimuthal angle in the first row (a), the second row (b) shows the low impact parameter ($D<75$~kpc) sub--sample, the third row (c) shows the blue ($B-K<1.5$) galaxy sub--sample and the fourth row (d) shows the edge--on ($i>60{\degree}$) galaxy sub--sample. Along the main diagonal, we show the metallicity measurements as purple circles and the metallicity upper limits as orange triangles. All other plots show the metallicity--azimuthal angle plot for the row sub--sample where columns (i), (ii), (iii) and (iv) are colored by {\HI} column density, inclination angle, impact parameter and galaxy color, respectively. 

In Figure \ref{fig:subsamp}(ai), we find that although the distribution of metallicities is clustered about the major and minor axes for pLLS and LLS, there is no trend between metallicity and azimuthal angle measurements. We do find that the pLLS and LLS typically occur for lower impact parameters with only one absorption system with $D>120$~kpc. We detect a range of inclination angles and galaxy colors for pLLS and LLS in our sample. This supports our conclusion that there is no CGM metallicity--azimuthal angle relationship for pLLS and LLS, driven by cold--accretion and outflows for our galaxy sample. 

In Figure \ref{fig:subsamp}(b), we show the sub--sample for low impact parameter ($D<75$~kpc) galaxies, where we detect a range of metallicities along both the major and minor axes. Many of the absorption systems with metallicity measurements at low impact parameter have high {\HI} column densities, which is representative of the anti--correlation found in Figure \ref{fig:HI_prop}(b). This is consistent with studies which have found that the strength of {\HI} absorption decreases with increasing impact parameter \citep{lanzetta95, tripp98, chen01a, rao11, curran16, prochaska17}. There is a large scatter in the galaxy inclination angles and colors along both major and minor axes for low impact parameter absorption systems.

In Figure \ref{fig:subsamp}(c), we show the sub--sample of blue ($B-K < 1.5$) galaxies and find that there is no relationship between the metallicity and azimuthal angle for this sub--sample. The scatter and detection of metallicity appears to be greatest along the major ($\Phi <30{\degree}$) and minor ($\Phi>60{\degree}$) axes, with only two metallicity measurements detected at median azimuthal angles ($30{\degree}<\Phi<60{\degree}$), which is consistent with the full sample observations. All but one metallicity measurement occurs at low ($D<120$~kpc) impact parameters, consistent with the range of impact parameters in our sample. Additionally, all but one metallicity measurement of blue, edge--on galaxies occur at high azimuthal angles. However, the small sample and large scatter in metallicity measurements means than we are unable to associate the absorbers with the expected metallicities of outflows in this situation. 

We also investigate the relationship of metallicity with azimuthal angle for galaxy-absorber pairs with high impact parameters. We take the mid-point of the impact parameters suggested by \citet{bordoloi11} ($50$~kpc) and \citet{lan18} ($100$~kpc) for which they observe a bimodal distribution of {\MgII} equivalent width with azimuthal angle at low impact parameters. Since we do not see a relationship between metallicity and impact parameter below $75$~kpc, we investigate the high impact parameter galaxy-absorber pairs. In Figure \ref{fig:highD}, we show the metallicity as a function of azimuthal angle for absorber-galaxy pairs with $D>75$~kpc. Circles and triangles represent metallicity measurements and metallicity upper limits, respectively. The color of the points indicate the {\HI} column density of the absorber. Interestingly, the high column density absorbers have low metallicities and are located along the minor axis where outflows are expected. However, a Kendall-$\tau$ rank correlation test, which accounts for metallicity upper limits, finds no significant evidence for a relationship between the metallicity and azimuthal angle ($1.5\sigma)$.

We describe the effect of galaxy inclination on the metallicity--azimuthal angle relationship in the next section, where the sub--sample for edge--on galaxies is shown in Figure \ref{fig:subsamp}(d). 

\begin{figure*}[t]
	\centering
	\includegraphics[width=\linewidth]{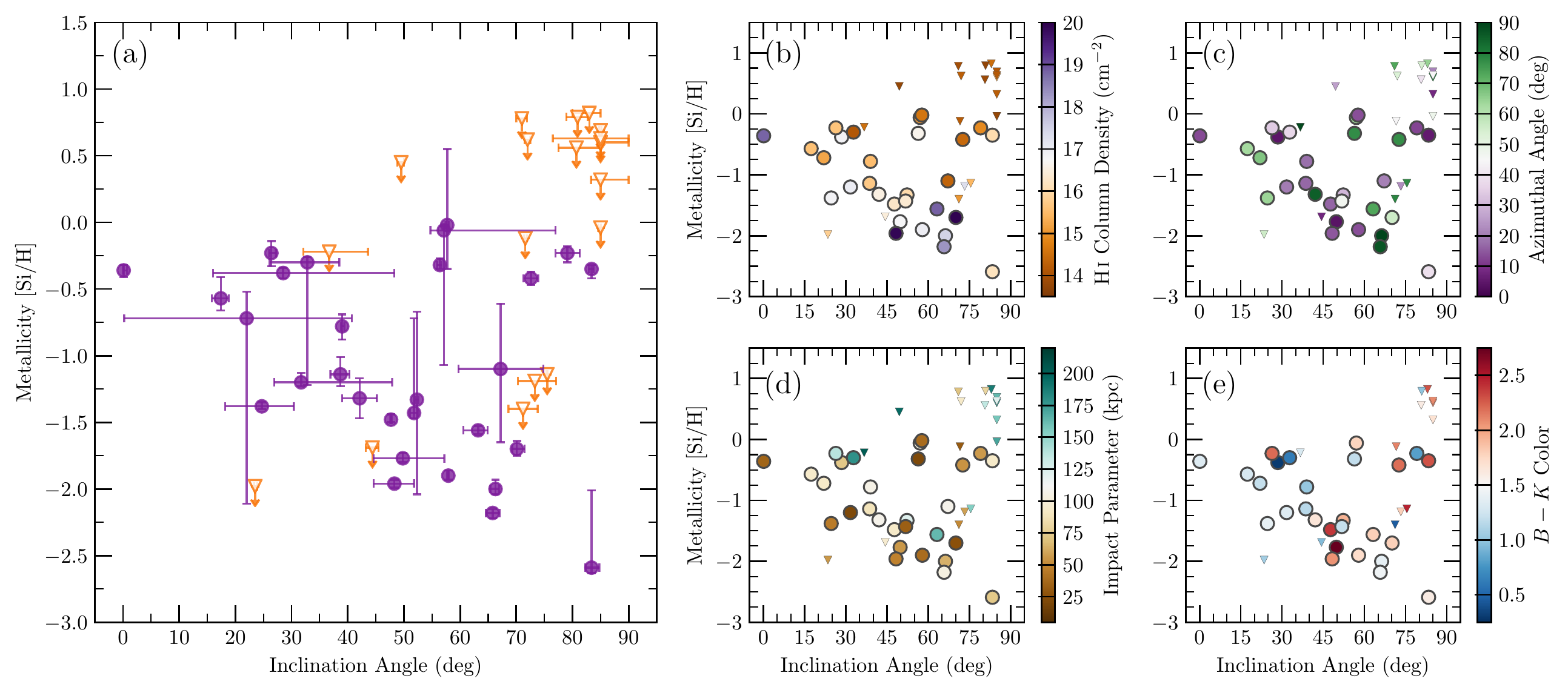}
	\caption{The relationship between metallicity and the inclination angle of the associated galaxies. Note that the major and minor axes correspond to azimuthal angles of $0{\degree}$ and $90{\degree}$, respectively. Similarly face-on and edge-on galaxies have inclination angles of $0{\degree}$ and $90{\degree}$, respectively. In panel (a) metallicity measurements are purple filled circles while the limits are orange open triangles. In the rest of the panels, the points are colored by the (b) {\HI} column density, (c) azimuthal angle, (d) impact parameter and (e) $B-K$ color while the metallicity measurements are circles and the upper limits are triangles.}
	\label{fig:met_inc}
\end{figure*}

\subsubsection{Metallicity and Inclination Angle}
It is possible that quasar sight--lines passing near low inclination galaxies could intersect both inflows and outflows regardless of the azimuthal angle. Thus, the value of the metallicity derived for these systems would represent an integrated metallicity with metal--poor absorption obscured by metal--rich CGM environments. However, the absorption near edge-on galaxies is less likely to pass though both outflows and inflows. Thus, a bimodal azimuthal distribution of the CGM metallicity would become more distinguishable for edge-on galaxies. 

To investigate this, we show the metallicity as a function of inclination angle in Figure \ref{fig:met_inc}(a), where metallicity measurements are purple circles and the upper limits are orange triangles. The data are reasonably flat with broad scatter. To determine if there is a difference between the metallicities detected for edge-on and face-on galaxies, we split the sample by $i=45{\degree}$. The median (mean) metallicities for $i<45$ and $i>45$, respectively are $-1.15\pm0.51$ ($-1.53\pm0.38$) and $-2.18\pm0.55$ ($-2.50\pm0.27$). There is no significant difference between the median ($1.0\sigma$) or mean ($1.5\sigma$) metallicities of high and low inclination angles. We also performed a Kolmogorov-Smirnov test on the metallicity measurements in the two samples. We find that the difference between the two samples is statistically insignificant ($2.5\sigma$), which indicates that the scatter in the CGM metallicity distribution obscures any potential trend between inclination angle and metallicity. 

We follow our investigation for the azimuthal angle by exploring the effect different physical parameters have on the relationship between metallicity and inclination angle. In Figure \ref{fig:met_inc}(b) we show the metallicity as a function of inclination angle colored by the {\HI} column density. Absorption systems with higher {\HI} column densities are found at lower metallicities, which is consistent with the suggestion of an anti-correlation between metallicity and {\HI} column density. However, the distribution of HI column densities and metallicities for edge-on and face-on galaxies is similar. Additionally, there is no significant metallicity bimodality for edge-on galaxies when we consider only LLS and pLLS.

In our simple model of the CGM, the azimuthal angle distribution of metallicities would become more apparent for edge-on galaxies. To investigate this, we show the metallicity as a function of inclination angle colored by azimuthal angle in Figure \ref{fig:met_inc}(c). For edge-on galaxies, $i>45{\degree}$, there is a large scatter in azimuthal angle for both high and low metallicities. This suggests that the distribution of CGM metallicities in azimuthal angle is complex and that metal--enriched gas is not only located along the minor axis, as seen in Figure \ref{fig:subsamp}(d). We find that the CGM metallicity has no apparent structure in azimuthal angle for edge-on galaxies, on average.

The effect of the impact parameter on the relationship between the metallicity and inclination angle is shown in Figure \ref{fig:met_inc}(d). Consistent with the results from Figure \ref{fig:subsamp}, there is no clear clustering with impact parameter for edge--on galaxies in the metallicity distribution. Even for absorbers at low impact parameters, there is no clear bimodality in CGM metallicity for edge-on galaxies.

Finally, we investigate the impact of galaxy color on the relationship between metallicity and inclination angle in Figure \ref{fig:met_inc}(e). \citet{kacprzak12b} found that the {\MgII} bimodality in azimuthal angle was strongest for blue star-forming galaxies. For $i>45{\degree}$, there is a wide range of galaxy colors at all metallicities. Thus, there is no evidence of a metallicity bimodality for blue, edge-on galaxies.

The edge--on galaxy sample in Figure \ref{fig:subsamp}(d) has metallicity measurements uniformly distributed across all azimuthal angles. However, the lowest metallicity measurements are found at median to high azimuthal angles, contrary to our expectations from a toy model. Interestingly, we find that the minor axis absorption systems for edge-on galaxies have higher {\HI} column densities than the major axis, which is similar to the results from \citet{bordoloi11} and \citet{lan18}. However, we note that the full sample has high {\HI} column densities present along both the major and minor axis. It is possible that multiple structures of both accreting gas and outflows contribute to higher {\HI} column densities for more inclined galaxies. The absorption for edge--on galaxies also primarily occurs for low impact parameter systems with only one metallicity measurement in the sub--sample having $D>120$~kpc. This is representative of the full sample distribution, shown in Figure \ref{fig:met_oth}(d). A range of galaxy colors are detected for edge--on galaxies.

\subsection{Ionization Parameter and Azimuthal Angle}

\begin{figure}[t]
	\centering
	\includegraphics[width=0.48\textwidth]{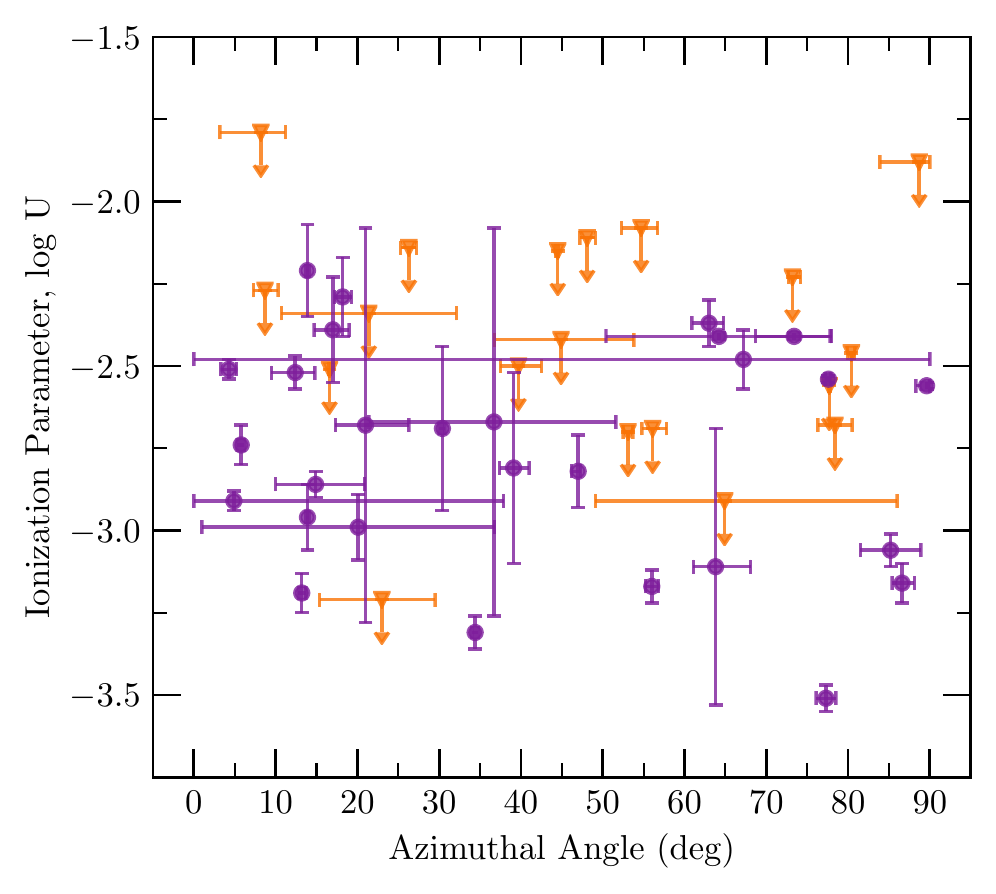}
	\caption{We show the ionization parameter, $\log U$, as a function of azimuthal angle. Note that the major and minor axes correspond to azimuthal angles of $0{\degree}$ and $90{\degree}$, respectively. Purple circles indicate measurements of the ionization parameter while orange triangles show upper limits. }
	\label{fig:usi}
\end{figure}

In the ionization modelling, it was assumed that the ionizing background was uniform. However, the ionization parameter could be higher along the minor axis where outflows occur, while accreting gas could shield the CGM along the major axis \citep{nielsenovi}. The ionization parameter as a function of azimuthal angle is presented in Figure \ref{fig:usi} where purple circles are  measurements of the ionization parameter and orange triangles are where we have upper limits on the ionization parameter. There is a spread of ionization parameter values along the major and minor axes, indicating that there is no orientation-dependence of the UV background or nearby ionizing sources. 

\section{Discussion}
\label{sec:discussion}

Our ``Multiphase Galaxy Halos'' Survey probed the relationship between the metallicity of the CGM and the orientation of $z < 0.7$, $L_{\ast}$ isolated galaxies. We investigated the metallicity of the CGM and examined its relationship CGM {\HI} column density and  galaxy orientation, impact parameter, color, halo mass and redshift.  A Hartigan dip test cannot rule out a unimodal distribution for the current sample, although it may be limited by sample size. If it is assumed that we observe a bimodal distribution, then the peaks are located at $[\text{Si}/\text{H}] = -0.4$ and $[\text{Si}/\text{H}] = -1.6$ compared to $[\text{X}/\text{H}] = -0.4$ and $[\text{X}/\text{H}] = -1.7$ from \citet{wotta18}. However, the peaks from \citet{wotta18} were calculated using pLLS while we used all {\HI} column density absorbers due to limited statistics. \citet{wotta18} also found that for ${\NHI} > 19~{\cms}$, the metallicity distribution becomes unimodal. Therefore, the higher {\HI} column density absorbers in our sample could prevent the detection of a bimodal distribution. However, restricting the sample to encompass only pLLS and LLS in Figure \ref{fig:subsamp} found that there was no significant difference between the metallicity distributions for major and minor axes. 

The COS-Halos galaxy-selected survey found a unimodal metallicity distribution, which has a similar {\HI} column density and metallicity range to ours and a mean of $[\text{Si}/\text{H}] = -0.51$ \citep{prochaska17}. The authors suggested that the lack of low metallicity pLLS systems in COS-Halos indicated that metal--poor pLLS absorbers are unlikely to be associated with $L_{\ast}$ galaxies. However, we have shown in Figure \ref{fig:HI_prop}(a) that low metallicity galaxy--absorber pairs have been found in our sample for galaxies with $10.77~M_{\odot}< \log M_h/M_{\odot}<12.51~M_{\odot} $.  Additionally, in Figure  \ref{fig:met_oth}(b), for fixed halo mass, we detect a range of metallicities consistent with that from the pLLS and LLS in \citet{wotta18} and \citet{prochaska17}. Therefore, the CGM of $L_{\ast}$ galaxies contains a range of metallicities and does not strongly contribute towards only the high metallicity branch in \citet{wotta18}. We note that we use the HM05 ionizing background in our metallicity calculations, while \citet{prochaska17} uses the HM12 background. However, we also observe a wide range of metallicities calculated using the HM12 background in our study ($-2.87 < \text{[Si/H]} < 1.43$) for the narrow halo mass range. This agrees with the conclusion that high metallicity CGM gas is not only associated with $L_{\ast}$ galaxies. However, it is possible that the CGM of sub-$L_{\ast}$ galaxies could contribute towards a lower metallicity peak. Future studies should investigate the relationship between the halo mass and the metallicity of the CGM. 

Interestingly, the zoom-in Feedback In Realistic Environments (FIRE) simulations suggest that for $z<1$ the metallicity distribution plateaus between $-1.3$ and $-0.5$ and is not bimodal, where inflows and outflows have the same distribution of metallicities \citep{hafen17, halfen18}. \citet{wotta18} suggests that the processes in the FIRE simulations which regulate the recycling of outflows and mixing of gas are too efficient since they do not produce a bimodality in pLLS and LLS. However, for the full sample of galaxy-absorber pairs with metallicity measurements, we are unable to rule out a unimodal metallicity distribution. 

We have further explored the relationship between the metallicity of the CGM and galaxy orientation. We find that there is no obvious relationship between the azimuthal angle of the galaxy and the CGM metallicity. While it appears that the range of CGM metallicities detected becomes wider for edge-on galaxies, the difference between edge-on and face-on CGM metallicities is insignificant. There is no significant difference between the CGM metallicities of the major and minor axes for edge-on galaxies, where a relationship between the metallicity and azimuthal angle would be easier to detect. Interestingly, \citet{peroux16} also found a large scatter ($2$~dex) in the galaxy--to--CGM metallicity ratio along the major axis. The presence of a range of CGM metallicities at both the major and minor axes indicates that the CGM is not easily described by a simple model with bipolar outflows and cold-mode planar accretion. 

Simulations of the CGM have indicated that outflows, which are metal--enriched, are more likely to be found along the minor axis \citep{brook11,stewart11,vandevoort+schaye12, shen13,peeples18, halfen18}. 
%Outflows can be generated from a variety of mechanisms including stellar feedback \citep{rubin14}, gas stripping from satellites \citep{lehner09}, recycled CGM gas \citep{oppenheimer10, ford13}, or intergalactic wind transfer \citep{angles16}. 
It is generally suggested that metal--rich outflows should have metallicities which are at least comparable to the associated galaxy's metallicity at the time of ejection \citep{fumagalli11, kimm11, shen13, wotta16}. The ISM metallicities for galaxies at fixed mass have relatively little scatter \citep[$<0.6$~dex;][]{erb06, mouhcine07, ellison09, scudder12, hughes13,  sanders15, kacprzak15b}. The large scatter ($2$~dex) in CGM gas metallicities detected in this sample for a fixed halo mass (Figure \ref{fig:met_oth}(b)) suggests there must be more complex phenomena occurring within the virial radii of galaxies compared to the ISM of galaxies.

In-falling gas in simulations is expected to align with the major axis of the galaxy and have low metallicities, which have only been enriched through Population III stars and dwarf galaxies. Simulations find that LLS inflows have metallicities of $\sim -1.5\pm0.3$ \citep{fumagalli11, vandevoort+schaye12,shen13}. This is consistent with the low metallicity branch found by \citet{lehner13} and \citet{wotta16, wotta18}. \citet{shen13} also found that in-falling gas contained a high proportion of recycled high metallicity gas from outflows. 

The relationship between the CGM metallicity and the galaxy orientation has been suggested as a method for determining if gas is accreting or outflowing \citep[e.g.,][]{kacprzak17} with metal--enriched outflows and low metallicity accreting filaments. Support for this relationship stems from the bimodality shown in the metallicity distribution of pLLS and LLS \citep{lehner13,wotta16,wotta18} and bimodalities in azimuthal angle detections of metals in the CGM with a concentration of metals towards the major and minor axes of galaxies \citep{bordoloi11,bordoloi14b, bouche12,kacprzak11morph, kacprzak12,kacprzak15, lan14, lan18}. Thus, it would be expected that the metallicity of the CGM should have some dependence on the orientation of the galaxy. We do not find a bimodal azimuthal angle distribution for our full sample (see Figure \ref{fig:AA_dist}. However, a significant peak found for systems with metallicity measurements was detected along the major axis. The strength of a peak in the azimuthal angle distribution along the minor axis would be highly correlated with the outflow opening angle since large opening angles could obscure a strong peak. Our sample also includes a range of galaxy inclination angles which would further dilute the signal. The presence of a significant peak along the major axis, despite the range of inclination angles, implies compact and narrow gas structures akin to filaments.

\begin{figure*}[t]
	\centering
	\includegraphics[width=\linewidth]{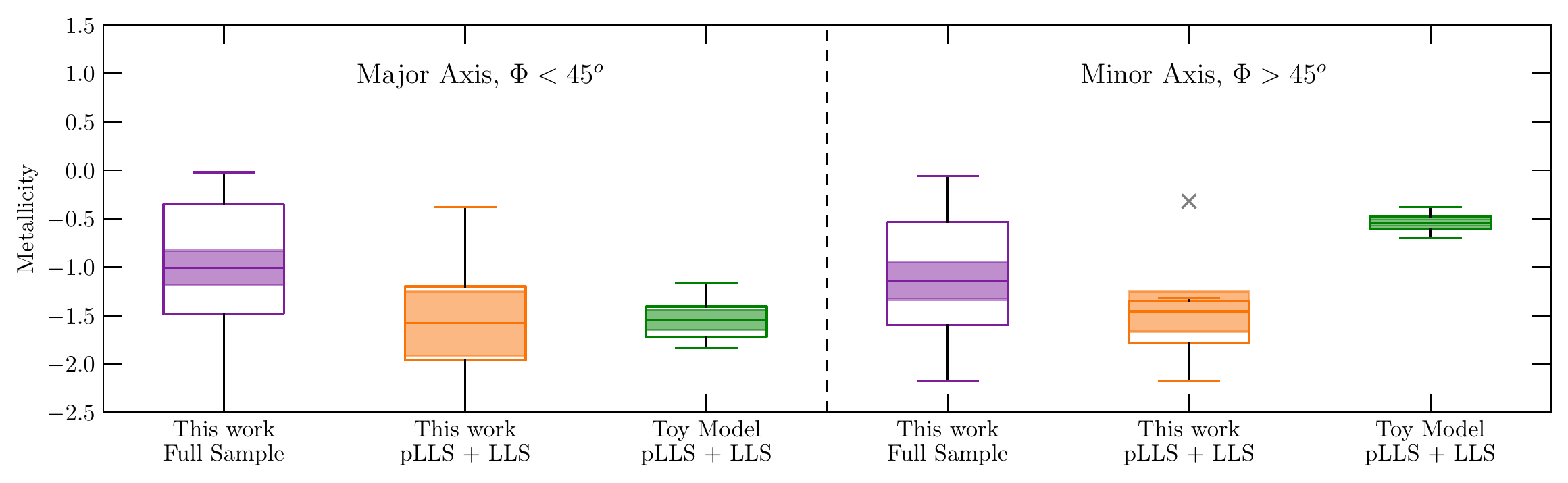}
	\caption{The purple box plots represent the full measured metallicity presented in this work, for major axis sight--lines on the left and minor axis sight--lines on the right. The orange box plots similarly represent the metallicity distribution along the major (left) and minor (right) axes for pLLS and LLS in our sample. For the green boxes, we assume that the pLLS high and low metallicity peaks found in \citet{wotta18} represent the metallicities expected from outflows and accretion, respectively. Therefore, these green boxes represent toy models of the CGM metallicities, where the low metallicity peak is associated with accretion along the major axis and the high metallicity peak is associated with outflows. Grey crosses indicate outliers. The box plots show the minimum, maximum, median and the $25\%$ and $75\%$ quartiles. The shaded regions show the error on the median.}
	\label{fig:comp_met}
\end{figure*}

To further explore the relationship between the CGM metallicity and azimuthal angle, we compare the metallicity measurements of the full sample, in purple, and the pLLS+LLS sample, in orange, on Figure \ref{fig:comp_met} for major and minor axis sight-lines. The metallicity distributions are shown as box plots with the error on the median represented by the shaded region. Along the major axis, the distributions of our full metallicity sample and the metallicities for pLLS+LLS are consistent. Similarly, the distributions of the CGM metallicity for our full and pLLS+LLS samples are consistent along the minor axis. However, it appears that we are less likely to detect high metallicity absorption systems where outflows are expected. There is no significant difference between the metallicity distributions along the major and minor axes for the full or limited {\HI} samples. 

Using the metallicity results from \citet{wotta18}, we construct predicted major and minor metallicity distributions based on our toy model. This allows us to determine if it is possible to statistically measure a bimodal relationship between the metallicity and azimuthal angle of our pLLS+LLS sample. We assumed that the low metallicity peak from \citet{wotta18} is associated with accreting gas along the major axis, while the high metallicity peak is associated with outflows along the minor axis. We randomly sample metallicities with replacement from each peak distribution, such that they have the same number of measurements for our pLLS+LLS sample. This process is bootstrapped 1000 times. We then plot the distributions in green on Figure \ref{fig:comp_met}. 

The means of the major and minor axis metallicity distributions for the toy model are significantly different ($3.2\sigma$). Assuming that the toy model is a true representation of the CGM, it is expected that we would detect a difference between the metallicity along the major and minor axes for our sample. The metallicity distributions for our full sample, pLLS+LLS sub--sample and the toy model are all consistent along the major axis, although the spreads of our metallicities are larger than the toy model. Along the minor axis, the toy model metallicity distribution appears higher than the measured metallicities from our pLLS+LLS sample, but the means are not significantly different ($2.9\sigma$). Both major and minor axis metallicity distributions in our survey (full and pLLS+LLS samples) are statistically consistent with the toy model metallicity predictions. Therefore, we are unable to use metallicity to distinguish between accretion and outflows at this time. This suggests that the spatial distribution of the CGM metallicity is not represented by a toy model with a clear separation between biconical outflows and accretion along the plane of the galaxy. Instead, the metallicity distributions are more consistent with recent simulation predictions from \citet{angles16} and \citet{ hafen17, halfen18} where the CGM is well mixed due to feedback processes and intergalactic wind transfer dominates over cold accretion. 

Low metallicity systems along the major axis of the galaxy are expected to indicate cold-mode accretion of cosmic filaments \citep[e.g.][]{stewart11,stewart13,stewart17,danovich12,danovich15}. However, the presence of low metallicity gas at all azimuthal angles could suggest that accreting filaments are not necessarily aligned with the major axis of the galaxy. Indeed, a study by \citet{kacprzak12} found evidence for cold-mode accretion along the minor axis of a $z=0.66$ galaxy. 

Simulations found that cold-mode accretion of gas will typically form filaments, where the alignment of the filaments is preferentially along the spin axis of the accreting stream rather than the galaxy axis. This would result in a misalignment of the low metallicity CGM accretion disk and the galactic plane. Additionally, \citet{angles16} studied accretion onto low redshift galaxies in the FIRE simulations and found that intergalactic transfer dominates accretion processes for $z<1$ $L_{\ast}$ galaxies. The gas filaments are then aligned to the axis between the galaxies, rather than the plane of the galaxy. Thus, accreting low metallicity gas may not have to be coplanar with the disk of the galaxy at low redshift. 

In addition, simulations find that in-falling gas aligned with the major axis is typically cold-mode accretion \citep{vandevoort11}. This mode of accretion is dominant for higher redshift ($z > 1$) halo masses of $M_h < 10^{12}$~$M_{\odot}$ and is more easily observed for pLLS \citep{fumagalli11, vandevoort+schaye12, vandevoort12, faucher15, hafen17, halfen18}. Therefore, we might not expect to see cold-mode accretion in our low redshift $z<1$, more massive galaxy halos $M_h \gtrsim 10^{12}$~$M_{\odot}$ where we have limited numbers of pLLS absorbers.

Interestingly, \citet{churchill13b} found that {\MgII} was present in the CGM of galaxies with halo masses of $M_h \gtrsim 10^{12}$~$M_{\odot}$. The presence of low ionization gas in massive halos where cold--mode accretion may have switched off could indicate efficient cooling of outflows or the presence of sub-halos. Either scenario points towards a more complex, clumpy structure of the CGM which is supported by the large scatter in the metallicity of absorbers for both major and minor axes. However, the presence of large metallicity scatter along the major axis does not preclude the possibility of cold accretion. However, the low metallicity gas would then have to be created by other mechanisms.

The presence of high metallicity systems for all azimuthal angles can be explained through feedback processes. At the peak of star formation ($z\approx 2$), winds generated by galactic outflows were stronger than they are in present--day galaxies. At this time, star formation was able to pollute the CGM with metals. Through feedback processes this gas could recycle through the CGM and back onto the galaxy within time scales of $<1$~Gyr, resulting in recycled metals existing at all orientations around the galaxy at the present time \citep{gnat07,oppenheimer13,oppenheimer16,angles16}. Similarly, observations of the CGM, including our sample, have found metal--enriched gas where accretion is expected \citep{peroux16,bouche16,muzahid16}. Therefore, we suggest that the structure of the CGM at low redshift is more mixed than expected from a toy model of the CGM.

The metallicities in this study have been calculated using integrated line-of-sight absorption profiles. However, the CGM is a multiphase environment where each quasar sight-line can potentially pass through different structural features. \citet{churchill15} and \citet{peeples18} found that for simulated absorption profiles, the gas occurs in a range of structures and physical locations. Along the line--of--sight, the CGM instead has metallicity structure across an absorption profile. The presence of any metals in the CGM along the sight--line will mask signatures of metal--poor gas at similar line--of--sight velocities even if the gas is at different physical locations. Thus, we are losing a tremendous amount of information by assuming a single metallicity per sight--line. Observationally, studies of individual absorbers have detected different metallicities within one absorber, suggesting that there can be a range of CGM properties along each sight--line \citep[e.g.][]{churchill12,crighton15,muzahid15, rosenwasser18, zahedy18}. Additionally, in the same absorber different metallicities have been derived for low and high ionization gas, suggesting that they trace different CGM structures \citep[e.g.][]{muzahid15}. In order to separate outflows from accretion, future studies should dissociate metal--enriched and metal--poor gas within an absorption system through multi-component and/or multi-phase modelling. However, {\HI} saturates at low column densities, making it difficult to determine how much hydrogen is associated with each metal line feature.

A general understanding of the relationship between the CGM metallicity and the galaxy orientation for surveys of multiple sight--lines towards many galaxies has not been possible thus far. While efforts have been made to select galaxy--absorber pairs where galaxy properties are similar, it is possible that the star--formation rate and merger histories, as well as prior AGN events of individual galaxies complicated the CGM \citep{heckman15, tumlinson17}. Additionally, the metallicity of the CGM is unlikely to exceed the galaxy metallicity. Therefore, studies, similar to that undertaken by \citet{peroux16}, need to take into account a range of galaxy properties, in addition to the orientation information. Instead, studies of the CGM using IFU observations could be extended to observe the spatial distribution of CGM metallicity throughout a specific halo \citep[e.g.,][]{lopez18}.  

A spatially distinct azimuthal distribution of CGM metallicity may be found at high redshifts where feedback processes have had less time to distribute high metallicity gas throughout the halo. At the peak of star formation, $z\approx2$, accretion and outflows should be enhanced \citep{vandevoort11}. Future studies may use the ability of \textit{James Webb Space Telescope} and the planned $30$m class telescopes to determine orientations of galaxies at the peak of star formation. Strong outflows and accretion could lead to a more distinct azimuthal distribution of CGM metallicity. Additionally, studies have started using IFU observations to map the structure of the CGM in emission at low redshifts \citep{martin16} as well as in absorption using large background lensed galaxies \citep{lopez18}. Future simulations and observations need to investigate how the velocity structure of an absorption profile relates to the combination of CGM geometry and metallicity for large samples.

Finally, we have neglected to consider the velocity structure of the absorption profiles in understanding the physical origin of the observed gas. However, an extensive body of work exists showing that kinematics are a powerful probe of both inflows and outflows regardless of whether the CGM metallicity was known \citep[e.g.][]{steidel02,tremonti07, weiner09, martin09,kacprzak10b, coil11, martin12, bouche12, gauthier12,rubin14,kacprzak14, bordoloi14a,magiicat5, schroetter16,ho17, martin19}. Future simulations and observations need to combine the information obtained from the absorption velocity structure, CGM geometry, and the CGM metallicities for large samples in order to have a more complete picture of baryon cycle processes.

%Larger equivalent widths along the minor axis could indicate either higher velocity spreads or larger column densities. Kinematic studies of {\MgII} absorption have shown that minor axis gas has large velocity spreads \citep{magiicat5} and the entrained gas flows away from galaxies at $100$--$1000$~{\kms} \citep[e.g.][]{tremonti07, weiner09, martin09, coil11, martin12, rubin14}. These absorption kinematics of minor axis systems are well-modeled by biconical outflows \citep{bouche12, gauthier12, kacprzak14, bordoloi14a, schroetter16,martin19}. At the same time, {\MgII} absorption velocity spreads are lower along the major axis \citep{magiicat5} and this gas exhibits  kinematics consistent with galaxy rotation, with a significant fraction of the gas having lower velocities \citep{steidel02, kacprzak10b, ho17,martin19}. The major axis gas kinematics are consistent with cold-mode gas accretion models \citep{stewart11, stewart17,danovich12, danovich15, steidel14, ho17}.

\section{Summary and Conclusions}
\label{sec:conclusions}
We used the ``Multiphase Galaxy Halos'' Survey to calculate the CGM metallicity of 47, $z<0.7$ isolated galaxy--absorber pairs. The metallicity of the CGM was then compared to other physical properties of the halo such as the  {\HI} column density and galaxy inclination, azimuthal angle, impact parameter, halo mass, color and redshift. 

Each ion in an absorption system was fit using the VPFIT software to calculate a column density. The column densities for the ions were then compared against a Cloudy ionization model using a MCMC analysis to calculate the metallicity. Galaxy inclinations and azimuthal angles were obtained by modelling the galaxies in \textit{HST} images with GIM2D. We then investigated the metallicity structure of the CGM. Our findings are:
\begin{enumerate}
    \item The {\HI} column density range of our sample is $13.8~\cms < {\NHI} < 19.9~\cms$ and decreases as the impact parameter increases, which is consistent with other CGM studies. 
    \item The absorption systems exhibit a uniform azimuthal distribution. However, the absorption systems with metallicity measurements are more likely to be detected along the major axis, but have no obvious peak in the PDF along the minor axis.
    \item The metallicity range for all absorption systems is $-2.6 < [\text{Si}/\text{H}] < 0.8$ with an average of $\langle [\text{Si}/\text{H}] \rangle = -1.3$. Although the metallicity distribution appears bimodal, we do not detect a significant bimodality ($0.4\sigma$). We also do not find a metallicity bimodality for pLLS and LLS due to the limited number of systems reported here.
    \item Metallicity is not anti-correlated with {\HI} column density. A Kendall-{$\tau$} rank correlation test, using metallicity measurements and upper limits, indicated that the significance of an anti-correlation is $0.1\sigma$. There is also no correlation between the CGM metallicity and the impact parameter.
    \item Despite observing a narrow range of halo mass ($10.77~M_{\odot} < \log M_{h}/M_{\odot} < 12.51~M_{\odot}$), we detect a wide range of metallicities and {\HI} column densities. Therefore, it is unlikely that pLLS and LLS, as well as low metallicity CGM gas, occur for low mass galaxies as suggested by \citet{prochaska17}.
    \item The metallicity is not related to the azimuthal angle. This relationship is also not influenced by sub-populations of {\HI} column density, inclination angle, impact parameter or $B-K$ color. The lack of correlation between the azimuthal angle and CGM metallicity in this low redshift study is not consistent with the simple model of the CGM. 
    \item Metallicity is not related to the galaxy inclination angle. Similarly to the azimuthal angle, the relationship is not influenced by sub-populations of the {\HI} column density, azimuthal angle, impact parameter or $B-K$ color. Even for edge-on galaxies, there is no indication of a relationship between the metallicity and azimuthal angle of the galaxy. We find no difference between the mean CGM metallicities of edge-on ($-1.53\pm0.38$) and face-on ($-2.50\pm0.27$) galaxies.
    \item The ionization parameter range for our sample is $-3.51 < \log U < -1.79$. The lack of azimuthal dependence of the ionization parameter shows that there is no ionization relation driven by outflows along the minor axis.
\end{enumerate}
In general, CGM metallicity does not appear to be dependent on the orientation of the associated galaxy. Thus, the structure of the CGM is not easily probed using the integrated line-of-sight metallicities. While we expected to find low metallicities in accreting gas along the major axis and high metallicity gas in outflows along the minor axis, we detect a range of metallicities for both infalling and outflowing gas. The presence of high metallicity gas at all azimuthal angles can be explained by feedback processes causing gas recycling at low redshifts, while low metallicity gas at all orientations suggests that cold streams of accreting filaments are not necessarily aligned with the galaxy plane at low redshifts or intergalactic transfer may dominate. In such a scenario, it may be possible to detect low metallicity gas at all azimuthal angles. The CGM at low redshift is complex, with a range of integrated line-of-sight metallicities at all azimuthal angle. We note though, that metallicity studies of individual galaxies may assist with the detection of accretion or outflows, where access to specific information about the galaxy star--formation history, past merger events and galaxy metallicity is available.
 
Our results are limited by the sample size. Larger surveys which combine CGM spectroscopy and galaxy imaging or IFU studies are required to understand the influence on properties such as the halo mass and redshift on the metallicity structure of the CGM. Furthermore, detailed modelling of absorption systems to determine the metallicity profile within each absorber may be able to identify structural features of the CGM. Additionally, future optical and infrared telescopes would allow galaxy orientations of higher redshift ($z = 2-3$) galaxies at the peak of star-formation to be determined, where stronger outflows and accretion signatures are expected to drive an azimuthal distribution of the CGM metallicity.

%%%%%%%%%%%%%%%%%%%%%%%%%%%%%%%%%%%%
\acknowledgments

We would like to thank John O'Meara for providing HIRES spectra and Nigel Mathes for reducing UVES spectra. We thank Nicolas Lehner for discussions on the UV ionizing background. We also thank Neil Crighton for the MCMC analysis software and Cloudy ionization training. Support for this research was provided by NASA through grants HST GO-13398 from the Space Telescope Science Institute, which is operated by the Association of Universities for Research in Astronomy, Inc., under NASA contract NAS5-26555. S.K.P acknowledges support through the Australian Government Research Training Program Scholarship. G.G.K, N.M.N, and M.T.M acknowledge the support of the Australian Research Council through the Discovery Project DP170103470. C.W.C. and J.C.C. acknowledges support by the National Science Foundation under Grant No. AST-1517816. Parts of this research were supported by the Australian Research Council Centre of Excellence for All Sky Astrophysics in 3 Dimensions (ASTRO 3D), through project number CE170100012. Some of the data presented herein were obtained at the W. M. Keck Observatory, which is operated as a scientific partnership among the California Institute of Technology, the University of California and the National Aeronautics and Space Administration. Observations were supported by Swinburne Keck programs 2014A\_W178E, 2014B\_W018E, 2015\_W018E and 2016A\_W056E. The Observatory was made possible by the generous financial support of the W. M. Keck Foundation.  The authors wish to recognize and acknowledge the very significant cultural role and reverence that the summit of Maunakea has always had within the indigenous Hawaiian community.  We are most fortunate to have the opportunity to conduct observations from this mountain.  Based on observations collected at the European Organisation for Astronomical Research in the Southern Hemisphere under ESO programmes listed in Table \ref{tab:obsgal}.

\bibliographystyle{apj}
\bibliography{refs}

\begin{appendices}
\section{HM05 and HM12 Comparison}
\label{app:appB}
\begin{figure}[h]
    \centering
    \includegraphics[width=\linewidth]{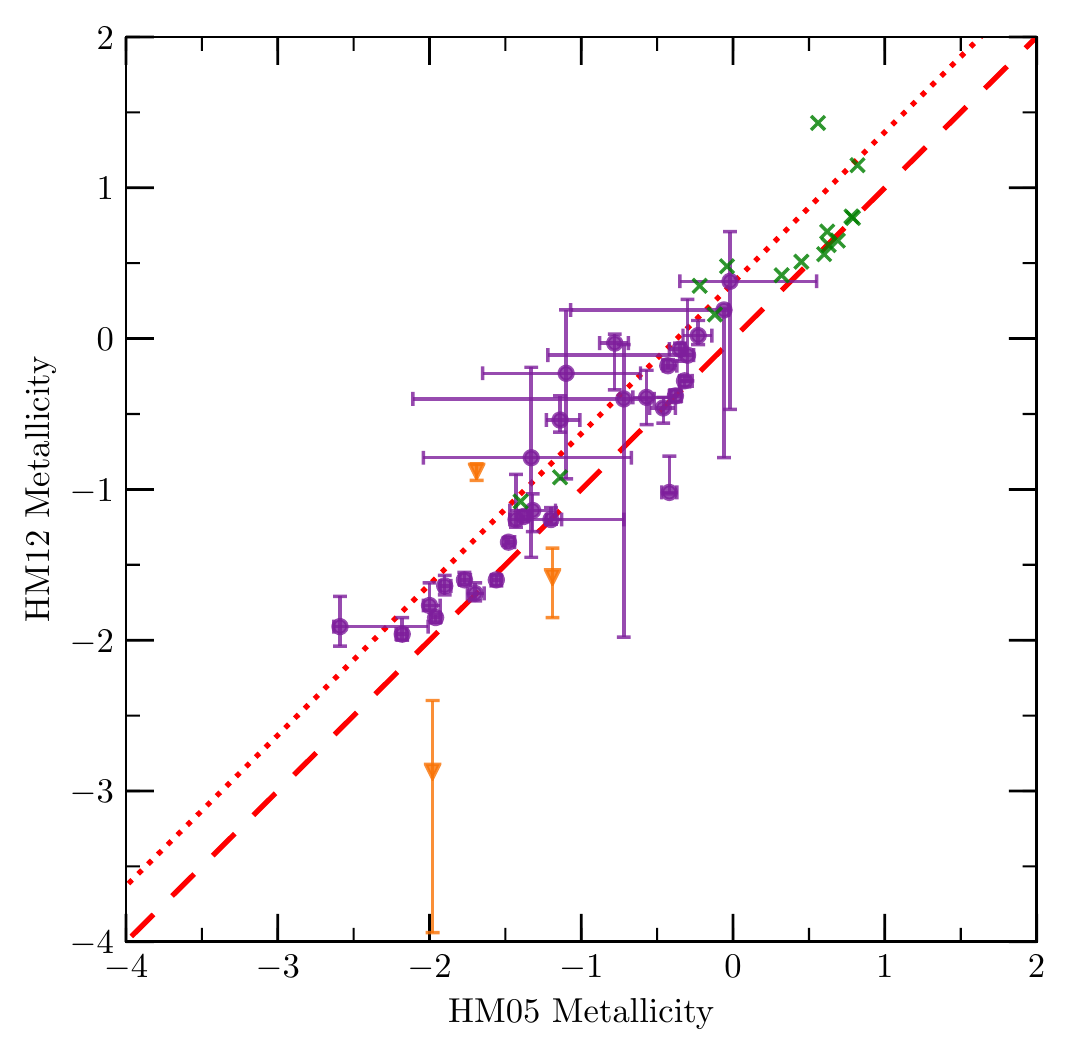}
    \caption{The metallicity calculated using the HM12 UV background as a function of the metallicity calculated using the HM05 UV ionizing background. The purple circles are where both metallicities are inferred by the models. The orange triangles are inferred metallicities for the HM12 UV background and upper limits for the metallicity from the HM05 UV background. The green crosses are metallicity upper limits for both ionizing backgrounds. The red dashed line shows where a 1-1 relationship lies, while the red dotted line shows the relationship derived by \citet{wotta18}.}
    \label{fig:met_comp}
\end{figure}

In Figure \ref{fig:met_comp}, we compare the metallicity measurements calculated using the HM05 and HM12 ionizing background in the Cloudy models. The purple circles indicate inferred metallicities for the HM05 and HM12 ionizing backgrounds, the orange triangles are modeled metallicities for HM12 background and a metallicity upper limit for the HM05 background. The green crosses are metallicity upper limits for both backgrounds. The red dashed line in the plot shows a 1-1 relationship, while the red dotted line shows the fit derived by \citet{wotta18}. The majority of the points are consistent with the lines. A Kolmogorov-Smirnov test on the two metallicity distributions finds no significant difference between the samples ($1.7\sigma$) despite most of the points lying above the $1-1$ relationship. However, there does seem to be some evidence that this metallicity difference depends on the {\HI} column density \citep{zahedy18,wotta18}. For consistency with the bimodal absorption selected surveys by \citet{lehner13,lehner18} and \citet{wotta16,wotta18}, we use the HM05 ionizing background.

\section{Absorption Profile Fits and Ionization Modelling Results}
\label{app:appA}
We present the results of the Voigt profile fits, the column densities used in the ionization modelling and the posterior distribution plots of the model parameters for each remaining galaxy--absorber pair. 

\begin{figure*}
	\centering
	\includegraphics[width=\linewidth]{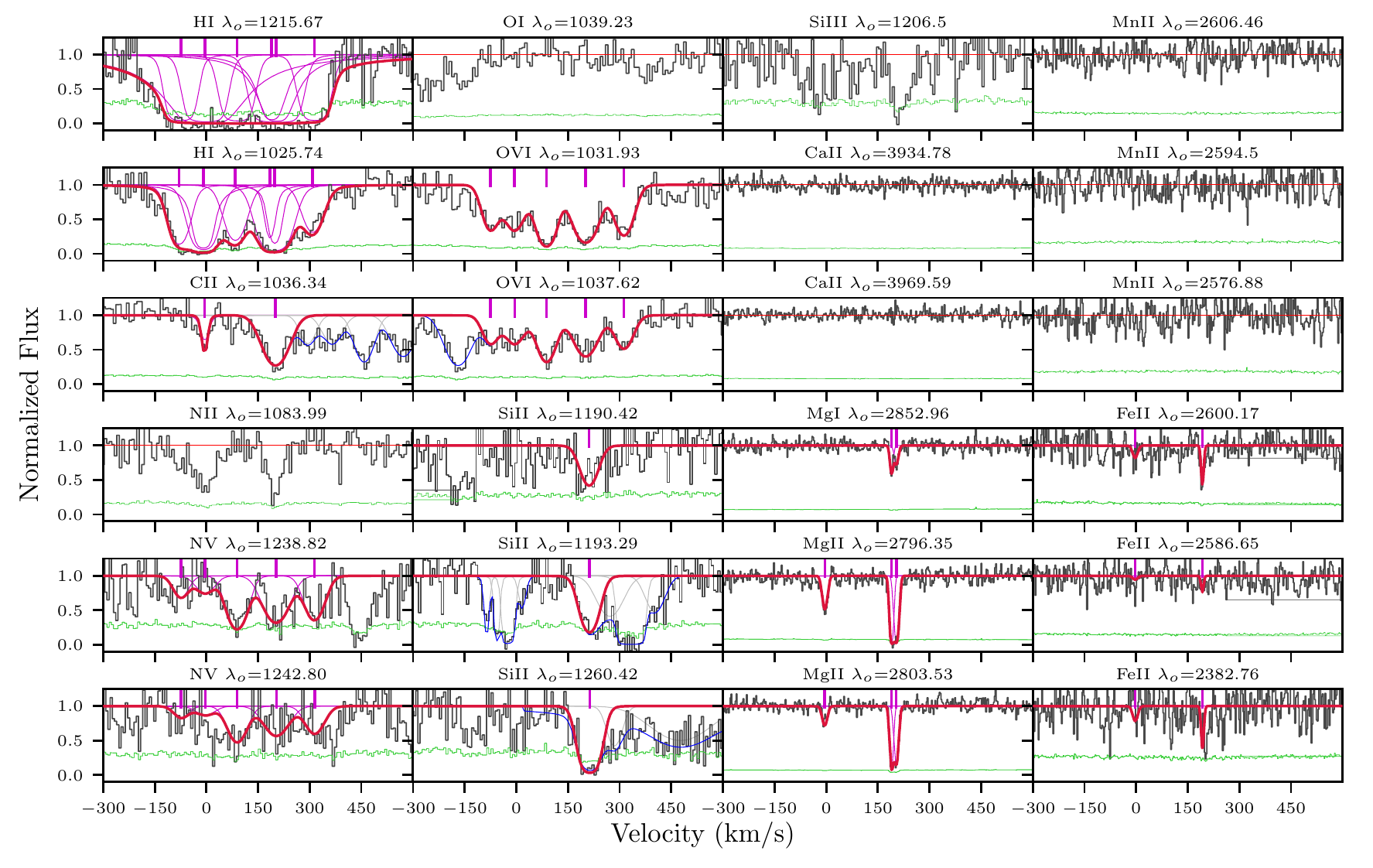}
	\caption{We show the {\HI}, {\CII}, {\OVI}, {\MgI}, {\MgII} and {\FeII} fits presented in \citet{muzahid15} for the system J$0125$, $z_{gal} = 0.398525$, using the same colors from Figure \ref{fig:Q0122_0.2119}. \citet{muzahid15} modelled the high and low ionization ions separately, matching {\HI} components to each low and high ionization components. We additionally model the {\SiII} and {\CII}. However the UV spectra is noisy compared to the optical spectra for {\MgII}. Hence, the same velocity structure was not able to be fitted to the data. The {\SiIII} column density upper limit was adopted from \citet{muzahid15}. The total {\OVI} fits from \citet{nielsenovi} are shown here for completeness, although they are not used in the ionization modelling. }
	\label{fig:J0125_3985}
\end{figure*}
\clearpage
\begin{deluxetable}{ccc}[hp]
	\tablecolumns{3}
	\tablewidth{\linewidth}
	\setlength{\tabcolsep}{0.06in}
	\tablecaption{J$0125$, $z_{gal} = 0.398525$ Column Densities\label{tab:J0125_3985}}
	\tablehead{
		\colhead{Ion}           	&
        \colhead{$\log N~({\cms})$}    &
		\colhead{$\log N$ Error~({\cms})}}
	\startdata
	{\HI   } &$ [18.85, 19.00]$  &   $\cdots$\\
{\CII  } &$ 14.59$  &   $0.04$\\
{\NV   } &$ 14.77$  &   $0.05$\\
{\OVI  } &$ 15.14$  &   $0.02$\\
{\SiII } &$ 13.75$  &   $0.08$\\
{\SiIII} &$ <12.80$  &   $\cdots$\\
{\MgI  } &$ 11.88$  &   $0.03$\\
{\MgII } &$ 13.48$  &   $0.02$\\
{\CaII } &$ <11.80$ &   $\cdots$\\
{\FeII } &$ 12.89$  &   $0.08$\\
{\MnII } &$ <12.65$ &   $\cdots$\\
{\NII  } &$ <13.86$ &   $\cdots$\\
{\OI   } &$ <14.67$ &   $\cdots$\\ [-10pt]
	\enddata
\end{deluxetable}
\begin{figure}[hp]
	\centering
	\includegraphics[width=\linewidth]{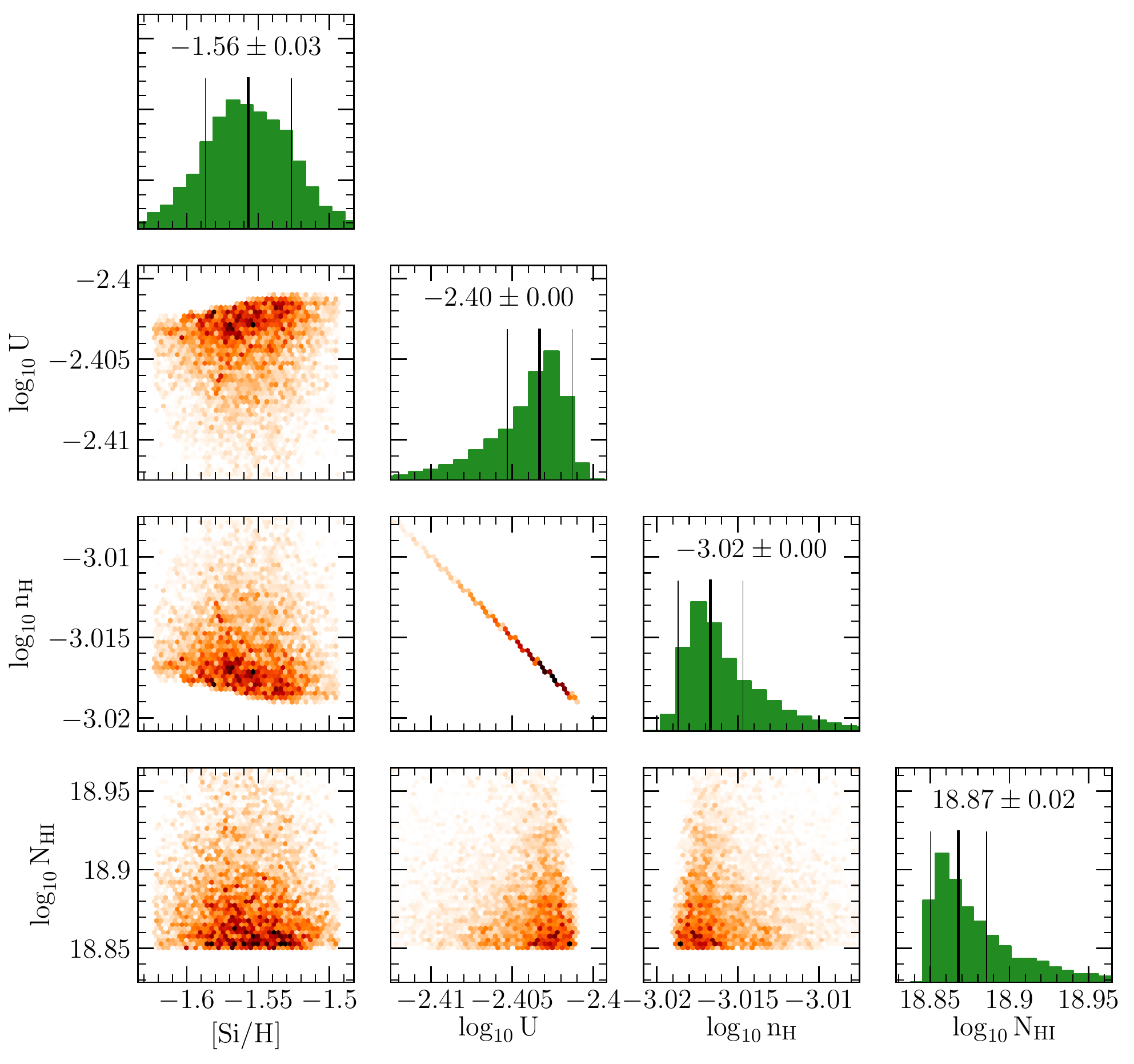}
	\caption{The posterior distribution profiles from the MCMC analysis of the Cloudy grids for J$0125$, $z_{gal} = 0.398525$, as for figure \ref{fig:Q0122_0.2119_par}.}
	\label{fig:J0125_3985_par}
\end{figure}
\newpage
\begin{deluxetable}{ccc}[hp]
	\tablecolumns{3}
	\tablewidth{\linewidth}
	\setlength{\tabcolsep}{0.06in}
	\tablecaption{J$0351$, $z_{gal} = 0.2617$ Column Densities\label{tab:J0351_0.2617}}
	\tablehead{
		\colhead{Ion}           	&
        \colhead{$\log N~({\cms})$}    &
		\colhead{$\log N$ Error~({\cms})}}
	\startdata
	{\HI}   & $14.51$   &$0.13$\\
{\CII}  & $<13.23$  &$\cdots$\\
{\CIII} & $<12.58$  &$\cdots$\\
{\NII}  & $<14.59$  &$\cdots$\\
{\NIII} & $<13.28$  &$\cdots$\\
{\NV}   & $<13.17$  &$\cdots$\\
{\OI}   & $<14.30$  &$\cdots$\\
{\SiII} & $<12.63$  &$\cdots$\\
{\SiIII}& $<12.06$  &$\cdots$\\
{\SiIV} & $<12.86$  &$\cdots$\\
{\CaII} & $<11.20$  &$\cdots$\\
{\MgI}  & $<11.36$  &$\cdots$\\
{\MgII} & $<11.88$  &$\cdots$\\[-5pt]

	\enddata
\end{deluxetable}
\begin{figure}[hp]
	\centering
	\includegraphics[width=\linewidth]{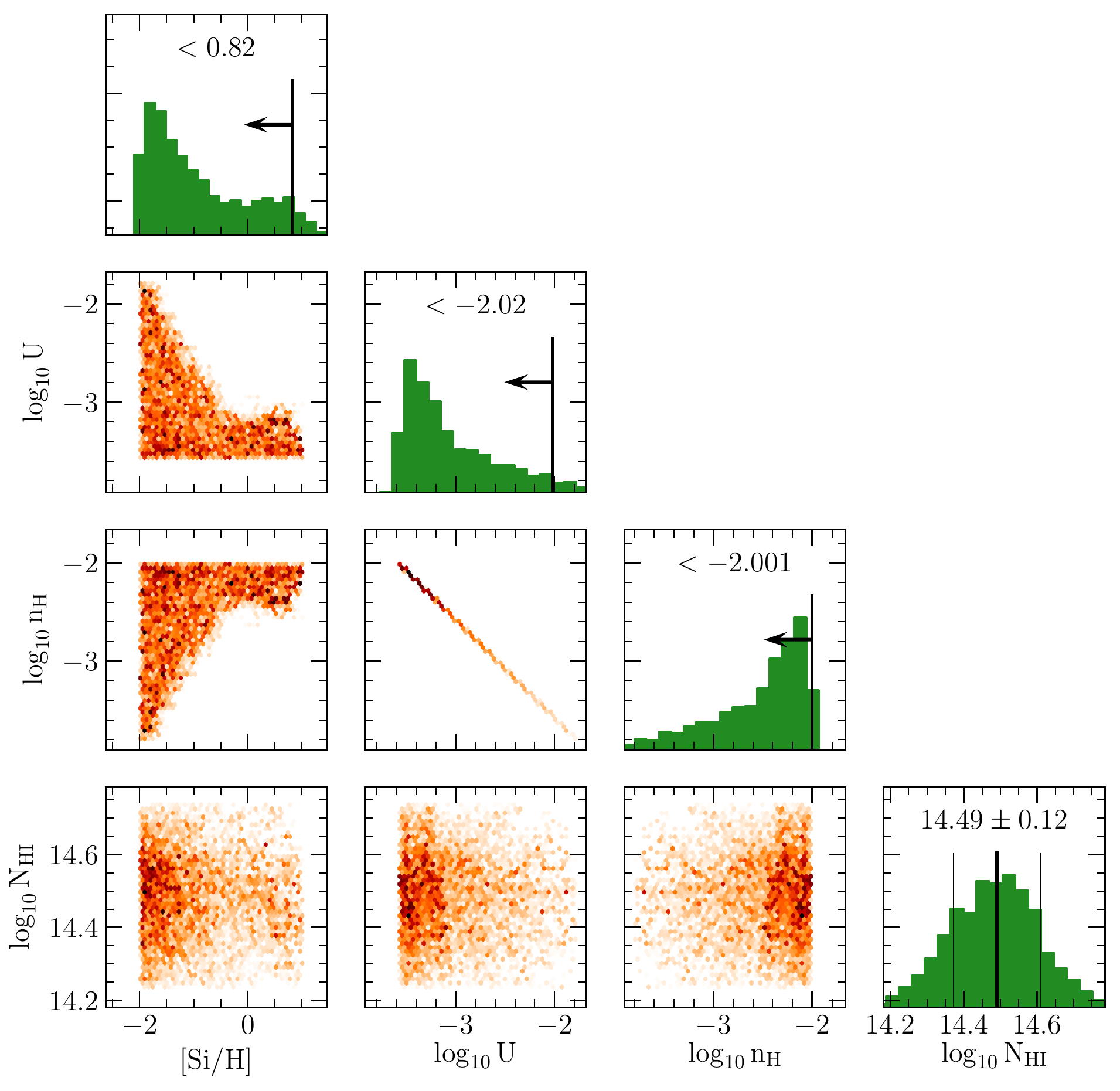}
	\caption{The posterior distribution profiles from the MCMC analysis of the Cloudy grids for J$0351$, $z_{gal} = 0.2617$, as for figure \ref{fig:Q0122_0.2119_par}.}
	\label{fig:J0351_2617_par}
\end{figure}
\begin{figure*}
	\centering
	\includegraphics[width=\linewidth]{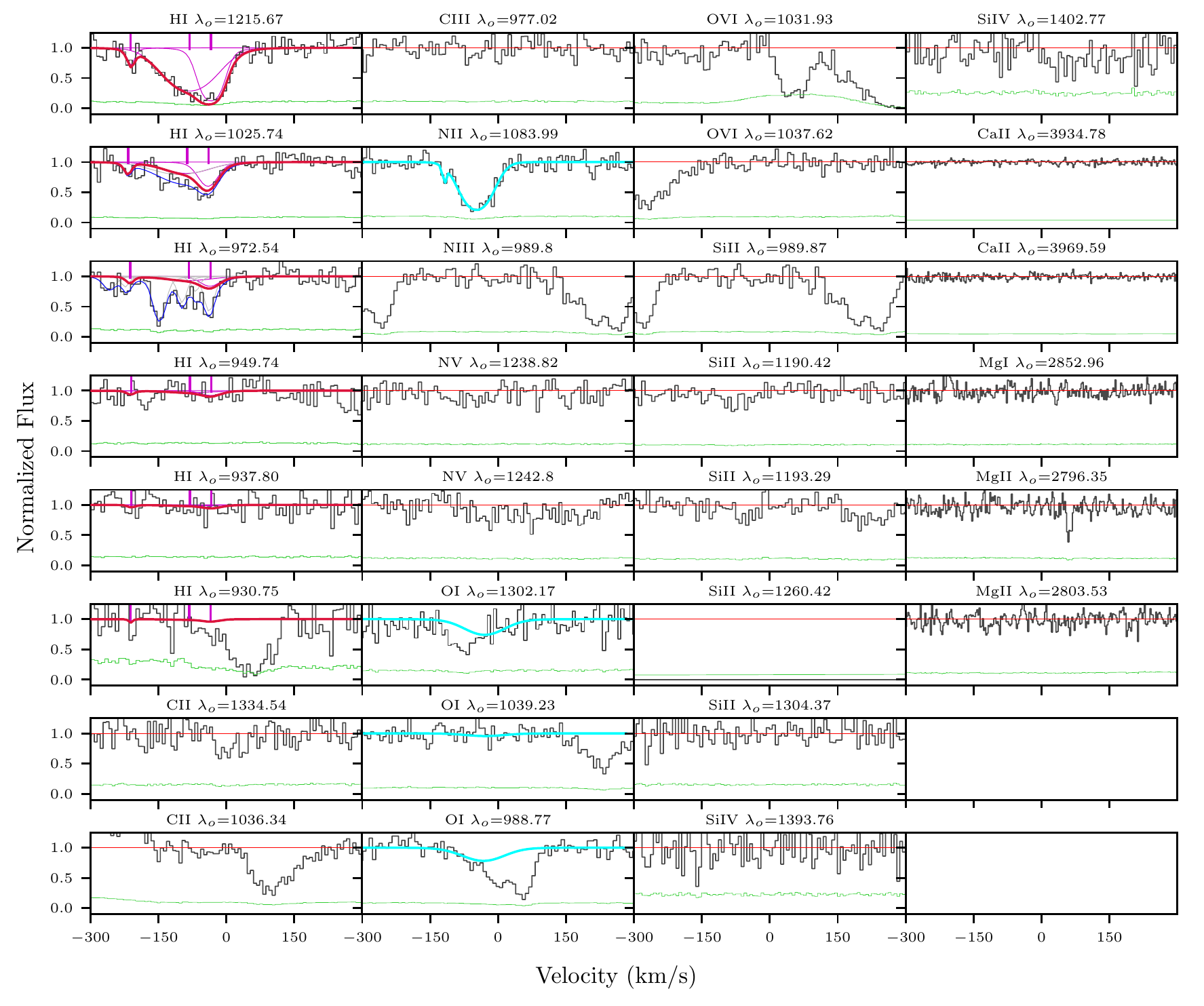}
	\caption{The fits for the system J$0351$, $z_{gal} = 0.2617$, as for figure \ref{fig:Q0122_0.2119}. The cyan fits for {NII} and {\OI} indicate where we have used the fit as a conservative upper limit on the column density, given that these lines are likely blended and we do not have supporting information to deblend them.}
	\label{fig:Q0349_0.2617}
\end{figure*}

\begin{figure*}[p]
	\centering
	\includegraphics[width=\linewidth]{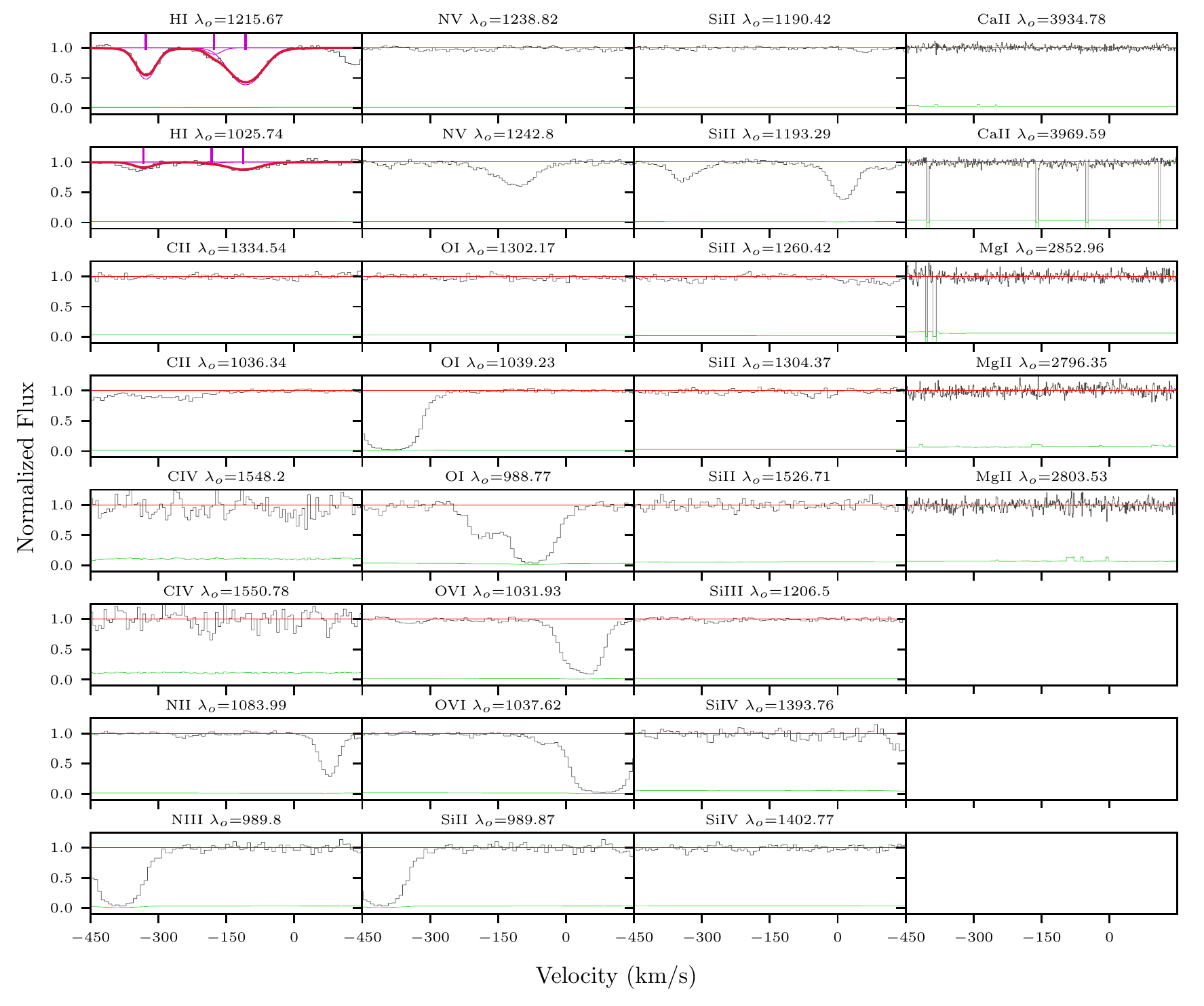}
	\caption{The fits for the system J$0407$, $z_{gal} = 0.1534$, as for figure \ref{fig:Q0122_0.2119}. Note that no metal lines are detected here.}
	\label{fig:J0407_1534}
\end{figure*}
\clearpage
\begin{deluxetable}{ccc}[hp]
	\tablecolumns{3}
	\tablewidth{\linewidth}
	\setlength{\tabcolsep}{0.06in}
	\tablecaption{J$0407$, $z_{gal} = 0.1534$ Column Densities\label{tab:J0407_1534}}
	\tablehead{
		\colhead{Ion}           	&
        \colhead{$\log N~({\cms})$}    &
		\colhead{$\log N$ Error~({\cms})}}
	\startdata
	{\HI}   & $13.79$   &$0.01$   \\
{\CII}  & $<12.44$   &$\cdots$\\
{\NII}  & $<12.43$   &$\cdots$\\
{\NIII} & $<12.85$   &$\cdots$\\
{\NV}   & $<12.17$   &$\cdots$\\
{\OI}   & $<13.00$   &$\cdots$\\
{\SiII} & $<11.49$   &$\cdots$\\
{\SiIII}& $<11.23$   &$\cdots$\\
{\SiIV} & $<12.17$   &$\cdots$\\
{\CaII} & $<11.03$   &$\cdots$\\
{\MgI}  & $<10.94$   &$\cdots$\\
{\MgII} & $<11.53$   &$\cdots$\\[-5pt]

	\enddata
\end{deluxetable}
\begin{figure}[hp]
	\centering
	\includegraphics[width=\linewidth]{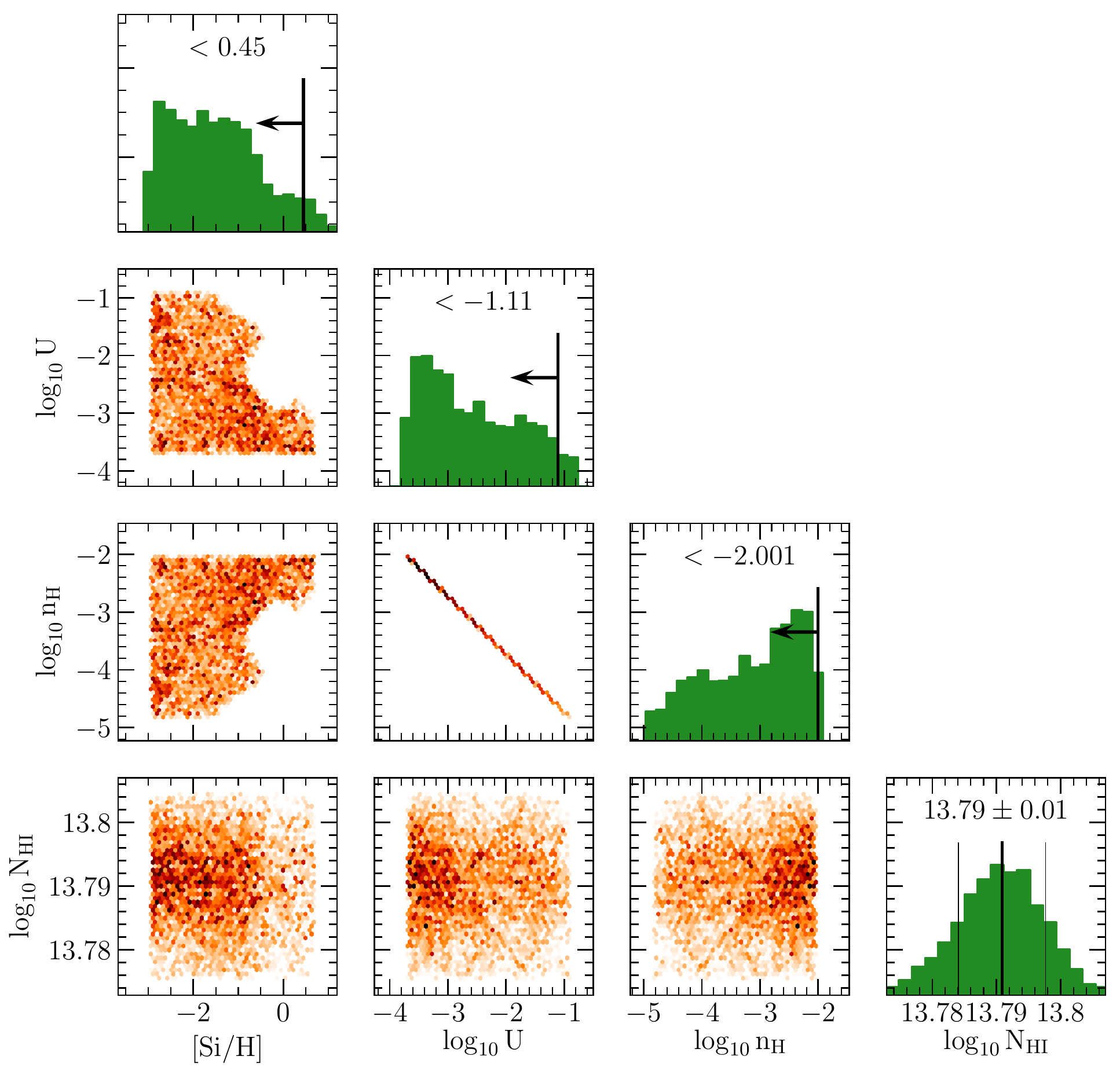}
	\caption{The posterior distribution profiles from the MCMC analysis of the Cloudy grids for J$0407$, $z_{gal} = 0.1534$, as for figure \ref{fig:Q0122_0.2119_par}.}
	\label{fig:J0407_1534_par}
\end{figure}
\newpage
\begin{deluxetable}{ccc}[hp]
	\tablecolumns{3}
	\tablewidth{\linewidth}
	\setlength{\tabcolsep}{0.06in}
	\tablecaption{J$0407$, $z_{gal} = 0.3422$ Column Densities\label{tab:J0407_3422}}
	\tablehead{
		\colhead{Ion}           	&
        \colhead{$\log N~({\cms})$}    &
		\colhead{$\log N$ Error~({\cms})}}
	\startdata
	{\HI}   & $13.78$   &$0.01$\\
{\CII}  & $<12.43$   &  $\cdots$\\
{\CIII} & $<11.64$   &  $\cdots$\\
{\NII}  & $<12.54$   &  $\cdots$\\
{\NIII} & $<12.51$   &  $\cdots$\\
{\NV}   & $<12.65$   &  $\cdots$\\
{\OI}   & $<13.24$   &  $\cdots$\\
{\SiII} & $<12.25$   &  $\cdots$\\
{\SiIII}& $<11.57$   &  $\cdots$\\
{\CaII} & $<10.65$   &  $\cdots$\\
{\MgI}  & $<10.65$   &  $\cdots$\\
{\MgII} & $<11.37$   &  $\cdots$\\
{\MnII} & $<12.77$   &  $\cdots$\\
{\FeII} & $<11.92$   &  $\cdots$\\[-5pt]
	\enddata
\end{deluxetable}
\begin{figure}[hp]
	\centering
	\includegraphics[width=\linewidth]{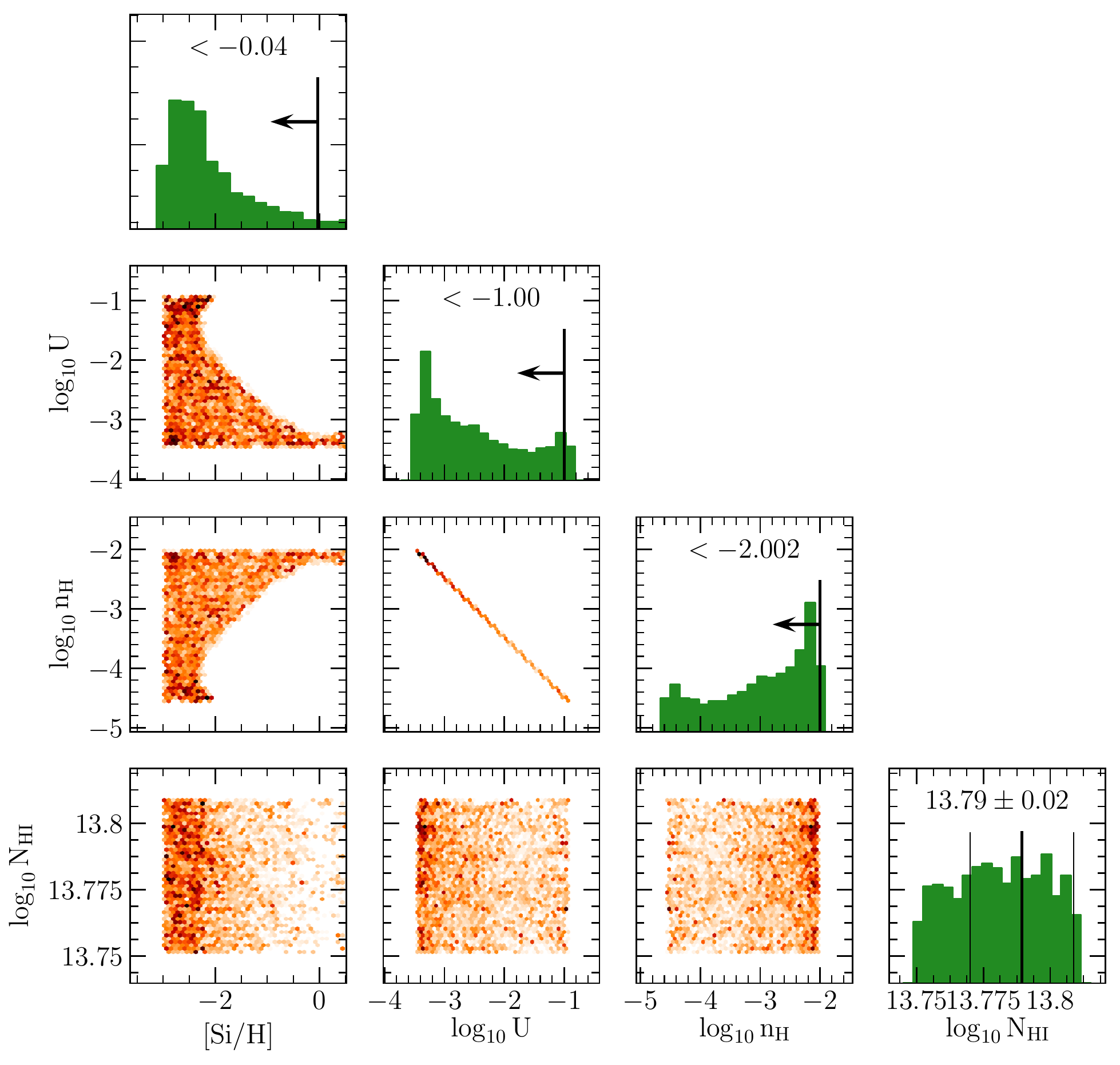}
	\caption{The posterior distribution profiles from the MCMC analysis of the Cloudy grids for J$0407$, $z_{gal} = 0.3422$, as for figure \ref{fig:Q0122_0.2119_par}. Note that {\OVI} is the only metal line detected here.}
	\label{fig:J0407_3422_par}
\end{figure}
\begin{figure*}
	\centering
	\includegraphics[width=\linewidth]{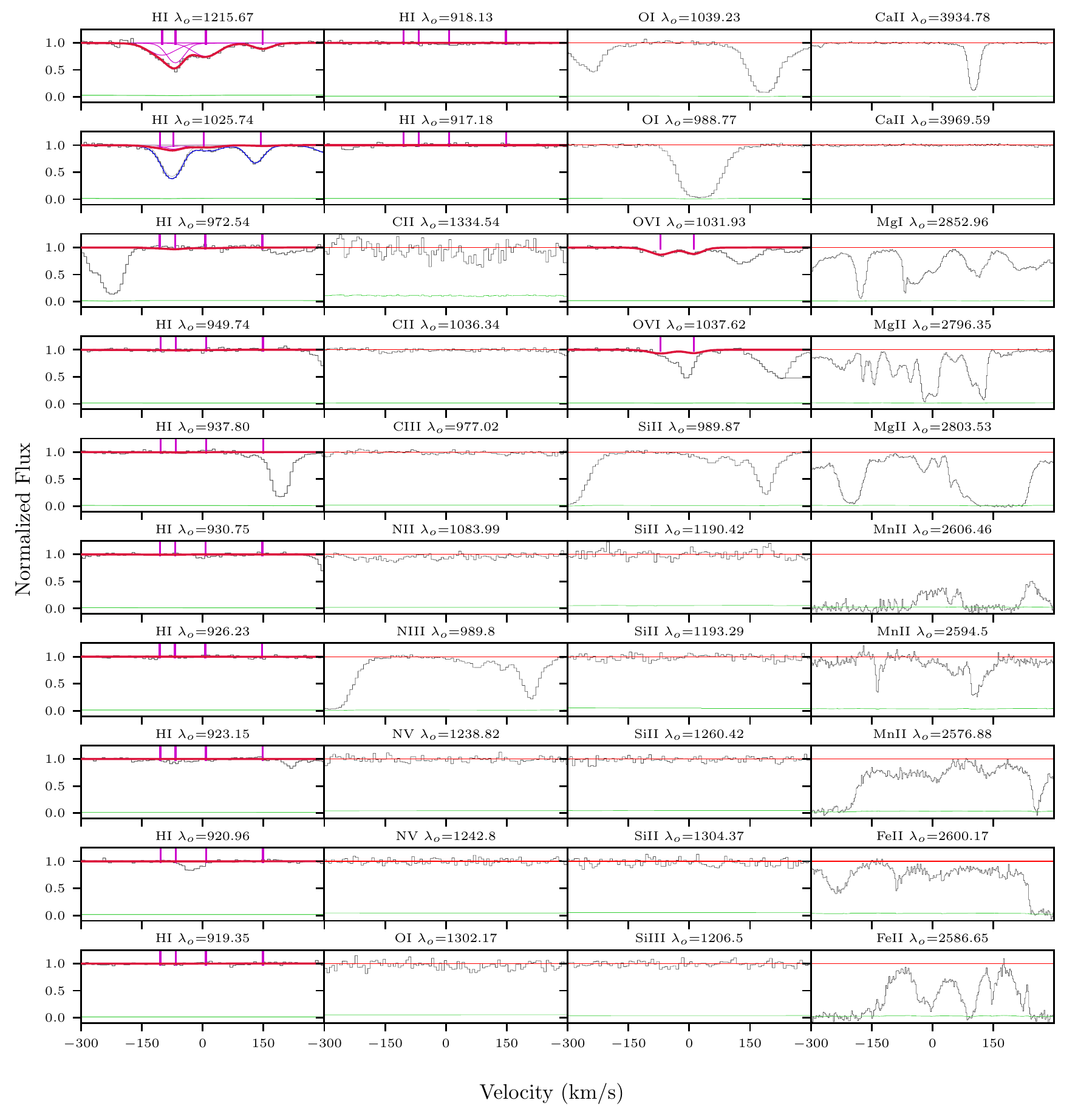}
	\caption{The fits for the system J$0407$, $z_{gal} = 0.3422$, as for figure \ref{fig:Q0122_0.2119}. The total {\OVI} fits from \citet{nielsenovi} are shown here for completeness, although they are not used in the ionization modelling.}
	\label{fig:J0407_3422}
\end{figure*}

\begin{figure*}
	\centering
	\includegraphics[width=\linewidth]{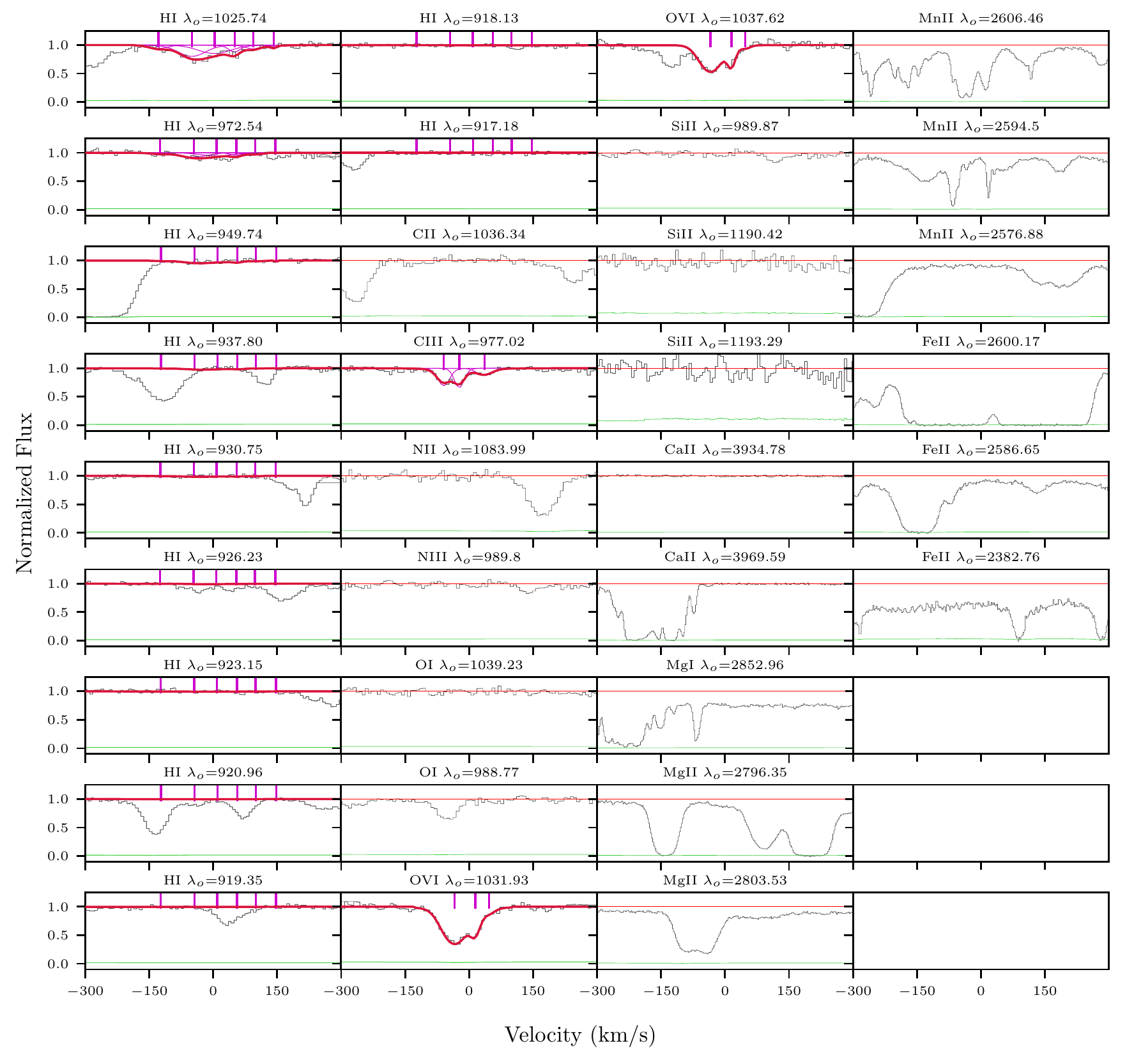}
	\caption{The fits for the system J$0407$, $z_{gal} = 0.495164$, as for figure \ref{fig:Q0122_0.2119}. The total {\OVI} fits from \citet{nielsenovi} are shown here for completeness, although they are not used in the ionization modelling.}
	\label{fig:J0407_4952}
\end{figure*}
\clearpage
\begin{deluxetable}{ccc}[hp]
	\tablecolumns{3}
	\tablewidth{\linewidth}
	\setlength{\tabcolsep}{0.06in}
	\tablecaption{J$0407$, $z_{gal} = 0.495164$ Column Densities\label{tab:J0407_4952}}
	\tablehead{
		\colhead{Ion}           	&
        \colhead{$\log N~({\cms})$}    &
		\colhead{$\log N$ Error~({\cms})}}
	\startdata
	{\HI}   & $14.34$    &$0.56$\\
{\CII}  & $<12.72$   &$\cdots$\\
{\CIII} & $13.20$    &$0.02$\\
{\NII}  & $<12.82$   &$\cdots$\\
{\NIII} & $<12.77$   &$\cdots$\\
{\OI}   & $<13.14$   &$\cdots$\\
{\SiII} & $<12.59$   &$\cdots$\\
{\CaII} & $<10.65$   &$\cdots$\\
{\MgI}  & $<10.61$   &$\cdots$\\
{\MgII} & $<10.87$   &$\cdots$\\
{\MnII} & $<11.26$   &$\cdots$\\
{\FeII} & $<11.86$   &$\cdots$\\[-5pt]

	\enddata
\end{deluxetable}
\begin{figure}[hp]
	\centering
	\includegraphics[width=\linewidth]{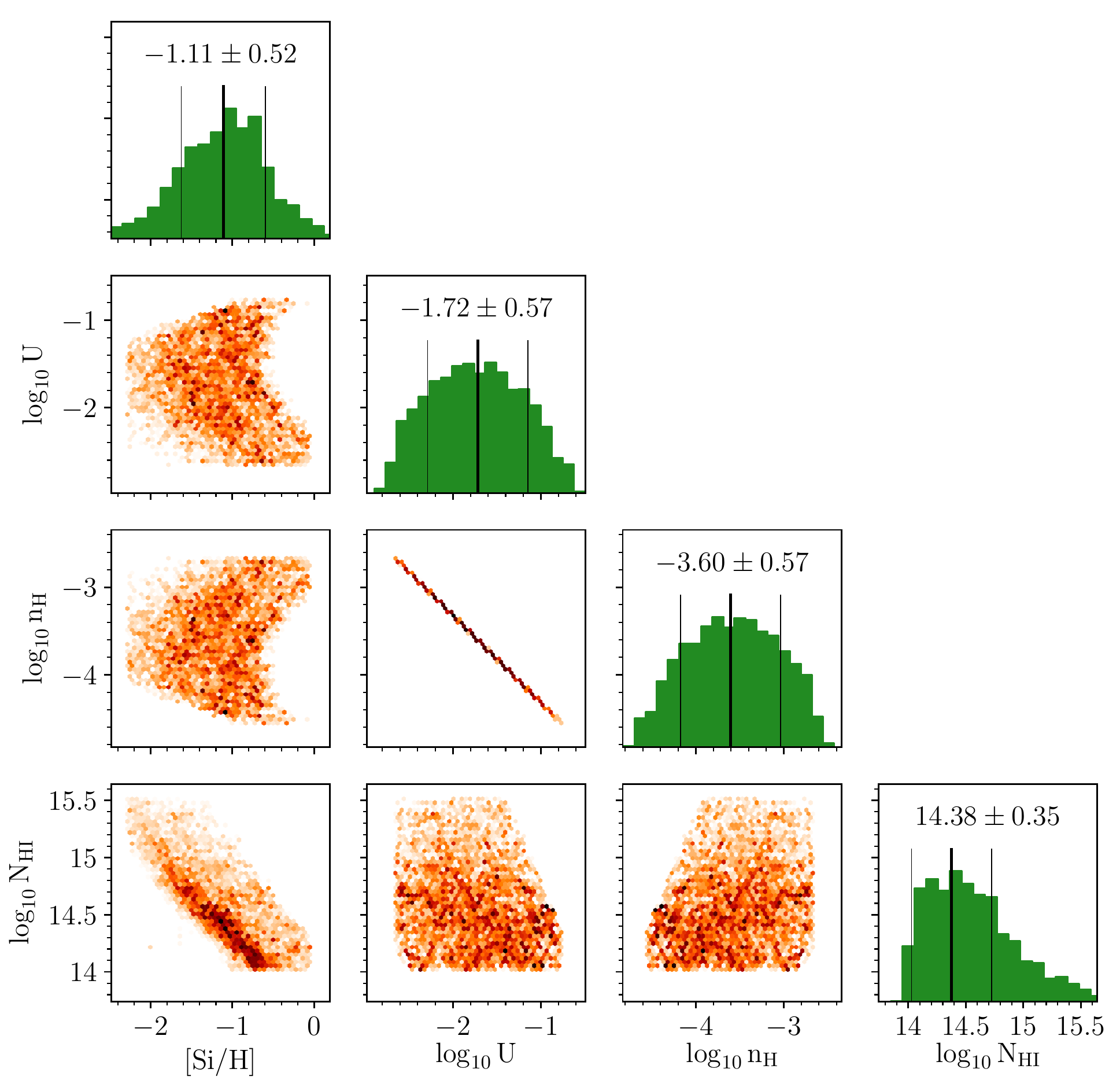}
	\caption{The posterior distribution profiles from the MCMC analysis of the Cloudy grids for J$0407$, $z_{gal} = 0.495164$, as for figure \ref{fig:Q0122_0.2119_par}.}
	\label{fig:J0407_4952_par}
\end{figure}
\newpage
\begin{deluxetable}{ccc}[hp]
	\tablecolumns{3}
	\tablewidth{\linewidth}
	\setlength{\tabcolsep}{0.06in}
	\tablecaption{J$0456$, $z_{gal} = 0.2784$ Column Densities\label{tab:J0456_2779}}
	\tablehead{
		\colhead{Ion}           	&
        \colhead{$\log N~({\cms})$}    &
		\colhead{$\log N$ Error~({\cms})}}
	\startdata
	{\HI}   & $[15.06, 19.00]$   &$\cdots$\\
{\CII}  & $<13.36$   &$\cdots$\\
{\NII}  & $<13.55$   &$\cdots$\\
{\OI}   & $<13.68$   &$\cdots$\\
{\SiII} & $<12.32$   &$\cdots$\\
{\SiIII}& $<12.04$   &$\cdots$\\
{\CaII} & $<10.96$   &$\cdots$\\[-5pt]

	\enddata
\end{deluxetable}
\begin{figure}[hp]
	\centering
	\includegraphics[width=\linewidth]{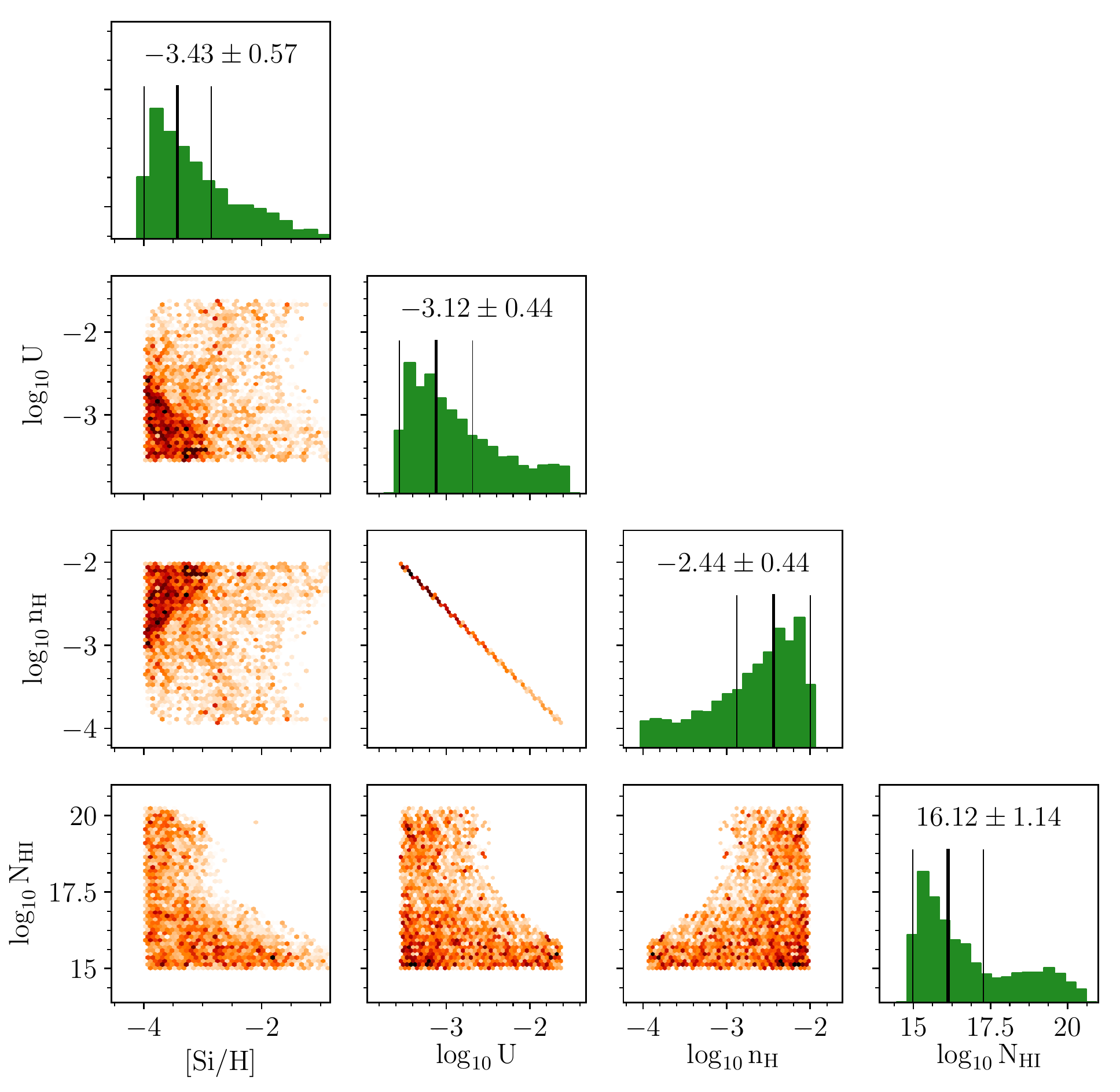}
	\caption{The posterior distribution profiles from the MCMC analysis of the Cloudy grids for J$0456$, $z_{gal} = 0.2784$, as for figure \ref{fig:Q0122_0.2119_par}. Note that no metal lines are detected here.}
	\label{fig:J0456_2779_par}
\end{figure}
\begin{figure*}[hp]
	\centering
	\includegraphics[width=\linewidth]{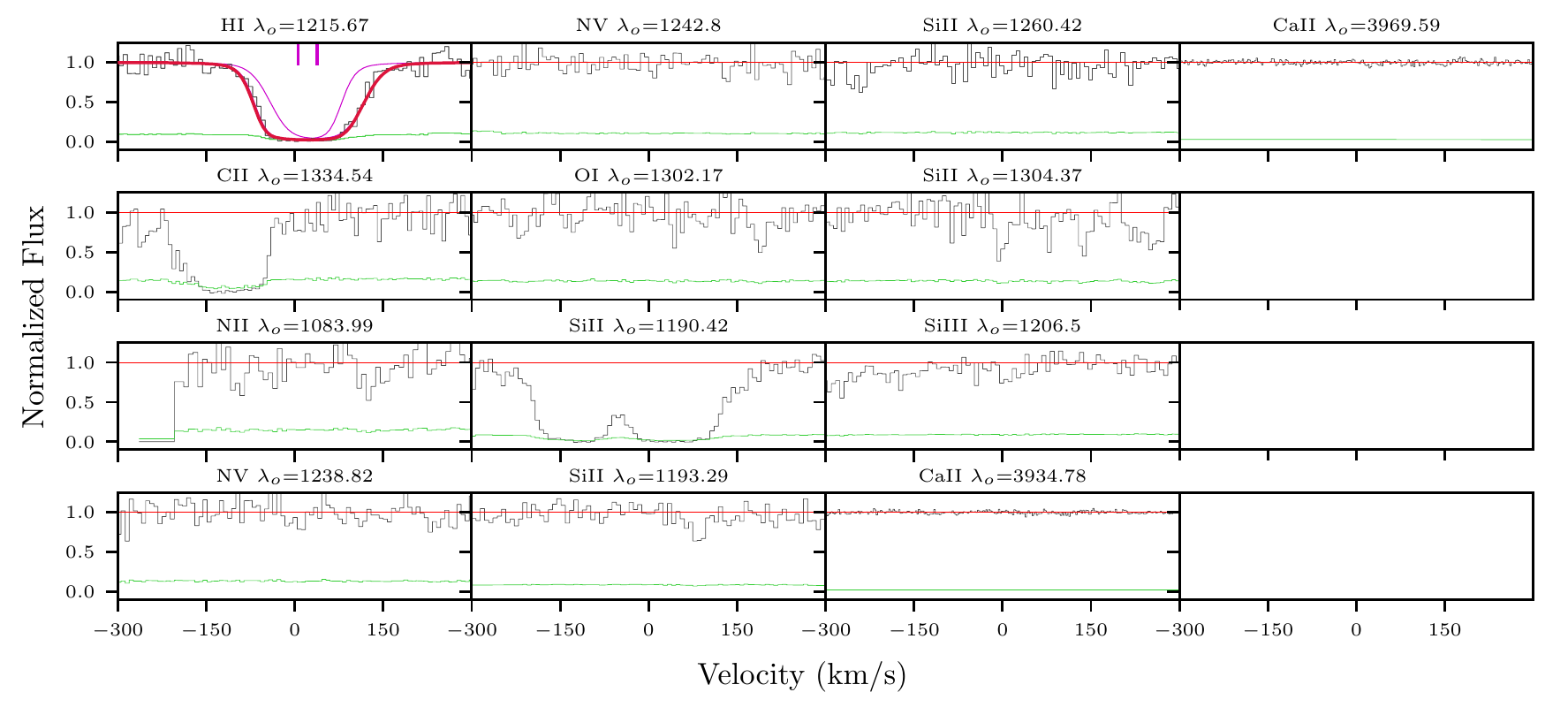}
	\caption{The fits for the system J$0456$, $z_{gal} = 0.2784$, as for figure \ref{fig:Q0122_0.2119}.}
	\label{fig:J0456_2779}
\end{figure*}

\begin{figure*}[hp]
	\centering
	\includegraphics[width=\linewidth]{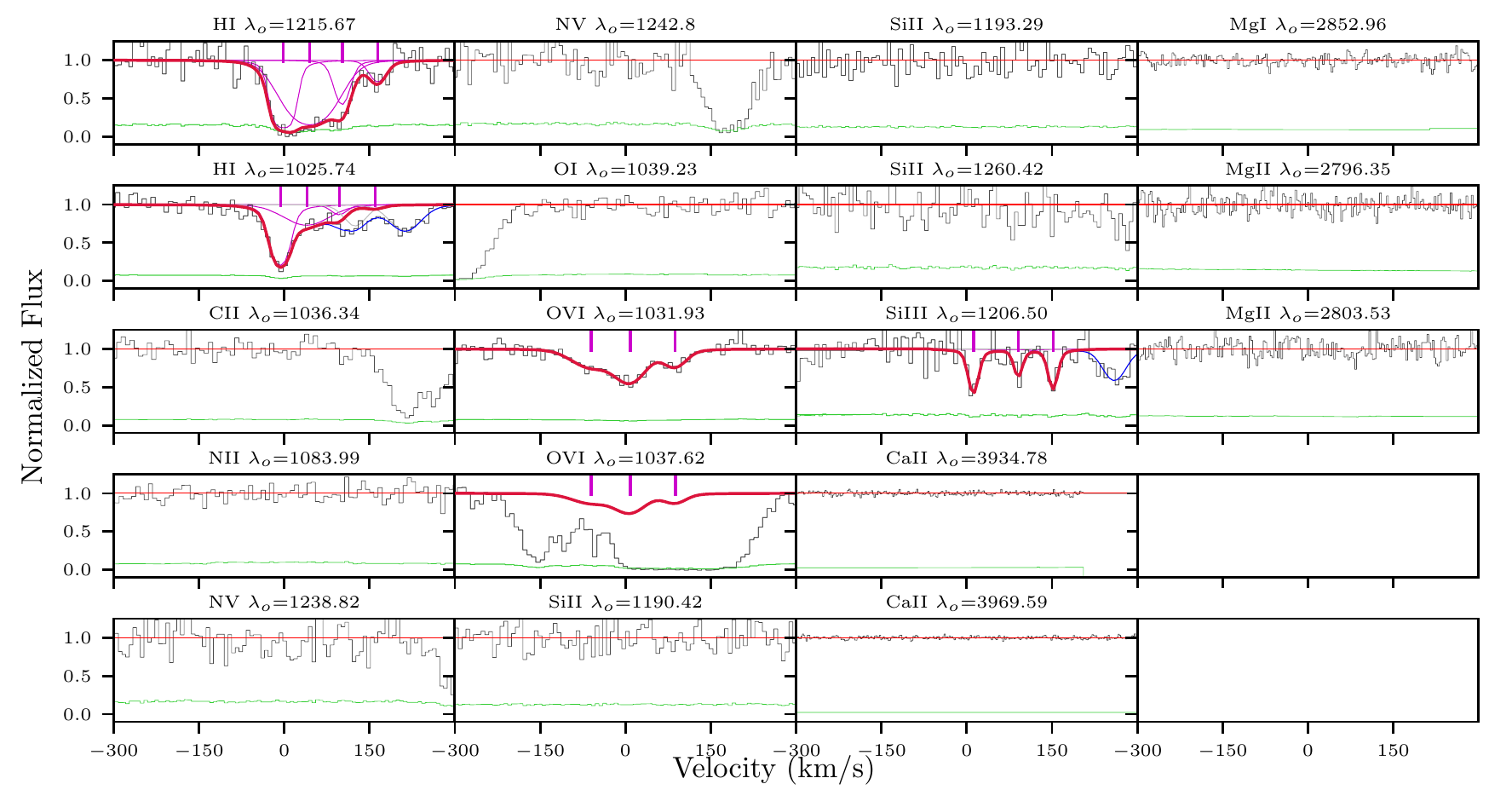}
	\caption{The fits for the system J$0456$, $z_{gal} = 0.381511$, as for figure \ref{fig:Q0122_0.2119}. The total {\OVI} fits from \citet{nielsenovi} are shown here for completeness, although they are not used in the ionization modelling.}
	\label{fig:J0456_3815}
\end{figure*}
\clearpage
\begin{deluxetable}{ccc}[hp]
	\tablecolumns{3}
	\tablewidth{\linewidth}
	\setlength{\tabcolsep}{0.06in}
	\tablecaption{J$0456$, $z_{gal} = 0.381511$ Column Densities\label{tab:J0456_3815}}
	\tablehead{
		\colhead{Ion}           	&
        \colhead{$\log N~({\cms})$}    &
		\colhead{$\log N$ Error~({\cms})}}
	\startdata
	{\HI}   & $15.10$   &$0.39$\\
{\CII}  & $<13.10$   &$\cdots$\\
{\NII}  & $<13.27$   &$\cdots$\\
{\NV}   & $<13.32$   &$\cdots$\\
{\OI}   & $<14.32$   &$\cdots$\\
{\SiII} & $<12.53$   &$\cdots$\\
{\SiIII}& $13.25$   &$0.55$\\
{\CaII} & $<11.05$   &$\cdots$\\
{\MgI}  & $<11.31$   &$\cdots$\\
{\MgII} & $<11.99$   &$\cdots$\\[-5pt]

	\enddata
\end{deluxetable}
\begin{figure}[hp]
	\centering
	\includegraphics[width=\linewidth]{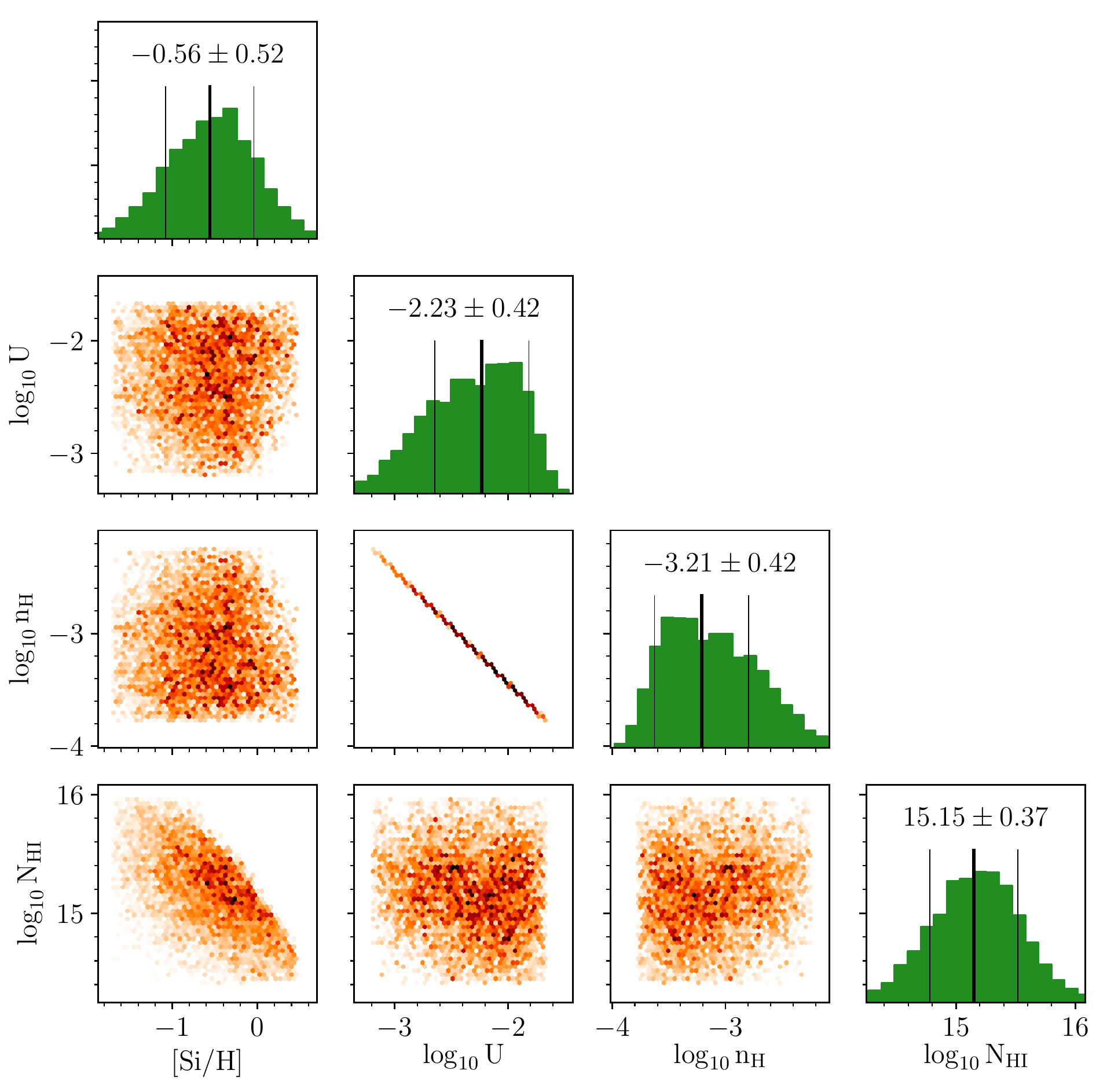}
	\caption{The posterior distribution profiles from the MCMC analysis of the Cloudy grids for J$0456$, $z_{gal} = 0.381511$, as for figure \ref{fig:Q0122_0.2119_par}.}
	\label{fig:J0456_3815_par}
\end{figure}
\newpage
\begin{deluxetable}{ccc}[hp]
	\tablecolumns{8}
	\tablewidth{\linewidth}
	\setlength{\tabcolsep}{0.06in}
	\tablecaption{J$0456$, $z_{gal} = 0.4838$ Column Densities\label{tab:J0456_4838}}
	\tablehead{
		\colhead{Ion}           	&
        \colhead{$\log N~({\cms})$}    &
		\colhead{$\log N$ Error~({\cms})}}
	\startdata
	{\HI}   & $[16.53,19.00]$   &$\cdots$\\
{\CII}  & $14.53$   &$0.03$\\
{\CIII} & $>14.34$   &$\cdots$\\
{\NII}  & $14.11$   &$0.05$\\
{\NIII} & $14.48$   &$0.03$\\
{\OI}   & $13.97$   &$\cdots$\\
{\SiII} & $13.34$   &$\cdots$\\
{\CaII} & $11.40$   &$0.04$\\
{\MgI}  & $11.84$   &$0.03$\\
{\MgII} & $13.60$   &$0.09$\\
{\MnII} & $12.64$   &$\cdots$\\
{\FeII} & $12.85$   &$0.82$\\[-5pt]

	\enddata
\end{deluxetable}
\begin{figure}[hp]
	\centering
	\includegraphics[width=\linewidth]{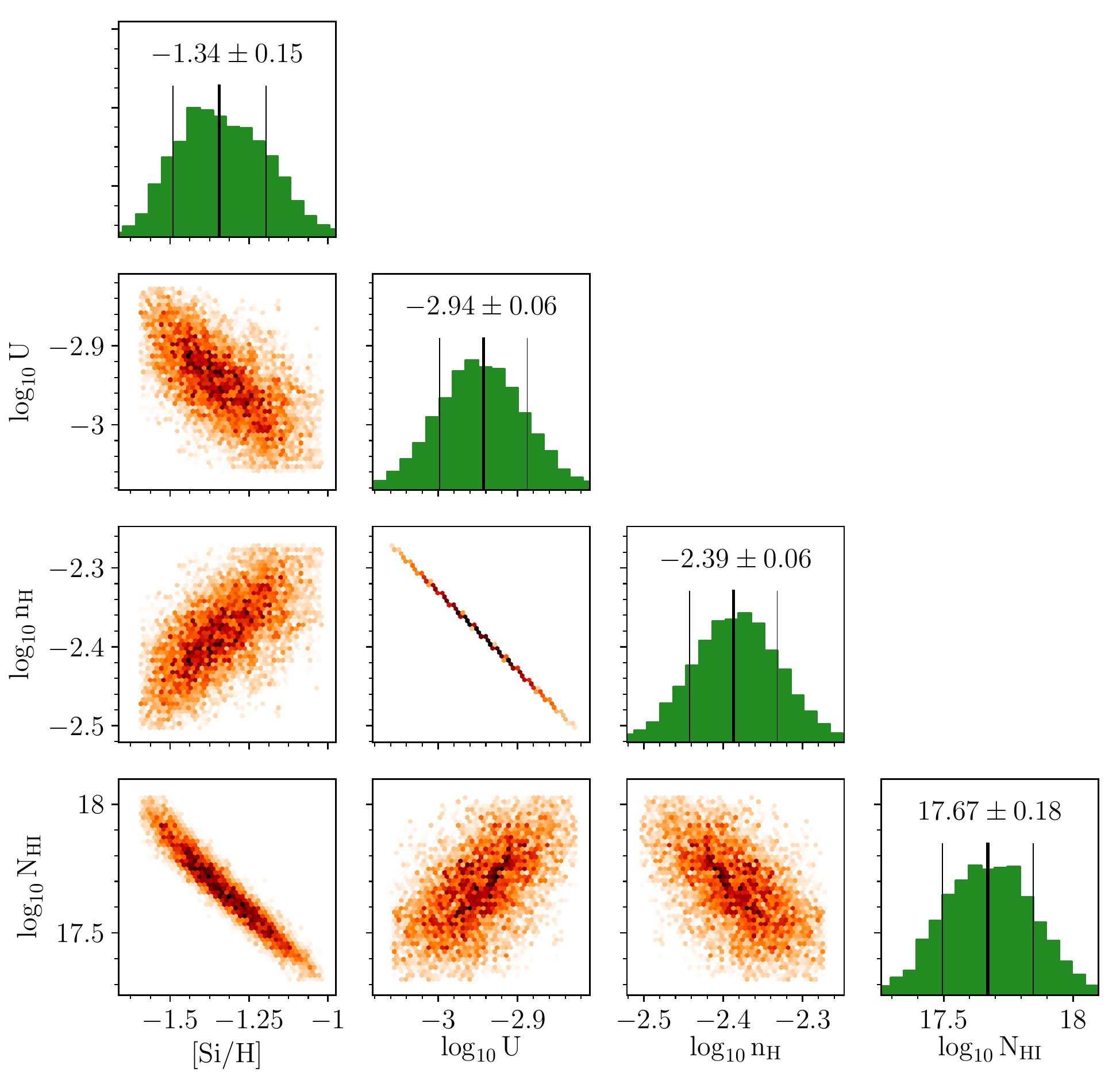}
	\caption{The posterior distribution profiles from the MCMC analysis of the Cloudy grids for J$0456$, $z_{gal} = 0.4838$, as for figure \ref{fig:Q0122_0.2119_par}.}
	\label{fig:J0456_4838_par}
\end{figure}
\begin{figure*}[hp]
	\centering
	\includegraphics[width=\linewidth]{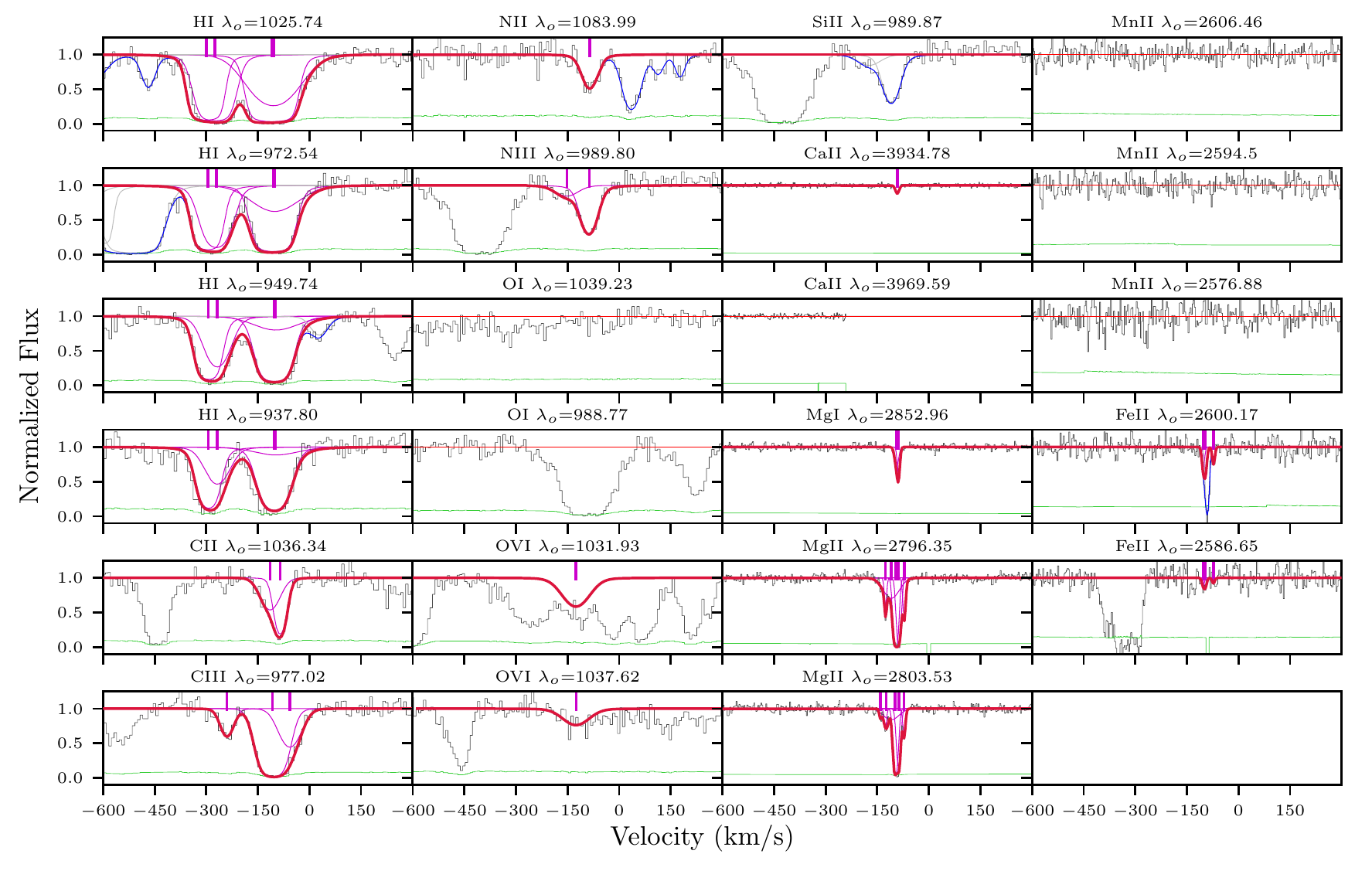}
	\caption{The fits for the system J$0456$, $z_{gal} = 0.4838$, as for figure \ref{fig:Q0122_0.2119}. The {\HI} Lyman series was moderately affected by the spectral shift in the COS spectrograph. In order to model all available {\HI} transitions, additionally components on the left {\HI} peak were required. The total {\OVI} fits from \citet{nielsenovi} are shown here for completeness, although they are not used in the ionization modelling. }
	\label{fig:J0456_4838}
\end{figure*}

\begin{figure*}[hp]
	\centering
	\includegraphics[width=\linewidth]{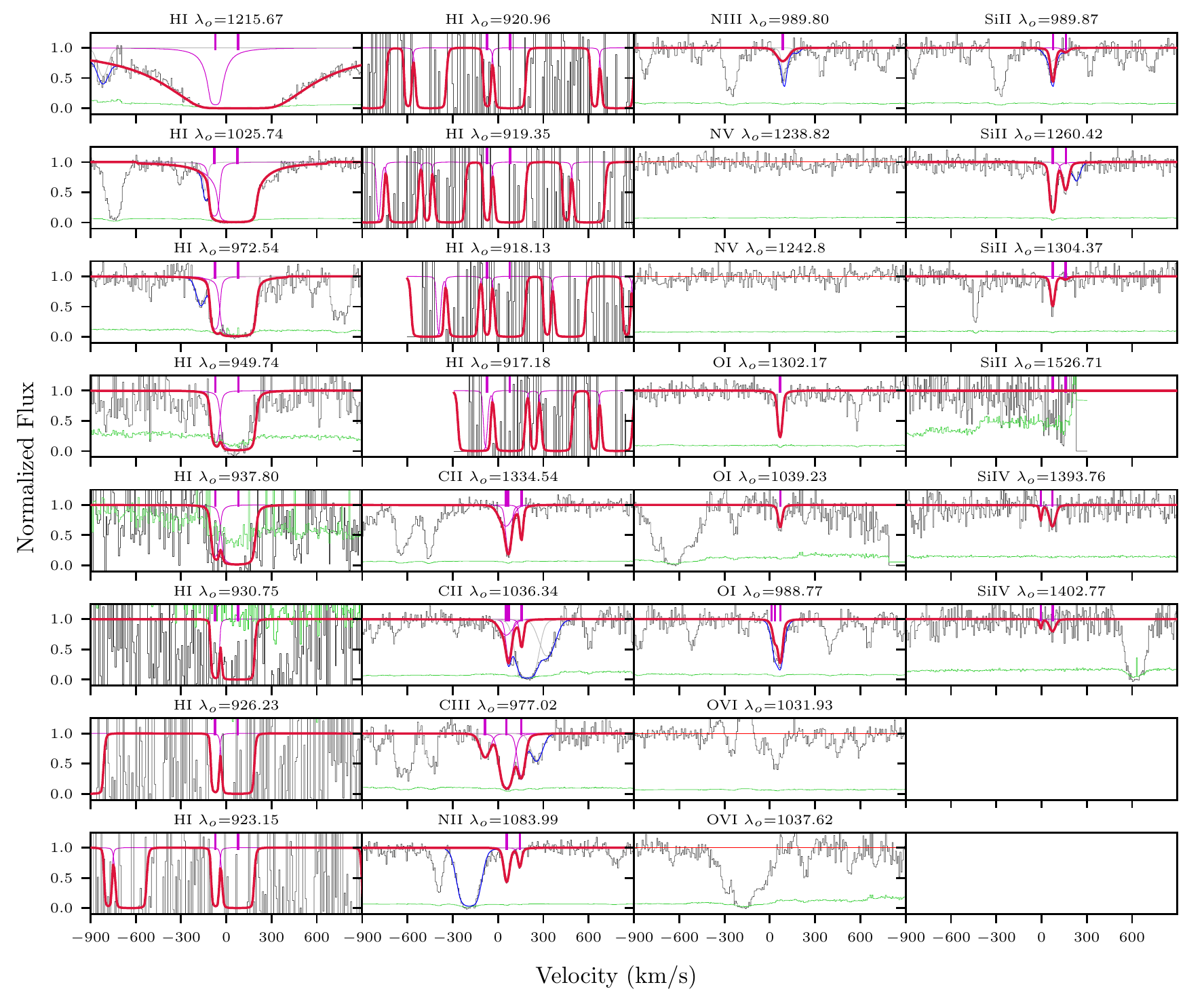}
	\caption{The fits for the system J$0853$, $z_{gal} = 0.1635$, as for figure \ref{fig:Q0122_0.2119}. The blend in the {\Lya} line is from the {\CIII} transition at $z_{gal} = 0.4402$, shown in blue.}
	\label{fig:J0853_1634}
\end{figure*}
\clearpage
\begin{deluxetable}{ccc}[hp]
	\tablecolumns{8}
	\tablewidth{\linewidth}
	\setlength{\tabcolsep}{0.06in}
	\tablecaption{J$0853$, $z_{gal} = 0.1635$ Column Densities\label{tab:J0853_1634}}
	\tablehead{
		\colhead{Ion}           	&
        \colhead{$\log N~({\cms})$}    &
		\colhead{$\log N$ Error~({\cms})}}
	\startdata
	{\HI}   & $19.93$   &$0.01$\\
{\CII}  & $14.58$   &$0.05$\\
{\CIII} & $14.31$   &$0.04$\\
{\NII}  & $14.28$   &$0.03$\\
{\NIII} & $13.88$   &$0.12$\\
{\NV}   & $<12.97$   &$\cdots$\\
{\OI}   & $14.91$   &$0.11$\\
{\SiII} & $14.04$   &$0.07$\\
{\SiIV} & $13.26$   &$0.19$\\[-5pt]

	\enddata
\end{deluxetable}
\begin{figure}[hp]
	\centering
	\includegraphics[width=\linewidth]{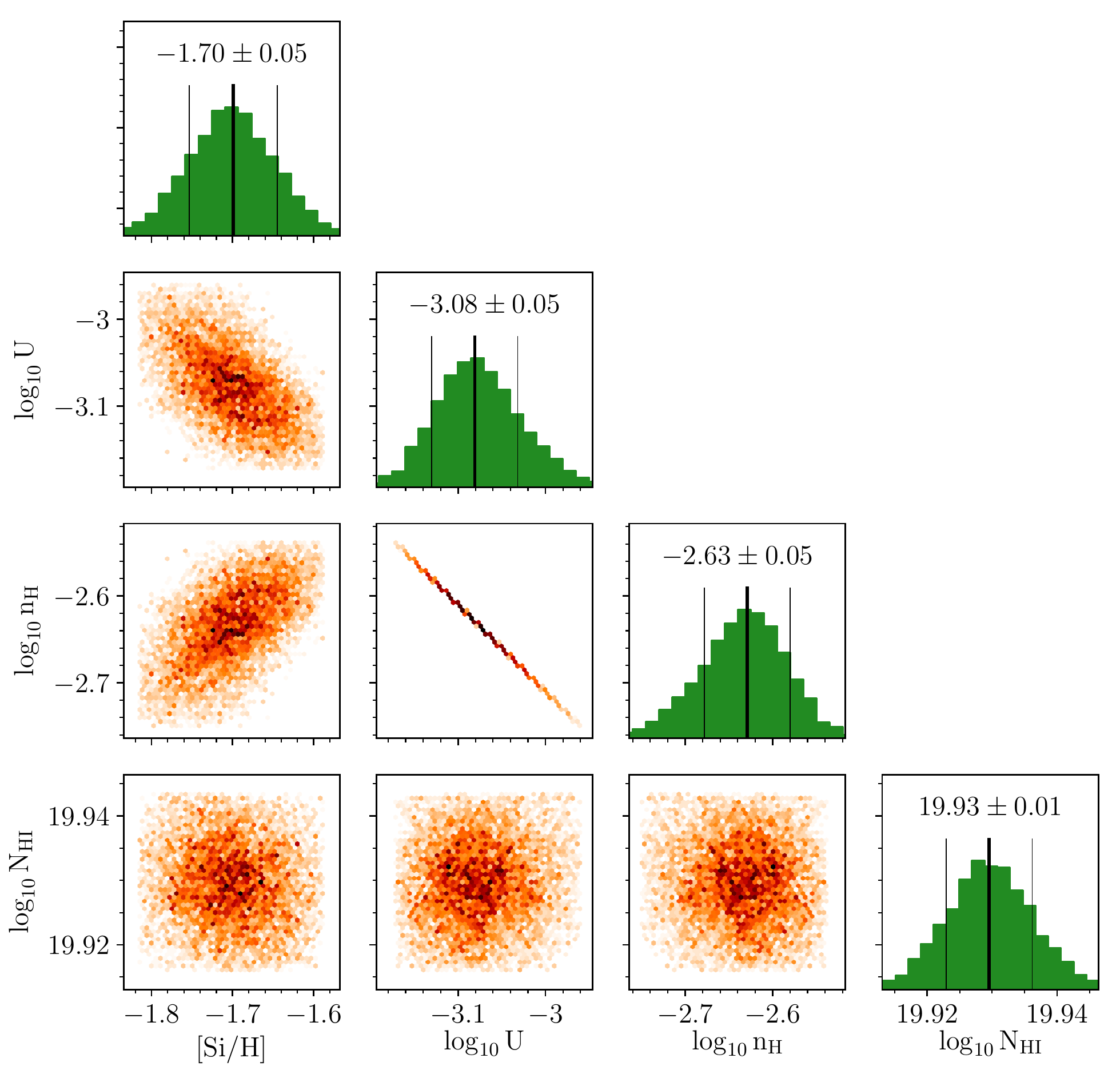}
	\caption{The posterior distribution profiles from the MCMC analysis of the Cloudy grids for J$0853$, $z_{gal} = 0.1635$, as for figure \ref{fig:Q0122_0.2119_par}.}
	\label{fig:J0853_1634_par}
\end{figure}
\newpage
\begin{deluxetable}{ccc}[hp]
	\tablecolumns{8}
	\tablewidth{\linewidth}
	\setlength{\tabcolsep}{0.06in}
	\tablecaption{J$0853$, $z_{gal} = 0.2766$ Column Densities\label{tab:J0853_2766}}
	\tablehead{
		\colhead{Ion}           	&
        \colhead{$\log N~({\cms})$}    &
		\colhead{$\log N$ Error~({\cms})}}
	\startdata
	{\HI}   & $14.15$   &$0.03$\\
{\CII}  & $<13.32$   &$\cdots$\\
{\CIII} & $13.09$   &$0.05$\\
{\NIII} & $<13.33$   &$\cdots$\\
{\NV}   & $<13.56$   &$\cdots$\\
{\OI}   & $<13.71$   &$\cdots$\\
{\SiII} & $<12.60$   &$\cdots$\\
{\SiIII}& $<12.12$   &$\cdots$\\[-5pt]

	\enddata
\end{deluxetable}
\begin{figure}[hp]
	\centering
	\includegraphics[width=\linewidth]{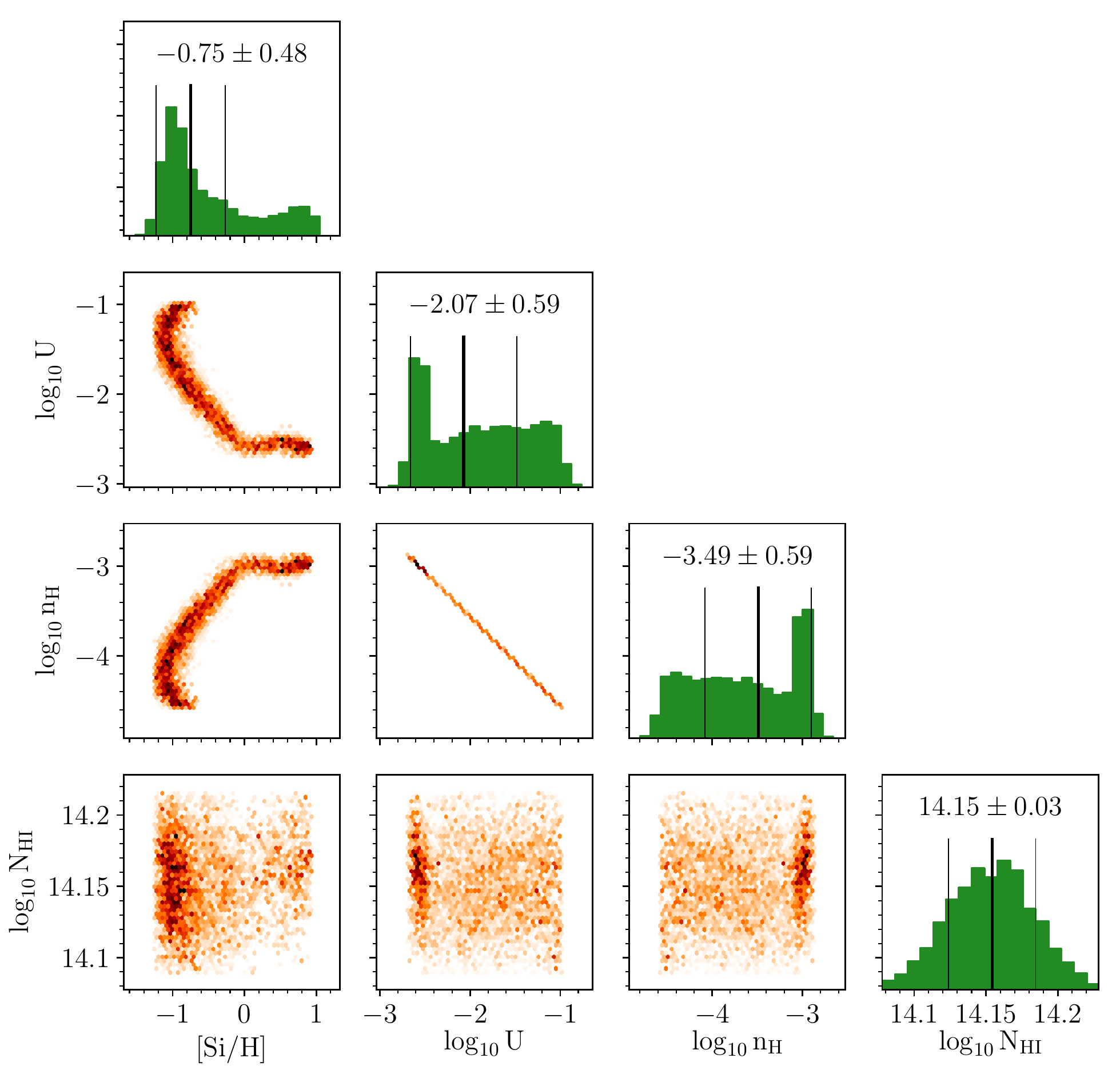}
	\caption{The posterior distribution profiles from the MCMC analysis of the Cloudy grids for J$0853$, $z_{gal} = 0.2766$, as for figure \ref{fig:Q0122_0.2119_par}.}
	\label{fig:J0853_2766_par}
\end{figure}
\begin{figure*}[hp]
	\centering
	\includegraphics[width=\linewidth]{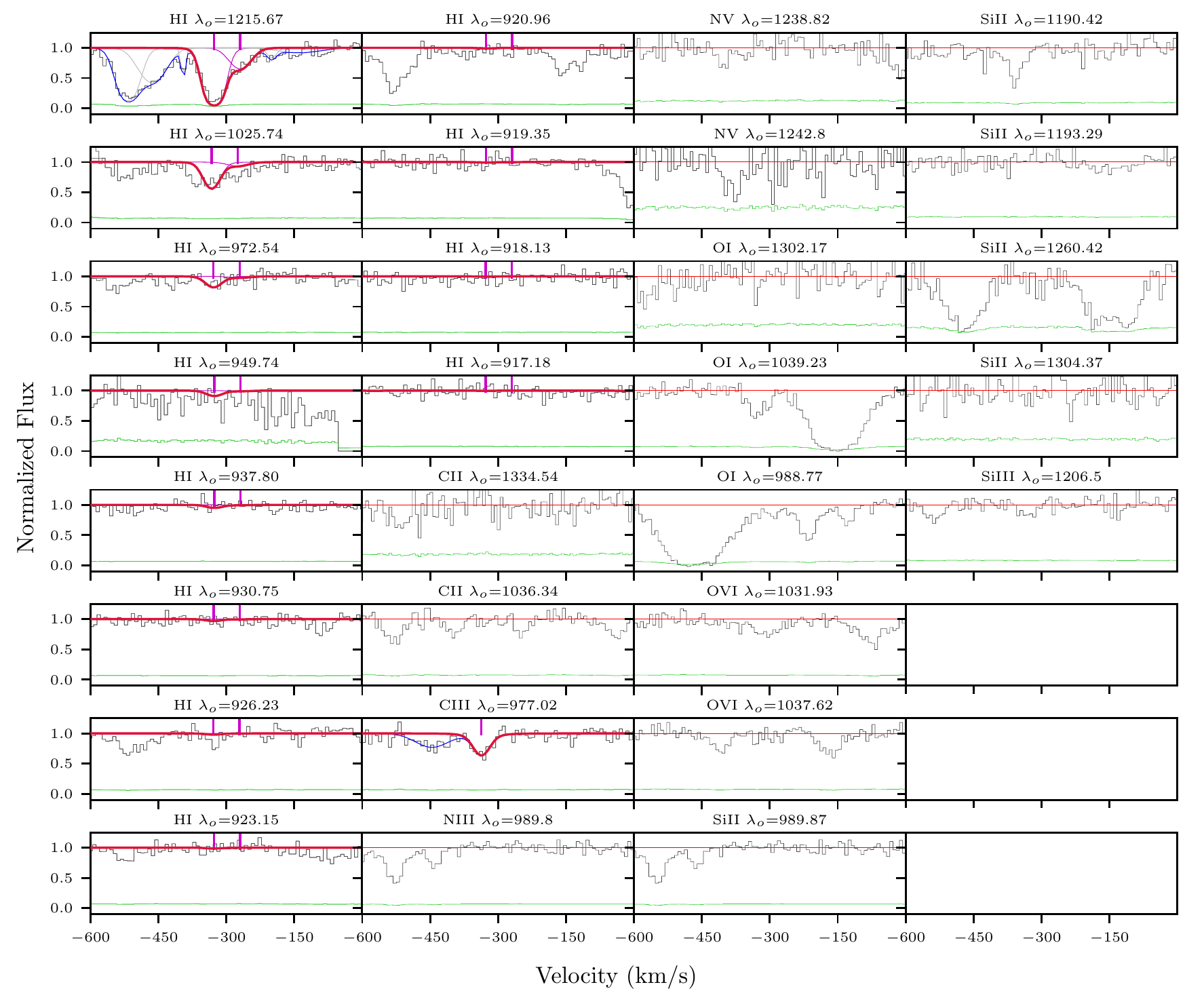}
	\caption{The fits for the system J$0853$, $z_{gal} = 0.2766$, as for figure \ref{fig:Q0122_0.2119}. Note that there is an unknown blended line (blue) with {\Lya}.}
	\label{fig:J0853_2766}
\end{figure*}

\begin{figure*}[hp]
	\centering
	\includegraphics[width=\linewidth]{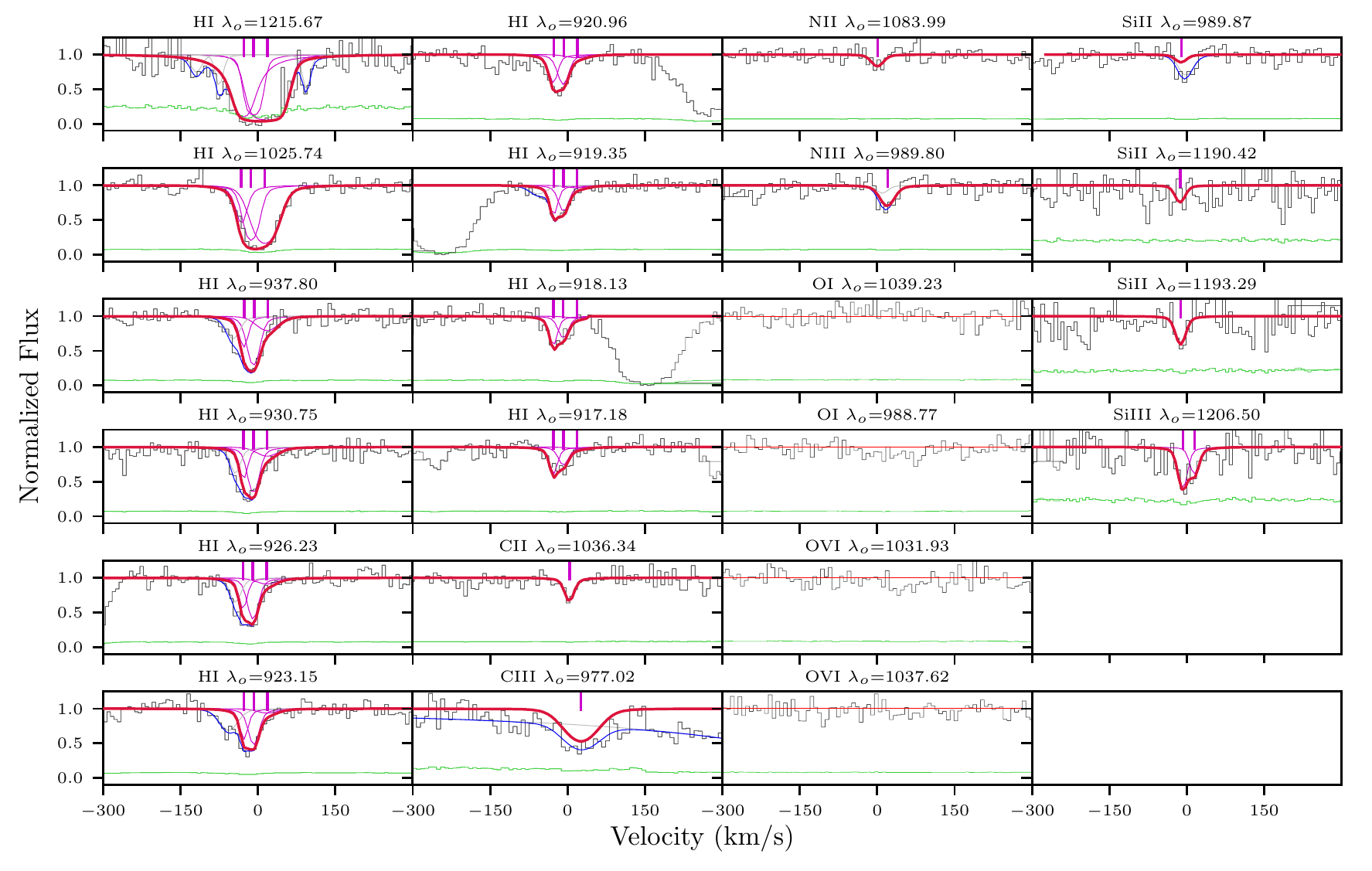}
	\caption{The fits for the system J$0853$, $z_{gal} = 0.4402$, as for figure \ref{fig:Q0122_0.2119}. The blend in the {\CIII} line in blue is from the DLA {\Lya} transition at $z_{gal} = 0.1634$.}
	\label{fig:J0853_4435}
\end{figure*}
\clearpage
\begin{deluxetable}{ccc}[hp]
	\tablecolumns{8}
	\tablewidth{\linewidth}
	\setlength{\tabcolsep}{0.06in}
	\tablecaption{J$0853$, $z_{gal} = 0.4402$ Column Densities\label{tab:J0853_4435}}
	\tablehead{
		\colhead{Ion}           	&
        \colhead{$\log N~({\cms})$}    &
		\colhead{$\log N$ Error~({\cms})}}
	\startdata
	{\HI}   & $17.30$   &$0.20$\\
{\CII}  & $13.68$   &$0.25$\\
{\CIII} & $13.45$   &$0.08$\\
{\NII}  & $13.32$   &$0.14$\\
{\NIII} & $13.68$   &$0.09$\\
{\OI}   & $<13.61$   &$\cdots$\\
{\SiII} & $12.97$   &$0.17$\\
{\SiIII}& $13.00$   &$0.40$\\[-5pt]

	\enddata
\end{deluxetable}
\begin{figure}[hp]
	\centering
	\includegraphics[width=\linewidth]{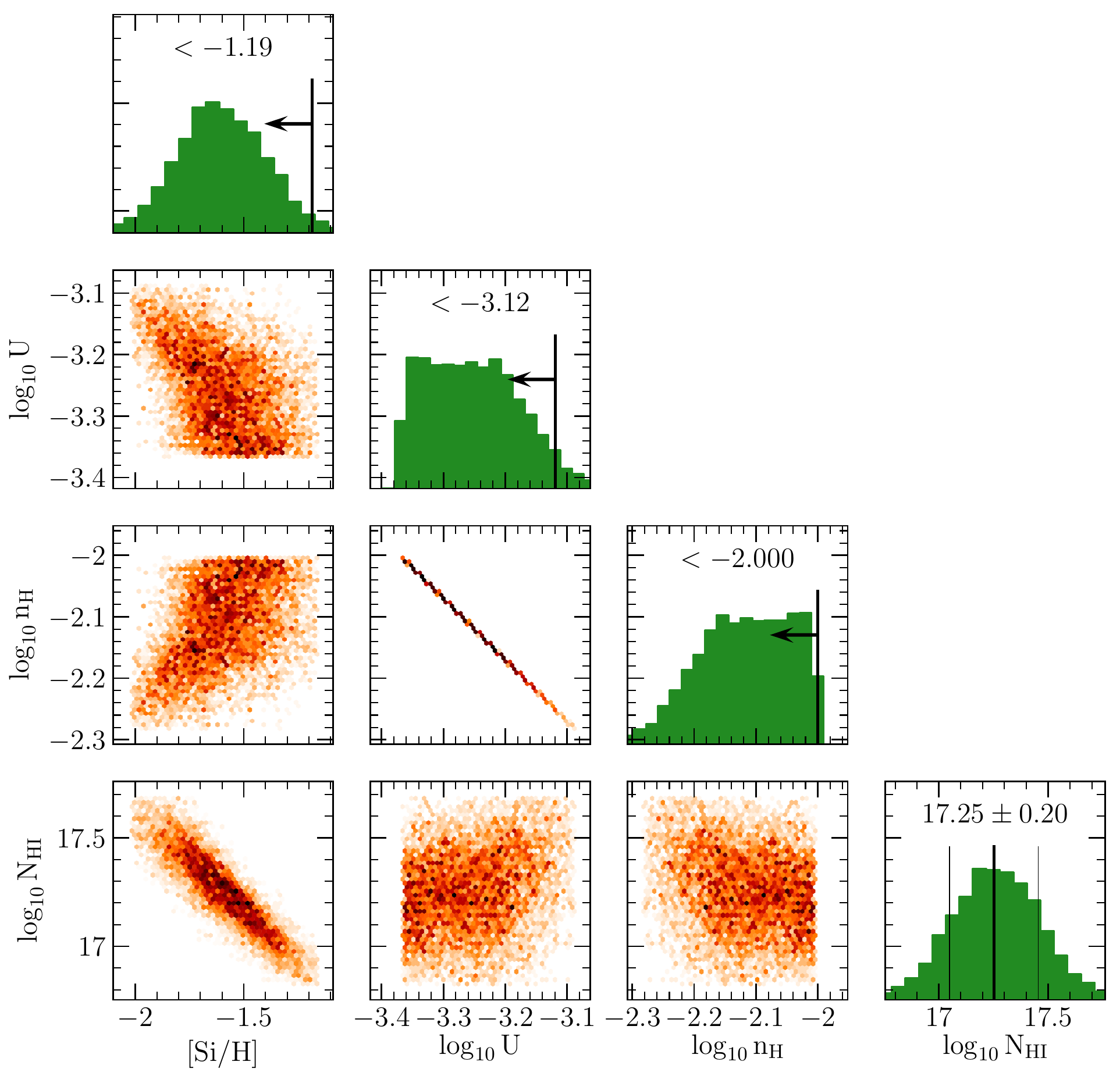}
	\caption{The posterior distribution profiles from the MCMC analysis of the Cloudy grids for J$0853$, $z_{gal} = 0.4402$, as for figure \ref{fig:Q0122_0.2119_par}.}
	\label{fig:J0853_4435_par}
\end{figure}
\newpage
\begin{deluxetable}{ccc}[hp]
	\tablecolumns{8}
	\tablewidth{\linewidth}
	\setlength{\tabcolsep}{0.06in}
	\tablecaption{J$0914$, $z_{gal} = 0.244312$ Column Densities\label{tab:J0914_2443}}
	\tablehead{
		\colhead{Ion}           	&
        \colhead{$\log N~({\cms})$}    &
		\colhead{$\log N$ Error~({\cms})}}
	\startdata
	{\HI}   & $15.55$   &$0.03$\\
{\CII}  & $13.52$   &$0.16$\\
{\CIII} & $<11.25$   &$\cdots$\\
{\NII}  & $<13.51$   &$\cdots$\\
{\NIII} & $13.89$   &$0.19$\\
{\NV}   & $<13.36$   &$\cdots$\\
{\OI}   & $<13.89$   &$\cdots$\\
{\SiII} & $12.52$   &$0.21$\\
{\SiIII}& $12.80$   &$0.09$\\
{\SiIV} & $<13.16$   &$\cdots$\\
{\CaII} & $<10.98$   &$\cdots$\\
{\MgI}  & $<10.67$   &$\cdots$\\
{\MgII} & $<11.13$   &$\cdots$\\
{\MnII} & $<11.41$   &$\cdots$\\
{\FeII} & $<11.59$   &$\cdots$\\[-5pt]

	\enddata
\end{deluxetable}
\begin{figure}[hp]
	\centering
	\includegraphics[width=\linewidth]{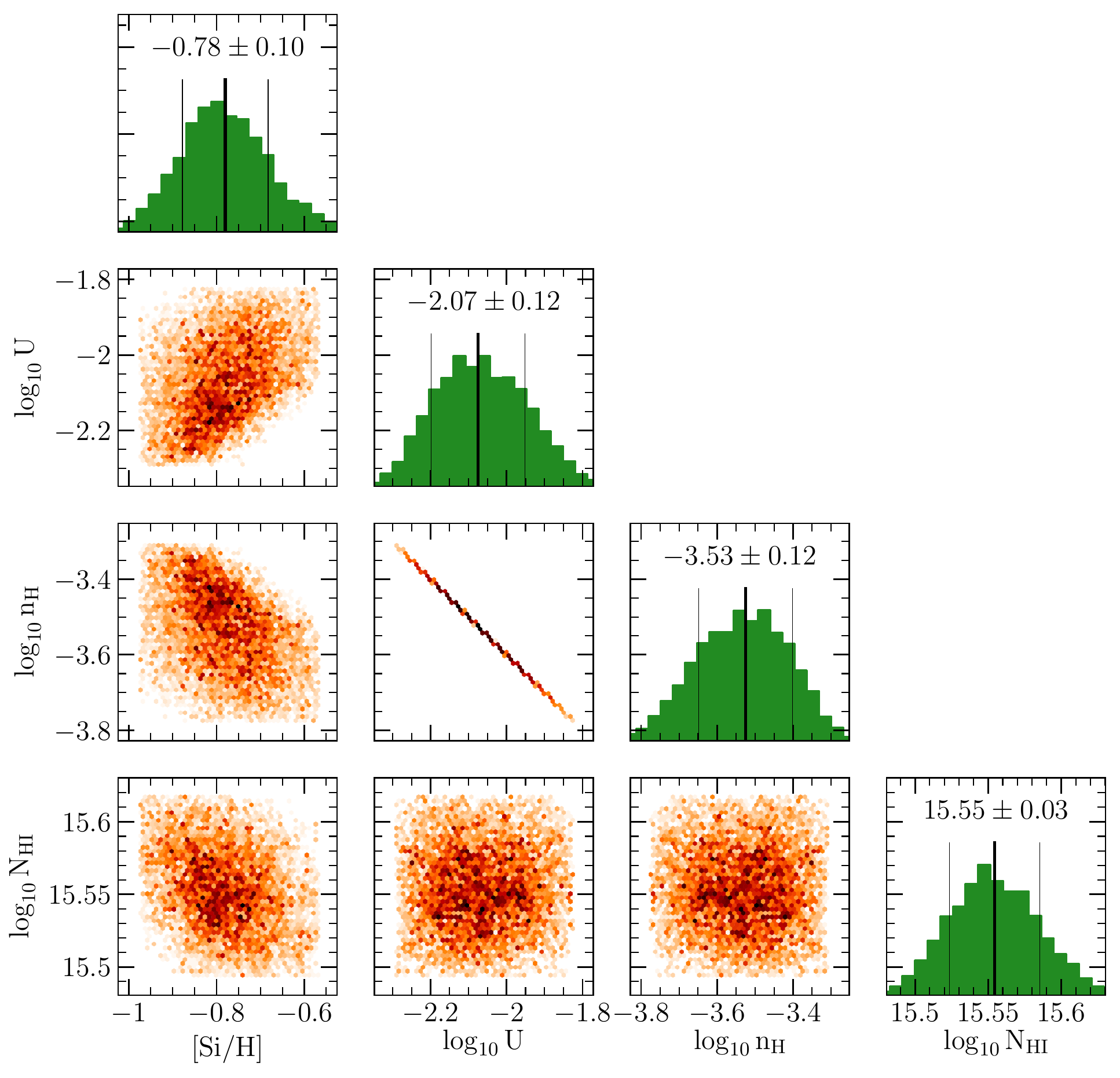}
	\caption{The posterior distribution profiles from the MCMC analysis of the Cloudy grids for J$0914$, $z_{gal} = 0.244312$, as for figure \ref{fig:Q0122_0.2119_par}.}
	\label{fig:J0914_2443_par}
\end{figure}
\begin{figure*}[hp]
	\centering
	\includegraphics[width=\linewidth]{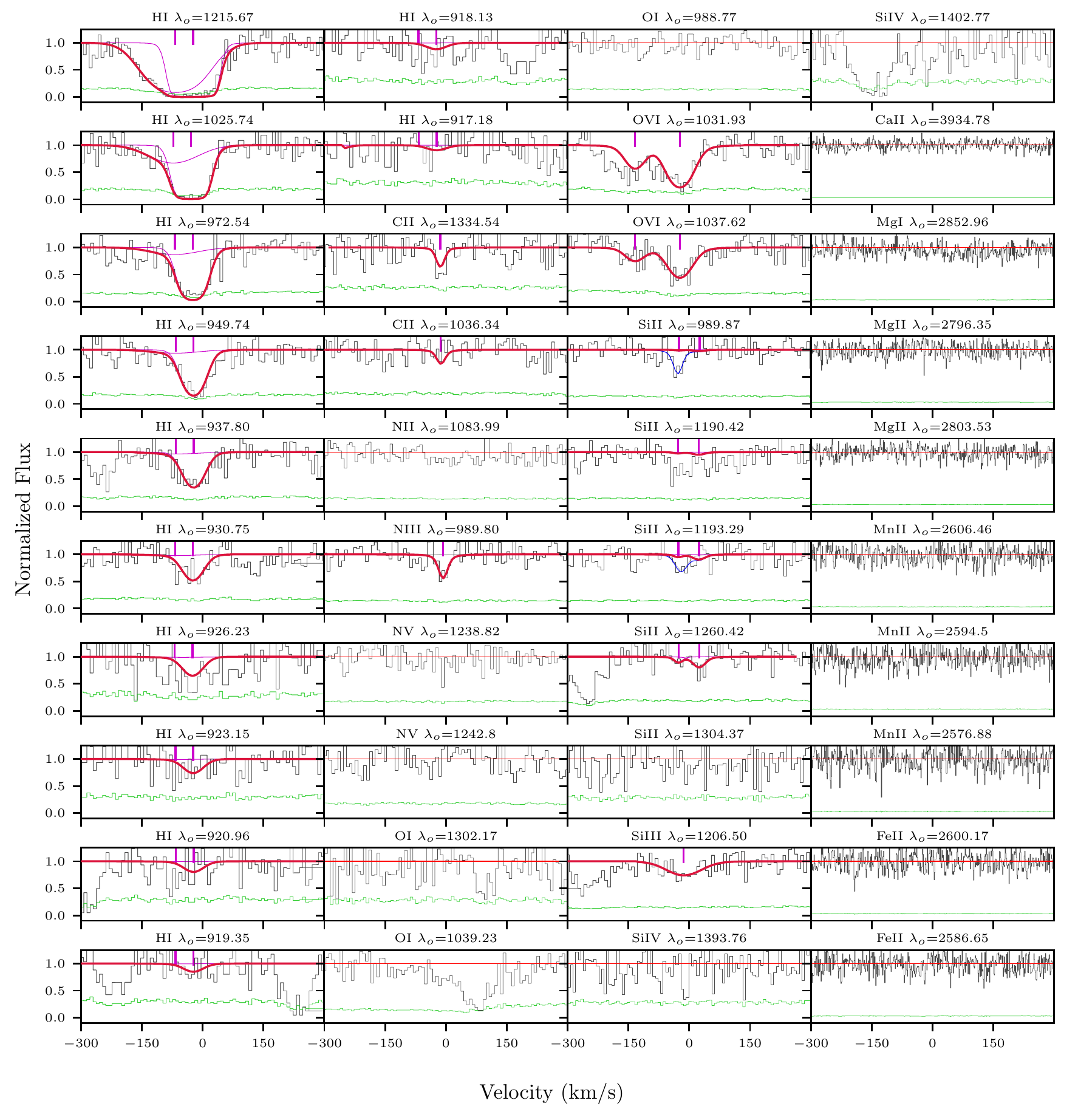}
	\caption{The fits for the system J$0914$, $z_{gal} = 0.244312$, as for figure \ref{fig:Q0122_0.2119}. The total {\OVI} fits from \citet{nielsenovi} are shown here for completeness, although they are not used in the ionization modelling.}
	\label{fig:J0914_2443}
\end{figure*}

\begin{figure*}[hp]
	\centering
	\includegraphics[width=\linewidth]{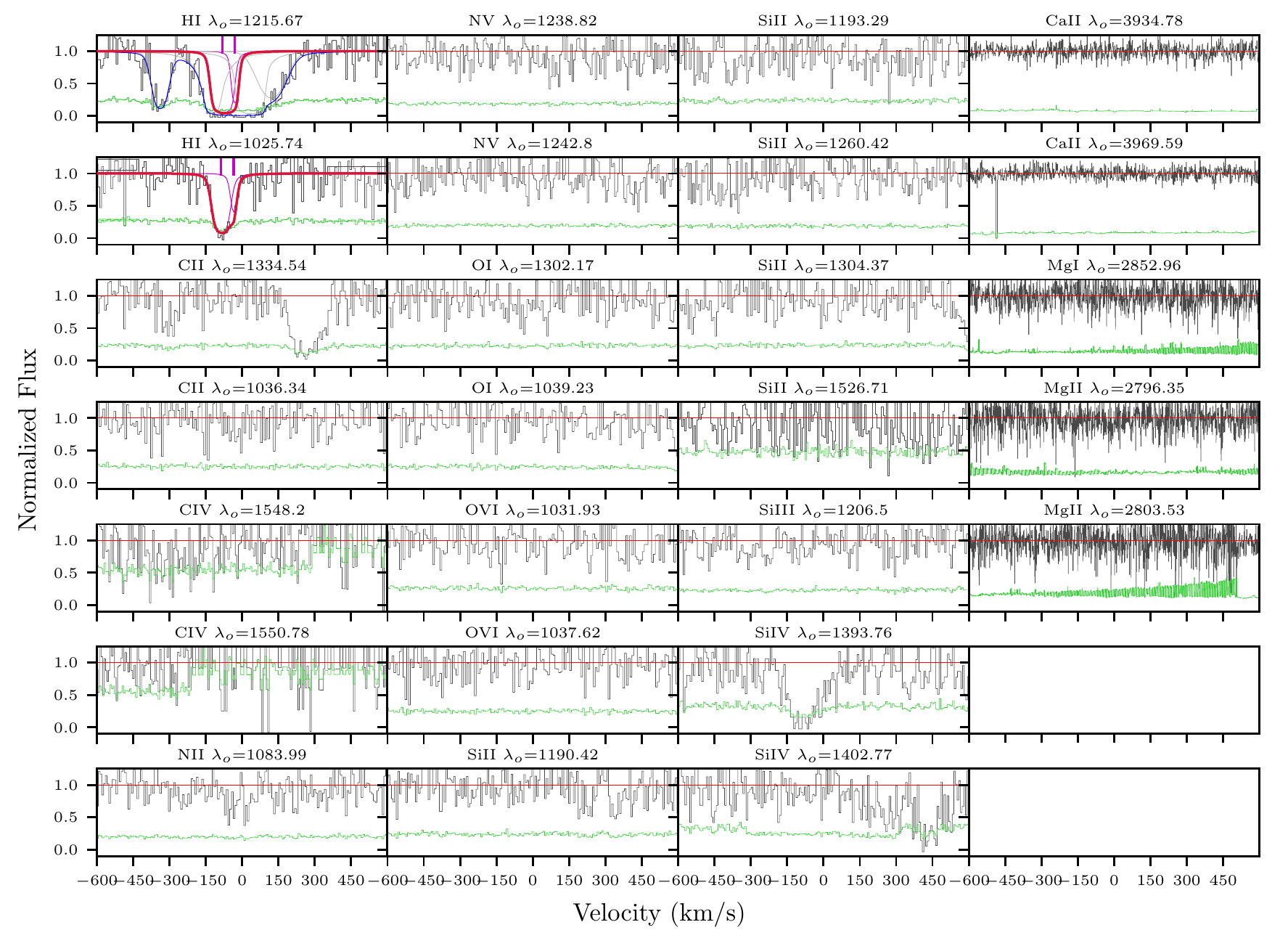}
	\caption{The fits for the system J$0943$, $z_{gal} = 0.1431$, as for figure \ref{fig:Q0122_0.2119}. Note that no metal lines are detected here.}
	\label{fig:J0914_1431}
\end{figure*}
\clearpage
\begin{deluxetable}{ccc}[hp]
	\tablecolumns{8}
	\tablewidth{\linewidth}
	\setlength{\tabcolsep}{0.06in}
	\tablecaption{J$0943$, $z_{gal} = 0.1431$ Column Densities\label{tab:J0943_1431}}
	\tablehead{
		\colhead{Ion}           	&
        \colhead{$\log N~({\cms})$}    &
		\colhead{$\log N$ Error~({\cms})}}
	\startdata
	{\HI}   & $[15.45, 17.20]$  &$\cdots$\\
{\CII}  & $<13.58$  &$\cdots$\\
{\NII}  & $<13.84$  &$\cdots$\\
{\NV}   & $<13.44$  &$\cdots$\\
{\OI}   & $<14.06$  &$\cdots$\\
{\SiII} & $<13.03$  &$\cdots$\\
{\SiIII}& $<12.48$  &$\cdots$\\
{\SiIV} & $<13.26$  &$\cdots$\\
{\CaII} & $<11.34$  &$\cdots$\\
{\MgI}  & $<11.30$  &$\cdots$\\
{\MgII} & $<11.87$  &$\cdots$\\[-5pt]

	\enddata
\end{deluxetable}
\begin{figure}[hp]
	\centering
	\includegraphics[width=\linewidth]{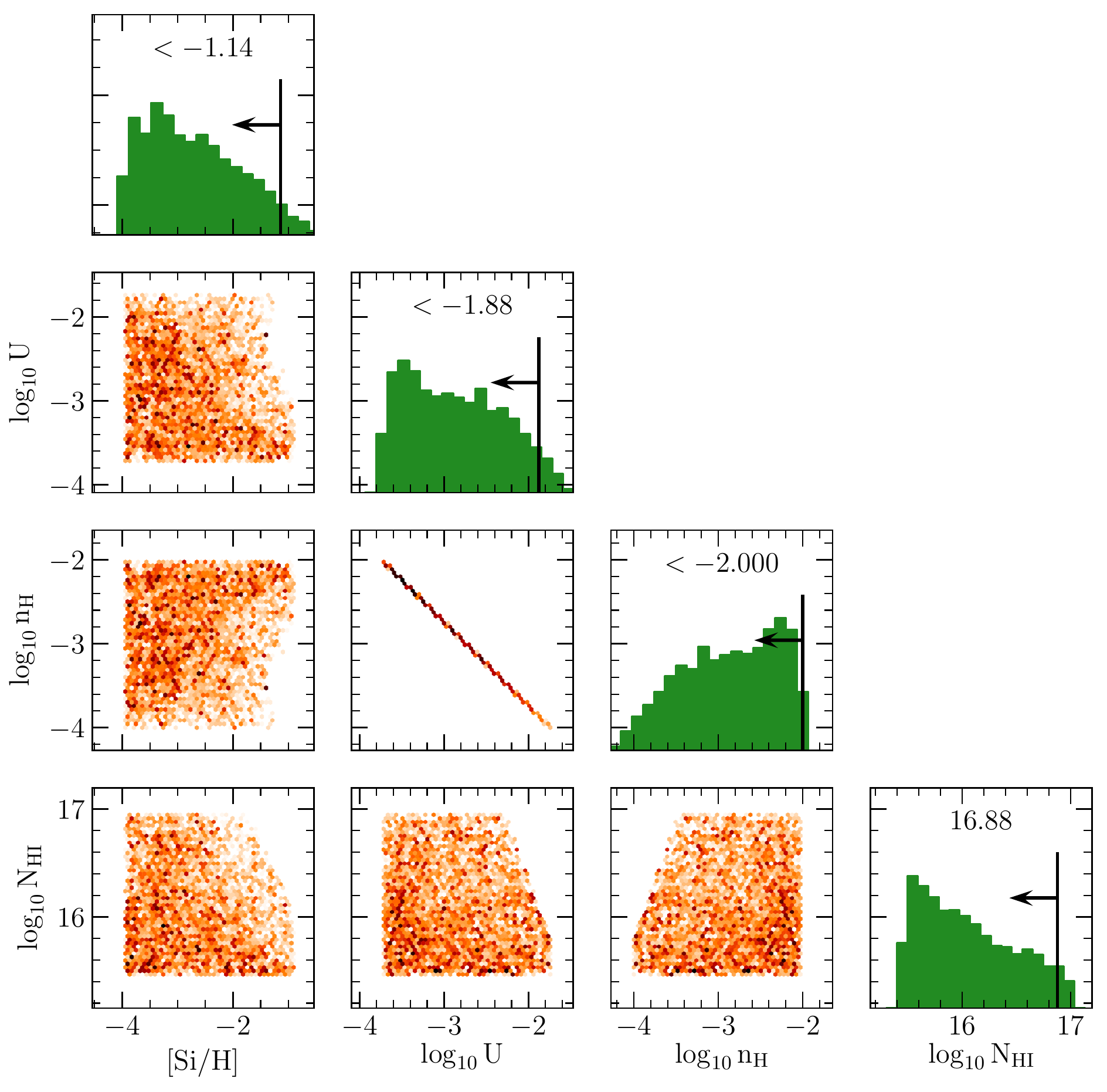}
	\caption{The posterior distribution profiles from the MCMC analysis of the Cloudy grids for J$0943$, $z_{gal} = 0.1431$, as for figure \ref{fig:Q0122_0.2119_par}.}
	\label{fig:J0943_1431_par}
\end{figure}
\newpage
\begin{deluxetable}{ccc}[hp]
	\tablecolumns{8}
	\tablewidth{\linewidth}
	\setlength{\tabcolsep}{0.06in}
	\tablecaption{J$0943$, $z_{gal} = 0.2284$ Column Densities\label{tab:J0943_2284}}
	\tablehead{
		\colhead{Ion}           	&
        \colhead{$\log N~({\cms})$}    &
		\colhead{$\log N$ Error~({\cms})}}
	\startdata
	{\HI}   & $16.03$   &$0.67$\\
{\CII}  & $<13.56$   &$\cdots$\\
{\CIII} & $14.18$   &$0.39$\\
{\NII}  & $<13.70$   &$\cdots$\\
{\NV}   & $<13.47$   &$\cdots$\\
{\OI}   & $<14.26$   &$\cdots$\\
{\SiII} & $<12.95$   &$\cdots$\\
{\SiIII}& $13.10$   &$0.09$\\
{\SiIV} & $<13.79$   &$\cdots$\\
{\CaII} & $<11.34$   &$\cdots$\\
{\MgI}  & $<11.21$   &$\cdots$\\
{\MgII} & $<11.67$   &$\cdots$\\
{\MnII} & $<12.17$   &$\cdots$\\
{\FeII} & $<12.29$   &$\cdots$\\

	\enddata
\end{deluxetable}
\begin{figure}[hp]
	\centering
	\includegraphics[width=\linewidth]{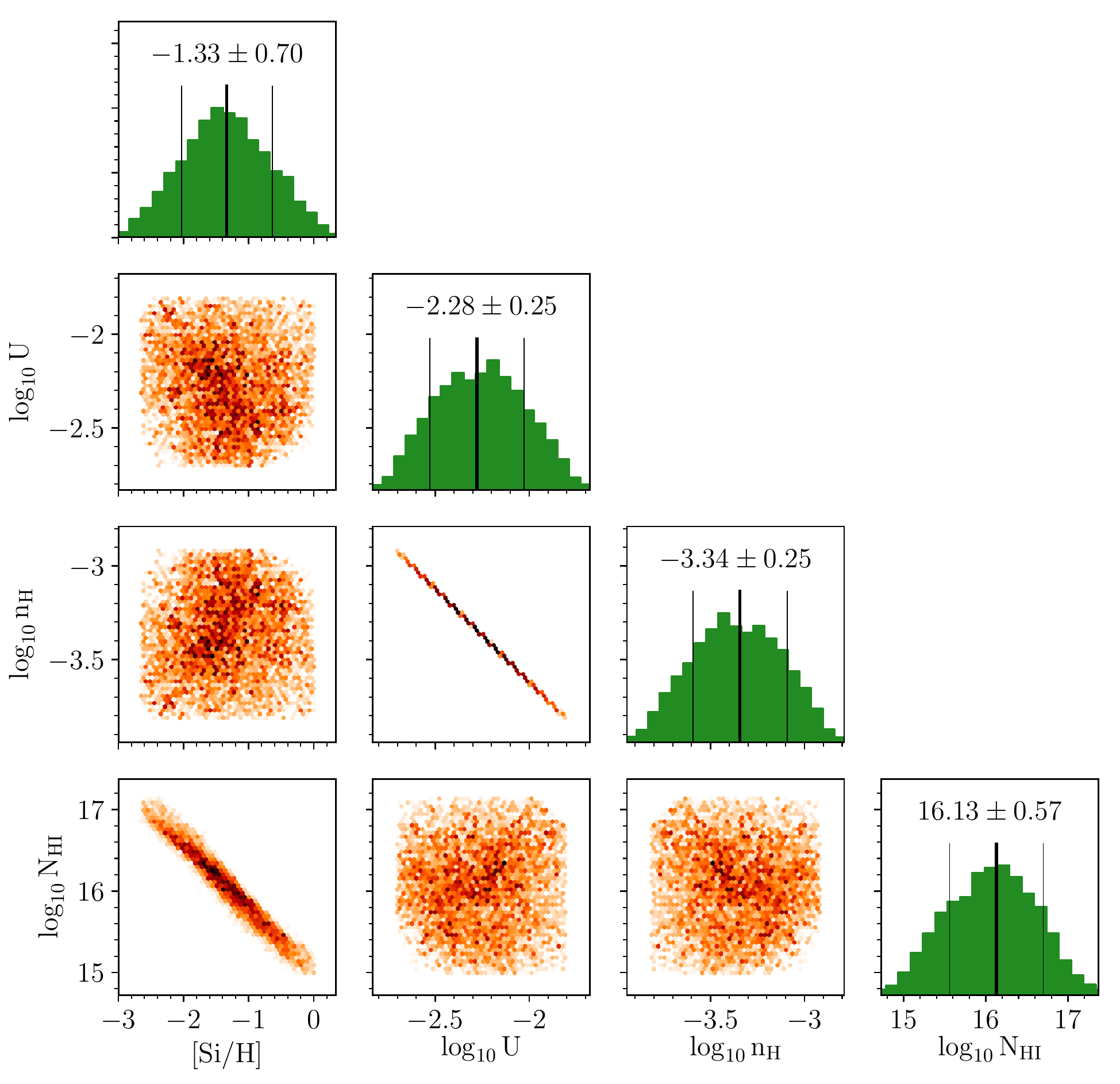}
	\caption{The posterior distribution profiles from the MCMC analysis of the Cloudy grids for J$0943$, $z_{gal} = 0.2284$, as for figure \ref{fig:Q0122_0.2119_par}.}
	\label{fig:J0943_2284_par}
\end{figure}
\begin{figure*}[hp]
	\centering
	\includegraphics[width=\linewidth]{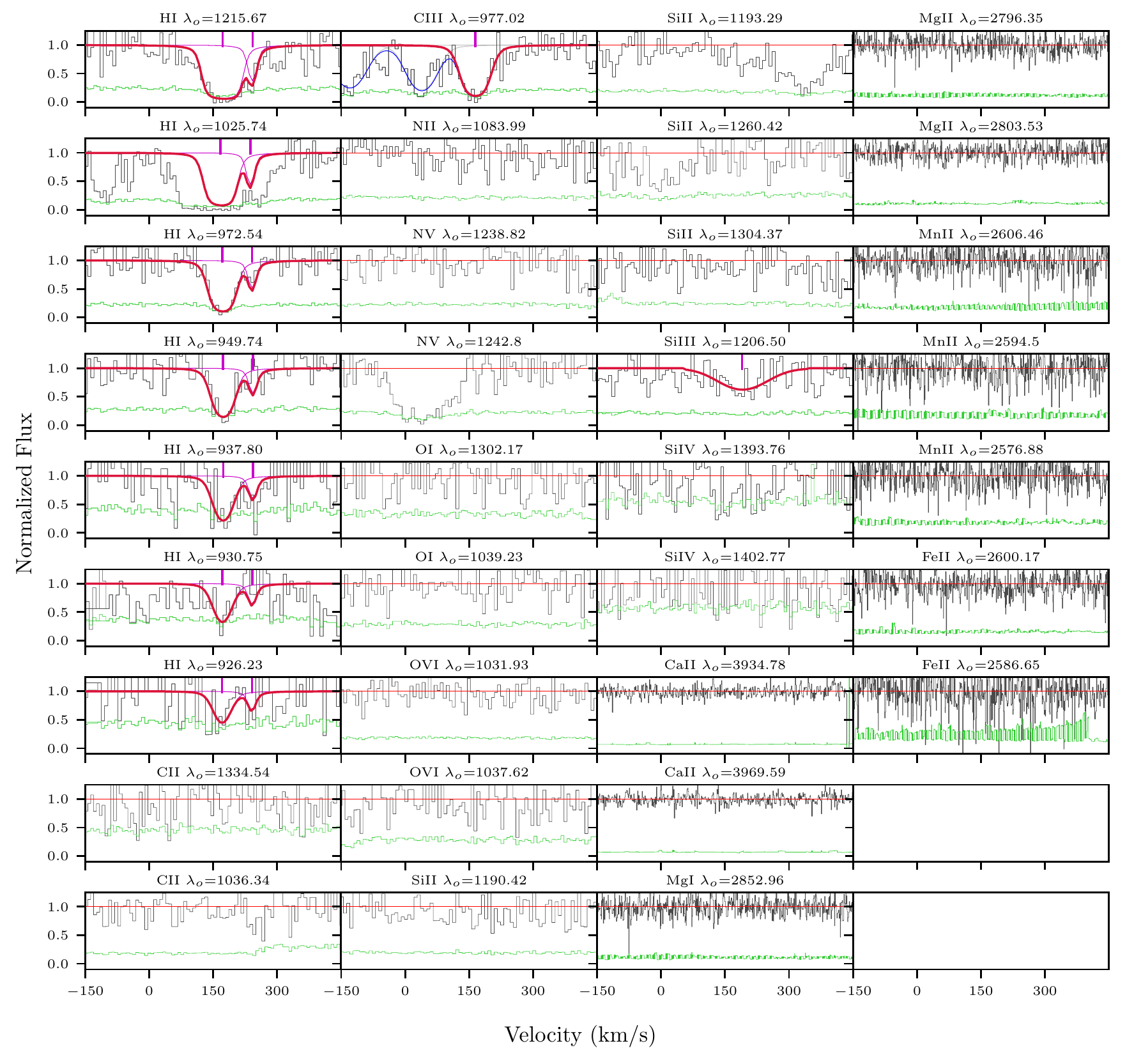}
	\caption{The fits for the system J$0943$, $z_{gal} = 0.2284$, as for figure \ref{fig:Q0122_0.2119}. Although there are absorption lines (blue) near {\CII}, we are confident that this line is real given the existance of {\SiIII} at the same redshift.}
	\label{fig:J0914_2284}
\end{figure*}

\begin{figure*}[hp]
	\centering
	\includegraphics[width=\linewidth]{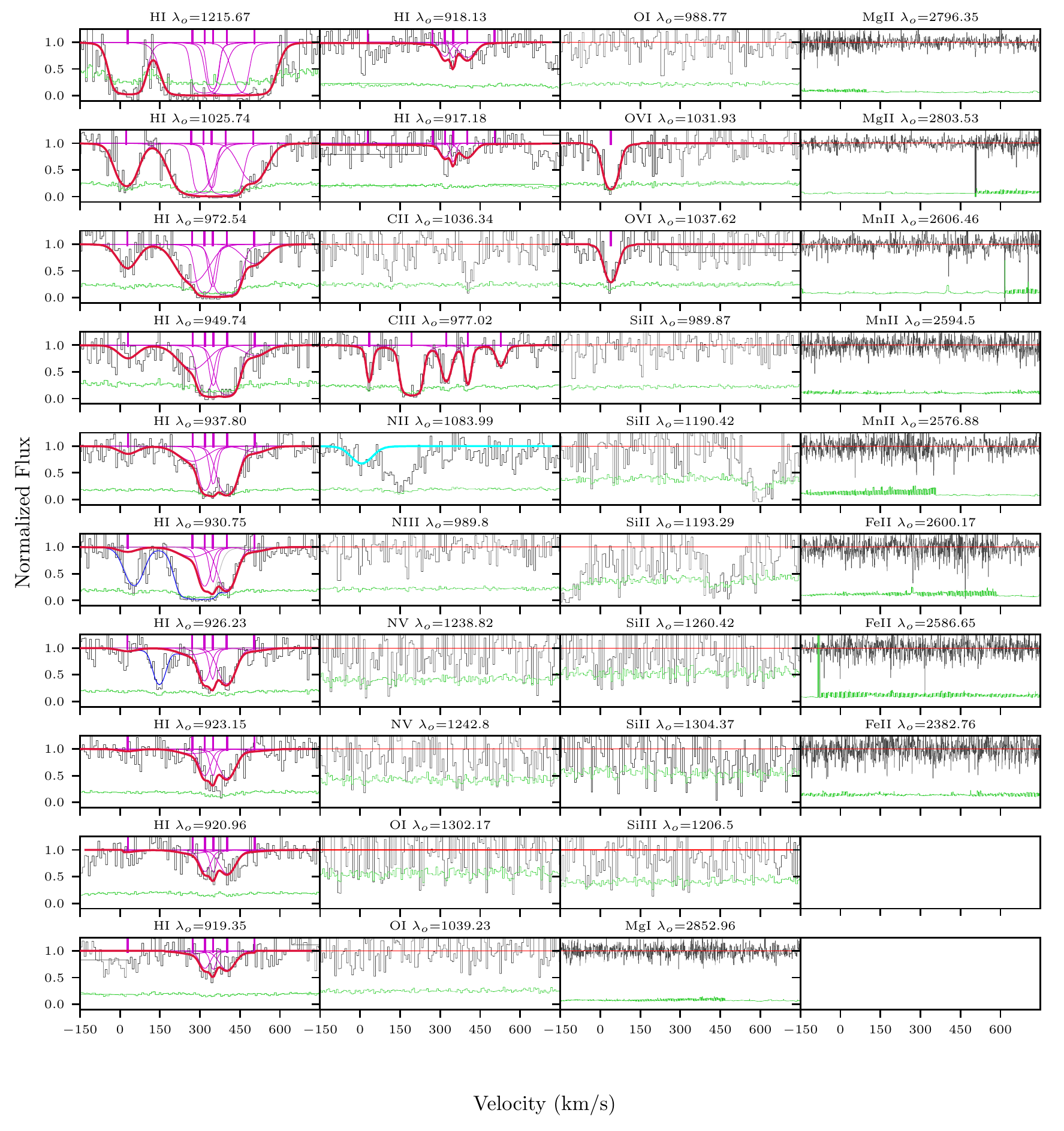}
	\caption{The fits for the system J$0943$, $z_{gal} = 0.353052$, as for figure \ref{fig:Q0122_0.2119}. The absorption profile where {\CII} is expected is the {\OVI} $1037$~{\AA} transition, so we calculate an upper limit from the spectra. The {\NII} absorption profile, in orange, is heavily blended. The column density from the profile fit is taken to the conservative upper limit on the column density. The total {\OVI} fits from \citet{nielsenovi} are shown here for completeness, although they are not used in the ionization modelling.}
	\label{fig:J0914_3531}
\end{figure*}
\clearpage
\begin{deluxetable}{ccc}[hp]
	\tablecolumns{8}
	\tablewidth{\linewidth}
	\setlength{\tabcolsep}{0.06in}
	\tablecaption{J$0943$, $z_{gal} = 0.353052$ Column Densities\label{tab:J0943_3531}}
	\tablehead{
		\colhead{Ion}           	&
        \colhead{$\log N~({\cms})$}    &
		\colhead{$\log N$ Error~({\cms})}}
	\startdata
	{\HI}   & $16.46$   &$0.03$\\
{\CII}  & $<14.44$   &$\cdots$\\
{\CIII} & $16.01$   &$1.92$\\
{\NII}  & $<14.09$   &$\cdots$\\
{\NIII} & $<13.93$   &$\cdots$\\
{\NV}   & $<14.04$   &$\cdots$\\
{\OI}   & $<14.29$   &$\cdots$\\
{\SiII} & $<13.35$   &$\cdots$\\
{\SiIII}& $<12.96$   &$\cdots$\\
{\MgI}  & $<11.24$   &$\cdots$\\
{\MgII} & $<11.82$   &$\cdots$\\
{\MnII} & $<12.22$   &$\cdots$\\
{\FeII} & $<12.39$   &$\cdots$\\[-5pt]

	\enddata
\end{deluxetable}
\begin{figure}[hp]
	\centering
	\includegraphics[width=\linewidth]{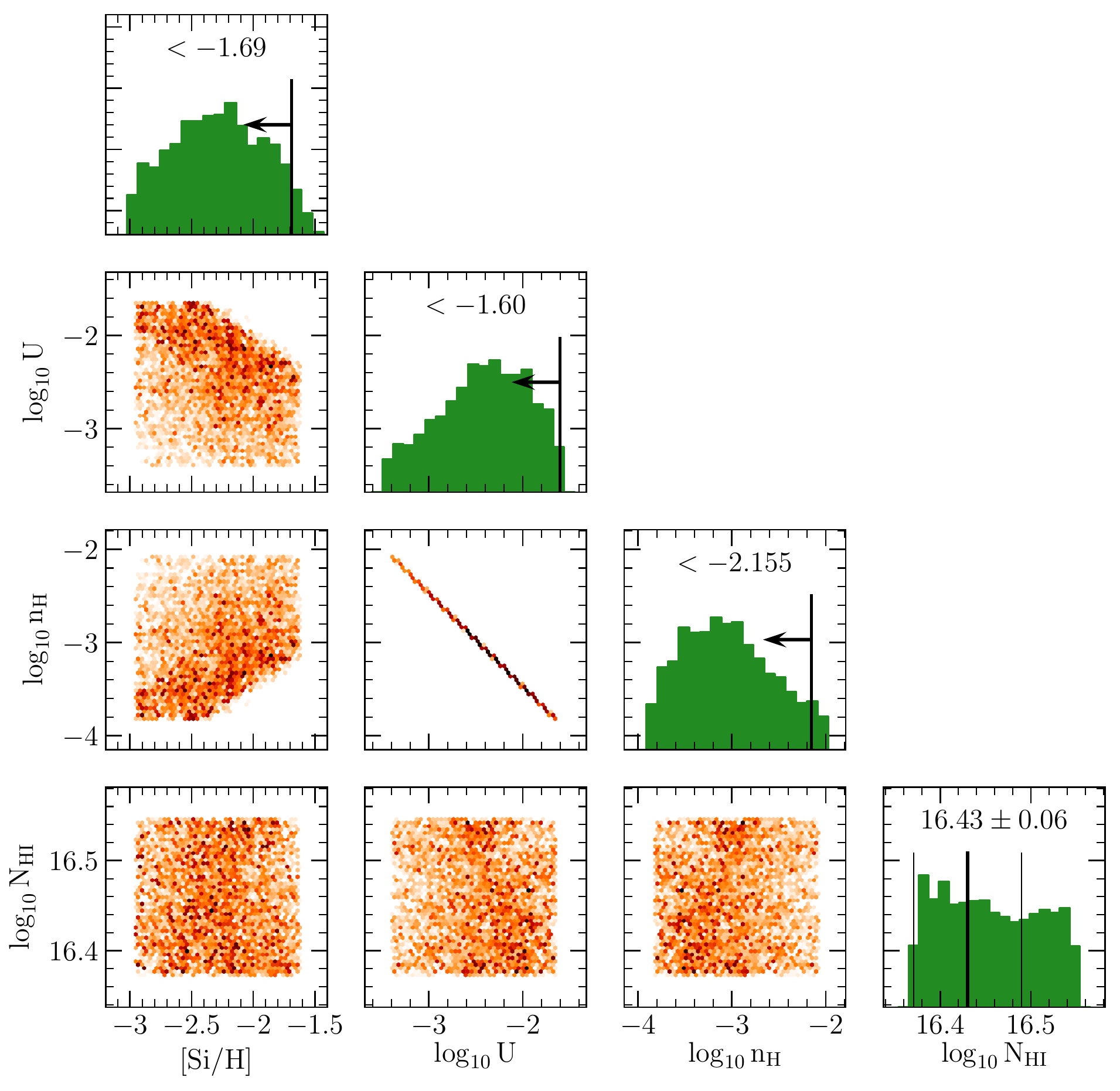}
	\caption{The posterior distribution profiles from the MCMC analysis of the Cloudy grids for J$0943$, $z_{gal} = 0.353052$, as for figure \ref{fig:Q0122_0.2119_par}.}
	\label{fig:J0943_3531_par}
\end{figure}
\newpage
\begin{deluxetable}{ccc}[hp]
	\tablecolumns{8}
	\tablewidth{\linewidth}
	\setlength{\tabcolsep}{0.06in}
	\tablecaption{J$0950$, $z_{gal} = 0.211866$ Column Densities\label{tab:J0950_2119}}
	\tablehead{
		\colhead{Ion}           	&
        \colhead{$\log N~({\cms})$}    &
		\colhead{$\log N$ Error~({\cms})}}
	\startdata
	{\HI}   & $[16.28, 19.00]$   &$\cdots$\\
{\CII}  & $14.71$   &$0.06$\\
{\CIII} & $>14.48$   &$\cdots$\\
{\NII}  & $14.59$   &$0.06$\\
{\NIII} & $>14.95$   &$\cdots$\\
{\NV}   & $13.66$   &$0.11$\\
{\OI}   & $14.35$   &$0.07$\\
{\SiII} & $14.30$   &$0.36$\\
{\SiIII}& $14.18$   &$0.33$\\
{\SiIV} & $13.98$   &$0.10$\\
{\CaII} & $<11.21$   &$\cdots$\\
{\MgI}  & $11.87$   &$0.07$\\
{\MgII} & $13.78$   &$0.34$\\
{\MnII} & $<12.51$   &$\cdots$\\
{\FeII} & $13.16$   &$0.07$\\

	\enddata
\end{deluxetable}
\begin{figure}[hp]
	\centering
	\includegraphics[width=\linewidth]{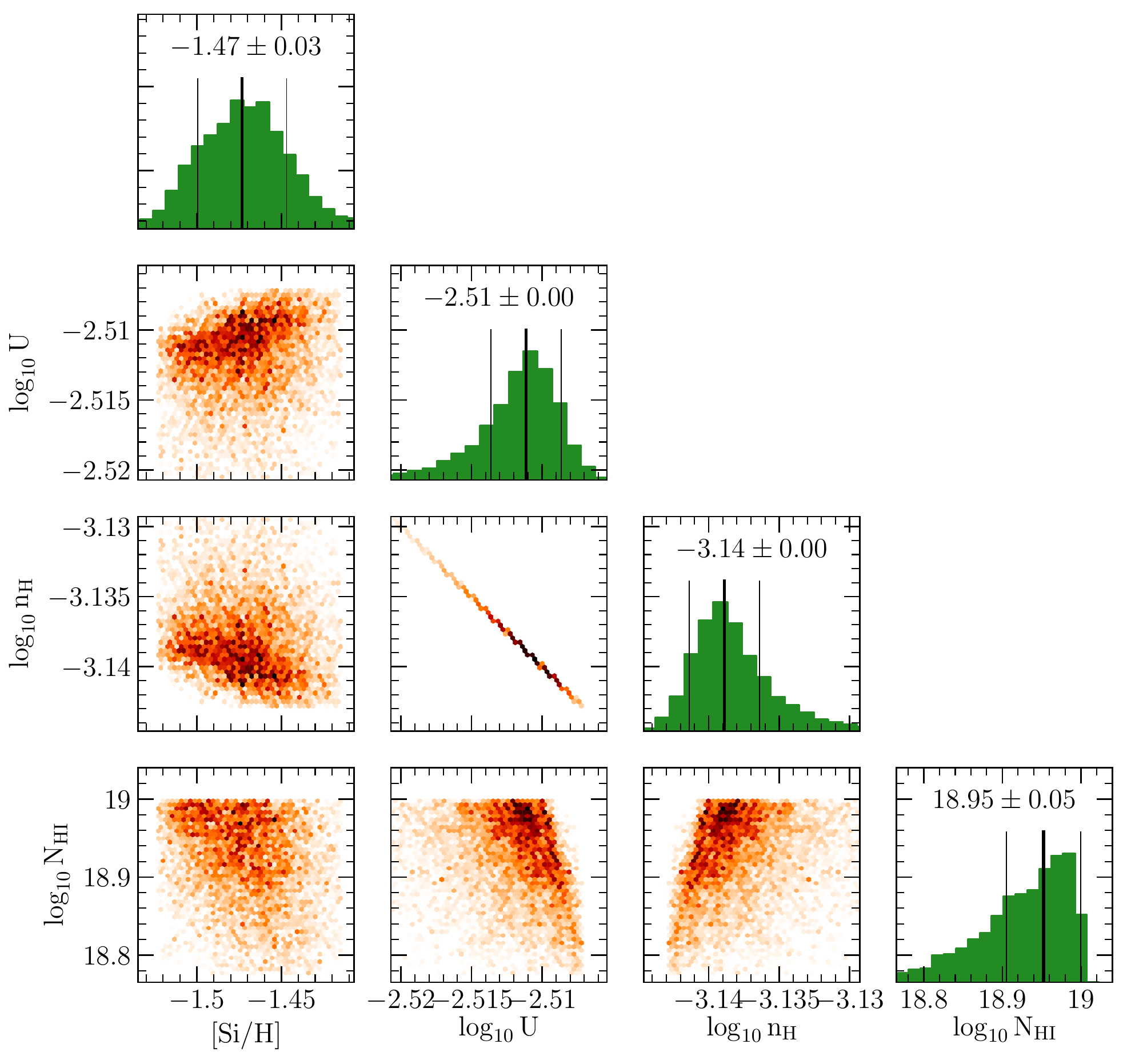}
	\caption{The posterior distribution profiles from the MCMC analysis of the Cloudy grids for J$0950$, $z_{gal} = 0.211866$, as for figure \ref{fig:Q0122_0.2119_par}.}
	\label{fig:J0950_2119_par}
\end{figure}
\begin{figure*}[hp]
	\centering
	\includegraphics[width=\linewidth]{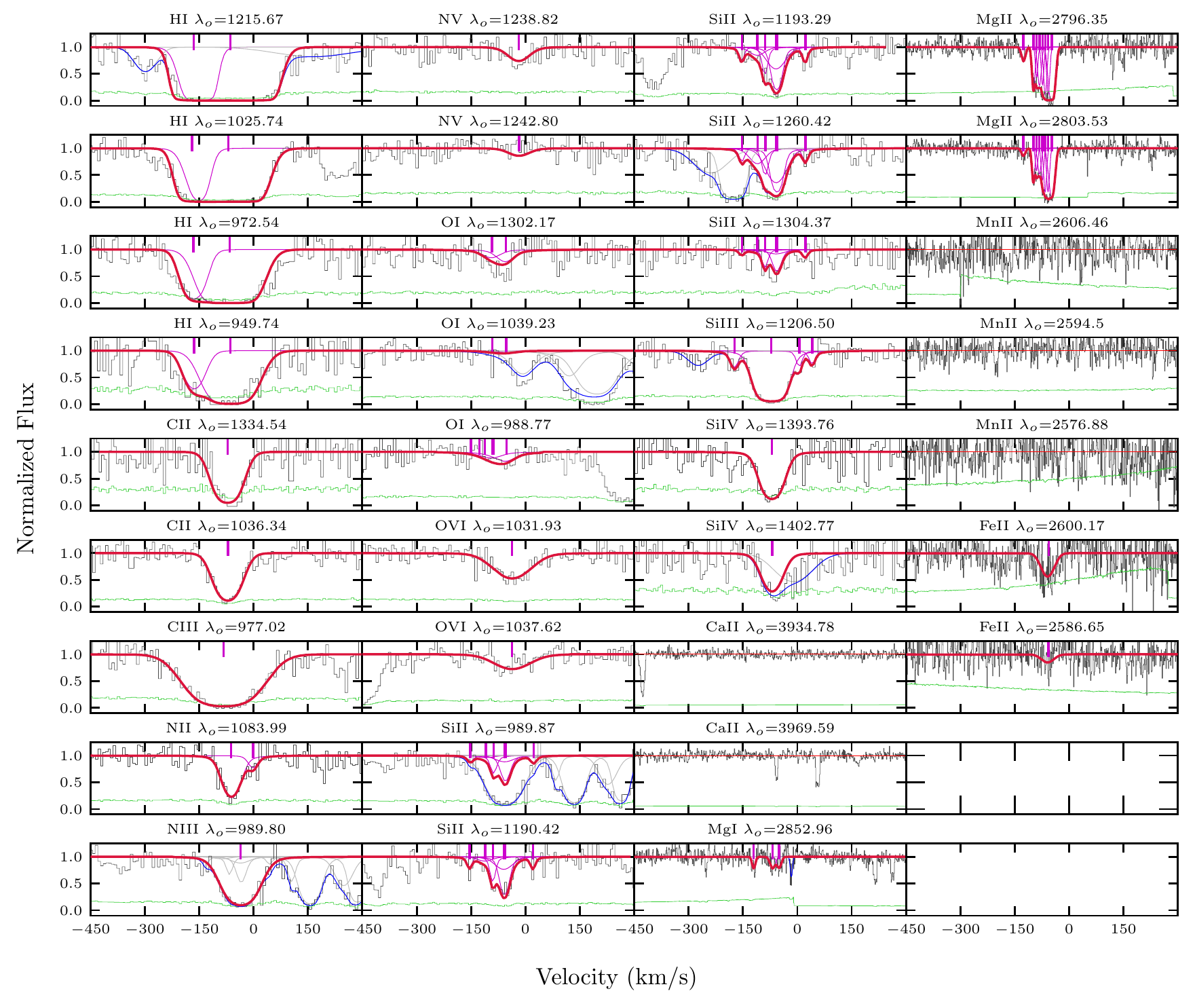}
	\caption{The fits for the system J$0950$, $z_{gal} = 0.211866$, as for figure \ref{fig:Q0122_0.2119}. As the {\CIII} and {\NIII} absorption profiles are saturated, the column densities from the fits are used as lower limits in the metallicity analysis. There are blends (blue) in the {\Lya} transition, although the {\HI} column density is constrained by the other lines. Similarly, the unknown blends in the {\OI} 1039~{\AA} and {\SiIV} 1402~{\AA} lines are constrained by the other {\OI} and {\SiIV} transitions. The {\NIII} and {\SiII} 989~{\AA} transitions are blended. However, the remaining {\SiII}, apart from the 1206~{\AA} transition with an unknown blend, constrain the column densities. The total {\OVI} fits from \citet{nielsenovi} are shown here for completeness, although they are not used in the ionization modelling.}
	\label{fig:J0950_2119}
\end{figure*}

\begin{figure*}[hp]
	\centering
	\includegraphics[width=\linewidth]{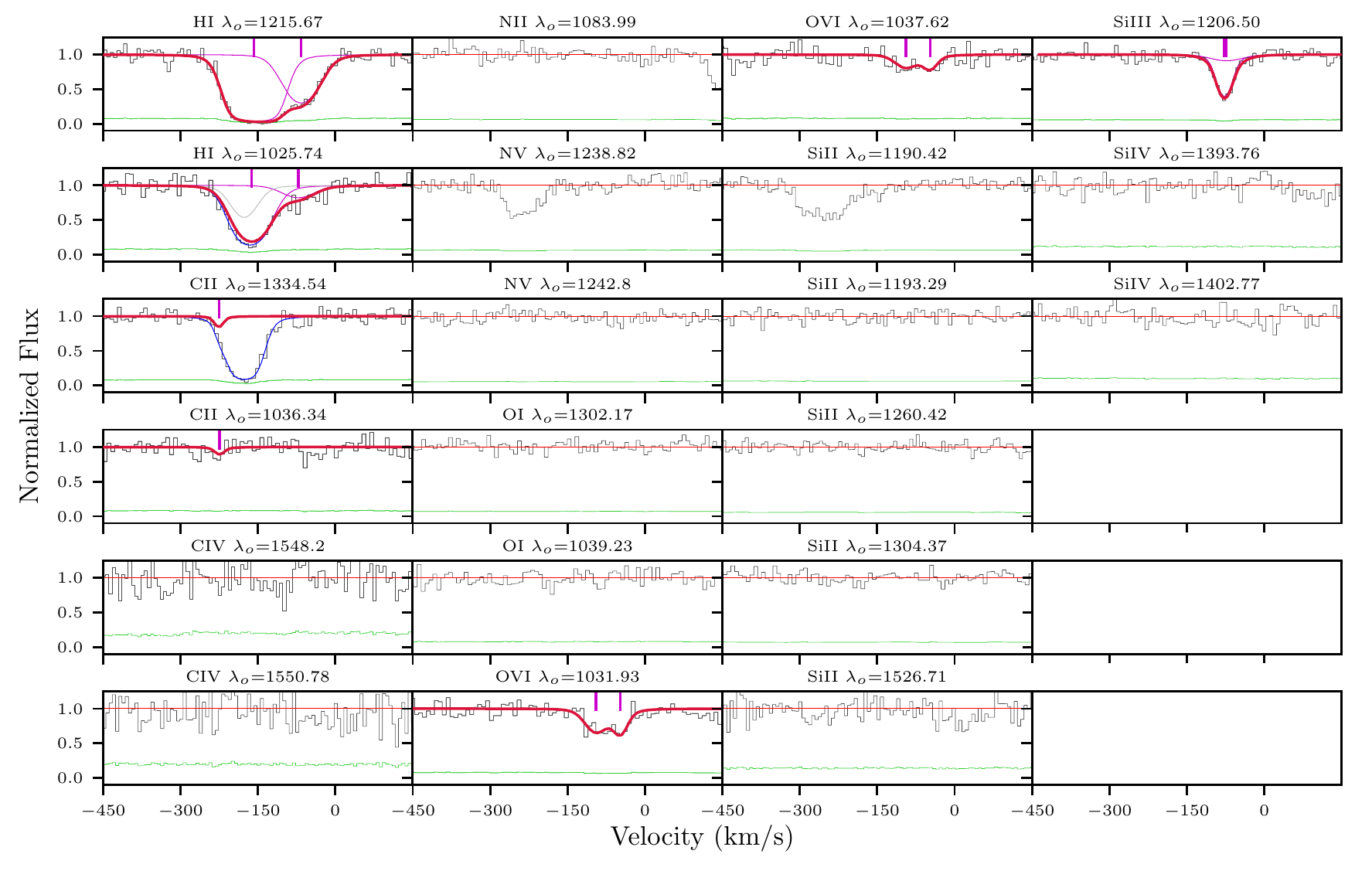}
	\caption{The fits for the system J$1004$, $z_{gal} = 0.1380$, as for figure \ref{fig:Q0122_0.2119}. Note that {\CII} 1334~{\AA} is blended with an unknown line. However, the column density is constrained by the {\CII} 1036~{\AA} line. The total {\OVI} fits from \citet{nielsenovi} are shown here for completeness, although they are not used in the ionization modelling.}
	\label{fig:J1004_1380}
\end{figure*}
\clearpage
\begin{deluxetable}{ccc}[hp]
	\tablecolumns{8}
	\tablewidth{\linewidth}
	\setlength{\tabcolsep}{0.06in}
	\tablecaption{J$1004$, $z_{gal} = 0.1380$ Column Densities\label{tab:J1004_1380}}
	\tablehead{
		\colhead{Ion}           	&
        \colhead{$\log N~({\cms})$}    &
		\colhead{$\log N$ Error~({\cms})}}
	\startdata
	{\HI}   & $14.91$   &$0.14$\\
{\CII}  & $13.07$   &$0.24$\\
{\NII}  & $<13.07$   &$\cdots$\\
{\NV}   & $<12.77$   &$\cdots$\\
{\OI}   & $<13.36$   &$\cdots$\\
{\SiII} & $<11.92$   &$\cdots$\\
{\SiIII}& $13.99$   &$0.03$\\
{\SiIV} & $<12.54$   &$\cdots$\\[-5pt]

	\enddata
\end{deluxetable}
\begin{figure}[hp]
	\centering
	\includegraphics[width=\linewidth]{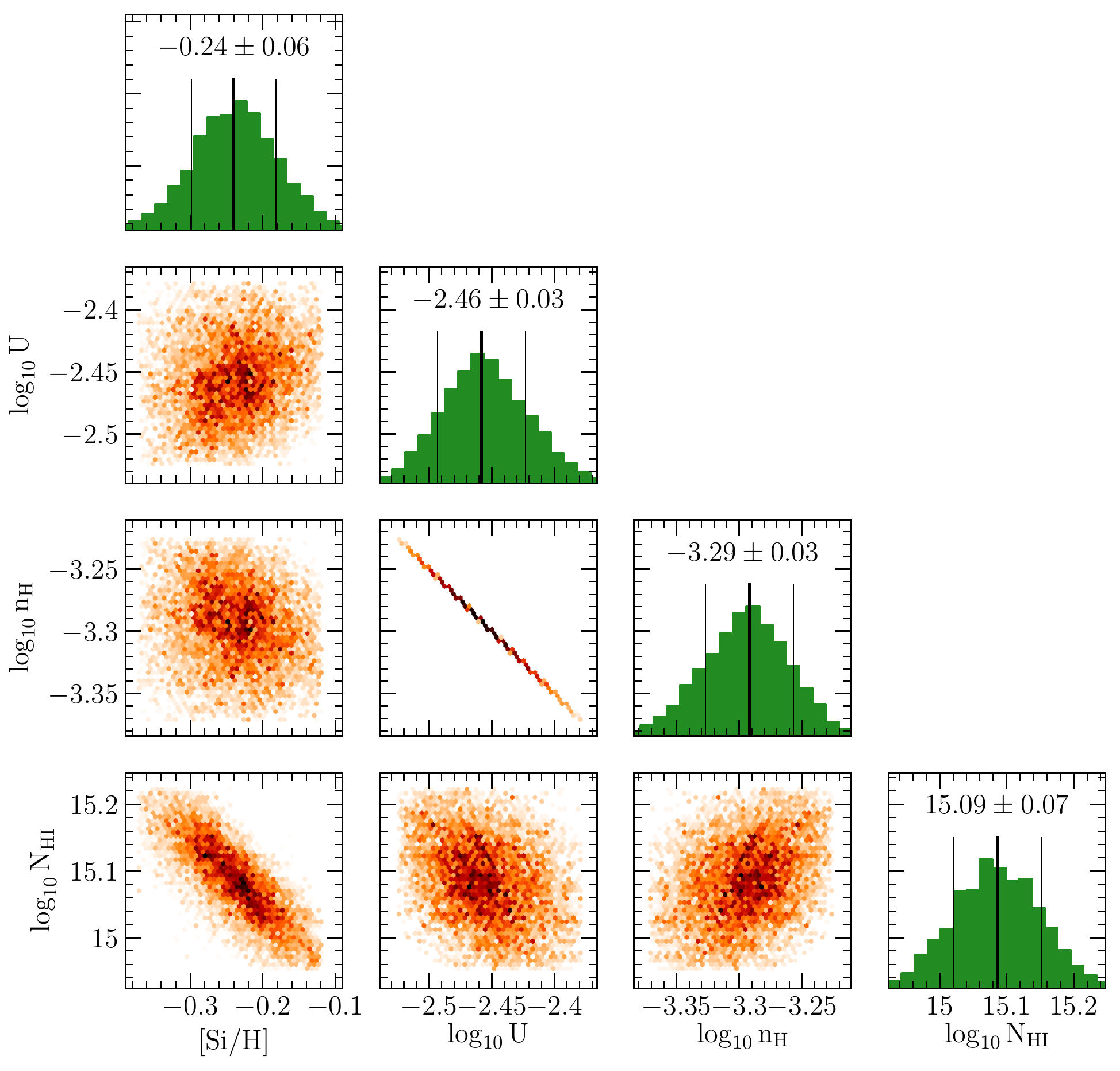}
	\caption{The posterior distribution profiles from the MCMC analysis of the Cloudy grids for J$1004$, $z_{gal} = 0.1380$, as for figure \ref{fig:Q0122_0.2119_par}.}
	\label{fig:J1004_1380_par}
\end{figure}
\newpage
\begin{deluxetable}{ccc}[hp]
	\tablecolumns{8}
	\tablewidth{\linewidth}
	\setlength{\tabcolsep}{0.06in}
	\tablecaption{J$1009$, $z_{gal} = 0.227855$ Column Densities\label{tab:J1009_2278}}
	\tablehead{
		\colhead{Ion}           	&
        \colhead{$\log N~({\cms})$}    &
		\colhead{$\log N$ Error~({\cms})}}
	\startdata
	{\HI}   & $[17.51, 19.00]$   &$\cdots$\\
{\CII}  & $14.55$   &$0.07$\\
{\NII}  & $<13.65$   &$\cdots$\\
{\NV}   & $<13.72$   &$\cdots$\\
{\OI}   & $<14.35$   &$\cdots$\\
{\SiII} & $<13.27$   &$\cdots$\\
{\SiIII}& $13.54$   &$0.27$\\
{\SiIV} & $13.57$   &$0.09$\\
{\CaII} & $<11.34$   &$\cdots$\\
{\MgI}  & $<11.22$   &$\cdots$\\
{\MgII} & $12.61$   &$0.05$\\
{\MnII} & $<12.28$   &$\cdots$\\
{\FeII} & $12.32$   &$0.11$\\[-5pt]

	\enddata
\end{deluxetable}
\begin{figure}[hp]
	\centering
	\includegraphics[width=\linewidth]{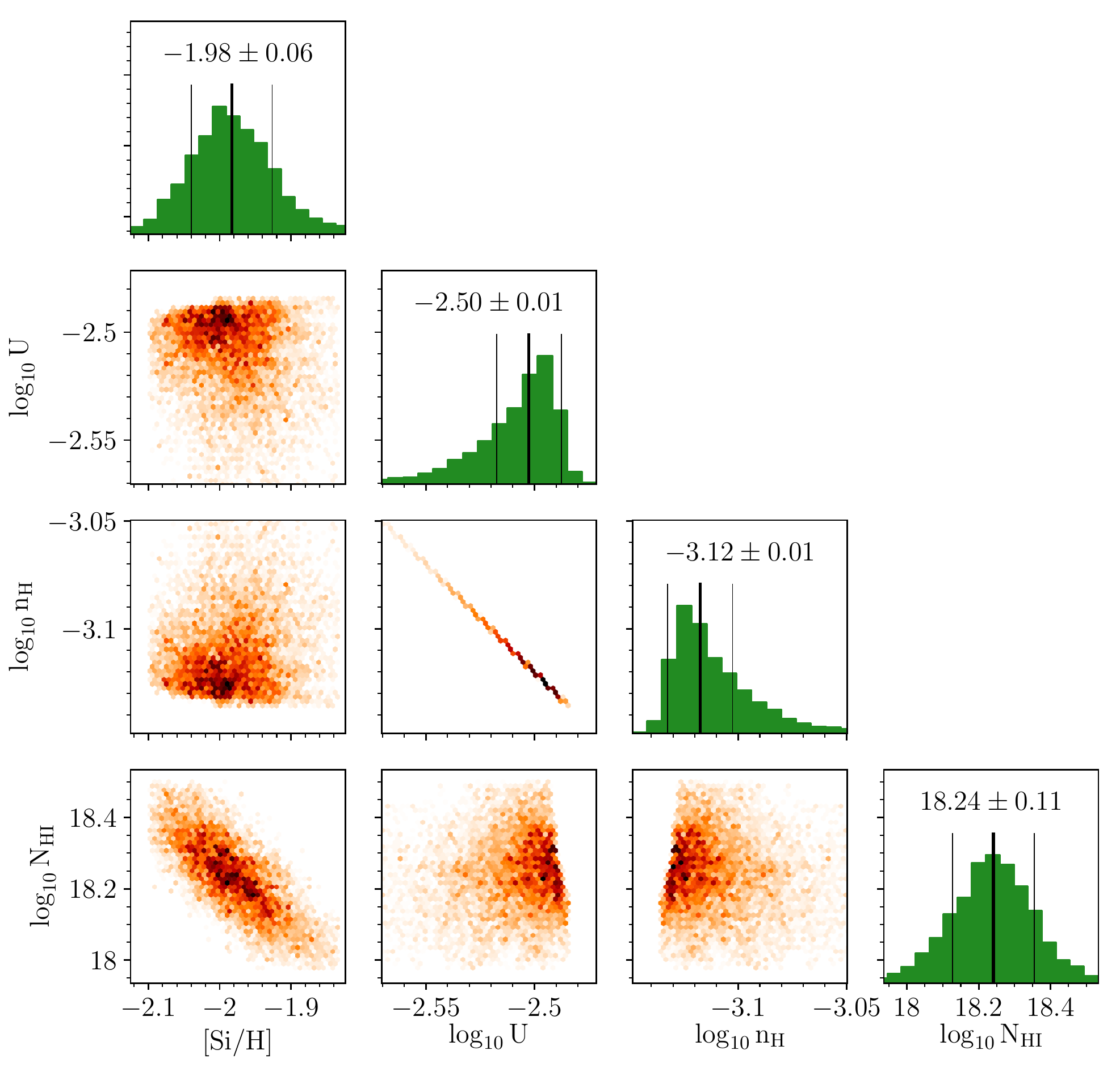}
	\caption{The posterior distribution profiles from the MCMC analysis of the Cloudy grids for J$1009$, $z_{gal} = 0.227855$, as for figure \ref{fig:Q0122_0.2119_par}.}
	\label{fig:J1009_2278_par}
\end{figure}
\begin{figure*}[hp]
	\centering
	\includegraphics[width=\linewidth]{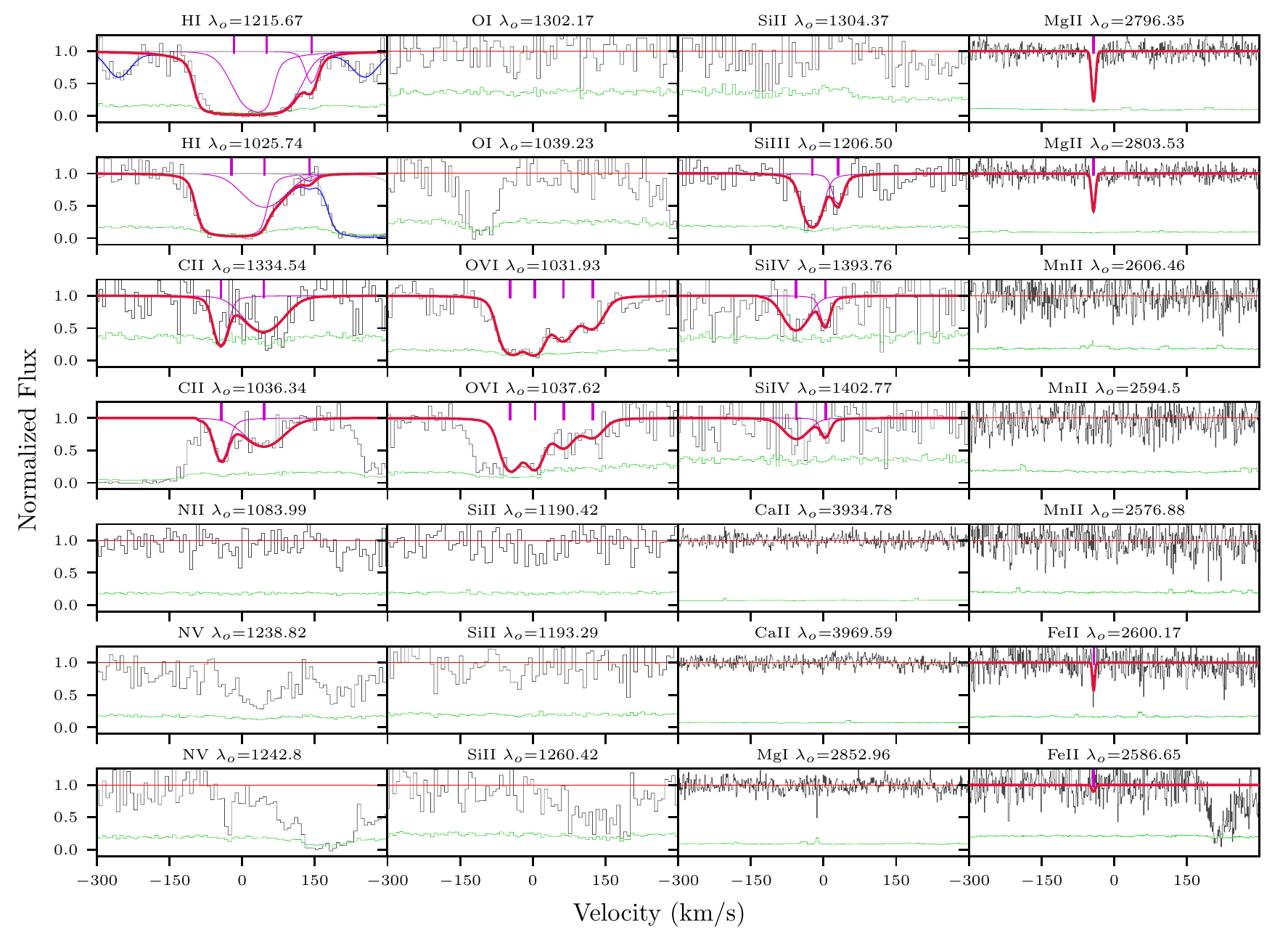}
	\caption{The fits for the system J$1009$, $z_{gal} = 0.227855$, as for figure \ref{fig:Q0122_0.2119}. The total {\OVI} fits from \citet{nielsenovi} are shown here for completeness, although they are not used in the ionization modelling.}
	\label{fig:J1009_2278}
\end{figure*}

\begin{figure*}[hp]
	\centering
	\includegraphics[width=\linewidth]{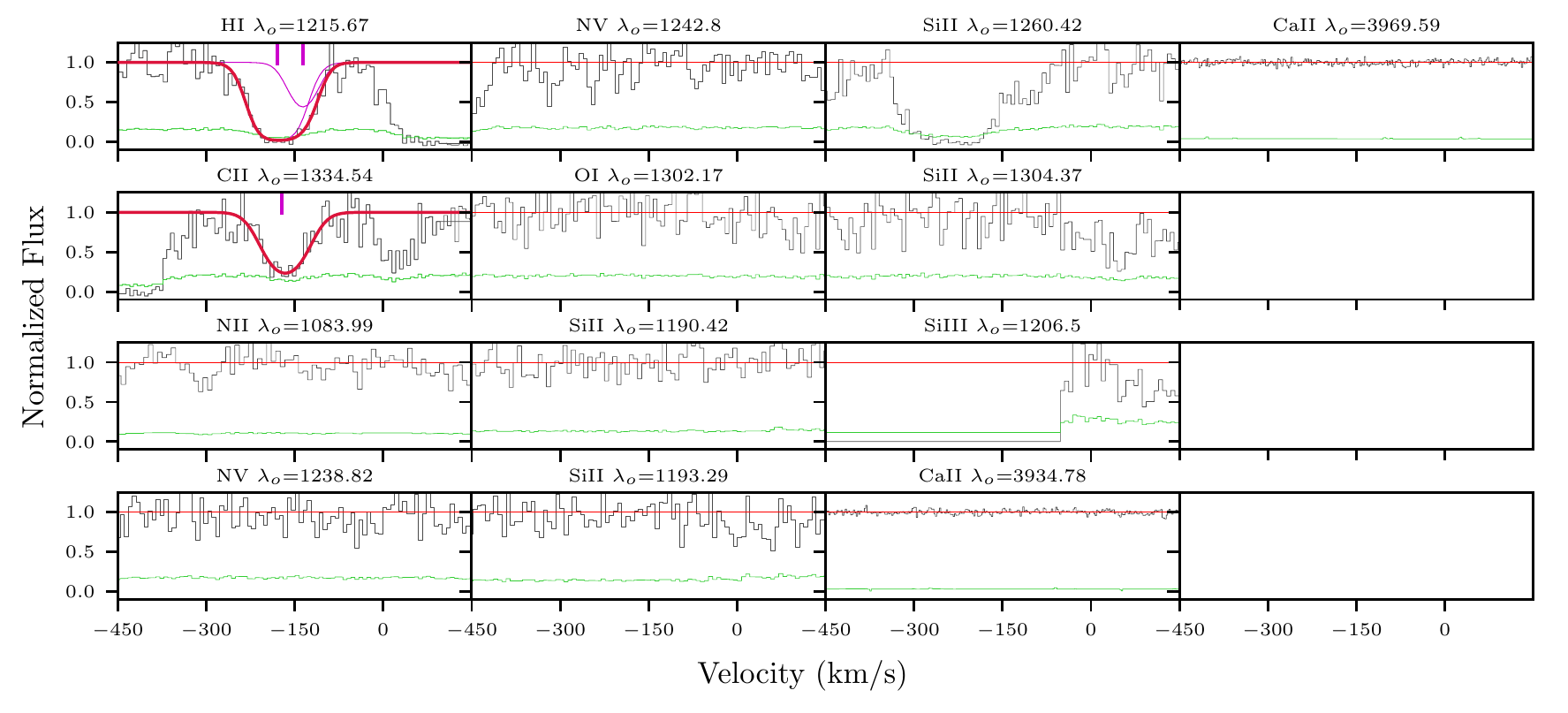}
	\caption{The fits for the system J$1041$, $z_{gal} = 0.3153$, as for figure \ref{fig:Q0122_0.2119}.}
	\label{fig:J1041_3153}
\end{figure*}
\clearpage
\begin{deluxetable}{ccc}[hp]
	\tablecolumns{8}
	\tablewidth{\linewidth}
	\setlength{\tabcolsep}{0.06in}
	\tablecaption{J$1041$, $z_{gal} = 0.3153$ Column Densities\label{tab:J1041_3153}}
	\tablehead{
		\colhead{Ion}           	&
        \colhead{$\log N~({\cms})$}    &
		\colhead{$\log N$ Error~({\cms})}}
	\startdata
	{\HI}   & $[14.43, 17.20]$   &$\cdots$\\
{\CII}  & $14.40$   &$0.05$\\
{\NII}  & $<13.34$   &$\cdots$\\
{\NV}   & $<13.36$   &$\cdots$\\
{\OI}   & $<13.91$   &$\cdots$\\
{\SiII} & $<12.60$   &$\cdots$\\
{\SiIII}& $<12.59$   &$\cdots$\\
{\CaII} & $<11.10$   &$\cdots$\\[-5pt]

	\enddata
\end{deluxetable}
\begin{figure}[hp]
	\centering
	\includegraphics[width=\linewidth]{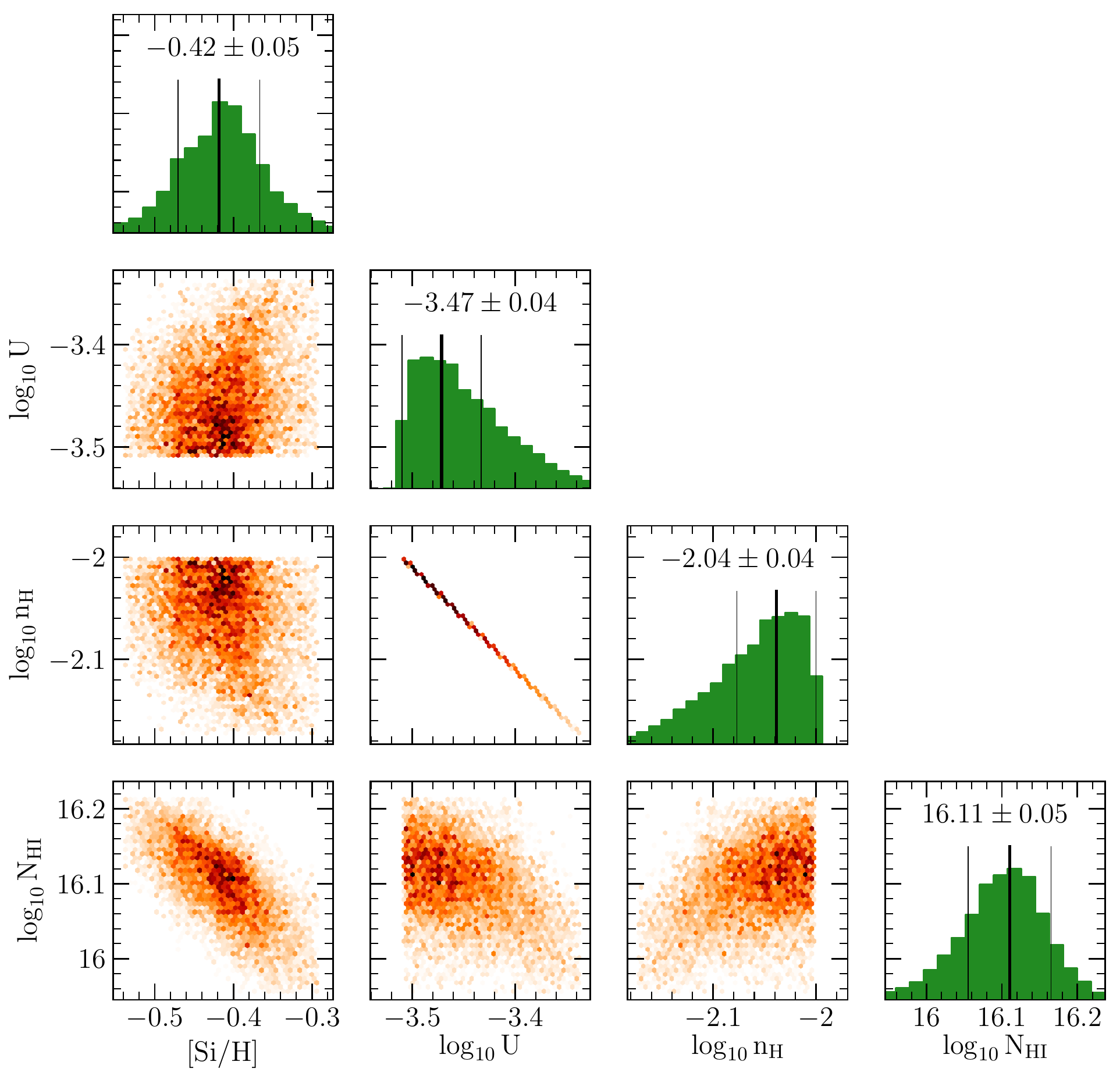}
	\caption{The posterior distribution profiles from the MCMC analysis of the Cloudy grids for J$1041$, $z_{gal} = 0.3153$, as for figure \ref{fig:Q0122_0.2119_par}.}
	\label{fig:J1041_3153_par}
\end{figure}
\newpage
\begin{deluxetable}{ccc}[hp]
	\tablecolumns{8}
	\tablewidth{\linewidth}
	\setlength{\tabcolsep}{0.06in}
	\tablecaption{J$1041$, $z_{gal} = 0.442173$ Column Densities\label{tab:J1041_4422}}
	\tablehead{
		\colhead{Ion}           	&
        \colhead{$\log N~({\cms})$}    &
		\colhead{$\log N$ Error~({\cms})}}
	\startdata
	{\HI}   & $[16.77, 19.00]$   &$\cdots$\\
{\CII}  & $15.05$   &$0.19$\\
{\CIII} & $>15.30$   &$\cdots$\\
{\NII}  & $14.47$   &$0.05$\\
{\NIII} & $14.80$   &$0.05$\\
{\OI}   & $<13.78$   &$\cdots$\\
{\SiII} & $13.81$   &$0.05$\\
{\SiIII}& $>14.14$   &$\cdots$\\
{\CaII} & $<11.04$   &$\cdots$\\
{\MgI}  & $11.59$   &$0.06$\\
{\MgII} & $13.67$   &$0.02$\\
{\MnII} & $<12.73$   &$\cdots$\\
{\FeII} & $<12.92$   &$\cdots$\\[-5pt]

	\enddata
\end{deluxetable}
\begin{figure}[hp]
	\centering
	\includegraphics[width=\linewidth]{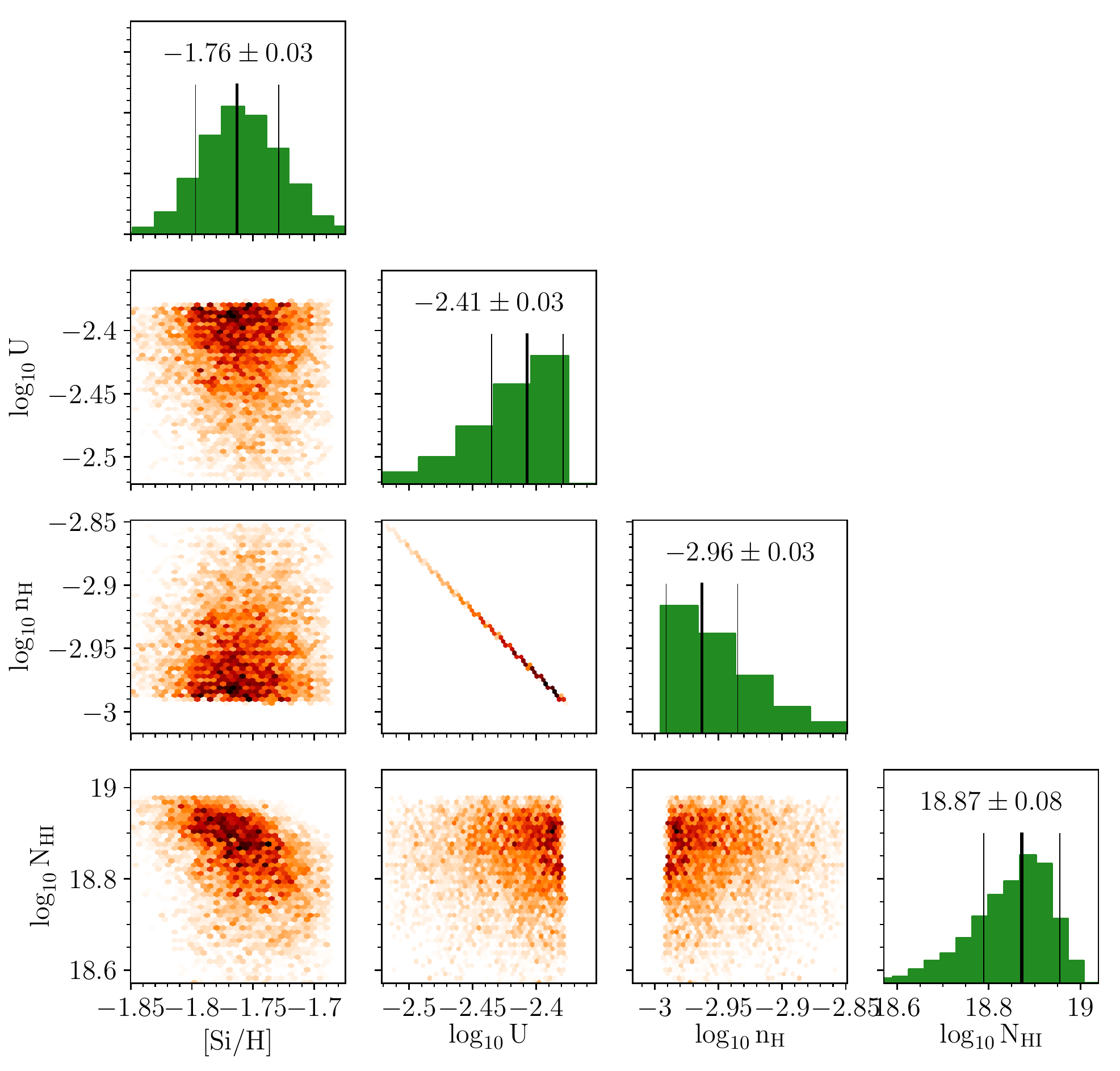}
	\caption{The posterior distribution profiles from the MCMC analysis of the Cloudy grids for J$1041$, $z_{gal} = 0.442173$, as for figure \ref{fig:Q0122_0.2119_par}.}
	\label{fig:J1041_4422_par}
\end{figure}
\begin{figure*}[hp]
	\centering
	\includegraphics[width=\linewidth]{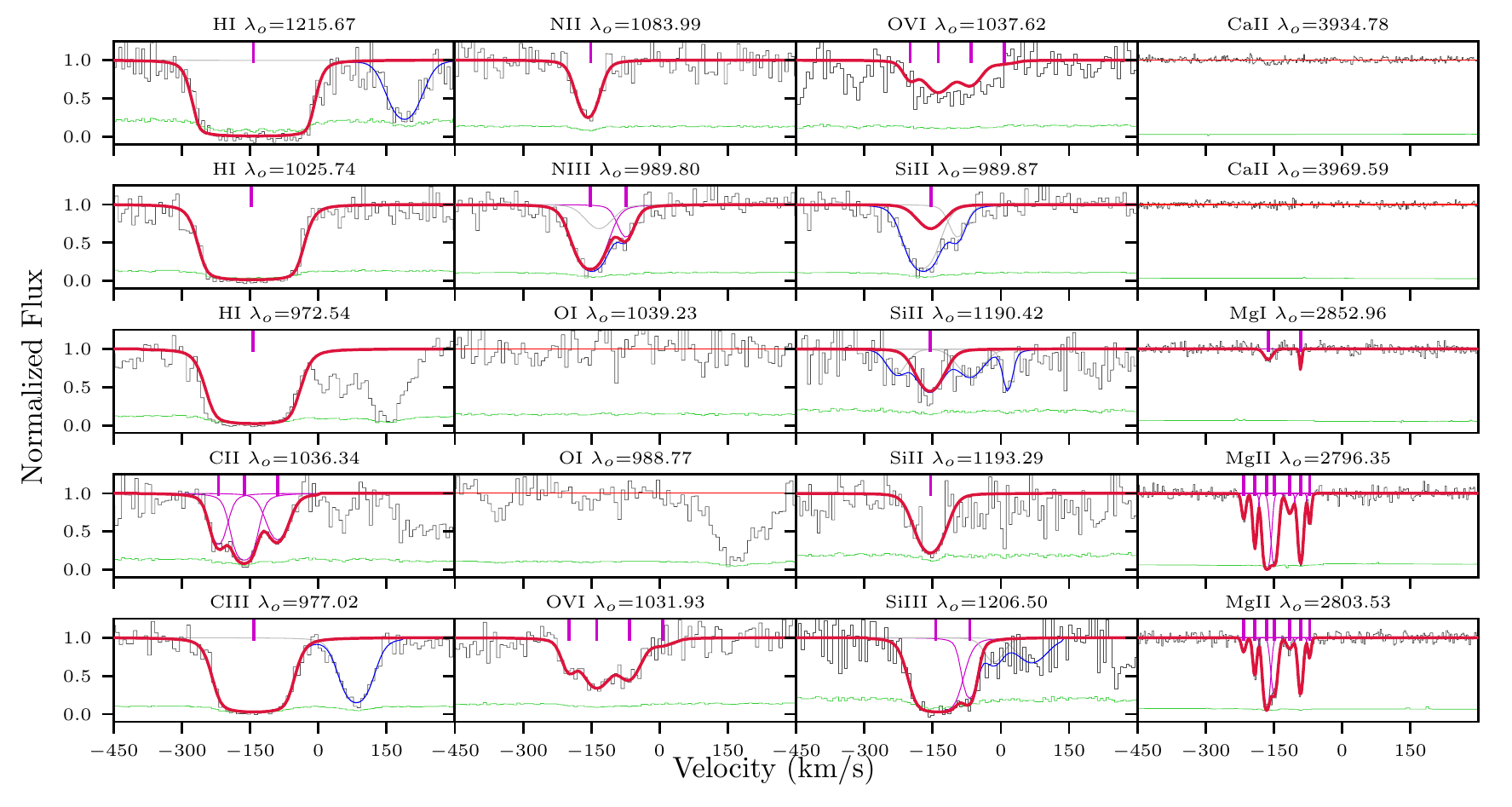}
	\caption{The fits for the system J$1041$, $z_{gal} = 0.442173$, as for figure \ref{fig:Q0122_0.2119}. The {\NIII} and {\SiII} 989~{\AA} lines are blended together, while there are unknown blends in the {\SiII} 1190~{\AA} transition. However, the {\SiII} 1193~{\AA} transition constrains the column densities. Additional unknown blend exits on the sides of the {\CIII} and {\SiIII} transitions, which were simple to remove. The total {\OVI} fits from \citet{nielsenovi} are shown here for completeness, although they are not used in the ionization modelling. }
	\label{fig:J1041_4422}
\end{figure*}

\begin{figure*}[hp]
	\centering
	\includegraphics[width=\linewidth]{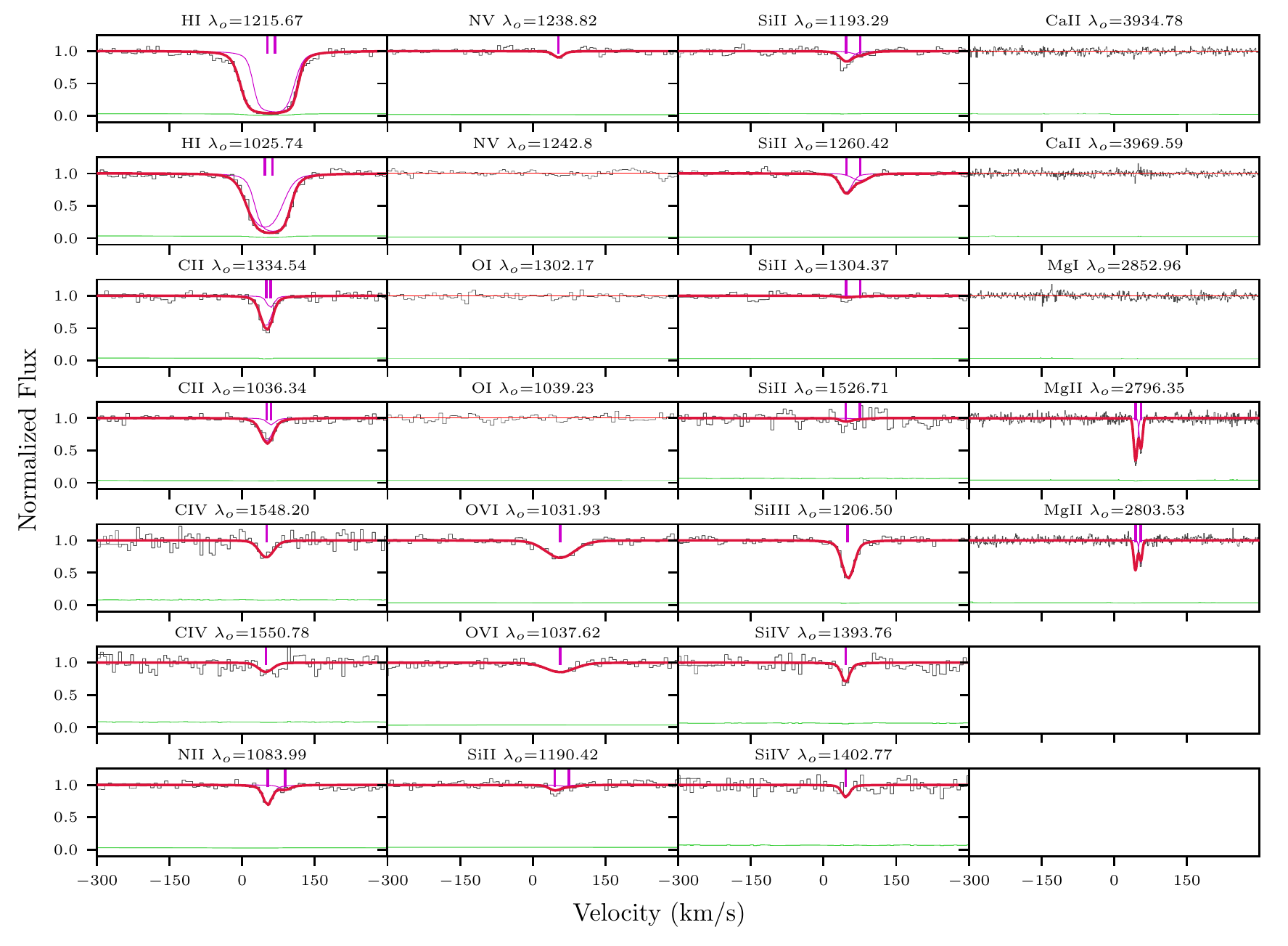}
	\caption{The fits for the system J$1119$, $z_{gal} = 0.1383$, as for figure \ref{fig:Q0122_0.2119}. The total {\OVI} fits from \citet{nielsenovi} are shown here for completeness, although they are not used in the ionization modelling.}
	\label{fig:J1119_1383}
\end{figure*}
\clearpage
\begin{deluxetable}{ccc}[hp]
	\tablecolumns{8}
	\tablewidth{\linewidth}
	\setlength{\tabcolsep}{0.06in}
	\tablecaption{J$1119$, $z_{gal} = 0.1383$ Column Densities\label{tab:J1119_1383}}
	\tablehead{
		\colhead{Ion}           	&
        \colhead{$\log N~({\cms})$}    &
		\colhead{$\log N$ Error~({\cms})}}
	\startdata
	{\HI}   & $15.64$   &$0.32$\\
{\CII}  & $13.99$   &$0.74$\\
{\NII}  & $13.72$   &$0.16$\\
{\OI}   & $<13.03$   &$\cdots$\\
{\SiII} & $12.66$   &$0.02$\\
{\SiIII}& $12.87$   &$0.03$\\
{\CaII} & $<10.90$   &$\cdots$\\
{\MgI}  & $<10.62$   &$\cdots$\\
{\MgII} & $12.72$   &$0.05$\\[-5pt]

	\enddata
\end{deluxetable}
\begin{figure}[hp]
	\centering
	\includegraphics[width=\linewidth]{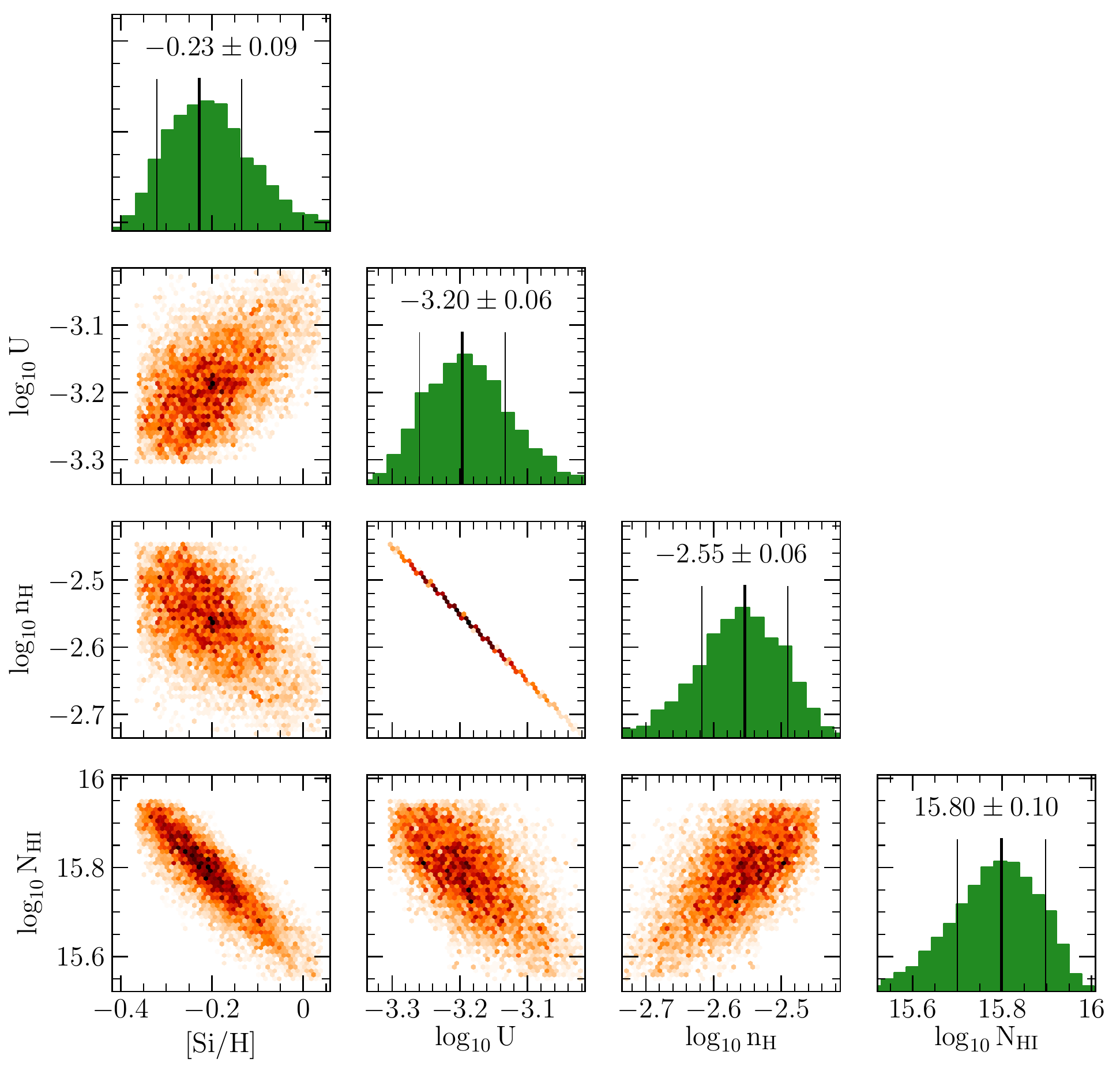}
	\caption{The posterior distribution profiles from the MCMC analysis of the Cloudy grids for J$1119$, $z_{gal} = 0.1383$, as for figure \ref{fig:Q0122_0.2119_par}.}
	\label{fig:J1119_1383_par}
\end{figure}
\newpage
\begin{deluxetable}{ccc}[hp]
	\tablecolumns{8}
	\tablewidth{\linewidth}
	\setlength{\tabcolsep}{0.06in}
	\tablecaption{J$1133$, $z_{gal} = 0.154599$ Column Densities\label{tab:J1133_1546}}
	\tablehead{
		\colhead{Ion}           	&
        \colhead{$\log N~({\cms})$}    &
		\colhead{$\log N$ Error~({\cms})}}
	\startdata
	{\HI}   & $[15.82, 17.20]$   &$\cdots$\\
{\CII}  & $<13.56$   &$\cdots$\\
{\NII}  & $<13.68$   &$\cdots$\\
{\NIII} & $<14.68$   &$\cdots$\\
{\NV}   & $<13.30$   &$\cdots$\\
{\OI}   & $<13.99$   &$\cdots$\\
{\SiII} & $<12.60$   &$\cdots$\\
{\SiIII}& $<12.43$   &$\cdots$\\
{\SiIV} & $<13.37$   &$\cdots$\\
{\CaII} & $<10.98$   &$\cdots$\\
{\MgI}  & $<10.68$   &$\cdots$\\
{\MgII} & $<11.15$   &$\cdots$\\[-5pt]

	\enddata
\end{deluxetable}
\begin{figure}[hp]
	\centering
	\includegraphics[width=\linewidth]{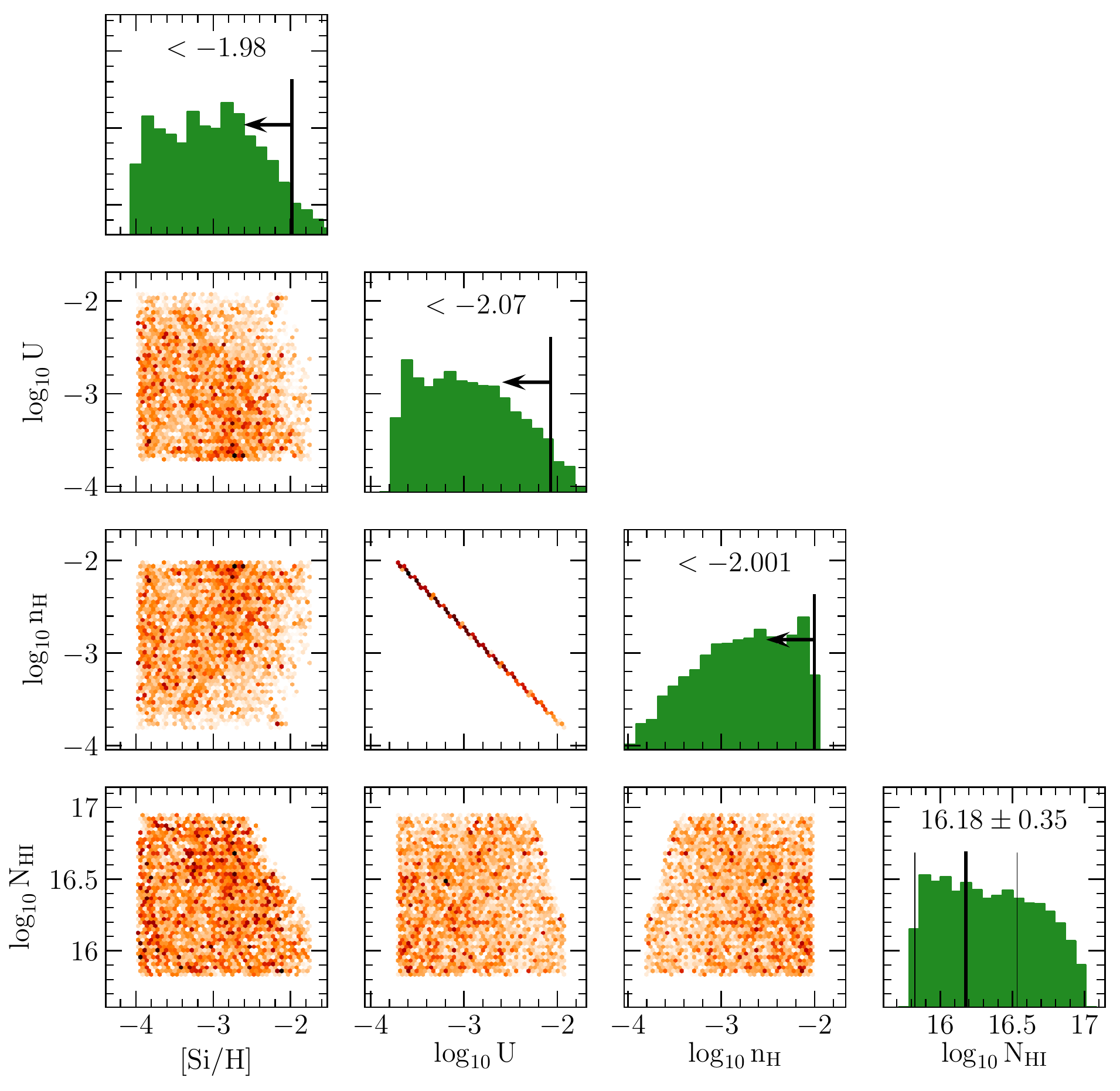}
	\caption{The posterior distribution profiles from the MCMC analysis of the Cloudy grids for J$1133$, $z_{gal} = 0.154599$, as for figure \ref{fig:Q0122_0.2119_par}.}
	\label{fig:J1133_1546_par}
\end{figure}
\begin{figure*}[hp]
	\centering
	\includegraphics[width=\linewidth]{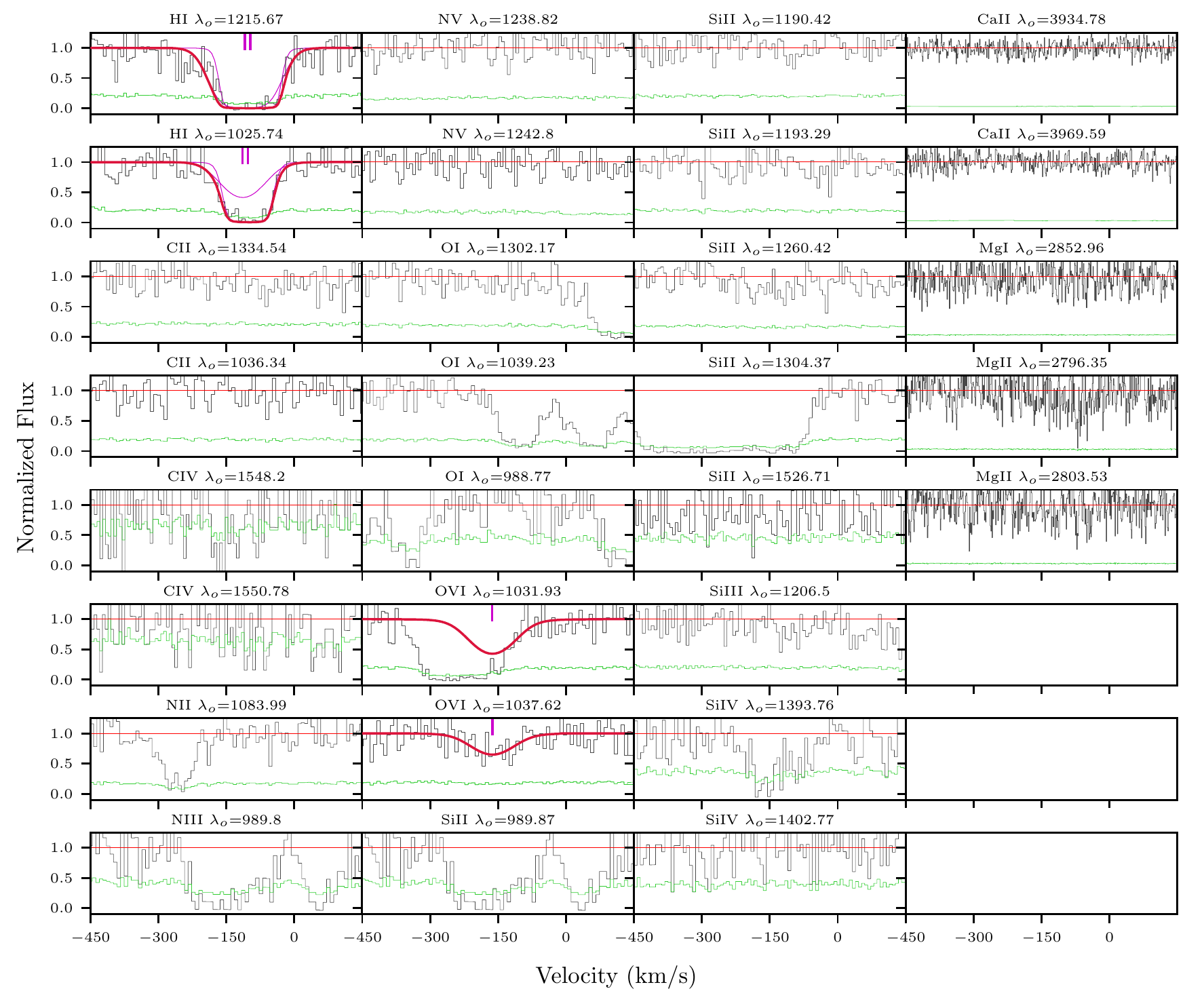}
	\caption{The fits for the system J$1133$, $z_{gal} = 0.154599$, as for figure \ref{fig:Q0122_0.2119}. The total {\OVI} fits from \citet{nielsenovi} are shown here for completeness, although they are not used in the ionization modelling.}
	\label{fig:J1133_1546}
\end{figure*}

\begin{figure*}[hp]
	\centering
	\includegraphics[width=\linewidth]{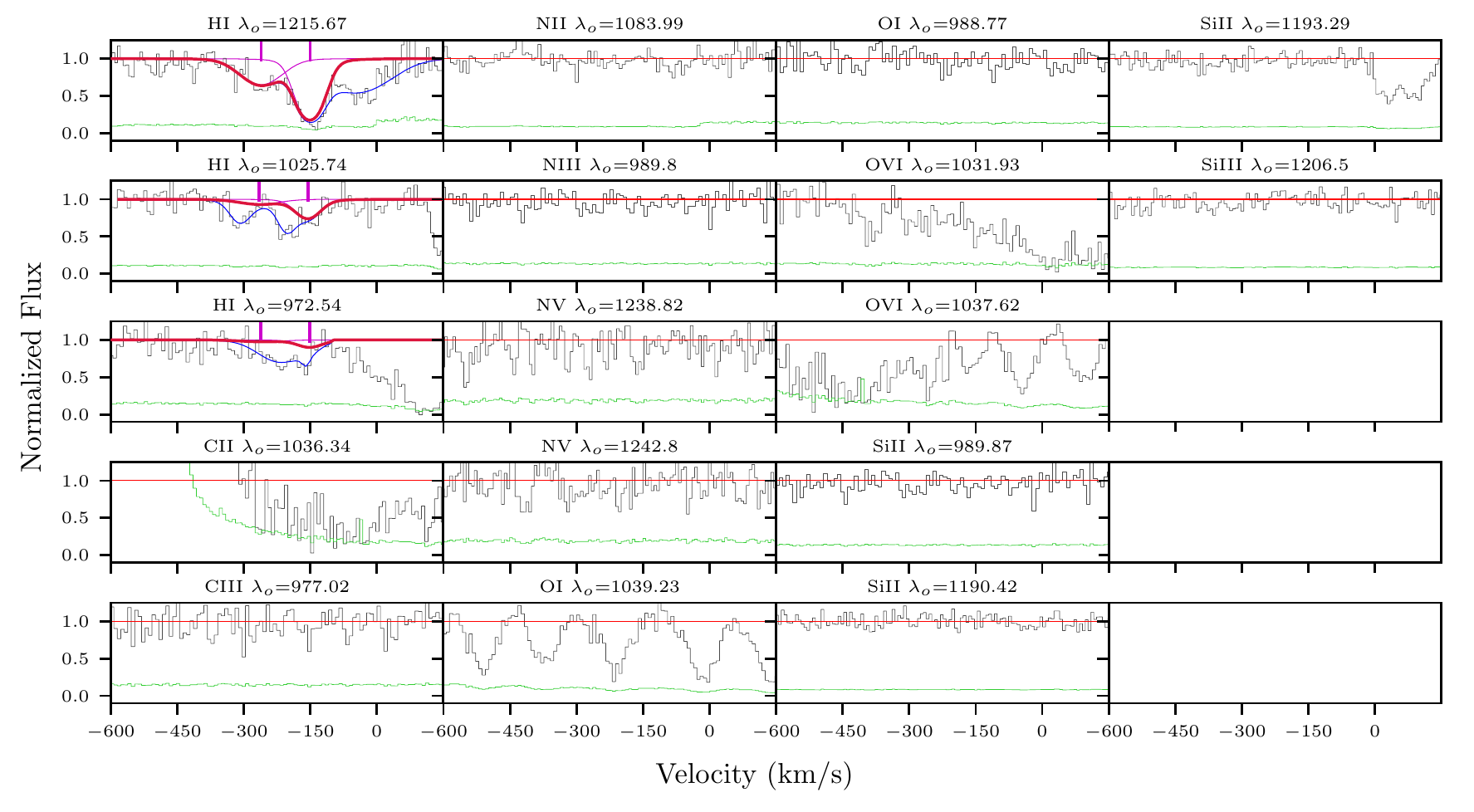}
	\caption{The fits for the system J$1139$, $z_{gal} = 0.1755$, as for figure \ref{fig:Q0122_0.2119}.}
	\label{fig:J1139_1755}
\end{figure*}
\clearpage
\begin{deluxetable}{ccc}[hp]
	\tablecolumns{8}
	\tablewidth{\linewidth}
	\setlength{\tabcolsep}{0.06in}
	\tablecaption{J$1139$, $z_{gal} = 0.1755$ Column Densities\label{tab:J1139_1755}}
	\tablehead{
		\colhead{Ion}           	&
        \colhead{$\log N~({\cms})$}    &
		\colhead{$\log N$ Error~({\cms})}}
	\startdata
	{\HI}   & $14.15$   &$0.05$\\
{\CII}  & $<14.18$   &$\cdots$\\
{\CIII} & $<12.66$   &$\cdots$\\
{\NII}  & $<13.46$   &$\cdots$\\
{\NIII} & $<13.52$   &$\cdots$\\
{\NV}   & $<13.41$   &$\cdots$\\
{\OI}   & $<13.89$   &$\cdots$\\
{\SiII} & $<12.57$   &$\cdots$\\
{\SiIII}& $<11.95$   &$\cdots$\\[-5pt]

	\enddata
\end{deluxetable}
\begin{figure}[hp]
	\centering
	\includegraphics[width=\linewidth]{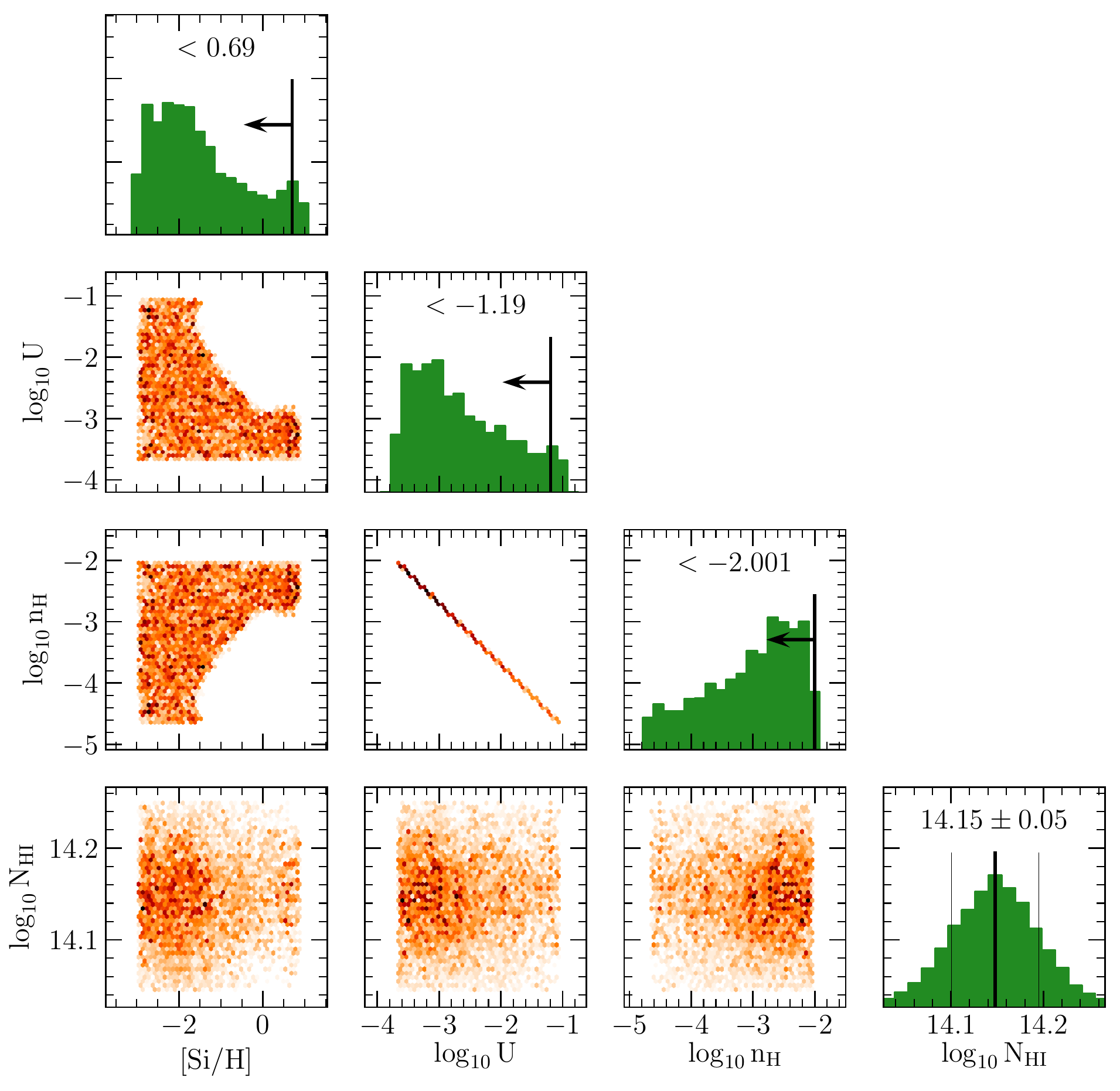}
	\caption{The posterior distribution profiles from the MCMC analysis of the Cloudy grids for J$1139$, $z_{gal} = 0.1755$, as for figure \ref{fig:Q0122_0.2119_par}.}
	\label{fig:J1139_1755_par}
\end{figure}
\newpage
\begin{deluxetable}{ccc}[hp]
	\tablecolumns{8}
	\tablewidth{\linewidth}
	\setlength{\tabcolsep}{0.06in}
	\tablecaption{J$1139$, $z_{gal} = 0.204194$ Column Densities\label{tab:J1139_2042}}
	\tablehead{
		\colhead{Ion}           	&
        \colhead{$\log N~({\cms})$}    &
		\colhead{$\log N$ Error~({\cms})}}
	\startdata
	{\HI}   & $[16.04,17.20]$   &$\cdots$\\
{\CII}  & $13.91$   &$0.06$\\
{\CIII} & $>14.41$   &$\cdots$\\
{\NII}  & $<13.04$   &$\cdots$\\
{\NIII} & $14.16$   &$0.05$\\
{\OI}   & $15.73$   &$2.60$\\
{\SiII} & $13.41$   &$0.06$\\
{\SiIII}& $13.88$   &$1.50$\\[-5pt]

	\enddata
\end{deluxetable}
\begin{figure}[hp]
	\centering
	\includegraphics[width=\linewidth]{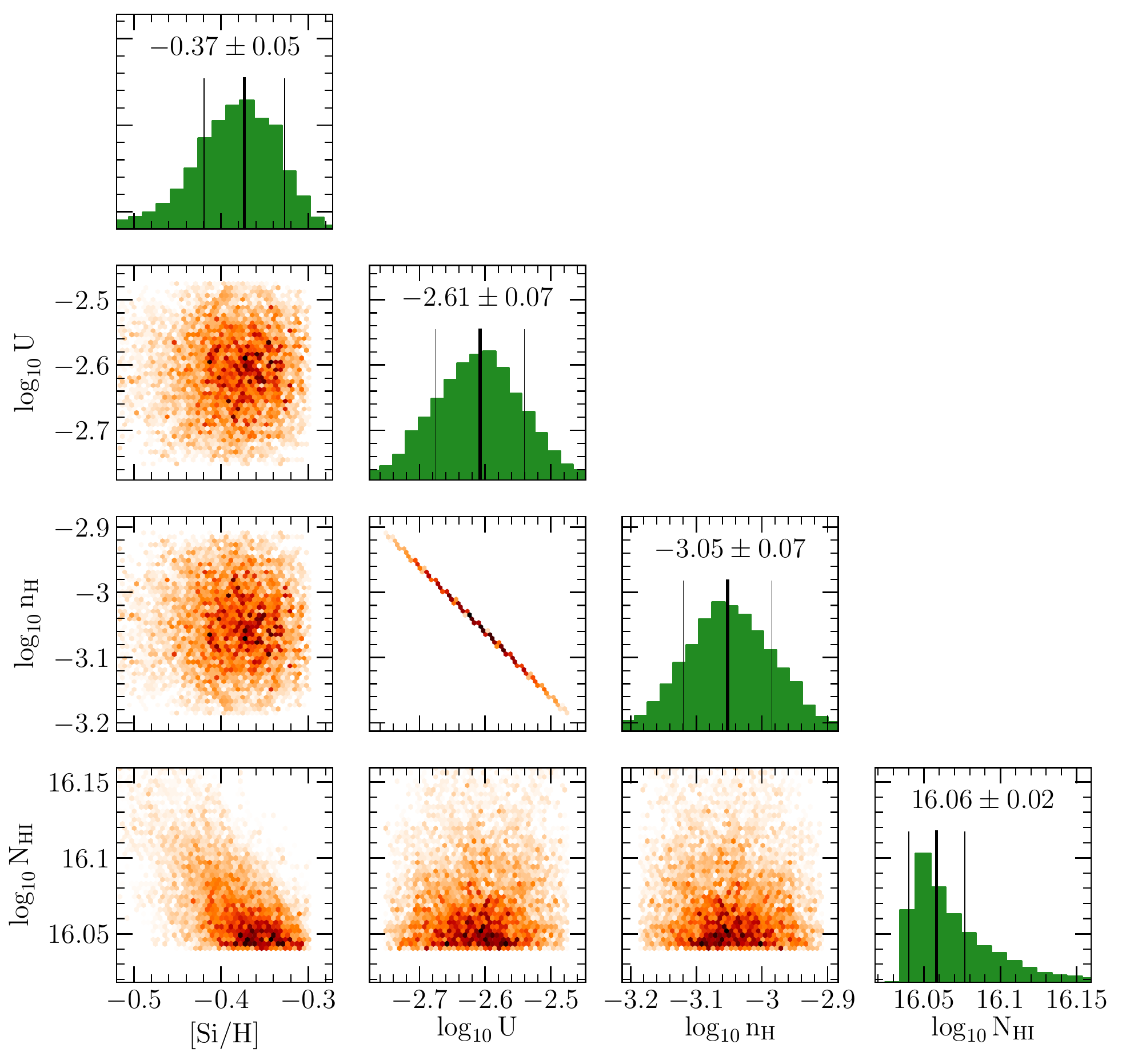}
	\caption{The posterior distribution profiles from the MCMC analysis of the Cloudy grids for J$1139$, $z_{gal} = 0.204194$, as for figure \ref{fig:Q0122_0.2119_par}.}
	\label{fig:J1139_2042_par}
\end{figure}
\begin{figure*}[hp]
	\centering
	\includegraphics[width=\linewidth]{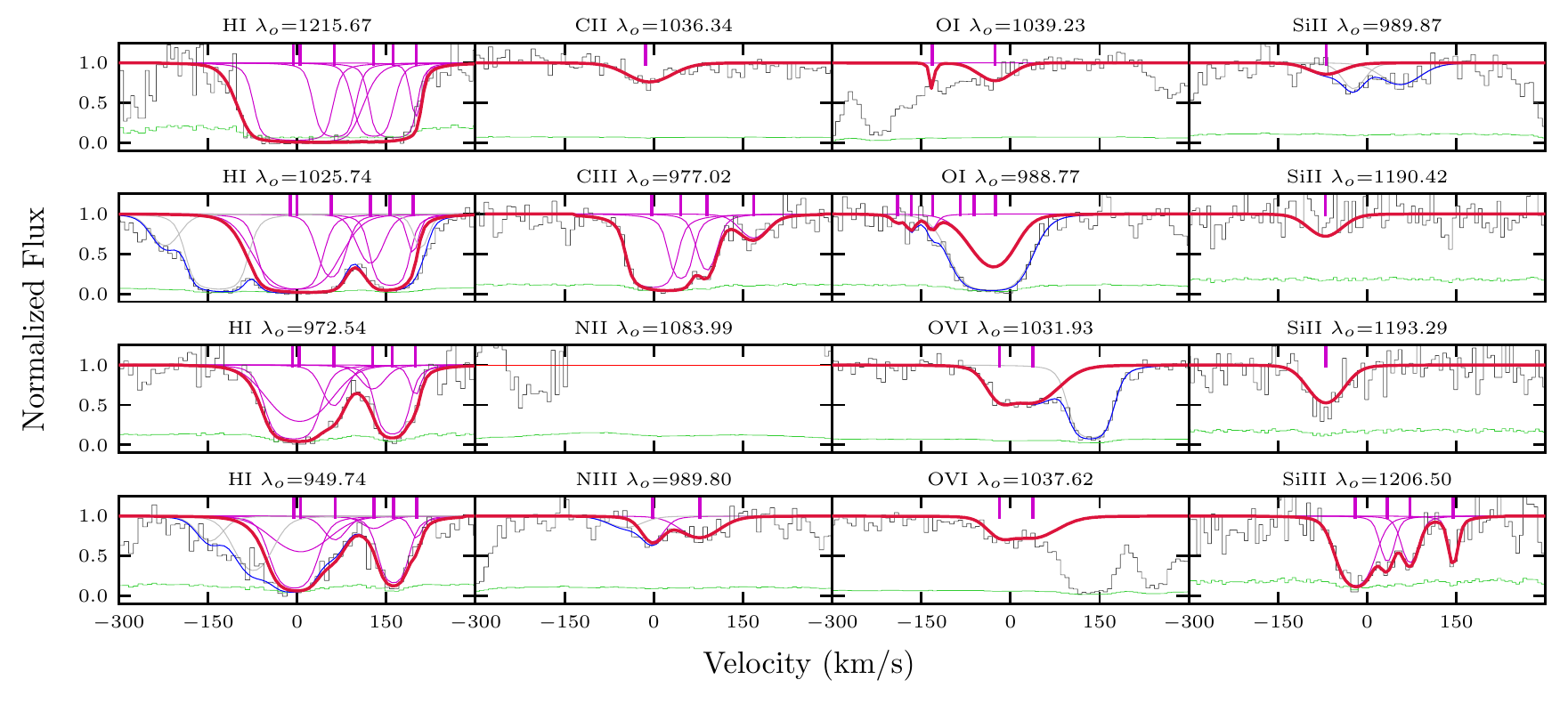}
	\caption{The fits for the system J$1139$, $z_{gal} = 0.204194$, as for figure \ref{fig:Q0122_0.2119}. There are blends present in the {\HI} 1025~{\AA} and 949~{\AA} transitions, while the remaining {\HI} lines constrain the column density. The {\SiII} and {\NIII} 989~{\AA} lines are blended together. The presence of {\SiII} 1990~{\AA} and 1993~{\AA} transitions constrain the column densities. The {\OI} 1039~{\AA} constrained the component at $v \sim -20$~{\kms}, while the {\OI} 988~{\AA} transition constrained the left component at $v \sim -140$~{\kms}.  The total {\OVI} fits and the blend in the {\OVI} $1031$~{\AA} line from \citet{nielsenovi} are shown here for completeness, although they are not used in the ionization modelling.  }
	\label{fig:J1139_2042}
\end{figure*}

\begin{figure*}[hp]
	\centering
	\includegraphics[width=\linewidth]{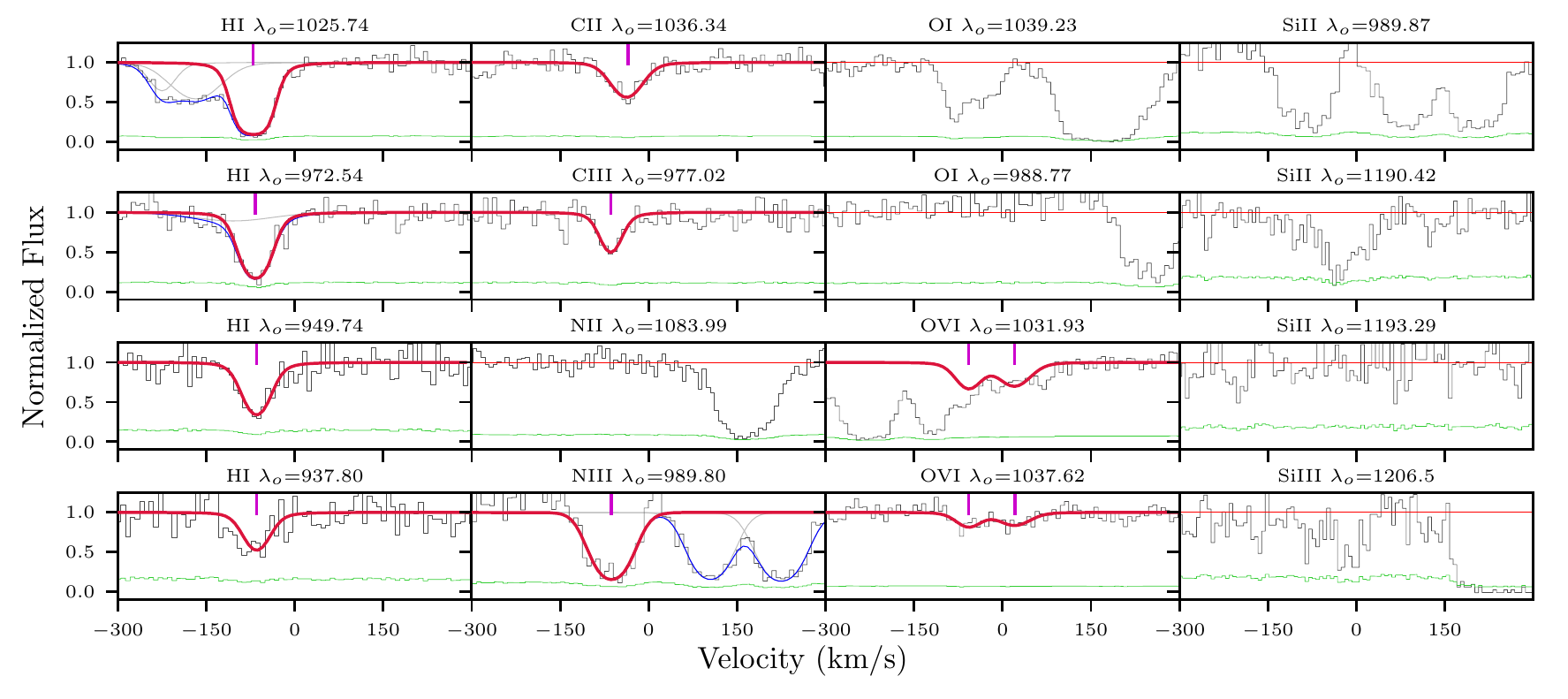}
	\caption{The fits for the system J$1139$, $z_{gal} = 0.212259$, as for figure \ref{fig:Q0122_0.2119}. Although there are unknown blends present in the {\HI} 1025~{\AA} and 972~{\AA} transitions, the {\HI} column density is constrained by the other {\HI} lines. We do not detect any {\SiII} absorption, although blends are present towards 989~{\AA} and 1190~{\AA} transitions. The {\SiII} and {\NIII} 989~{\AA} lines are blended together, although the non-detection of {\SiII} in other transitions suggests that the absorption is primarily due to {\NIII}. The total {\OVI} fits from \citet{nielsenovi} are shown here for completeness, although they are not used in the ionization modelling.}
	\label{fig:J1139_2123}
\end{figure*}
\clearpage
\begin{deluxetable}{ccc}[hp]
	\tablecolumns{8}
	\tablewidth{\linewidth}
	\setlength{\tabcolsep}{0.06in}
	\tablecaption{J$1139$, $z_{gal} = 0.212259$ Column Densities\label{tab:J1139_2123}}
	\tablehead{
		\colhead{Ion}           	&
        \colhead{$\log N~({\cms})$}    &
		\colhead{$\log N$ Error~({\cms})}}
	\startdata
	{\HI}   & $15.33$  &$0.04$\\
{\CII}  & $<14.10$  &$\cdots$\\
{\NII}  & $<13.24$  &$\cdots$\\
{\NIII} & $<14.63$  &$0.04$\\
{\OI}   & $<13.73$  &$\cdots$\\
{\SiII} & $<12.93$  &$\cdots$\\
{\SiIII}& $<12.60$  &$\cdots$\\[-5pt]

	\enddata
\end{deluxetable}
\begin{figure}[hp]
	\centering
	\includegraphics[width=\linewidth]{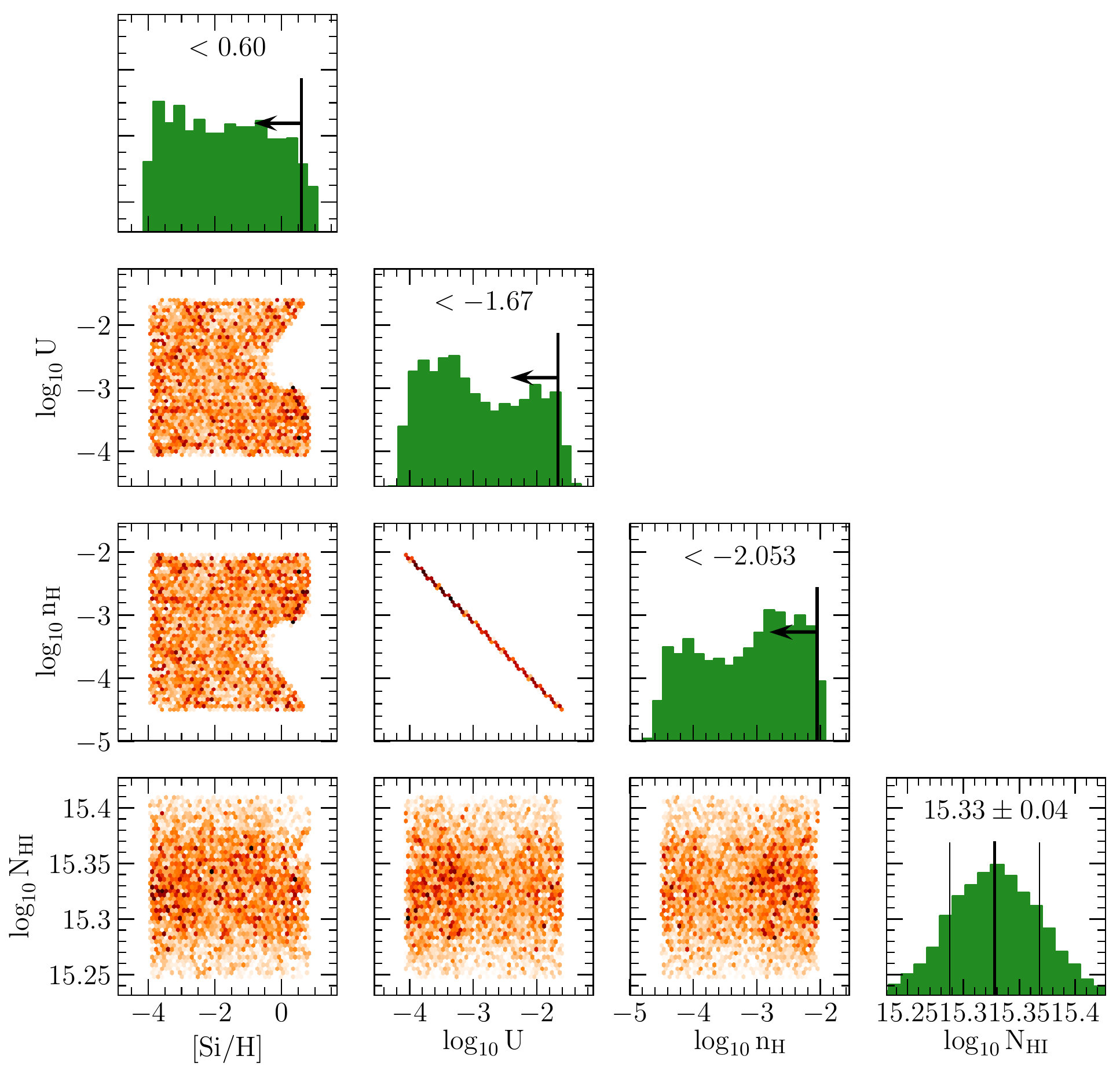}
	\caption{The posterior distribution profiles from the MCMC analysis of the Cloudy grids for J$1139$, $z_{gal} = 0.212259$, as for figure \ref{fig:Q0122_0.2119_par}.}
	\label{fig:J1139_2123_par}
\end{figure}
\newpage
\begin{deluxetable}{ccc}[hp]
	\tablecolumns{8}
	\tablewidth{\linewidth}
	\setlength{\tabcolsep}{0.06in}
	\tablecaption{J$1139$, $z_{gal} = 0.219724$ Column Densities\label{tab:J1139_2197}}
	\tablehead{
		\colhead{Ion}           	&
        \colhead{$\log N~({\cms})$}    &
		\colhead{$\log N$ Error~({\cms})}}
	\startdata
	{\HI}   & $14.20$&$0.07$\\
{\CII}  & $<13.09$   &$\cdots$\\
{\CIII} & $<12.62$   &$\cdots$\\
{\NII}  & $<13.21$   &$\cdots$\\
{\NIII} & $<13.41$   &$\cdots$\\
{\OI}   & $<13.77$   &$\cdots$\\
{\SiII} & $<12.91$   &$\cdots$\\[-5pt]

	\enddata
\end{deluxetable}
\begin{figure}[hp]
	\centering
	\includegraphics[width=\linewidth]{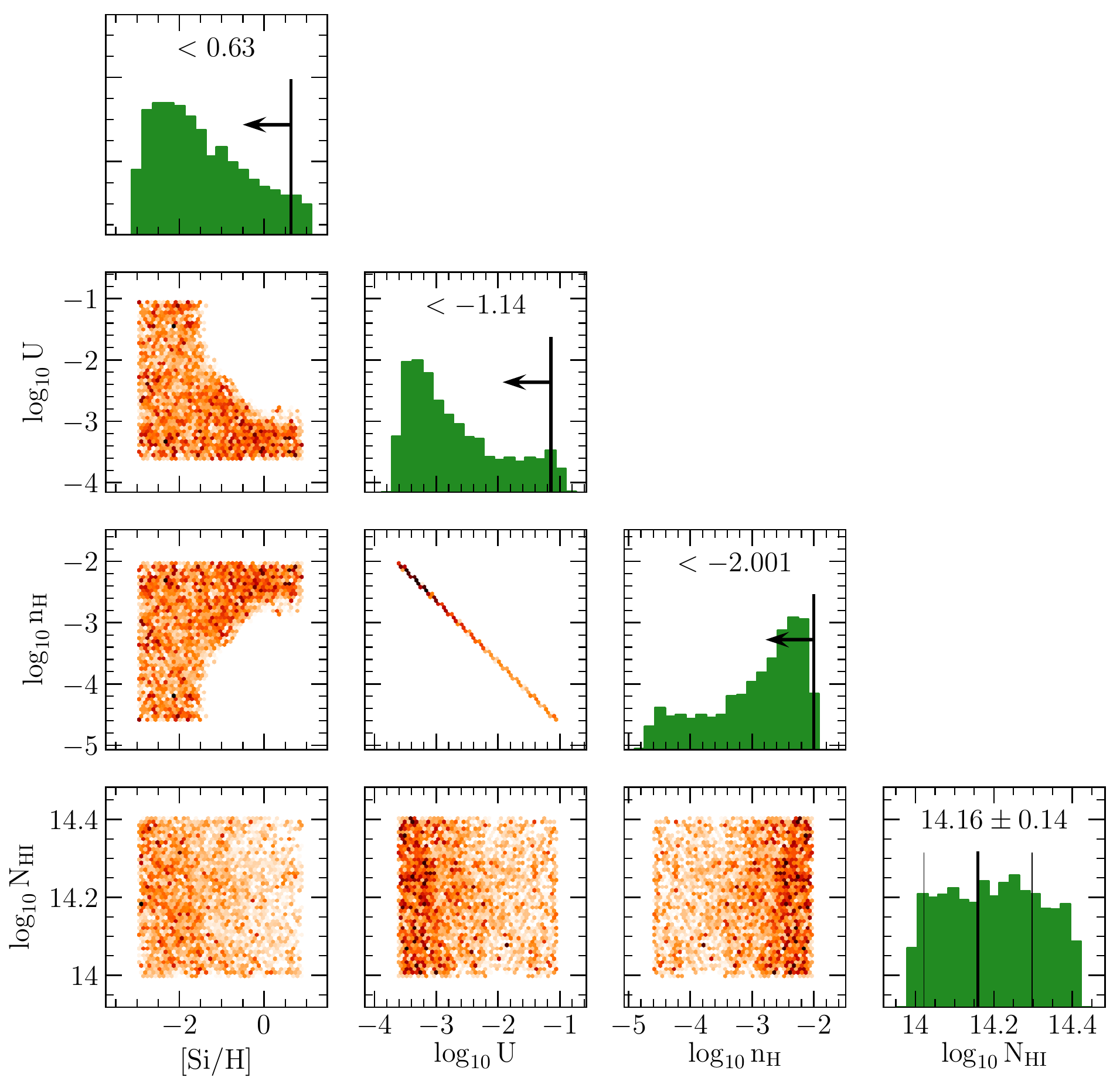}
	\caption{The posterior distribution profiles from the MCMC analysis of the Cloudy grids for J$1139$, $z_{gal} = 0.219724$, as for figure \ref{fig:Q0122_0.2119_par}.}
	\label{fig:J1139_2197_par}
\end{figure}
\begin{figure*}[hp]
	\centering
	\includegraphics[width=\linewidth]{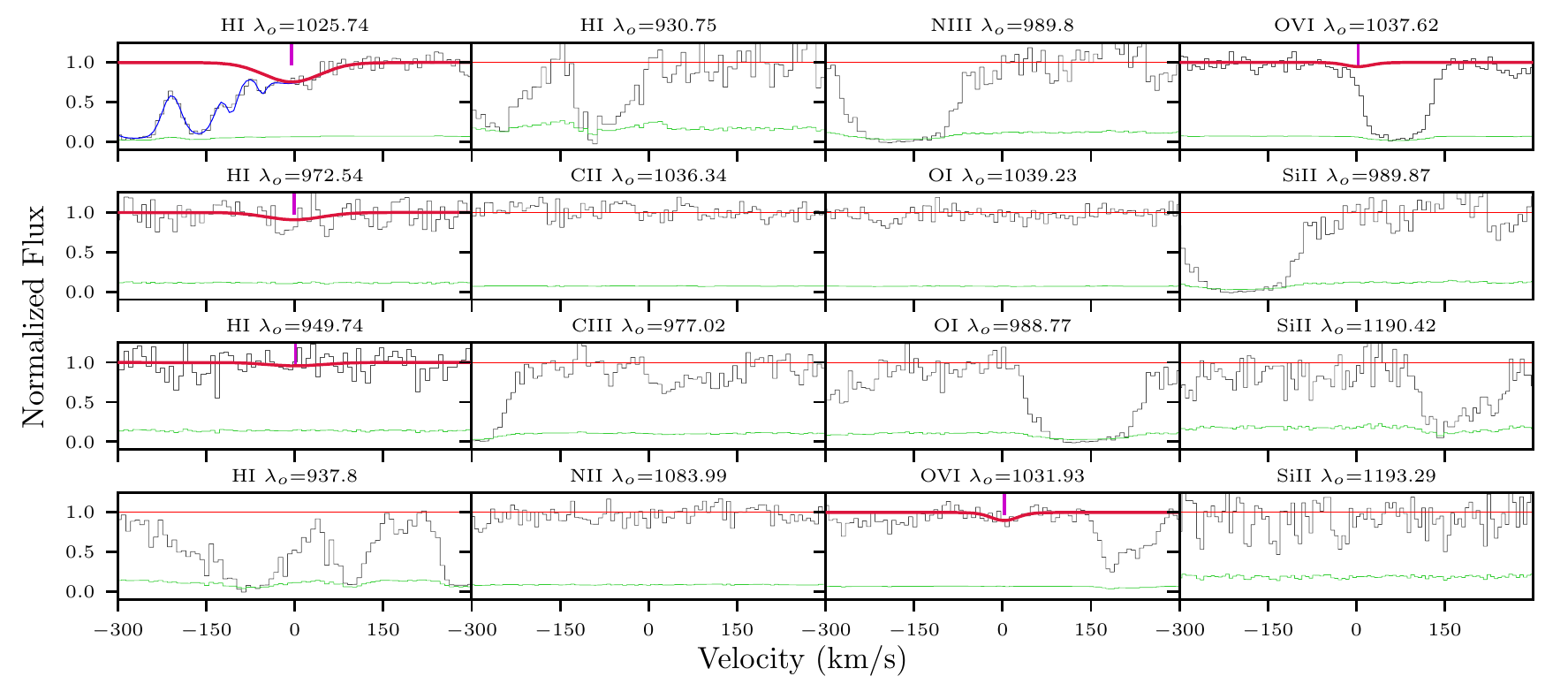}
	\caption{The fits for the system J$1139$, $z_{gal} = 0.219724$, as for figure \ref{fig:Q0122_0.2119}. There are blends present in the {\HI} 1025~{\AA} transition. However, the other {\HI} lines constrain the column density. There are no other metals, apart from {\OVI} detected here. The total {\OVI} fits from \citet{nielsenovi} are shown here for completeness, although they are not used in the ionization modelling.}
	\label{fig:J1139_2197}
\end{figure*}

\begin{figure*}[hp]
	\centering
	\includegraphics[width=\linewidth]{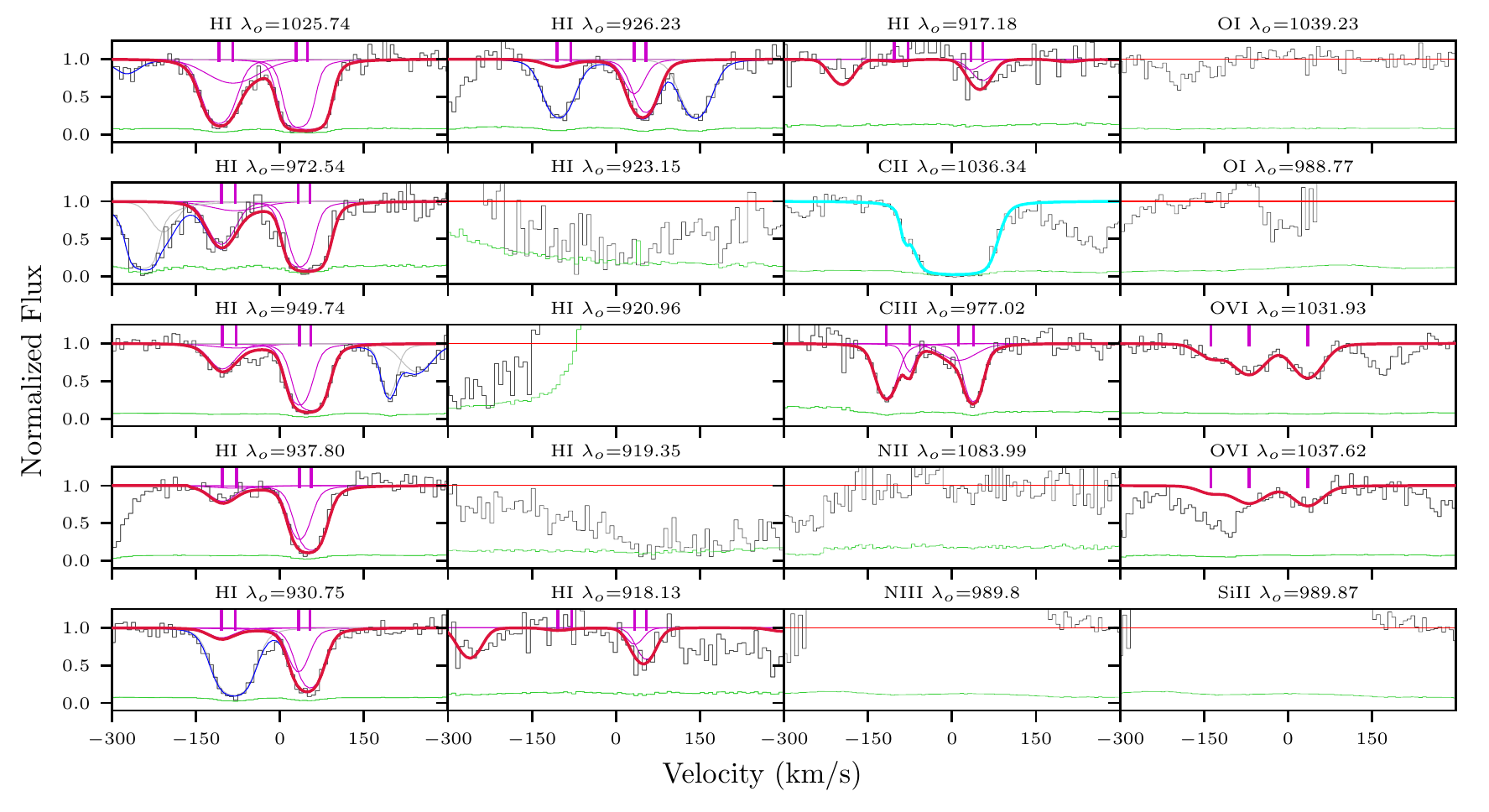}
	\caption{The fits for the system J$1139$, $z_{gal} = 0.319255$, as for figure \ref{fig:Q0122_0.2119}. The blends fitted in the {\HI} 1025~{\AA} and 949~{\AA} transitions were done to improve the model and do no overlap with the {\HI} absorption. However, the blends in the {\HI} 972~{\AA}, 930~{\AA} and 926~{\AA} lines were significant. The {\HI} column density was constrained using the 937~{\AA} transition. The {\CII} 1036~{\AA} transition is havily blended. However, in the absence of additional data the modelled transition had to be used as an upper limit on the column density. The total {\OVI} fits from \citet{nielsenovi} are shown here for completeness, although they are not used in the ionization modelling.}
	\label{fig:J1139_3193}
\end{figure*}
\clearpage
\begin{deluxetable}{ccc}[hp]
	\tablecolumns{8}
	\tablewidth{\linewidth}
	\setlength{\tabcolsep}{0.06in}
	\tablecaption{J$1139$, $z_{gal} = 0.319255$ Column Densities\label{tab:J1139_3193}}
	\tablehead{
		\colhead{Ion}           	&
        \colhead{$\log N~({\cms})$}    &
		\colhead{$\log N$ Error~({\cms})}}
	\startdata
	{\HI}   & $16.19$   &$0.03$\\
{\CII}  & $<15.60$   &$\cdots$\\
{\CIII} & $14.09$   &$0.16$\\
{\NII}  & $<13.73$   &$\cdots$\\
{\NIII} & $<13.27$   &$\cdots$\\
{\OI}   & $<14.41$   &$\cdots$\\
{\SiII} & $<13.16$   &$\cdots$\\[-5pt]
	\enddata
\end{deluxetable}
\begin{figure}[hp]
	\centering
	\includegraphics[width=\linewidth]{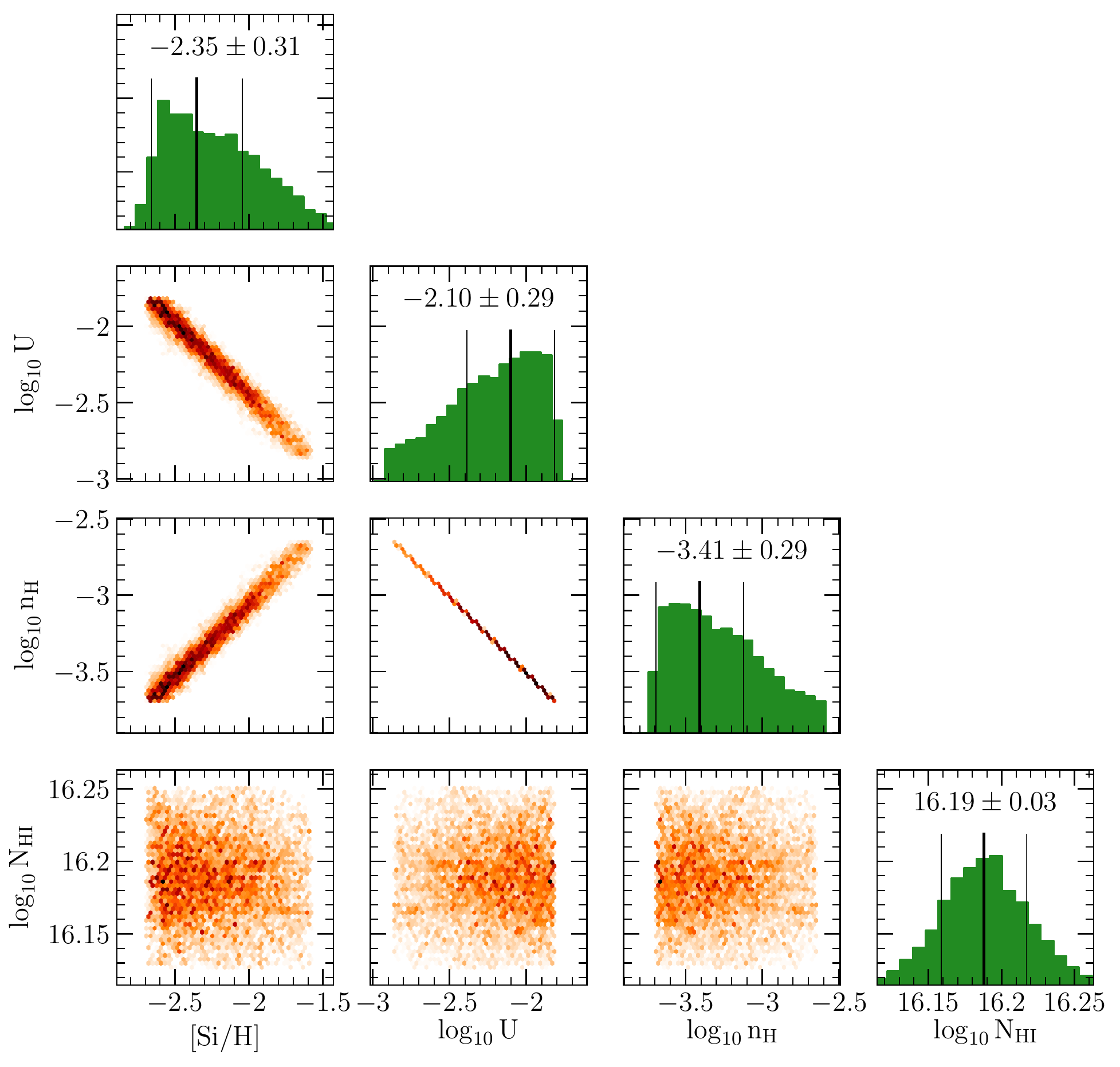}
	\caption{The posterior distribution profiles from the MCMC analysis of the Cloudy grids for J$1139$, $z_{gal} = 0.319255$, as for figure \ref{fig:Q0122_0.2119_par}.}
	\label{fig:J1139_3193_par}
\end{figure}
\newpage
\begin{deluxetable}{ccc}[hp]
	\tablecolumns{8}
	\tablewidth{\linewidth}
	\setlength{\tabcolsep}{0.06in}
	\tablecaption{J$1219$, $z_{gal} = 0.1241$ Column Densities\label{tab:J1219_1241}}
	\tablehead{
		\colhead{Ion}           	&
        \colhead{$\log N~({\cms})$}    &
		\colhead{$\log N$ Error~({\cms})}}
	\startdata
	{\HI}   & $15.25$   &$0.03$\\
{\CII}  & $<13.19$   &$\cdots$\\
{\NII}  & $<13.48$   &$\cdots$\\
{\NV}   & $<13.01$   &$\cdots$\\
{\OI}   & $<14.48$   &$\cdots$\\
{\SiII} & $<12.11$   &$\cdots$\\
{\SiIII}& $12.81$   &$0.04$\\
{\SiIV} & $13.68$   &$0.75$\\[-5pt]

	\enddata
\end{deluxetable}
\begin{figure}[hp]
	\centering
	\includegraphics[width=\linewidth]{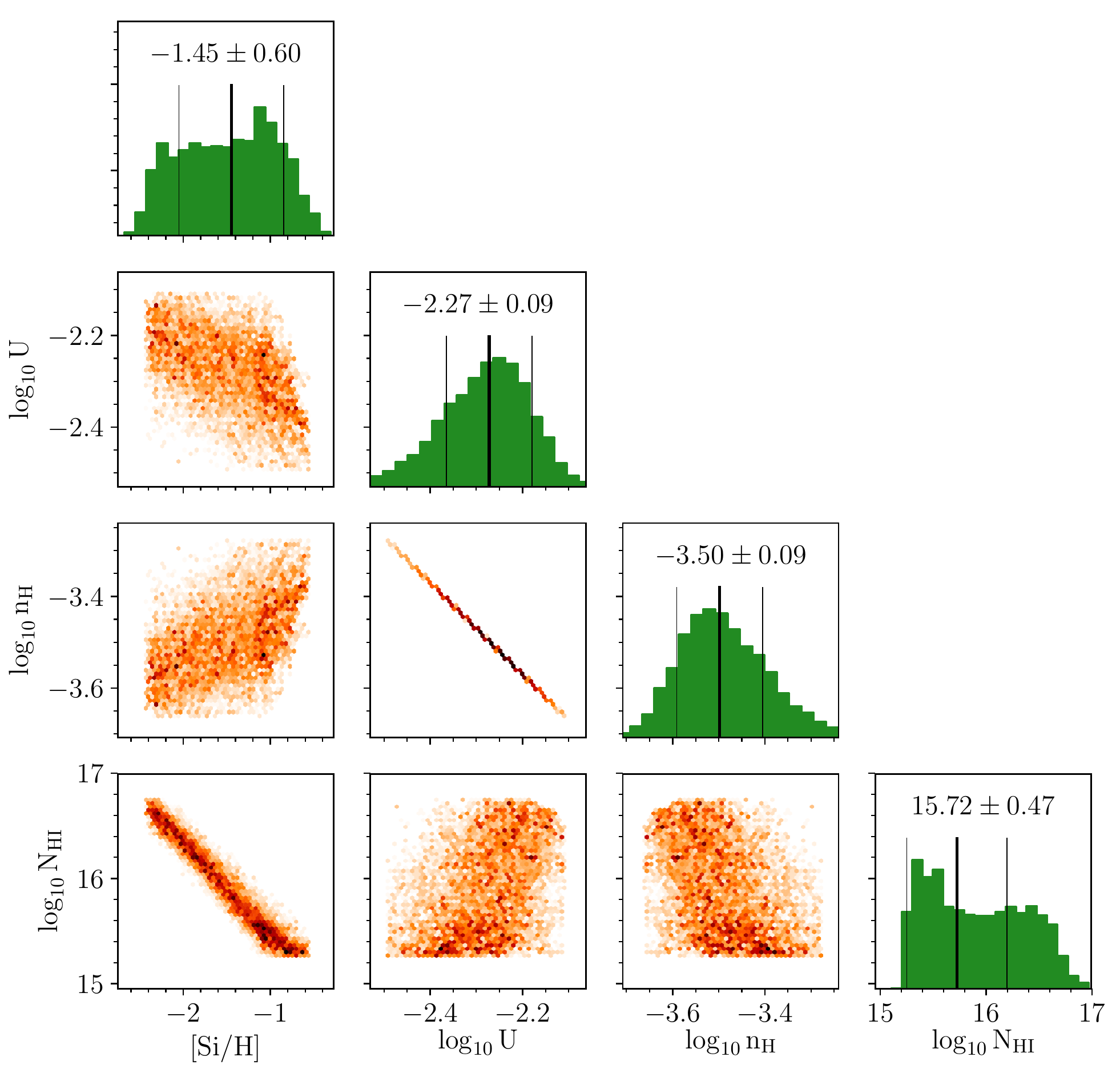}
	\caption{The posterior distribution profiles from the MCMC analysis of the Cloudy grids for J$1219$, $z_{gal} = 0.1241$, as for figure \ref{fig:Q0122_0.2119_par}.}
	\label{fig:J1219_1241_par}
\end{figure}
\begin{figure*}[hp]
	\centering
	\includegraphics[width=\linewidth]{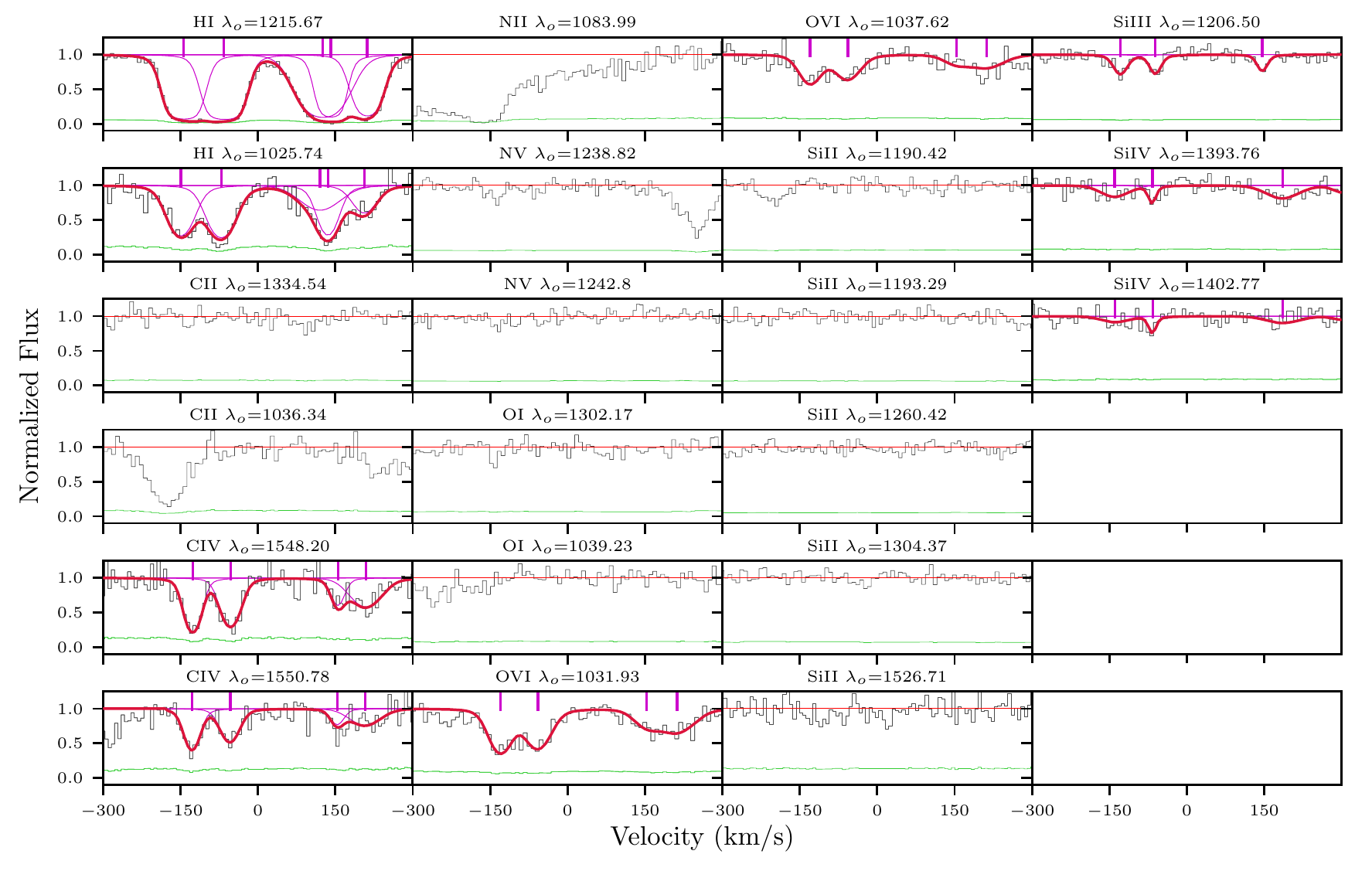}
	\caption{The fits for the system J$1219$, $z_{gal} = 0.1241$, as for figure \ref{fig:Q0122_0.2119}. The total {\OVI} fits from \citet{nielsenovi} are shown here for completeness, although they are not used in the ionization modelling.}
	\label{fig:J1219_1241}
\end{figure*}

\begin{figure*}[hp]
	\centering
	\includegraphics[width=\linewidth]{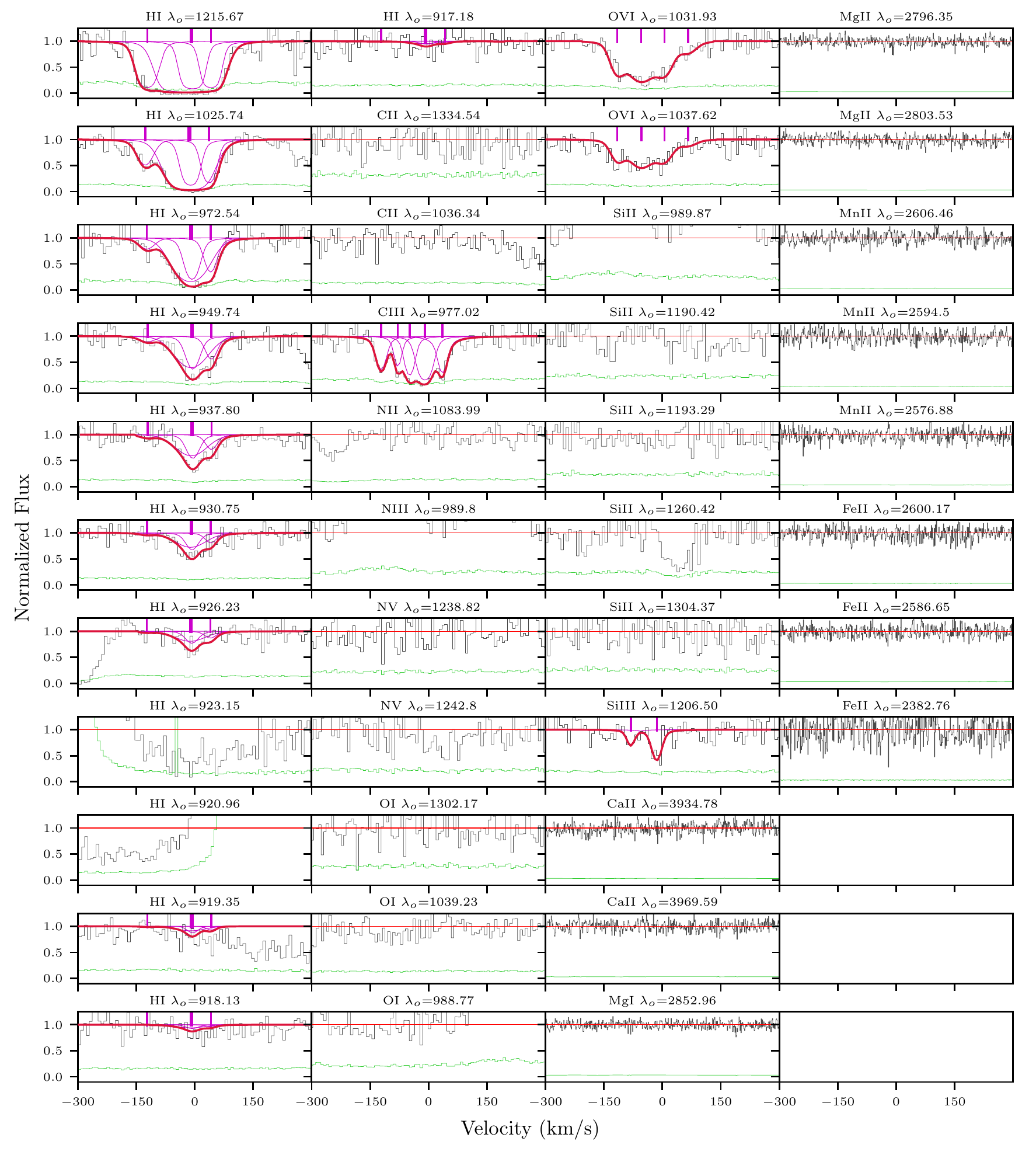}
	\caption{The fits for the system J$1233$, $z_{gal} = 0.318757$, as for figure \ref{fig:Q0122_0.2119}. The total {\OVI} fits from \citet{nielsenovi} are shown here for completeness, although they are not used in the ionization modelling.}
	\label{fig:J1233_3188}
\end{figure*}
\clearpage
\begin{deluxetable}{ccc}[hp]
	\tablecolumns{8}
	\tablewidth{\linewidth}
	\setlength{\tabcolsep}{0.06in}
	\tablecaption{J$1233$, $z_{gal} = 0.318757$ Column Densities\label{tab:J1233_3188}}
	\tablehead{
		\colhead{Ion}           	&
        \colhead{$\log N~({\cms})$}    &
		\colhead{$\log N$ Error~({\cms})}}
	\startdata
	{\HI}   & $15.72$   &$0.02$\\
{\CII}  & $<13.41$   &$\cdots$\\
{\CIII} & $15.84$   &$0.62$\\
{\NII}  & $<13.39$   &$\cdots$\\
{\NIII} & $13.66$   &$0.19$\\
{\NV}   & $<13.81$   &$\cdots$\\
{\OI}   & $<14.11$   &$\cdots$\\
{\SiII} & $<12.72$   &$\cdots$\\
{\SiIII}& $12.99$   &$0.12$\\
{\CaII} & $<10.98$   &$\cdots$\\
{\MgI}  & $<10.66$   &$\cdots$\\
{\MgII} & $<11.14$   &$\cdots$\\
{\MnII} & $<11.43$   &$\cdots$\\
{\FeII} & $<11.52$   &$\cdots$\\[-5pt]

	\enddata
\end{deluxetable}
\begin{figure}[hp]
	\centering
	\includegraphics[width=\linewidth]{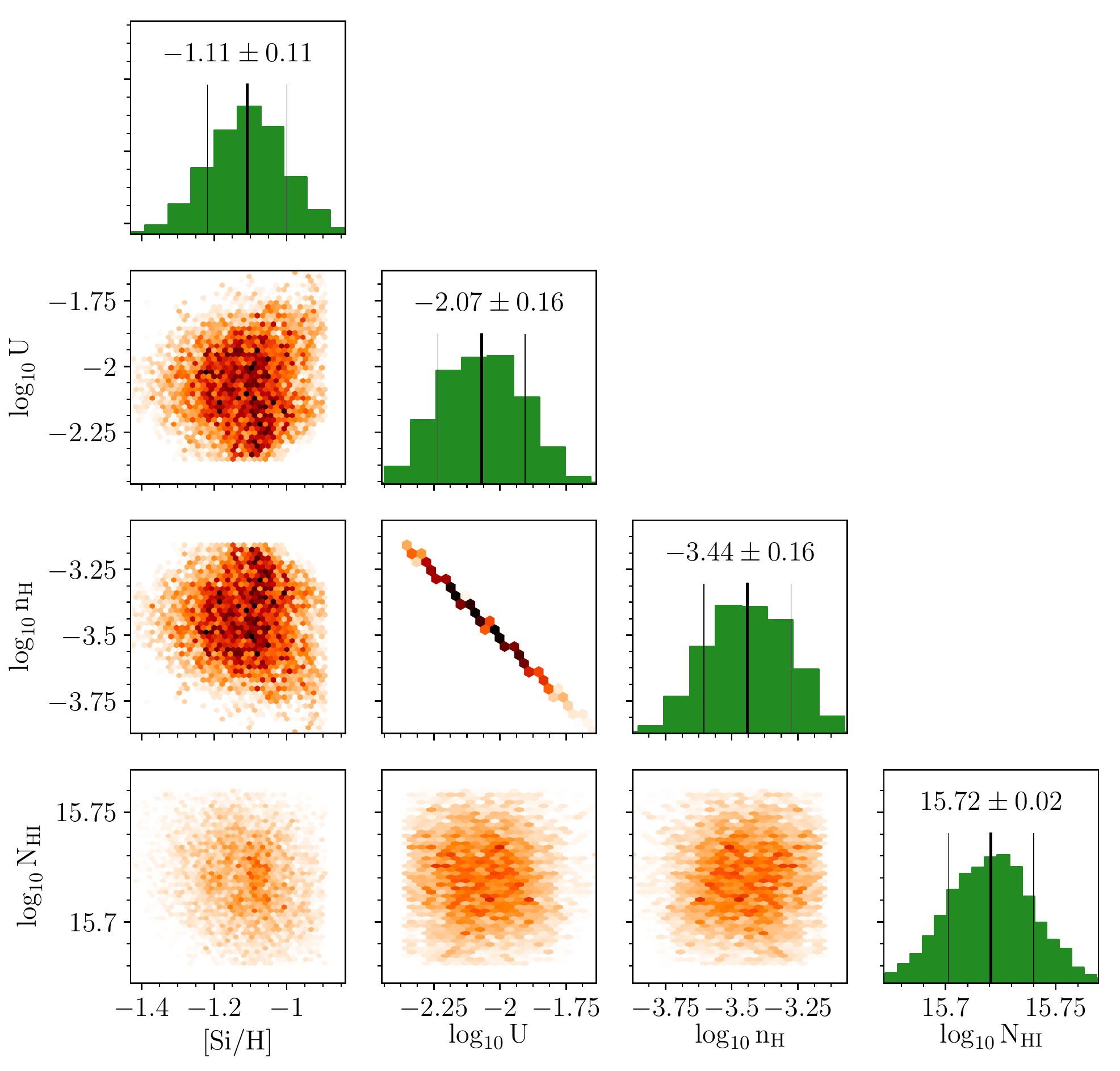}
	\caption{The posterior distribution profiles from the MCMC analysis of the Cloudy grids for J$1233$, $z_{gal} = 0.318757$, as for figure \ref{fig:Q0122_0.2119_par}.}
	\label{fig:J1233_3188_par}
\end{figure}
\newpage
\begin{deluxetable}{ccc}[hp]
	\tablecolumns{8}
	\tablewidth{\linewidth}
	\setlength{\tabcolsep}{0.06in}
	\tablecaption{J$1241$, $z_{gal} = 0.205267$ Column Densities\label{tab:J1241_2053}}
	\tablehead{
		\colhead{Ion}           	&
        \colhead{$\log N~({\cms})$}    &
		\colhead{$\log N$ Error~({\cms})}}
	\startdata
	{\HI}   & $[16.63, 19.00]$   &$\cdots$\\
{\CII}  & $>15.22$   &$\cdots$\\
{\CIII} & $>15.47$   &$\cdots$\\
{\NII}  & $<13.19$   &$\cdots$\\
{\NIII} & $>16.27$   &$\cdots$\\
{\NV}   & $13.81$   &$0.08$\\
{\OI}   & $<13.76$   &$\cdots$\\
{\SiII} & $14.53$   &$0.10$\\
{\SiIII}& $>14.13$   &$\cdots$\\
{\SiIV} & $14.79$   &$2.44$\\
{\CaII} & $<11.58$   &$\cdots$\\
{\MgI}  & $12.03$   &$0.08$\\
{\MgII} & $<14.26$   &$\cdots$\\
{\MnII} & $<12.96$   &$\cdots$\\
{\FeII} & $<13.61$   &$\cdots$\\[-5pt]

	\enddata
\end{deluxetable}
\begin{figure}[hp]
	\centering
	\includegraphics[width=\linewidth]{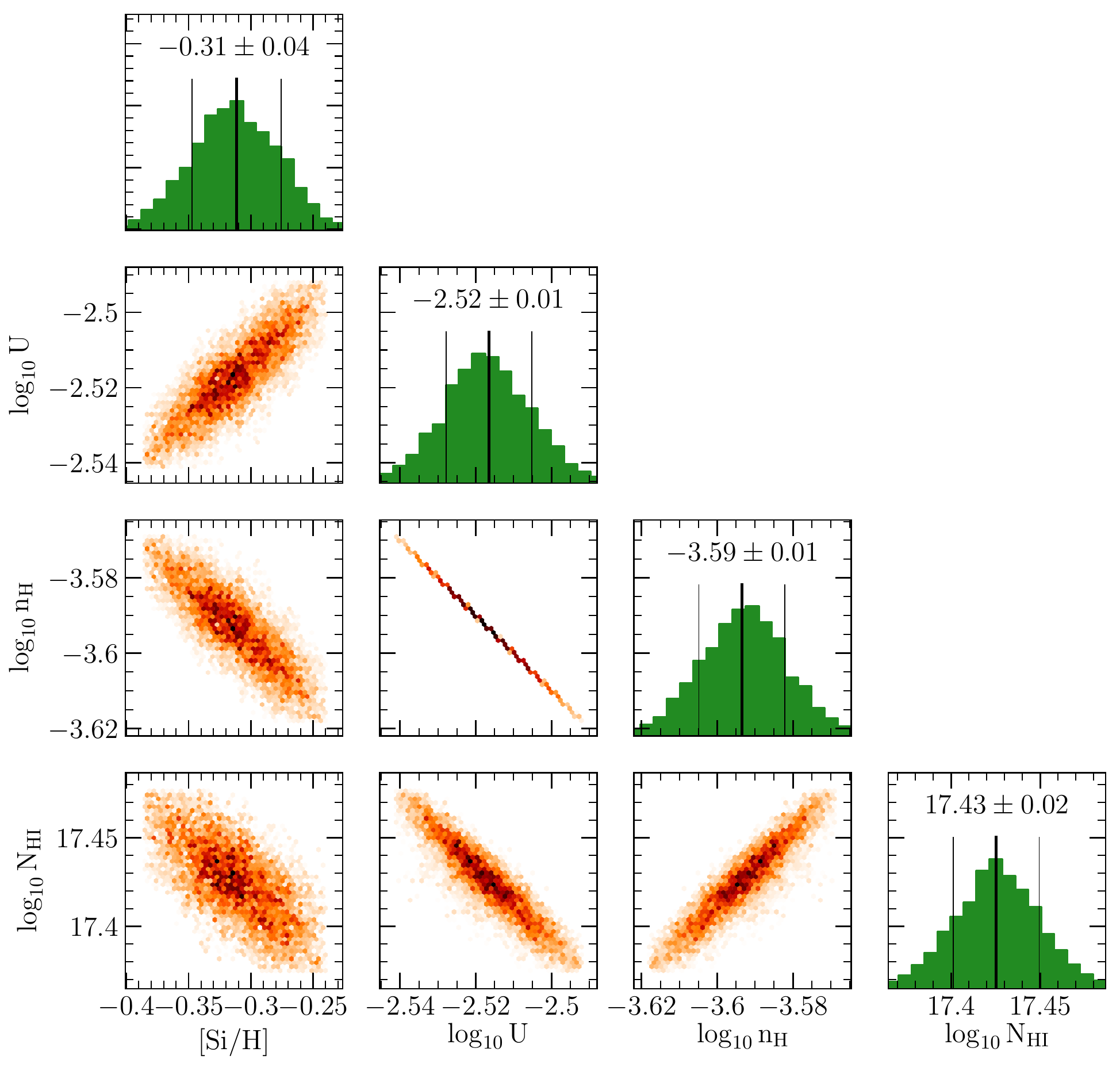}
	\caption{The posterior distribution profiles from the MCMC analysis of the Cloudy grids for J$1241$, $z_{gal} = 0.205267$, as for figure \ref{fig:Q0122_0.2119_par}.}
	\label{fig:J1241_2053_par}
\end{figure}
\begin{figure*}[hp]
	\centering
	\includegraphics[width=\linewidth]{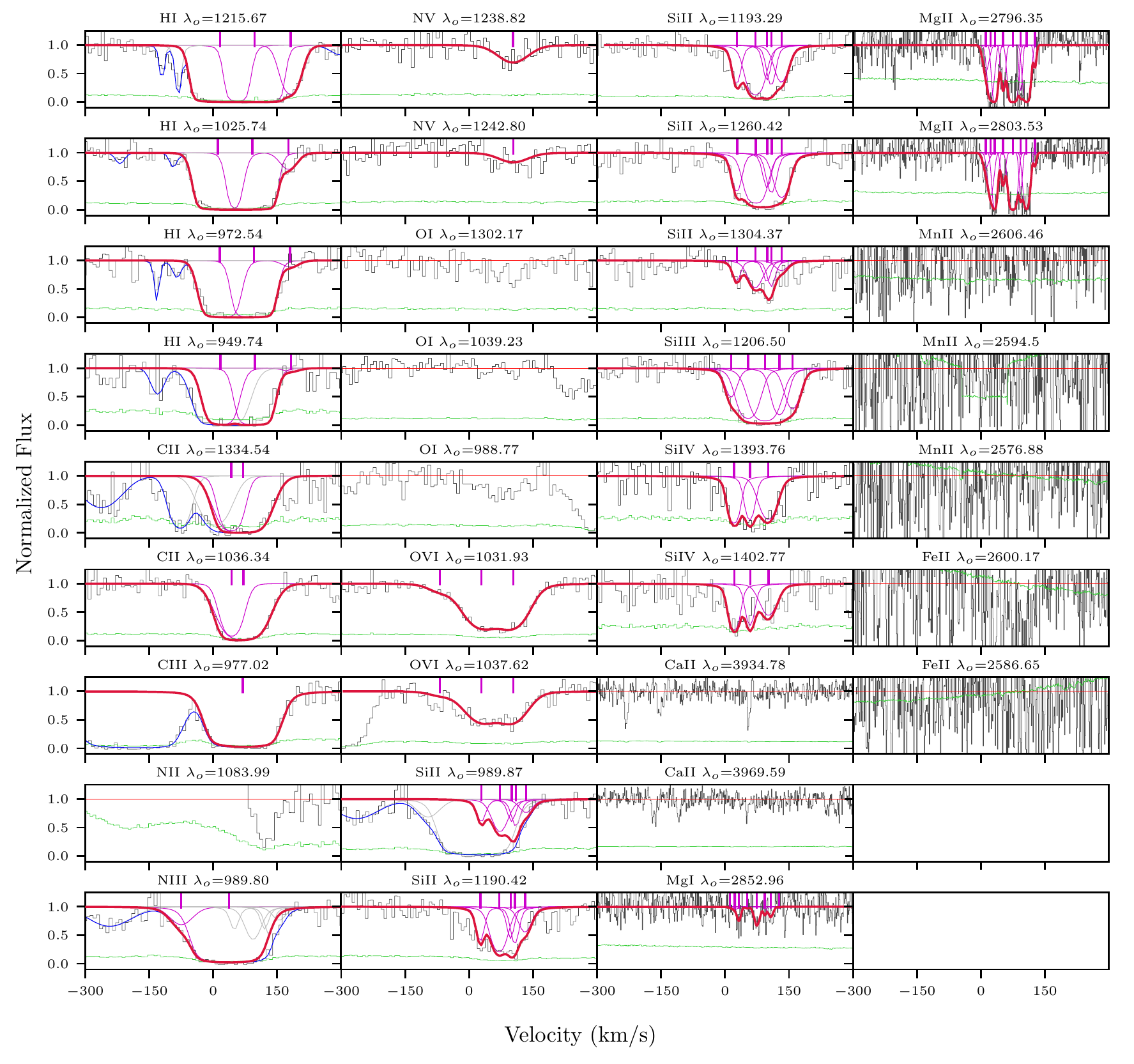}
	\caption{The fits for the system J$1241$, $z_{gal} = 0.205267$, as for figure \ref{fig:Q0122_0.2119}. There are unknown blends on the left of the {\HI} absorption. The positions of the blends were not consistent with deuterium absorption. However, the small strength of the blends, particularly in the {\HI} 1025~{\AA} and 972~{\AA} lines, meant that a reasonable upper limits on the {\HI} column density was obtained. The {\SiII} and {\NIII} 989~{\AA} lines are blended together. The presence of the other {\SiII} transitions constrained the column densities. The total {\OVI} fits from \citet{nielsenovi} are shown here for completeness, although they are not used in the ionization modelling.}
	\label{fig:J1241_2053}
\end{figure*}

\begin{figure*}[hp]
	\centering
	\includegraphics[width=\linewidth]{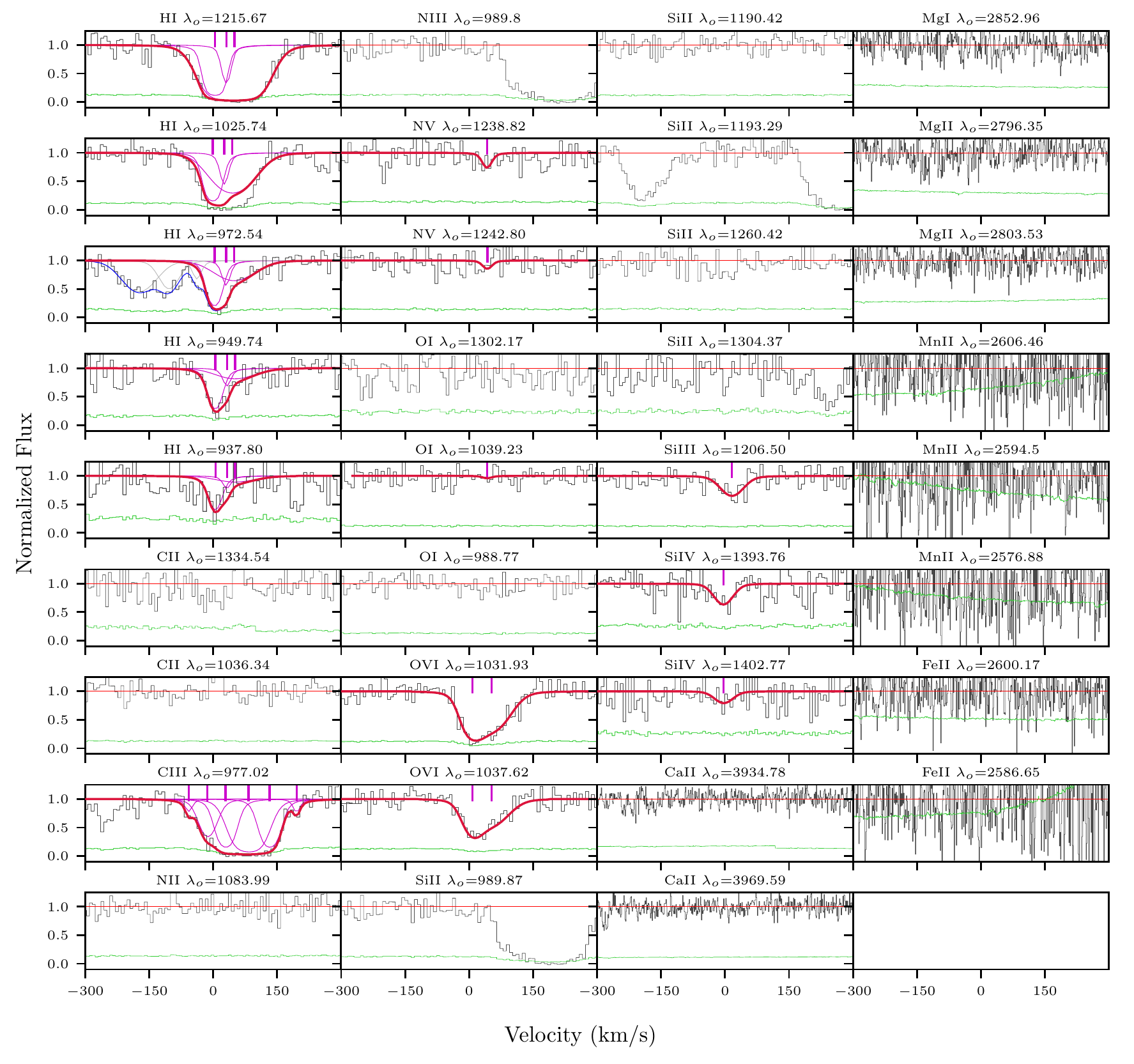}
	\caption{The fits for the system J$1241$, $z_{gal} = 0.217905$, as for figure \ref{fig:Q0122_0.2119}. Although there was unknown blend in the {\HI} 972~{\AA} transition, the presence of other {\HI} lines was suffieient to constrain the column density. The total {\OVI} fits from \citet{nielsenovi} are shown here for completeness, although they are not used in the ionization modelling.}
	\label{fig:J1241_2179}
\end{figure*}
\clearpage
\begin{deluxetable}{ccc}[hp]
	\tablecolumns{8}
	\tablewidth{\linewidth}
	\setlength{\tabcolsep}{0.06in}
	\tablecaption{J$1241$, $z_{gal} = 0.217905$ Column Densities\label{tab:J1241_2179}}
	\tablehead{
		\colhead{Ion}           	&
        \colhead{$\log N~({\cms})$}    &
		\colhead{$\log N$ Error~({\cms})}}
	\startdata
	{\HI}   & $15.59$   &$0.12$\\
{\CII}  & $<13.37$   &$\cdots$\\
{\CIII} & $>14.66$   &$\cdots$\\
{\NII}  & $<13.44$   &$\cdots$\\
{\NIII} & $<13.50$   &$\cdots$\\
{\OI}   & $<13.86$   &$\cdots$\\
{\SiII} & $<12.49$   &$\cdots$\\
{\SiIII}& $12.77$   &$0.09$\\
{\SiIV} & $13.15$   &$0.12$\\
{\CaII} & $<11.74$   &$\cdots$\\
{\MgI}  & $<11.66$   &$\cdots$\\
{\MgII} & $<12.20$   &$\cdots$\\
{\MnII} & $<13.07$   &$\cdots$\\
{\FeII} & $<12.98$   &$\cdots$\\[-5pt]

	\enddata
\end{deluxetable}
\begin{figure}[hp]
	\centering
	\includegraphics[width=\linewidth]{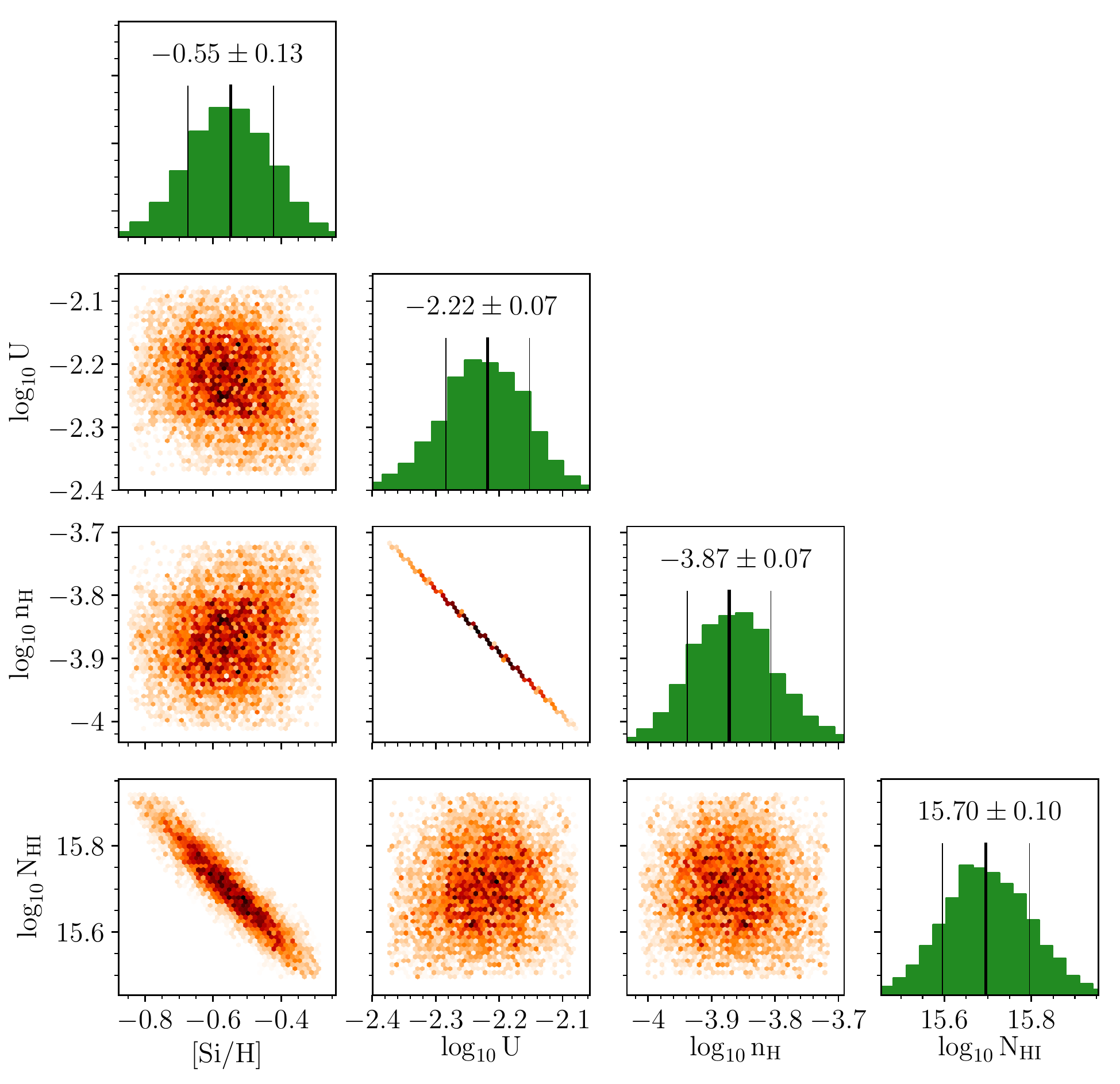}
	\caption{The posterior distribution profiles from the MCMC analysis of the Cloudy grids for J$1241$, $z_{gal} = 0.217905$, as for figure \ref{fig:Q0122_0.2119_par}.}
	\label{fig:J1241_2179_par}
\end{figure}
\newpage
\begin{deluxetable}{ccc}[hp]
	\tablecolumns{8}
	\tablewidth{\linewidth}
	\setlength{\tabcolsep}{0.06in}
	\tablecaption{J$1244$, $z_{gal} = 0.5504$ Column Densities\label{tab:J1244_5504}}
	\tablehead{
		\colhead{Ion}           	&
        \colhead{$\log N~({\cms})$}    &
		\colhead{$\log N$ Error~({\cms})}}
	\startdata
	{\HI}   & $[17.00, 19.00]$   &$\cdots$\\
{\CII}  & $14.82$   &$0.08$\\
{\CIII} & $>15.23$   &$\cdots$\\
{\NII}  & $14.50$   &$0.09$\\
{\NIII} & $>14.90$   &$\cdots$\\
{\OI}   & $<14.83$   &$\cdots$\\
{\SiII} & $<14.23$   &$\cdots$\\
{\CaII} & $11.73$   &$0.05$\\
{\MgI}  & $11.83$   &$0.04$\\
{\MgII} & $13.55$   &$0.05$\\
{\MnII} & $<12.40$   &$\cdots$\\
{\FeII} & $13.60$   &$0.05$\\

	\enddata
\end{deluxetable}
\begin{figure}[hp]
	\centering
	\includegraphics[width=\linewidth]{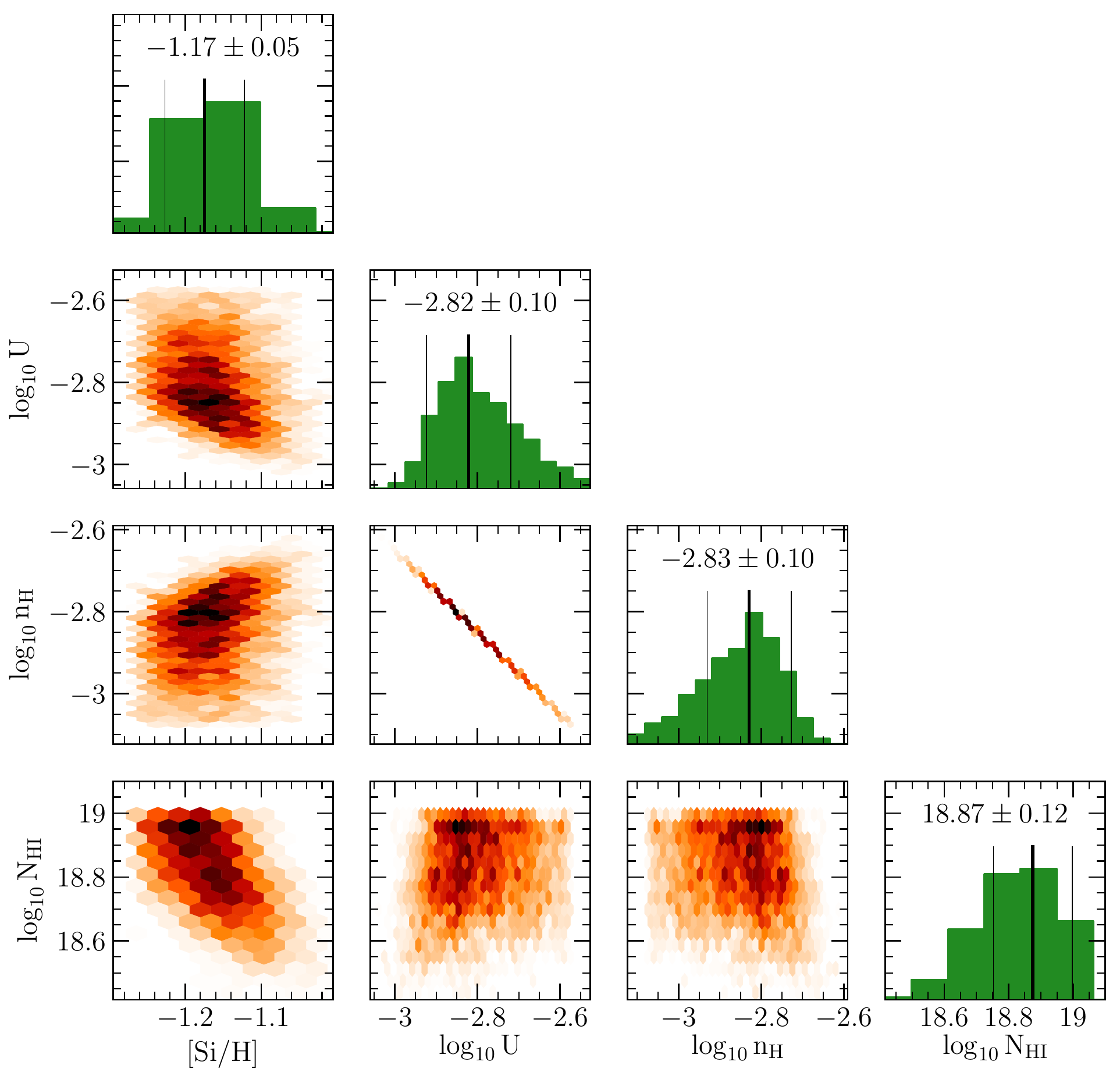}
	\caption{The posterior distribution profiles from the MCMC analysis of the Cloudy grids for J$1244$, $z_{gal} = 0.5504$, as for figure \ref{fig:Q0122_0.2119_par}.}
	\label{fig:J1244_5504_par}
\end{figure}
\begin{figure*}[hp]
	\centering
	\includegraphics[width=\linewidth]{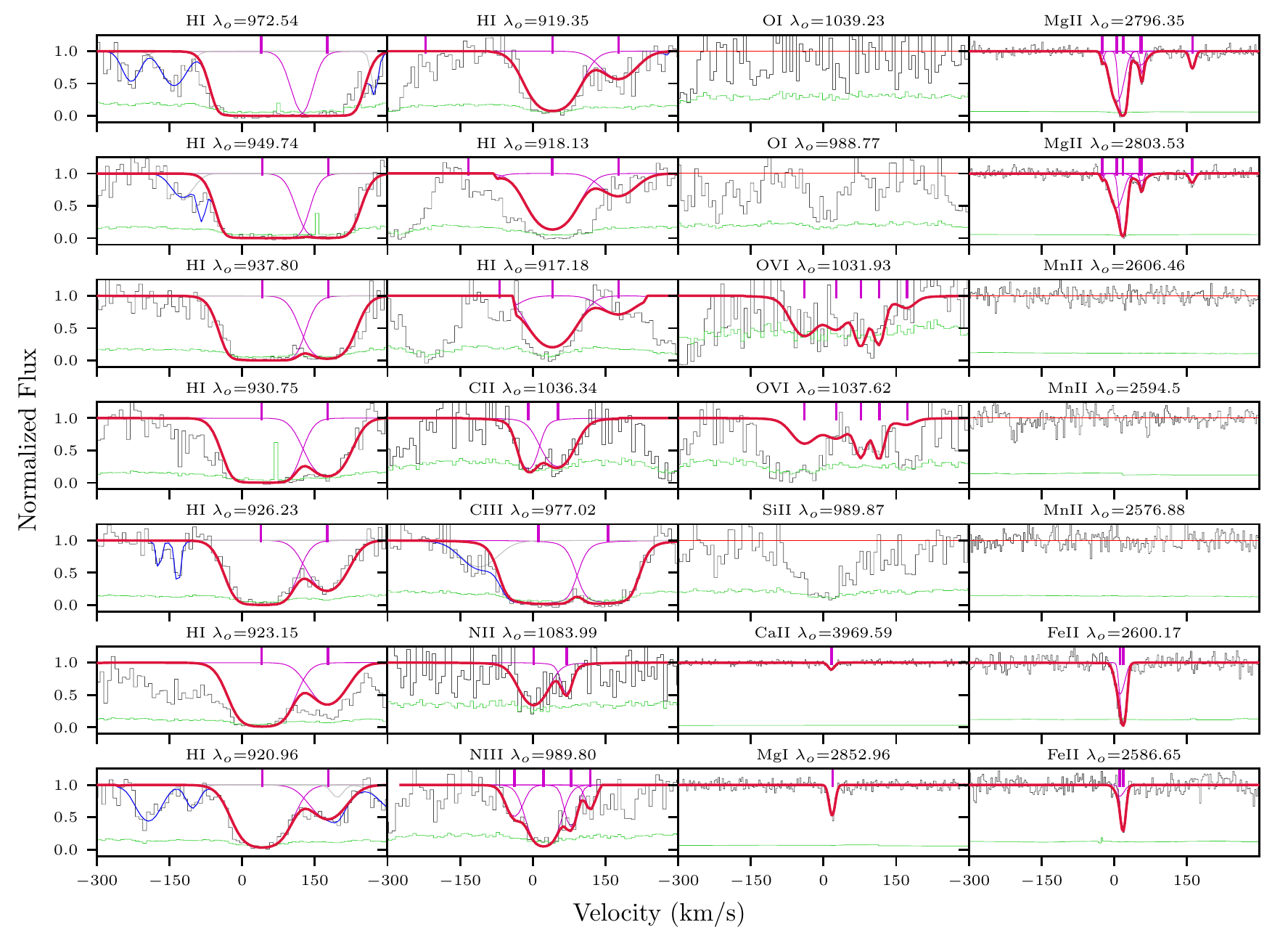}
	\caption{The fits for the system J$1244$, $z_{gal} = 0.5504$, as for figure \ref{fig:Q0122_0.2119}. Although there are blends present in the {\HI} 972~{\AA}, 949~{\AA}, 926~{\AA} and 920~{\AA} transitions, the other {\HI} lines are able to constrain the upper limit on the column density. Additionally, there is a blend on the left of the {\CII} 977~{\AA} line. However, it is sufficiently offset from the absorption, such that it has minimal impact on obtaining an upper limit on the column density. The total {\OVI} fits from \citet{nielsenovi} are shown here for completeness, although they are not used in the ionization modelling.}
	\label{fig:J1244_5504}
\end{figure*}

\begin{figure*}[hp]
	\centering
	\includegraphics[width=\linewidth]{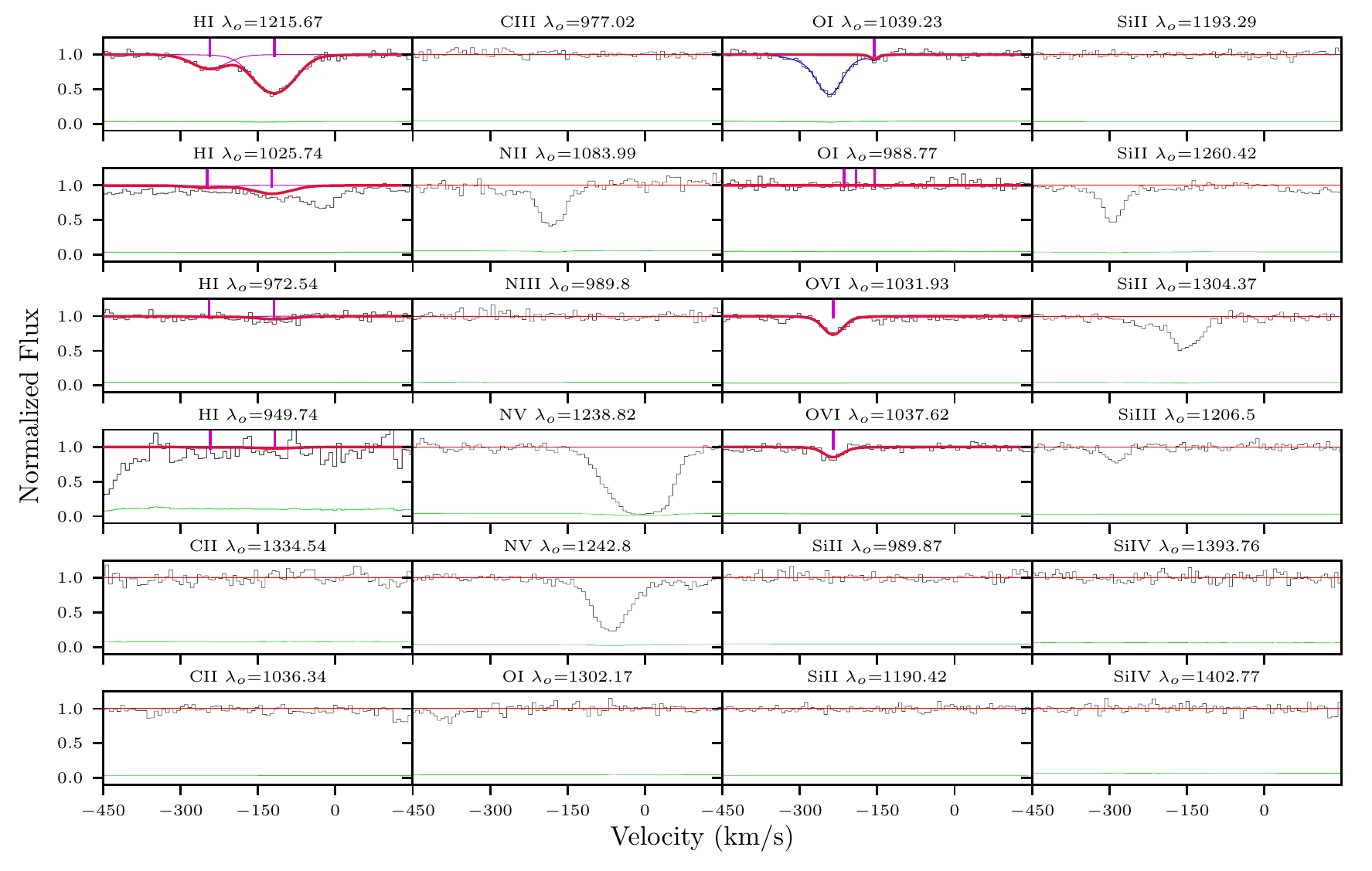}
	\caption{The fits for the system J$1301$, $z_{gal} = 0.1967$, as for figure \ref{fig:Q0122_0.2119}. There is an unknown blend for the {\OI} 1039~{\AA} transition. However, the presence of other {\OI} transitions is sufficient to constrain the column density. The total {\OVI} fits from \citet{nielsenovi} are shown here for completeness, although they are not used in the ionization modelling.}
	\label{fig:J1301_1967}
\end{figure*}
\clearpage
\begin{deluxetable}{ccc}[hp]
	\tablecolumns{8}
	\tablewidth{\linewidth}
	\setlength{\tabcolsep}{0.06in}
	\tablecaption{J$1301$, $z_{gal} = 0.1967$ Column Densities\label{tab:J1301_1967}}
	\tablehead{
		\colhead{Ion}           	&
        \colhead{$\log N~({\cms})$}    &
		\colhead{$\log N$ Error~({\cms})}}
	\startdata
	{\HI}   & $13.86$   &$0.01$\\
{\CII}  & $<12.74$   &$\cdots$\\
{\CIII} & $<12.08$   &$\cdots$\\
{\NII}  & $<13.01$   &$\cdots$\\
{\NIII} & $<12.94$   &$\cdots$\\
{\NV}   & $<12.94$   &$\cdots$\\
{\OI}   & $<13.18$   &$\cdots$\\
{\SiII} & $<12.30$   &$\cdots$\\
{\SiIII}& $<11.46$   &$\cdots$\\
{\SiIV} & $<12.24$   &$\cdots$\\[-5pt]
	\enddata
\end{deluxetable}
\begin{figure}[hp]
	\centering
	\includegraphics[width=\linewidth]{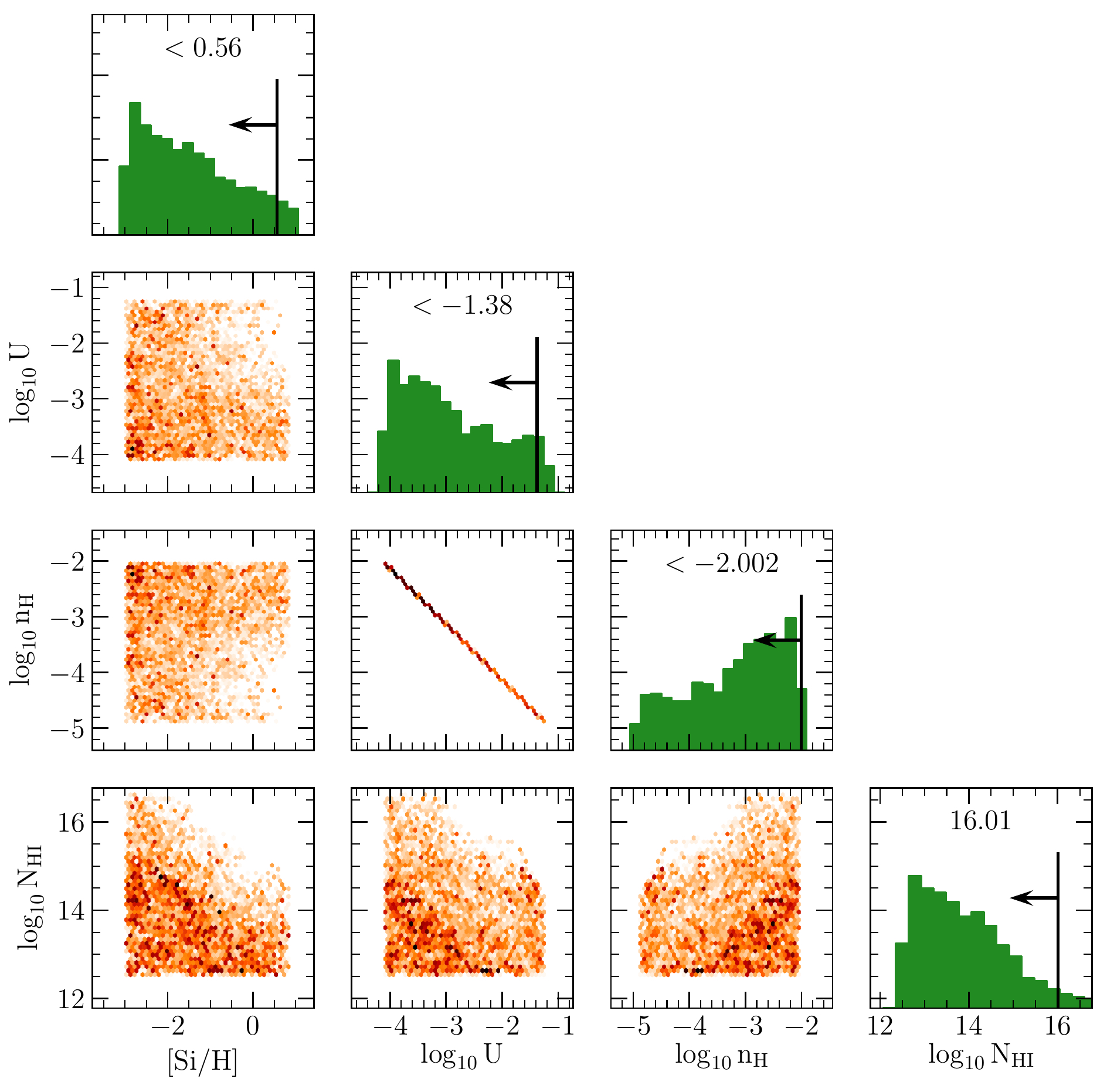}
	\caption{The posterior distribution profiles from the MCMC analysis of the Cloudy grids for J$1301$, $z_{gal} = 0.1967$, as for figure \ref{fig:Q0122_0.2119_par}.}
	\label{fig:J1301_1967_par}
\end{figure}
\newpage
\begin{deluxetable}{ccc}[hp]
	\tablecolumns{8}
	\tablewidth{\linewidth}
	\setlength{\tabcolsep}{0.06in}
	\tablecaption{J$1319$, $z_{gal} = 0.6610$ Column Densities\label{tab:J1319_6610}}
	\tablehead{
		\colhead{Ion}           	&
        \colhead{$\log N~({\cms})$}    &
		\colhead{$\log N$ Error~({\cms})}}
	\startdata
	{\HI}   & $18.30$   &$0.30$\\
{\CII}  & $<13.34$   &$\cdots$\\
{\CIII} & $14.35$   &$0.07$\\
{\NIII} & $<13.32$   &$\cdots$\\
{\OI}   & $<13.73$   &$\cdots$\\
{\SiII} & $<13.24$   &$\cdots$\\
{\MgI}  & $11.37$   &$0.04$\\
{\MgII} & $13.17$   &$0.02$\\
{\MnII} & $<11.80$   &$\cdots$\\
{\FeII} & $13.07$   &$0.02$\\[-5pt]

	\enddata
\end{deluxetable}
\begin{figure}[hp]
	\centering
	\includegraphics[width=\linewidth]{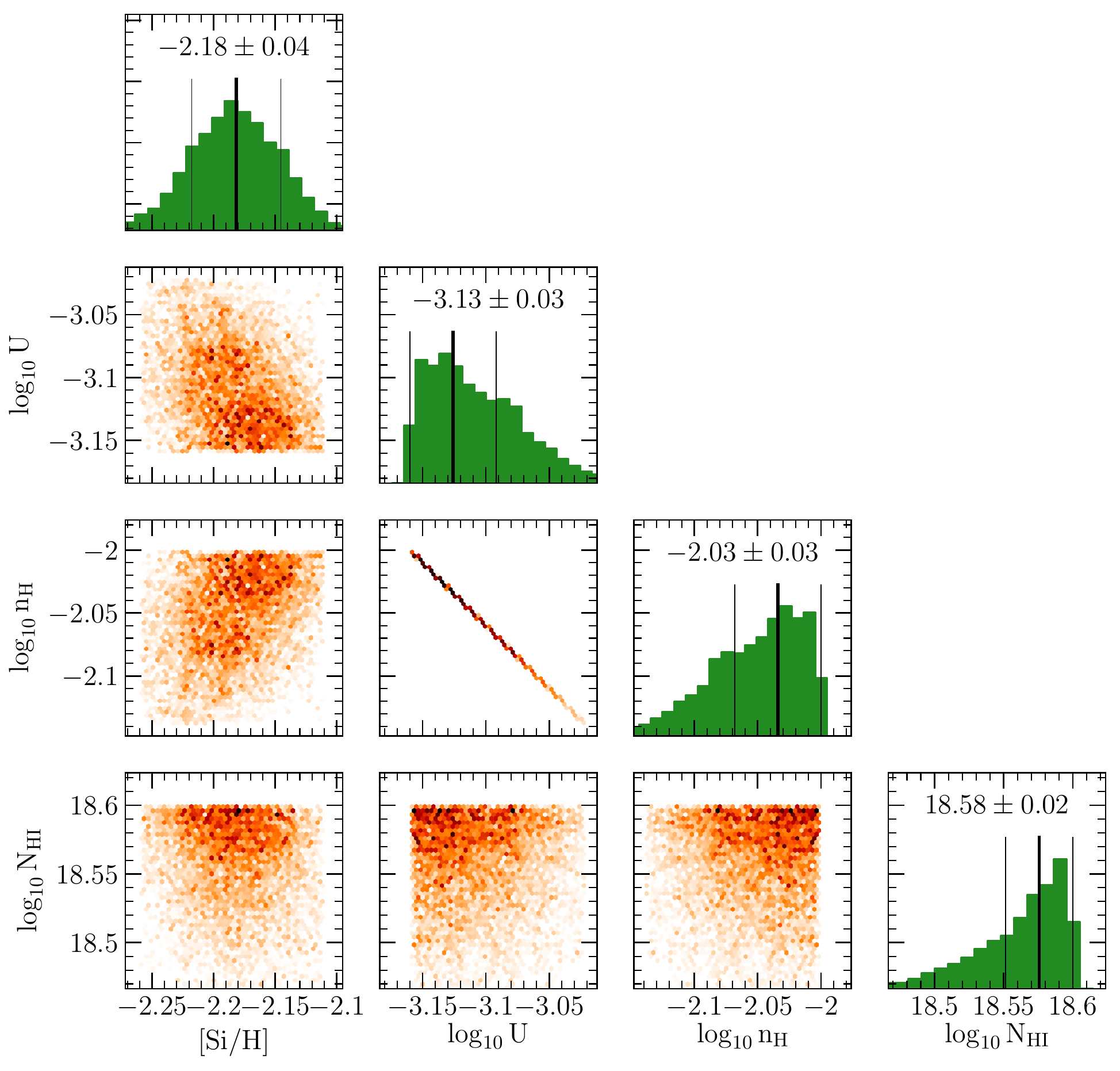}
	\caption{The posterior distribution profiles from the MCMC analysis of the Cloudy grids for J$1319$, $z_{gal} = 0.6610$, as for figure \ref{fig:Q0122_0.2119_par}.}
	\label{fig:J1319_6610_par}
\end{figure}
\begin{figure*}[hp]
	\centering
	\includegraphics[width=\linewidth]{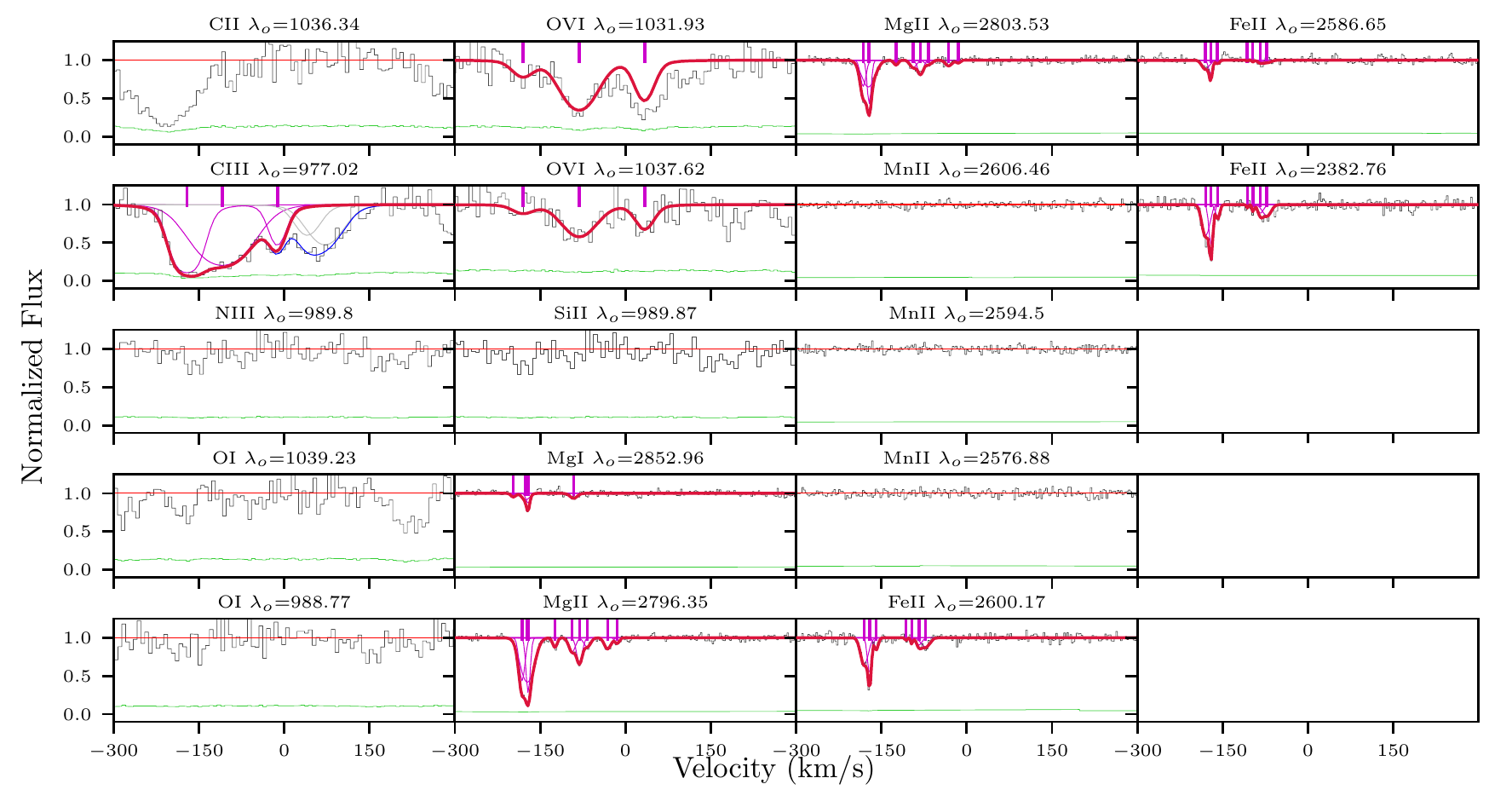}
	\caption{The fits for the system J$1319$, $z_{gal} = 0.6610$, as for figure \ref{fig:Q0122_0.2119}. The {\HI} column density measurement was taken from \citet{kacprzak12b} where it was obtained by fitting the Lyman break. There is a blend on the right of the {\CIII} transition. However, it is sufficiently distinct from the transition, such that the column density is well constrained.  The total {\OVI} fits from \citet{nielsenovi} are shown here for completeness, although they are not used in the ionization modelling.}
	\label{fig:J1319_6610}
\end{figure*}

\begin{figure*}[hp]
	\centering
	\includegraphics[width=\linewidth]{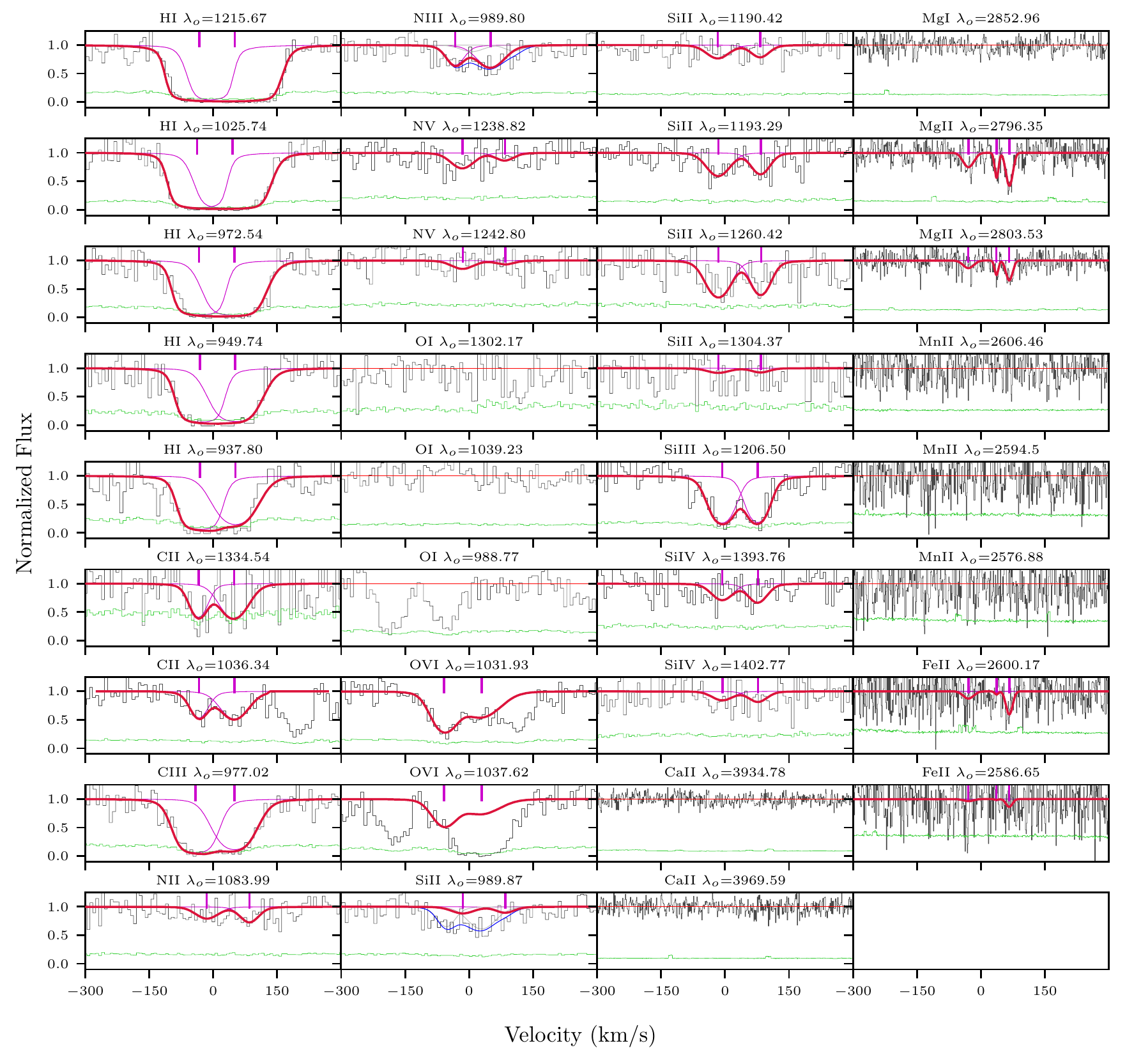}
	\caption{The fits for the system J$1322$, $z_{gal} = 0.214431$, as for figure \ref{fig:Q0122_0.2119}. The {\SiII} and {\NIII} 989~{\AA} lines are blended together. The presence of other {\SiII} constrains the column densities. The total {\OVI} fits from \citet{nielsenovi} are shown here for completeness, although they are not used in the ionization modelling.}
	\label{fig:J1322_2144}
\end{figure*}
\clearpage
\begin{deluxetable}{ccc}[hp]
	\tablecolumns{8}
	\tablewidth{\linewidth}
	\setlength{\tabcolsep}{0.06in}
	\tablecaption{J$1322$, $z_{gal} = 0.214431$ Column Densities\label{tab:J1322_2144}}
	\tablehead{
		\colhead{Ion}           	&
        \colhead{$\log N~({\cms})$}    &
		\colhead{$\log N$ Error~({\cms})}}
	\startdata
	{\HI}   & $[16.97, 19.00]$   &$\cdots$\\
{\CII}  & $14.45$   &$0.03$\\
{\NII}  & $14.03$   &$0.10$\\
{\OI}   & $<14.19$   &$\cdots$\\
{\SiII} & $13.53$   &$0.07$\\
{\SiIII}& $13.74$   &$0.06$\\
{\CaII} & $<11.43$   &$\cdots$\\
{\MgI}  & $<11.34$   &$\cdots$\\
{\MgII} & $12.83$   &$0.03$\\
{\MnII} & $<12.56$   &$\cdots$\\
{\FeII} & $12.93$   &$0.11$\\[-5pt]

	\enddata
\end{deluxetable}
\begin{figure}[hp]
	\centering
	\includegraphics[width=\linewidth]{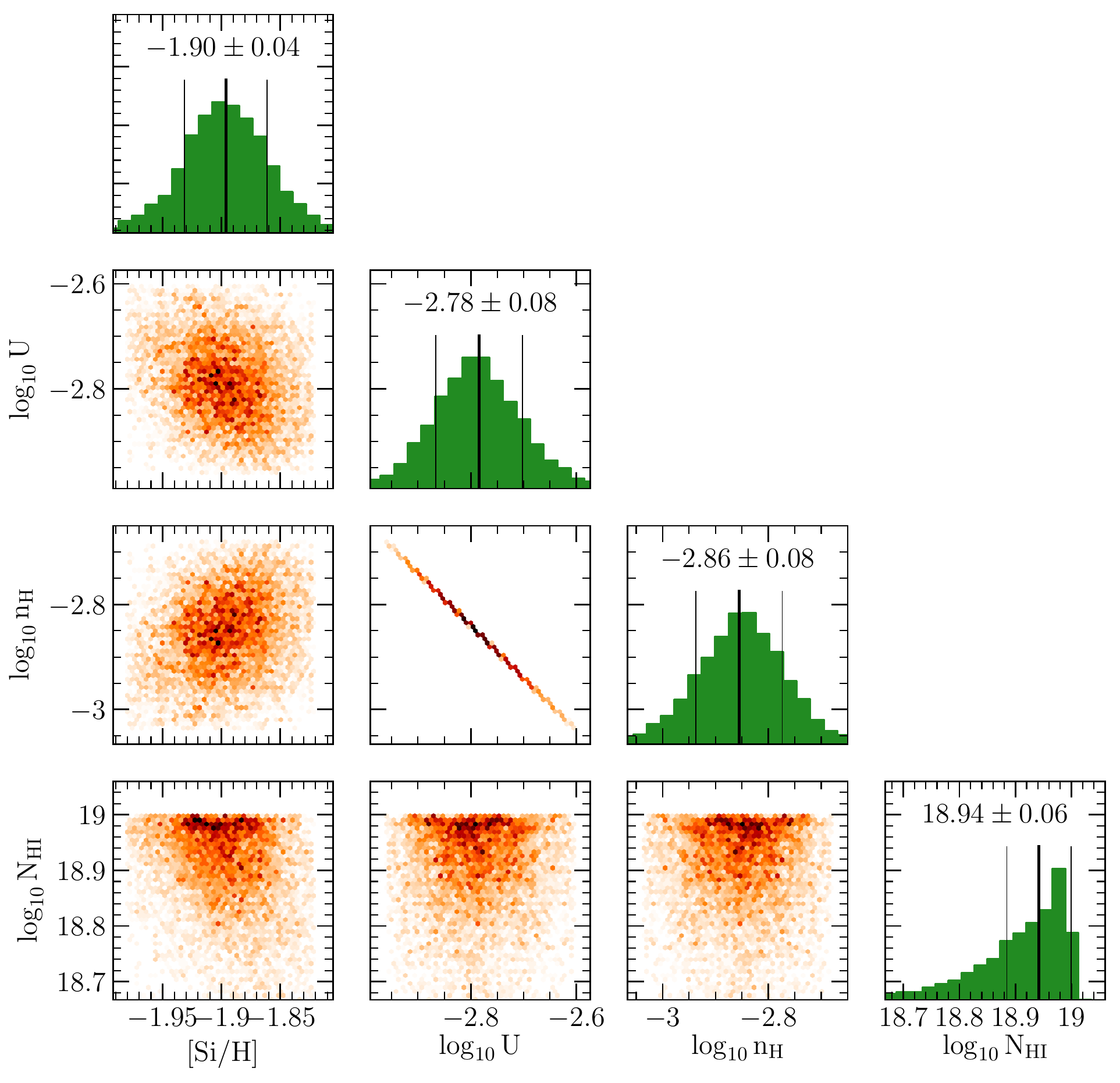}
	\caption{The posterior distribution profiles from the MCMC analysis of the Cloudy grids for J$1322$, $z_{gal} = 0.214431$, as for figure \ref{fig:Q0122_0.2119_par}.}
	\label{fig:J1322_2144_par}
\end{figure}
\newpage
\begin{deluxetable}{ccc}[hp]
	\tablecolumns{8}
	\tablewidth{\linewidth}
	\setlength{\tabcolsep}{0.06in}
	\tablecaption{J$1342$, $z_{gal} = 0.0708$ Column Densities\label{tab:J1342_0708}}
	\tablehead{
		\colhead{Ion}           	&
        \colhead{$\log N~({\cms})$}    &
		\colhead{$\log N$ Error~({\cms})}}
	\startdata
	{\HI}   & $14.61$   &$0.47$\\
{\CII}  & $<14.68$   &$\cdots$\\
{\CIV}  & $14.19$    &$0.05$\\
{\NII}  & $<13.94$   &$\cdots$\\
{\NV}   & $<13.71$   &$\cdots$\\
{\OI}   & $<14.46$   &$\cdots$\\
{\SiII} & $<13.98$   &$\cdots$\\
{\SiIII}& $<13.63$   &$\cdots$\\
{\SiIV} & $13.49$   &$0.10$\\
{\CaII} & $<10.98$   &$\cdots$\\
{\MgI}  & $<10.95$   &$\cdots$\\[-5pt]

	\enddata
\end{deluxetable}
\begin{figure}[hp]
	\centering
	\includegraphics[width=\linewidth]{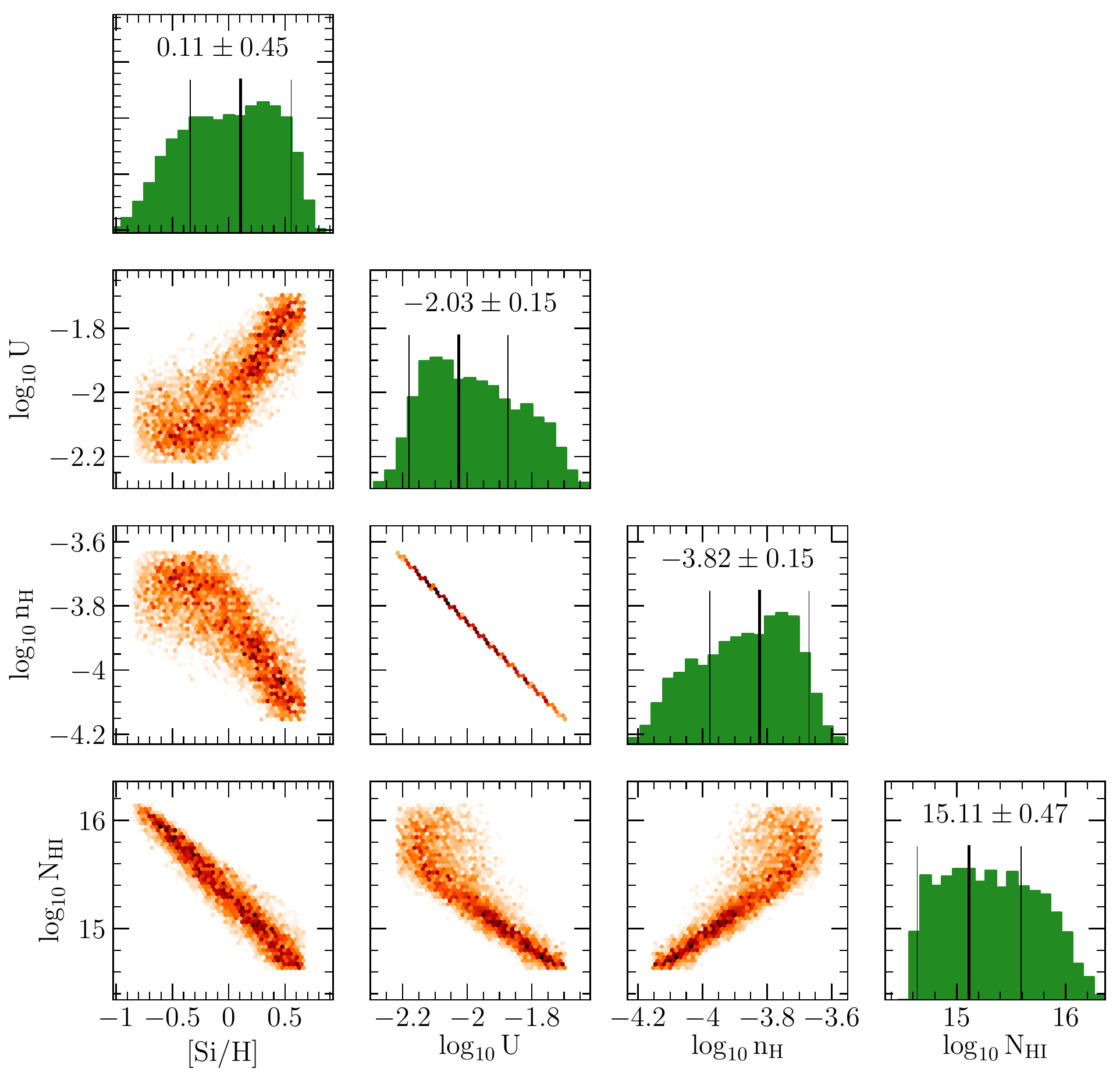}
	\caption{The posterior distribution profiles from the MCMC analysis of the Cloudy grids for J$1342$, $z_{gal} = 0.0708$, as for figure \ref{fig:Q0122_0.2119_par}.}
	\label{fig:J1342_0708_par}
\end{figure}
\begin{figure*}[hp]
	\centering
	\includegraphics[width=\linewidth]{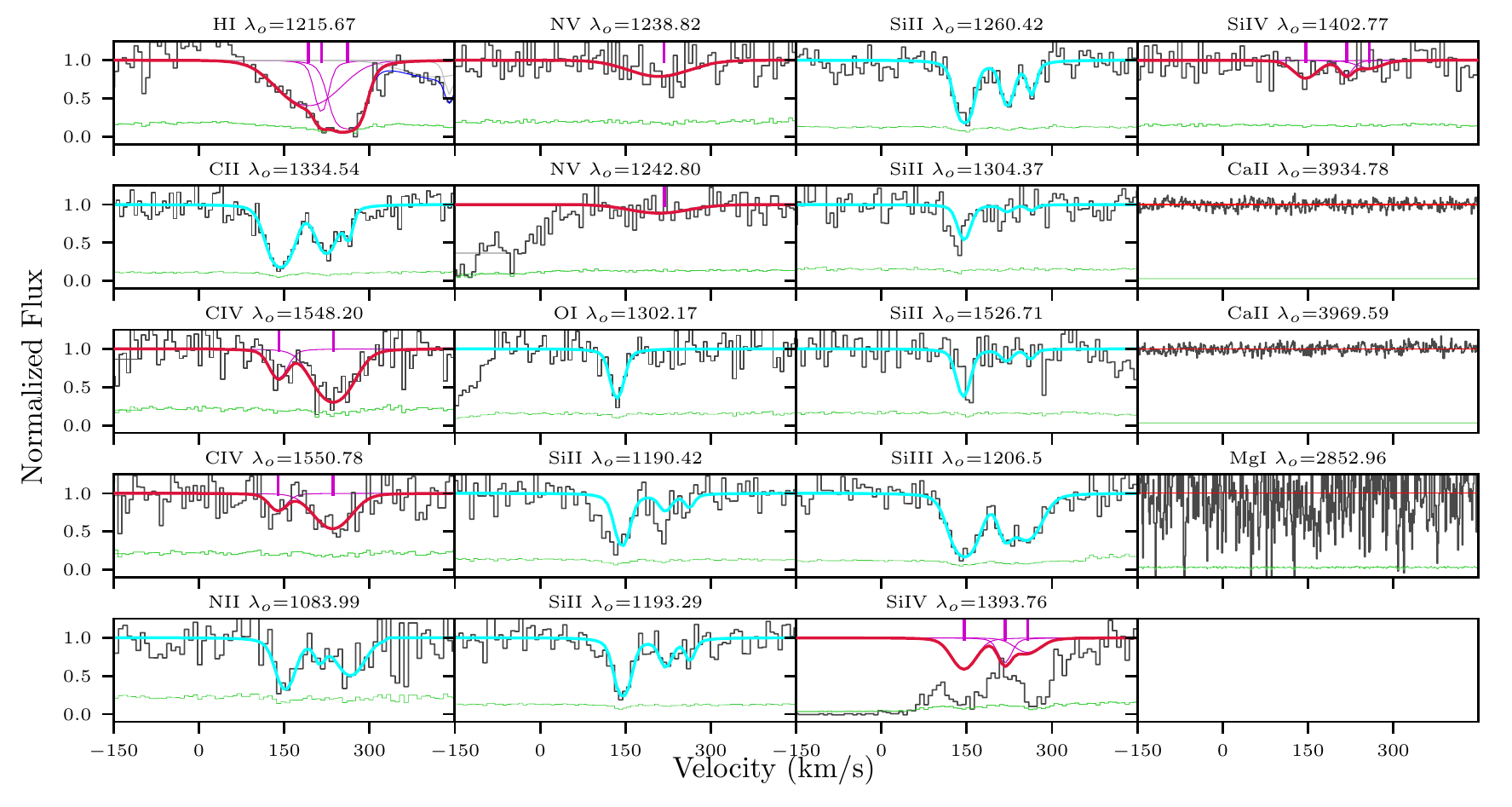}
	\caption{The fits for the system J$1342$, $z_{gal} = 0.0708$, as for figure \ref{fig:Q0122_0.2119}. We have used the column densities from the {\CII}, {\NII}, {\OI}, {\SiII} and {\SiIII} line models as upper limits in the metallicity calculation. The main absorption peak for each of these ions is coincident with only weak {\HI} absorption and is typically very strong. The inclusion of these ions in the metallicity calculation results in a super-solar metallicity ([Si/H]$>1$), which is unlikely to occur in the CGM.}
	\label{fig:J1342_0708}
\end{figure*}

\begin{figure*}[hp]
	\centering
	\includegraphics[width=\linewidth]{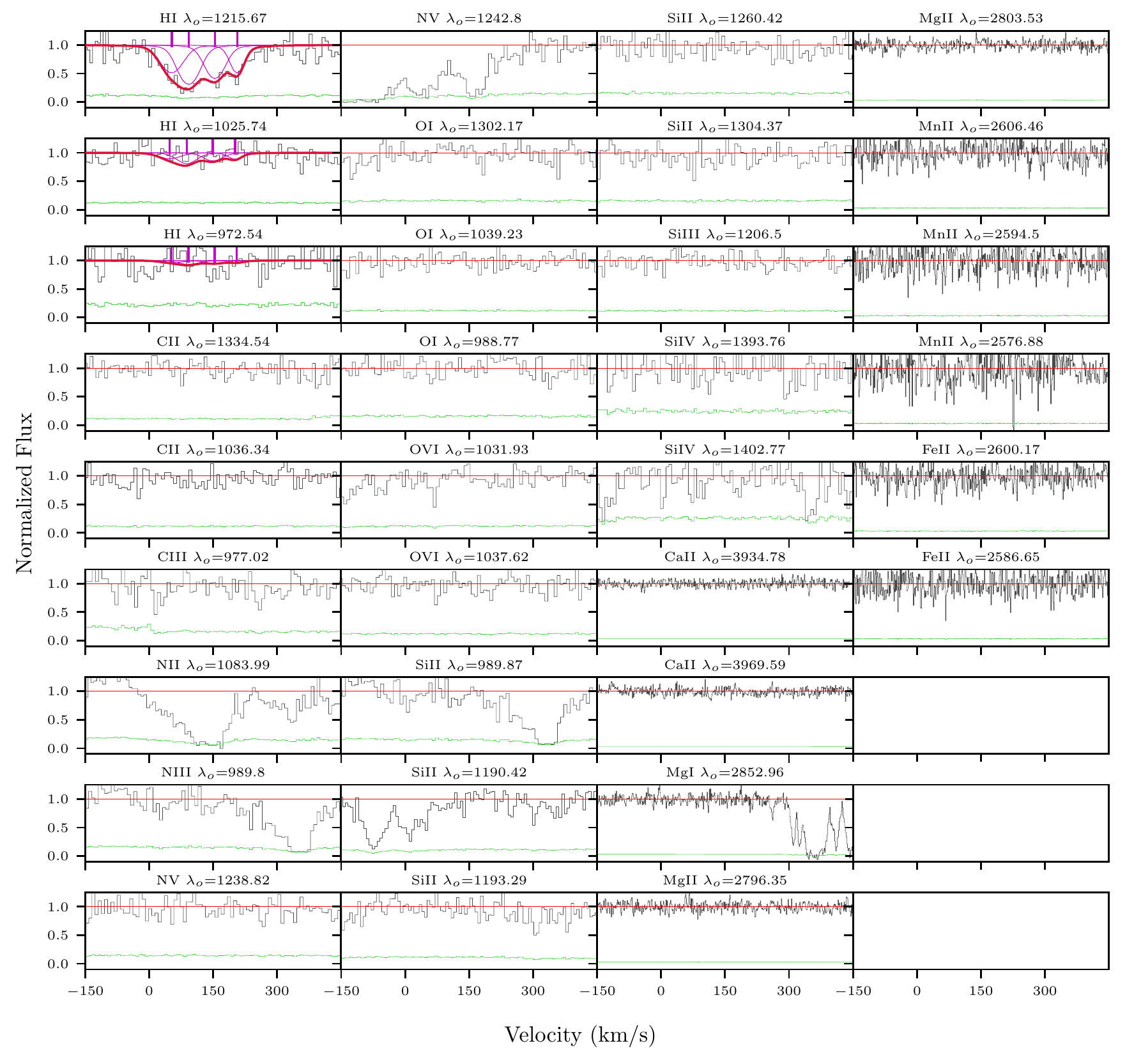}
	\caption{The fits for the system J$1342$, $z_{gal} = 0.2013$, as for figure \ref{fig:Q0122_0.2119}.}
	\label{fig:J1342_2013}
\end{figure*}
\clearpage
\begin{deluxetable}{ccc}[hp]
	\tablecolumns{8}
	\tablewidth{\linewidth}
	\setlength{\tabcolsep}{0.06in}
	\tablecaption{J$1342$, $z_{gal} = 0.2013$ Column Densities\label{tab:J1342_2013}}
	\tablehead{
		\colhead{Ion}           	&
        \colhead{$\log N~({\cms})$}    &
		\colhead{$\log N$ Error~({\cms})}}
	\startdata
	{\HI}   & $14.22$   &$0.03$\\
{\CII}  & $<13.18$   &$\cdots$\\
{\CIII} & $<12.73$   &$\cdots$\\
{\NII}  & $<14.16$   &$\cdots$\\
{\NIII} & $<13.61$   &$\cdots$\\
{\NV}   & $<13.23$   &$\cdots$\\
{\OI}   & $<13.73$   &$\cdots$\\
{\SiII} & $<12.54$   &$\cdots$\\
{\SiIII}& $<12.10$   &$\cdots$\\
{\SiIV} & $<13.03$   &$\cdots$\\
{\CaII} & $<10.98$   &$\cdots$\\
{\MgI}  & $<10.67$   &$\cdots$\\
{\MgII} & $<11.15$   &$\cdots$\\
{\MnII} & $<11.43$   &$\cdots$\\
{\FeII} & $<11.61$   &$\cdots$\\[-5pt]

	\enddata
\end{deluxetable}
\begin{figure}[hp]
	\centering
	\includegraphics[width=\linewidth]{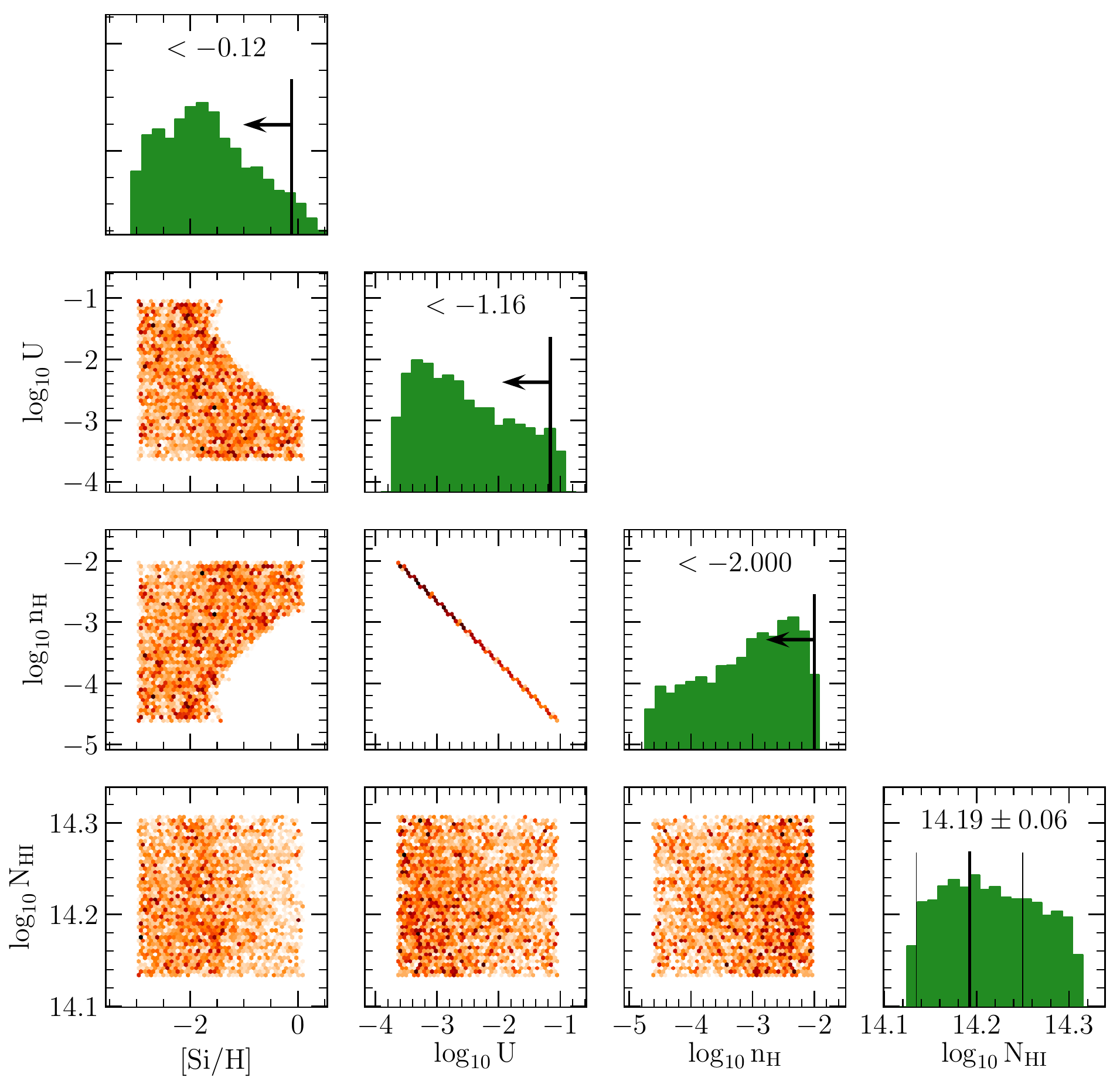}
	\caption{The posterior distribution profiles from the MCMC analysis of the Cloudy grids for J$1342$, $z_{gal} = 0.2013$, as for figure \ref{fig:Q0122_0.2119_par}.}
	\label{fig:J1342_2013_par}
\end{figure}
\newpage
\begin{deluxetable}{ccc}[hp]
	\tablecolumns{8}
	\tablewidth{\linewidth}
	\setlength{\tabcolsep}{0.06in}
	\tablecaption{J$1342$, $z_{gal} = 0.227042$ Column Densities\label{tab:J1342_2270}}
	\tablehead{
		\colhead{Ion}           	&
        \colhead{$\log N~({\cms})$}    &
		\colhead{$\log N$ Error~({\cms})}}
	\startdata
	{\HI}   & $18.83$   &$0.05$\\
{\CII}  & $15.46$   &$0.12$\\
{\CIII} & $15.28$   &$0.23$\\
{\NII}  & $15.07$   &$0.12$\\
{\NV}   & $<13.37$   &$\cdots$\\
{\OI}   & $15.45$   &$0.09$\\
{\SiII} & $14.72$   &$0.09$\\
{\SiIII}& $14.65$   &$0.22$\\
{\SiIV} & $13.64$   &$0.07$\\
{\CaII} & $12.49$   &$0.05$\\
{\MgI}  & $12.58$   &$0.12$\\
{\MgII} & $15.00$   &$0.16$\\
{\MnII} & $12.30$   &$0.06$\\
{\FeII} & $15.00$   &$0.17$\\[-5pt]

	\enddata
\end{deluxetable}
\begin{figure}[hp]
	\centering
	\includegraphics[width=\linewidth]{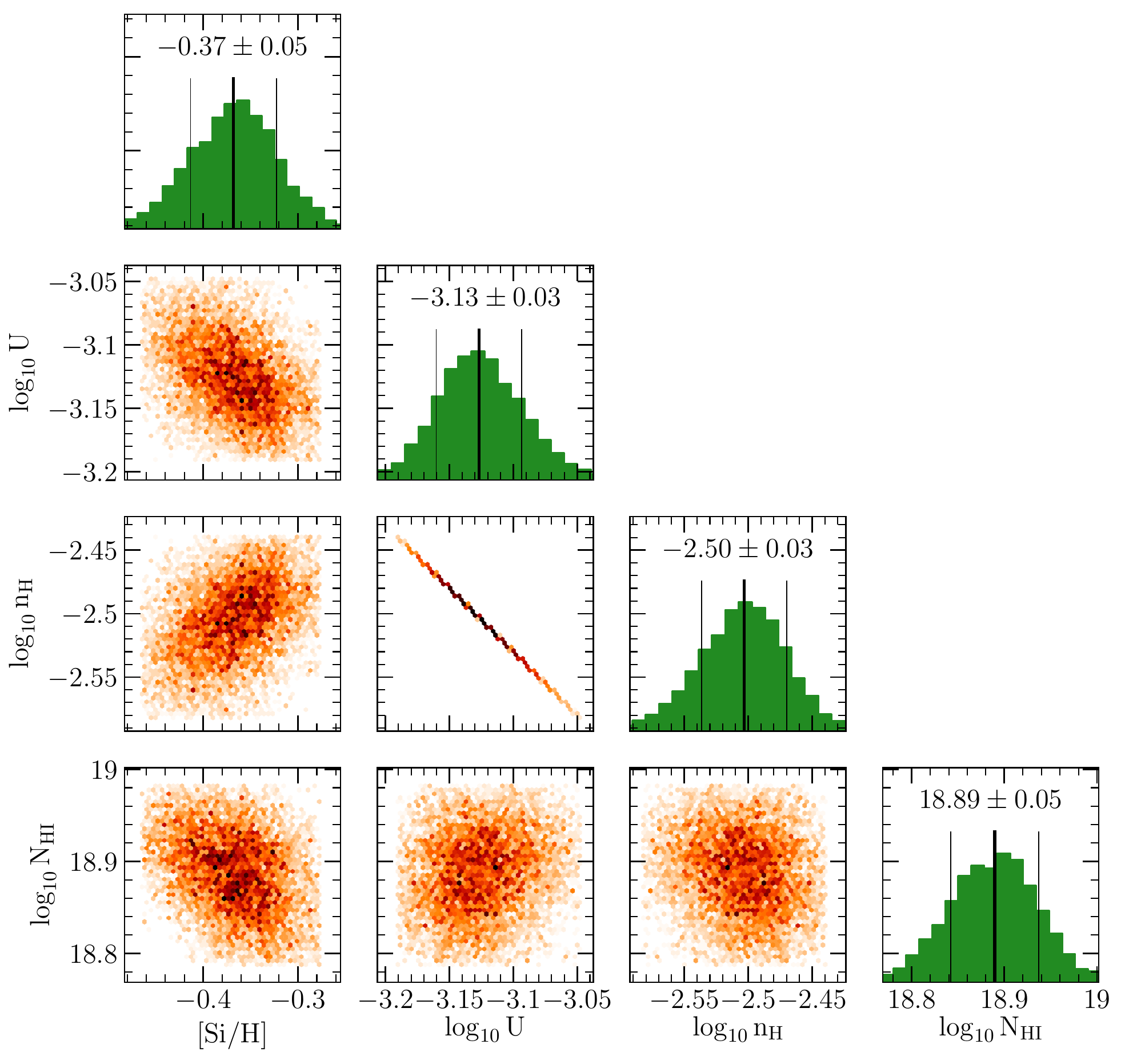}
	\caption{The posterior distribution profiles from the MCMC analysis of the Cloudy grids for J$1342$, $z_{gal} = 0.227042$, as for figure \ref{fig:Q0122_0.2119_par}.}
	\label{fig:J1342_2270_par}
\end{figure}
\begin{figure*}[hp]
	\centering
	\includegraphics[width=\linewidth]{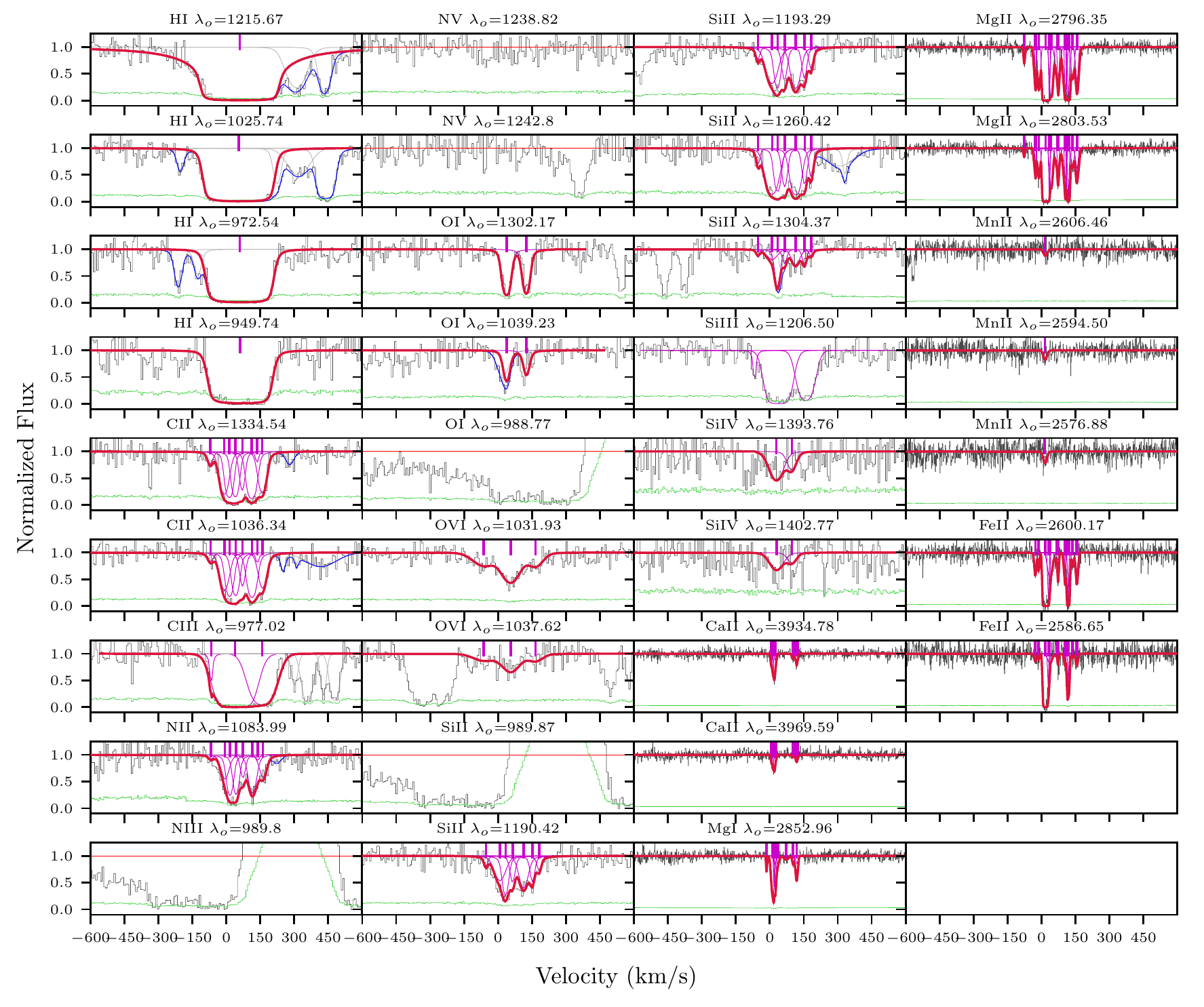}
	\caption{The fits for the system J$1342$, $z_{gal} = 0.227042$, as for figure \ref{fig:Q0122_0.2119}. The {\HI} column density is well constrained by the presence of damping wings, despite unidentified blends in the {\HI} 1215~{\AA}, 1025~{\AA} and 972~{\AA} transitions. There are blends in the {\CII}, {\CIII} and {\SiII} transitions. However, the offset of the blends is sufficient that the column density of the ions can be constrained. The {\OI} 1039~{\AA} transition was also blended with an unknown line. However, the column density was constrained using the {\OI} transition.  The total {\OVI} fits from \citet{nielsenovi} are shown here for completeness, although they are not used in the ionization modelling.}
	\label{fig:J1342_2270}
\end{figure*}

\begin{figure*}[hp]
	\centering
	\includegraphics[width=\linewidth]{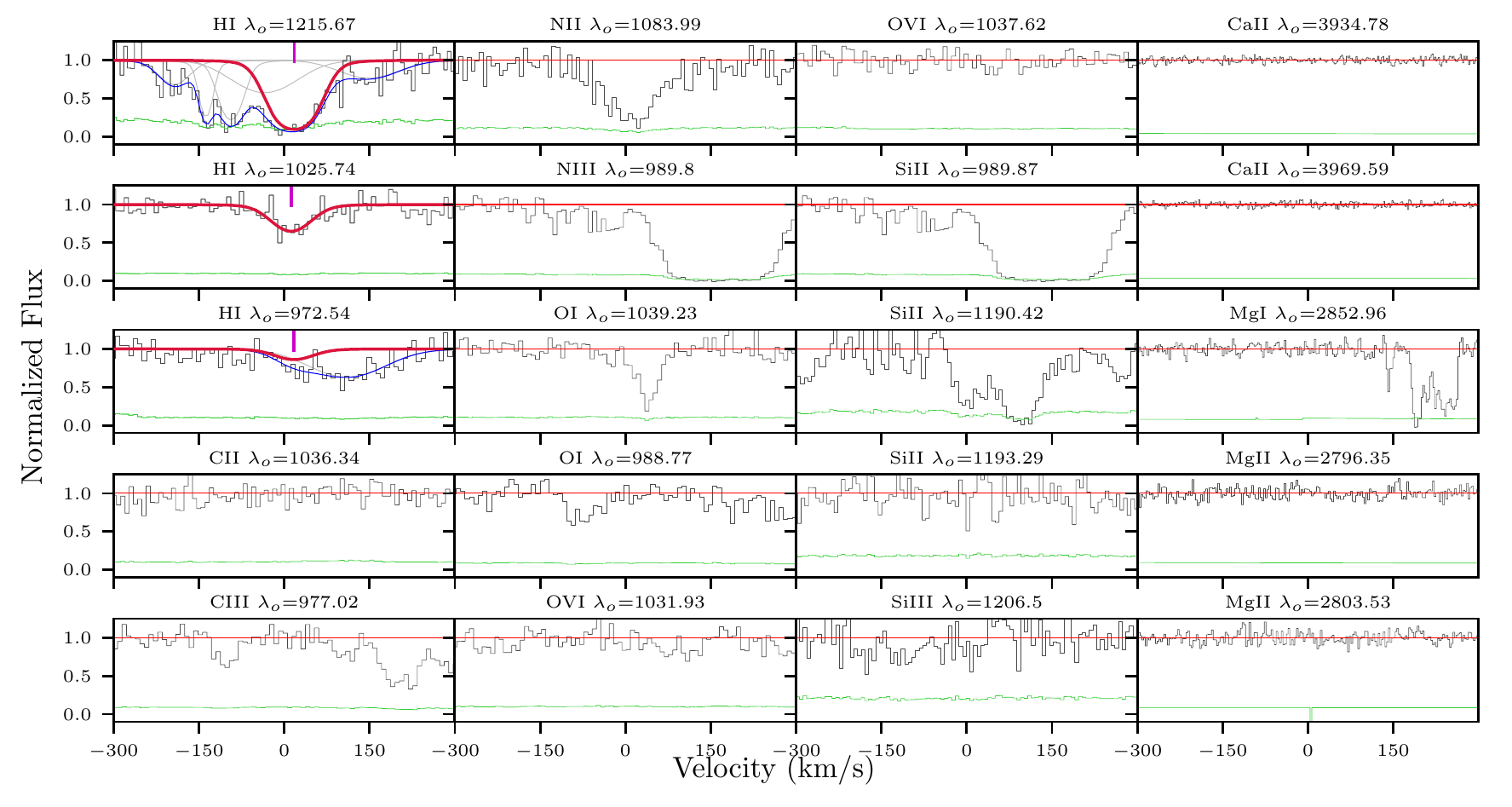}
	\caption{The fits for the system J$1357$, $z_{gal} = 0.4295$, as for figure \ref{fig:Q0122_0.2119}. The {\HI} column density was constrained using the 1025~{\AA} line, despite blends in the 1215~{\AA} and 972~{\AA} transitions. No other metals were detected here.}
	\label{fig:J1357_4303}
\end{figure*}
\clearpage
\begin{deluxetable}{ccc}[hp]
	\tablecolumns{8}
	\tablewidth{\linewidth}
	\setlength{\tabcolsep}{0.06in}
	\tablecaption{J$1357$, $z_{gal} = 0.4295$ Column Densities\label{tab:J1357_4303}}
	\tablehead{
		\colhead{Ion}           	&
        \colhead{$\log N~({\cms})$}    &
		\colhead{$\log N$ Error~({\cms})}}
	\startdata
	{\HI}   & $14.25$   &$0.05$\\
{\CII}  & $<13.27$   &$\cdots$\\
{\CIII} & $<12.44$   &$\cdots$\\
{\NII}  & $<13.79$   &$\cdots$\\
{\NIII} & $<13.34$   &$\cdots$\\
{\OI}   & $<13.69$   &$\cdots$\\
{\SiII} & $<12.89$   &$\cdots$\\
{\SiIII}& $<12.47$   &$\cdots$\\
{\CaII} & $<11.17$   &$\cdots$\\
{\MgI}  & $<11.29$   &$\cdots$\\
{\MgII} & $<11.77$   &$\cdots$\\[-5pt]

	\enddata
\end{deluxetable}
\begin{figure}[hp]
	\centering
	\includegraphics[width=\linewidth]{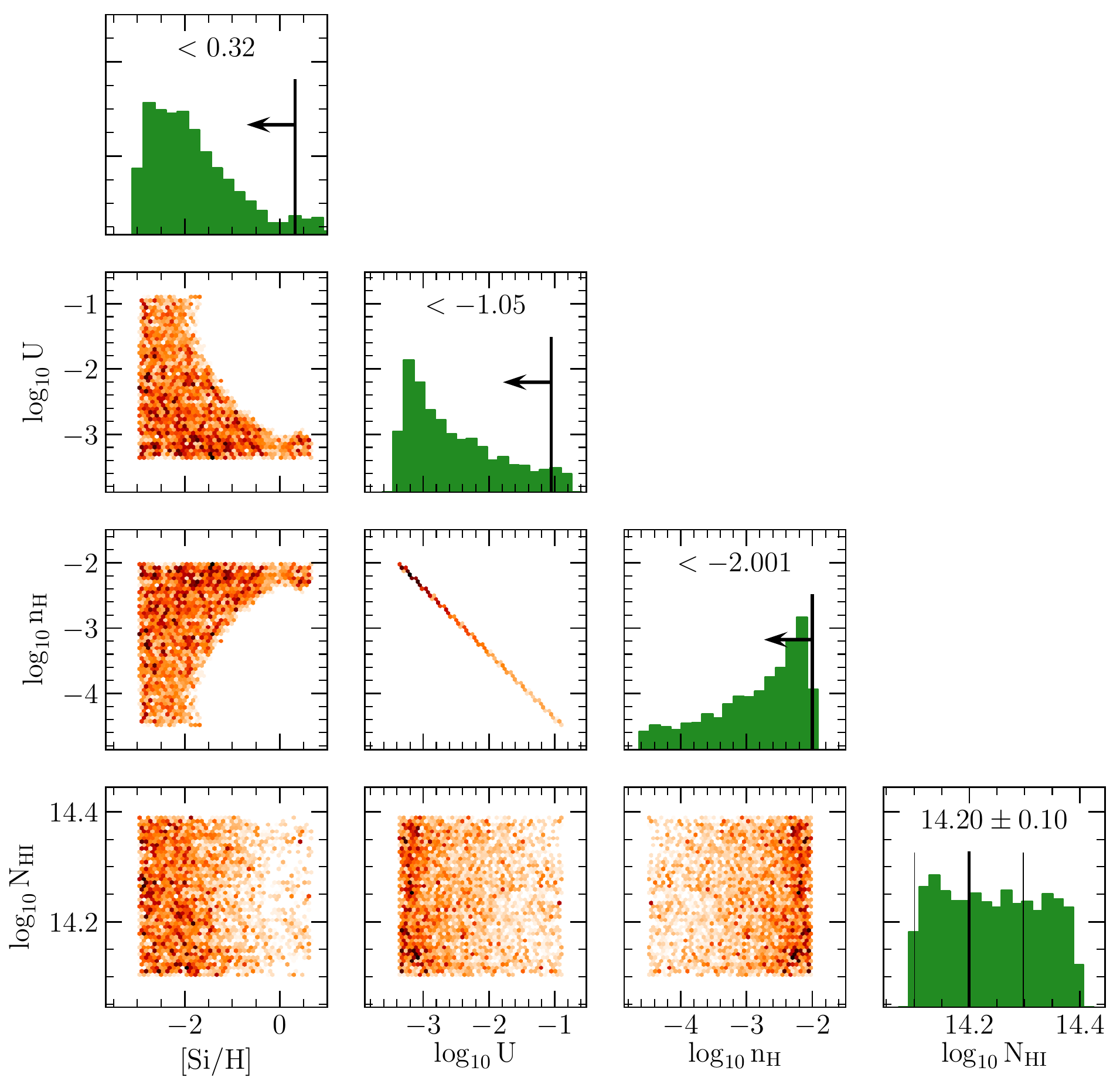}
	\caption{The posterior distribution profiles from the MCMC analysis of the Cloudy grids for J$1357$, $z_{gal} = 0.4295$, as for figure \ref{fig:Q0122_0.2119_par}.}
	\label{fig:J1357_4303_par}
\end{figure}
\newpage
\begin{deluxetable}{ccc}[hp]
	\tablecolumns{8}
	\tablewidth{\linewidth}
	\setlength{\tabcolsep}{0.06in}
	\tablecaption{J$1357$, $z_{gal} = 0.4592$ Column Densities\label{tab:J1357_4566}}
	\tablehead{
		\colhead{Ion}           	&
        \colhead{$\log N~({\cms})$}    &
		\colhead{$\log N$ Error~({\cms})}}
	\startdata
	{\HI}   & $[16.87, 19.00]$   &$\cdots$\\
{\CII}  & $>15.10$   &$\cdots$\\
{\CIII} & $>15.04$   &$\cdots$\\
{\NII}  & $>14.72$   &$\cdots$\\
{\NIII} & $>15.30$   &$\cdots$\\
{\OI}   & $<13.74$   &$\cdots$\\
{\SiII} & $15.31$   &$0.23$\\
{\CaII} & $<11.08$   &$\cdots$\\
{\MgI}  & $11.63$   &$0.14$\\
{\MgII} & $13.74$   &$0.01$\\[-5pt]

	\enddata
\end{deluxetable}
\begin{figure}[hp]
	\centering
	\includegraphics[width=\linewidth]{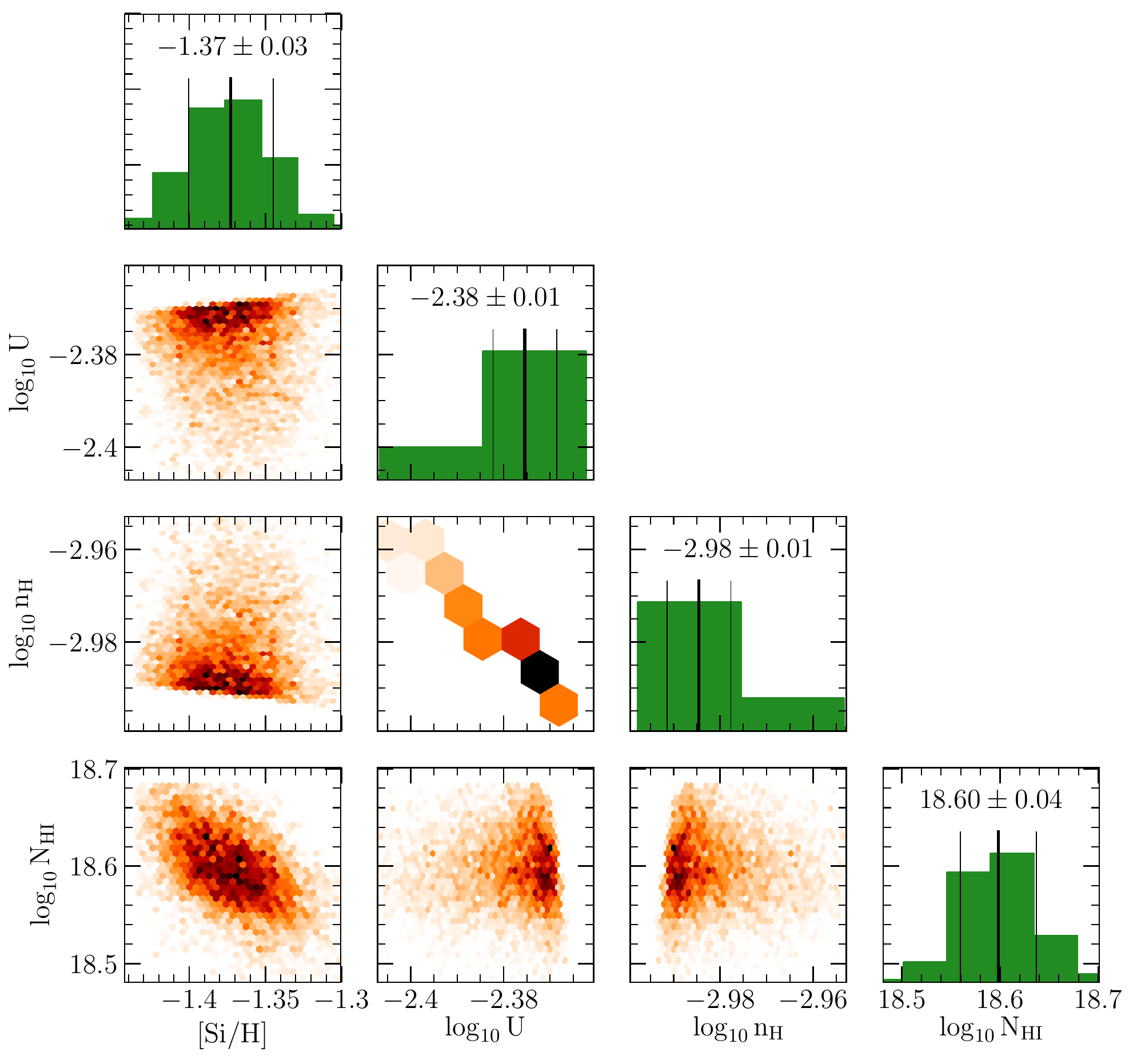}
	\caption{The posterior distribution profiles from the MCMC analysis of the Cloudy grids for J$1357$, $z_{gal} = 0.4592$, as for figure \ref{fig:Q0122_0.2119_par}.}
	\label{fig:J1357_4566_par}
\end{figure}
\begin{figure*}[hp]
	\centering
	\includegraphics[width=\linewidth]{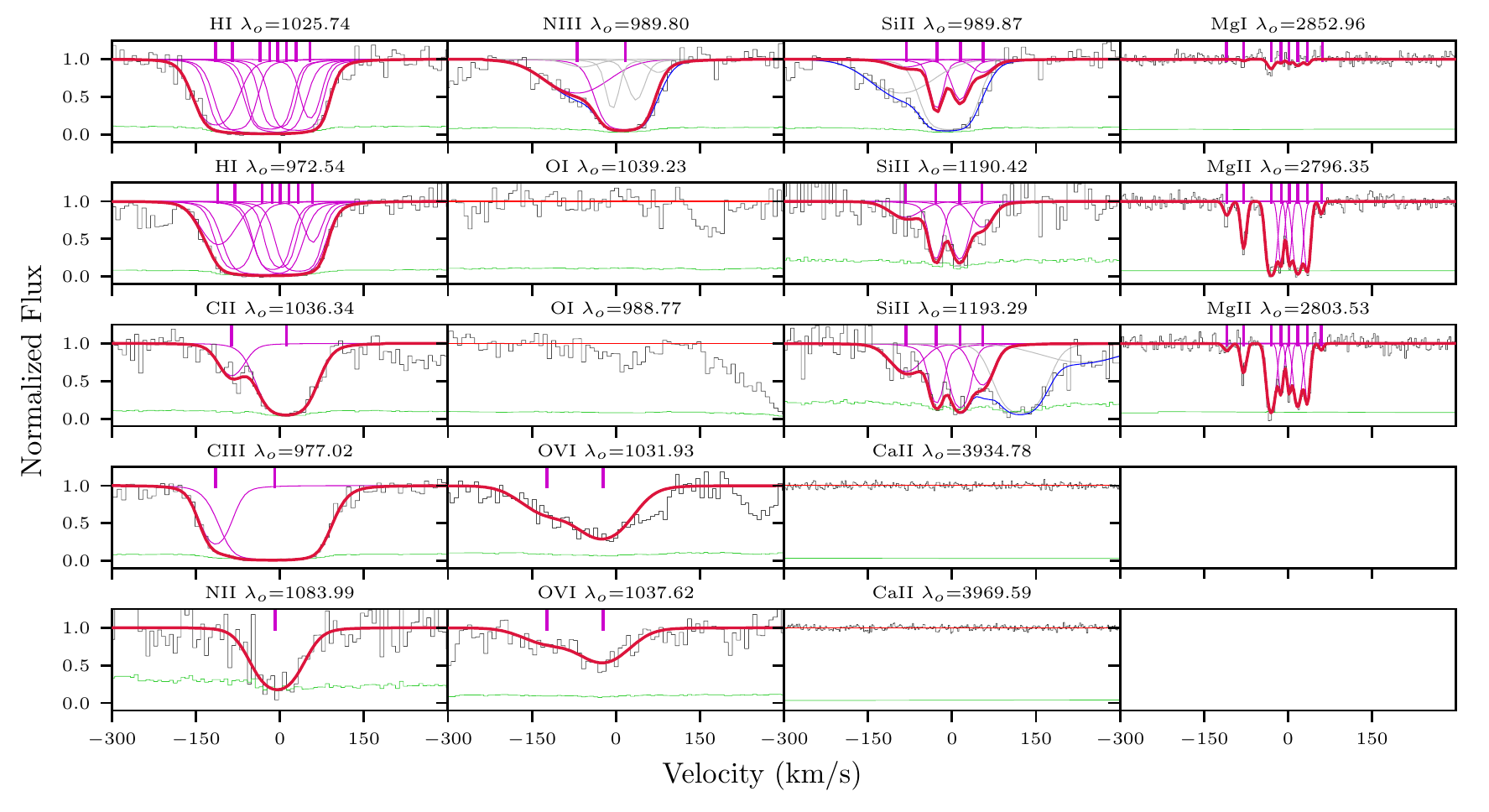}
	\caption{The fits for the system J$1357$, $z_{gal} = 0.4592$, as for figure \ref{fig:Q0122_0.2119}. The {\SiII} and {\NIII} 989~{\AA} lines are blended together. Additionally, an unknown blend was present in the {\SiII} 1193~{\AA} transition. The presence of the {\SiII} 1990~{\AA} transition constrains the column densities. The total {\OVI} fits from \citet{nielsenovi} are shown here for completeness, although they are not used in the ionization modelling.}
	\label{fig:J1357_4566}
\end{figure*}

\begin{figure*}[hp]
	\centering
	\includegraphics[width=\linewidth]{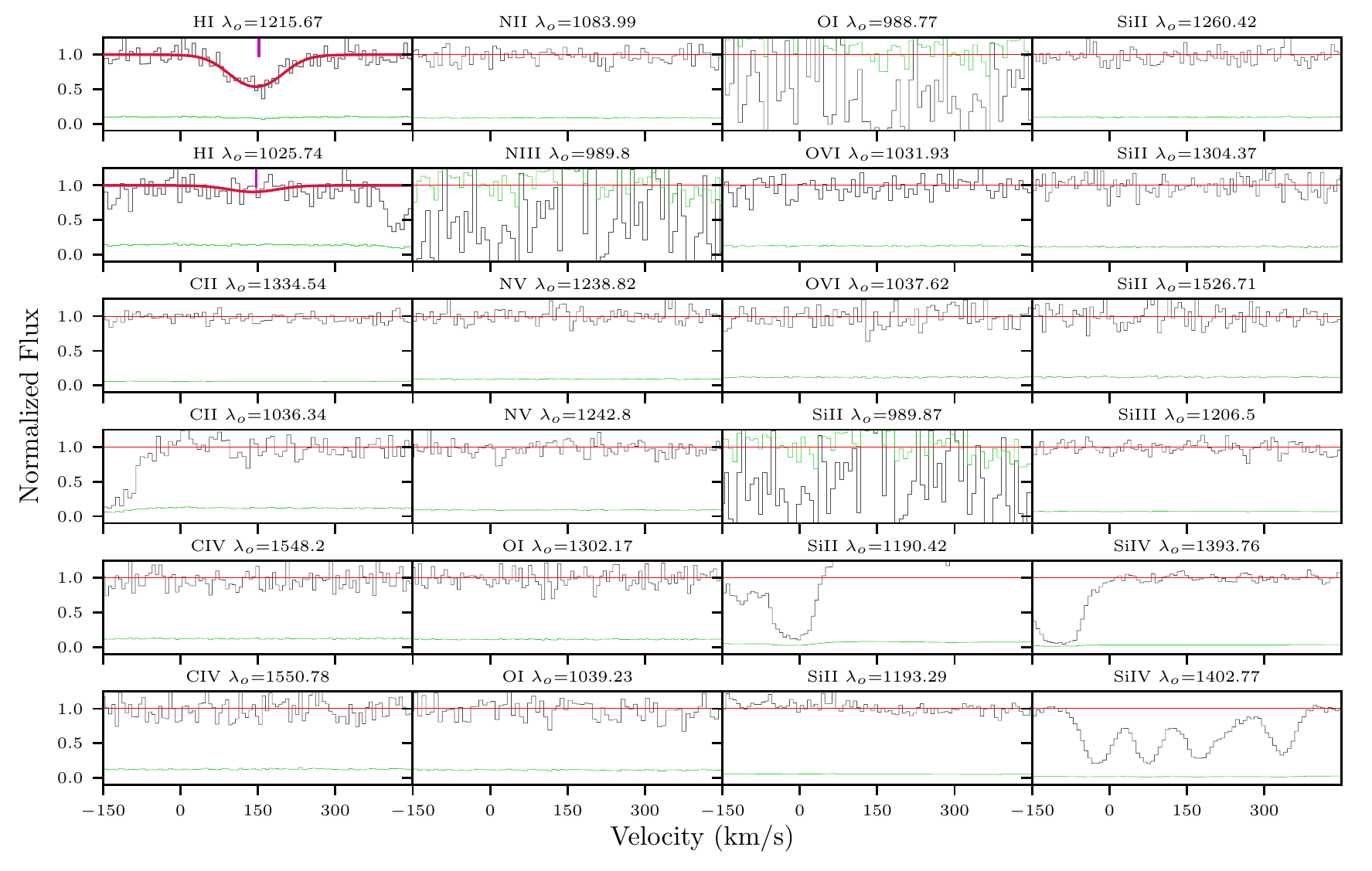}
	\caption{The fits for the system J$1547$, $z_{gal} = 0.0949$, as for figure \ref{fig:Q0122_0.2119}. There were no other metals detected here.}
	\label{fig:J1547_0965}
\end{figure*}
\clearpage
\begin{deluxetable}{ccc}[hp]
	\tablecolumns{8}
	\tablewidth{\linewidth}
	\setlength{\tabcolsep}{0.06in}
	\tablecaption{J$1547$, $z_{gal} = 0.0949$ Column Densities\label{tab:J1547_0965}}
	\tablehead{
		\colhead{Ion}           	&
        \colhead{$\log N~({\cms})$}    &
		\colhead{$\log N$ Error~({\cms})}}
	\startdata
	{\HI}   & $13.75$   &$0.03$\\
{\CII}  & $<12.88$   &$\cdots$\\
{\NII}  & $<13.22$   &$\cdots$\\
{\NIII} & $<15.66$   &$\cdots$\\
{\NV}   & $<12.97$   &$\cdots$\\
{\OI}   & $<13.57$   &$\cdots$\\
{\SiII} & $<12.19$   &$\cdots$\\
{\SiIII}& $<11.87$   &$\cdots$\\
{\SiIV} & $<11.95$   &$\cdots$\\[-5pt]

	\enddata
\end{deluxetable}
\begin{figure}[hp]
	\centering
	\includegraphics[width=\linewidth]{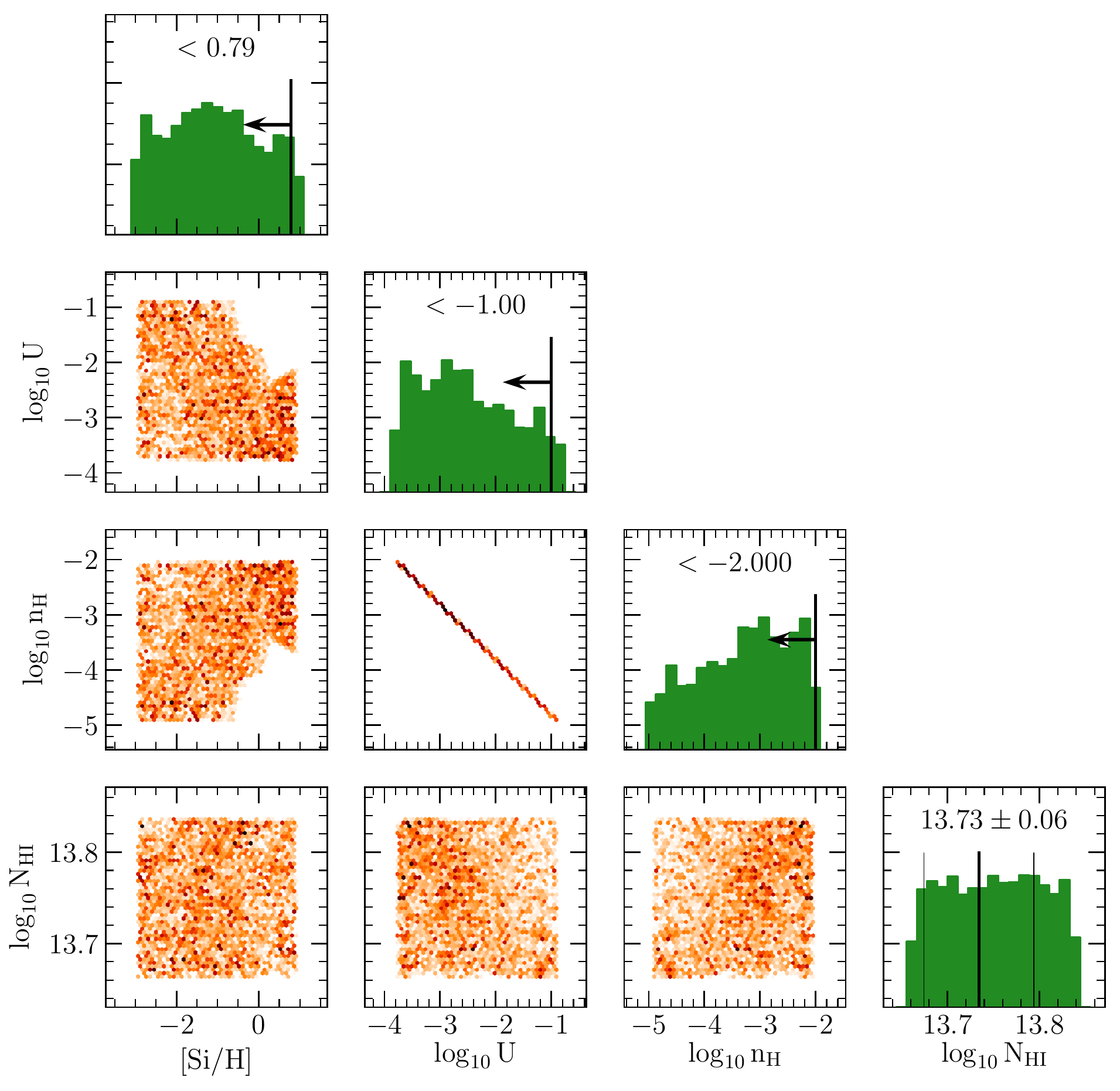}
	\caption{The posterior distribution profiles from the MCMC analysis of the Cloudy grids for J$1547$, $z_{gal} = 0.0949$, as for figure \ref{fig:Q0122_0.2119_par}.}
	\label{fig:J1547_0965_par}
\end{figure}
\newpage
\begin{deluxetable}{ccc}[hp]
	\tablecolumns{8}
	\tablewidth{\linewidth}
	\setlength{\tabcolsep}{0.06in}
	\tablecaption{J$1555$, $z_{gal} = 0.189201$ Column Densities\label{tab:J1555_1894}}
	\tablehead{
		\colhead{Ion}           	&
        \colhead{$\log N~({\cms})$}    &
		\colhead{$\log N$ Error~({\cms})}}
	\startdata
	{\HI}   & $[16.37, 19.00]$   &$\cdots$\\
{\CII}  & $14.56$   &$0.04$\\
{\CIII} & $>15.46$   &$\cdots$\\
{\NII}  & $14.25$   &$0.09$\\
{\NIII} & $15.62$   &$1.50$\\
{\NV}   & $<13.45$   &$\cdots$\\
{\OI}   & $<14.08$   &$\cdots$\\
{\SiII} & $14.05$   &$0.70$\\
{\SiIII}& $13.98$   &$0.93$\\
{\SiIV} & $13.80$   &$0.08$\\
{\CaII} & $<12.06$   &$\cdots$\\
{\MgI}  & $<11.77$   &$\cdots$\\
{\MgII} & $14.38$   &$0.40$\\
{\MnII} & $<12.63$   &$\cdots$\\
{\FeII} & $<12.62$   &$\cdots$\\[-5pt]

	\enddata
\end{deluxetable}
\begin{figure}[hp]
	\centering
	\includegraphics[width=\linewidth]{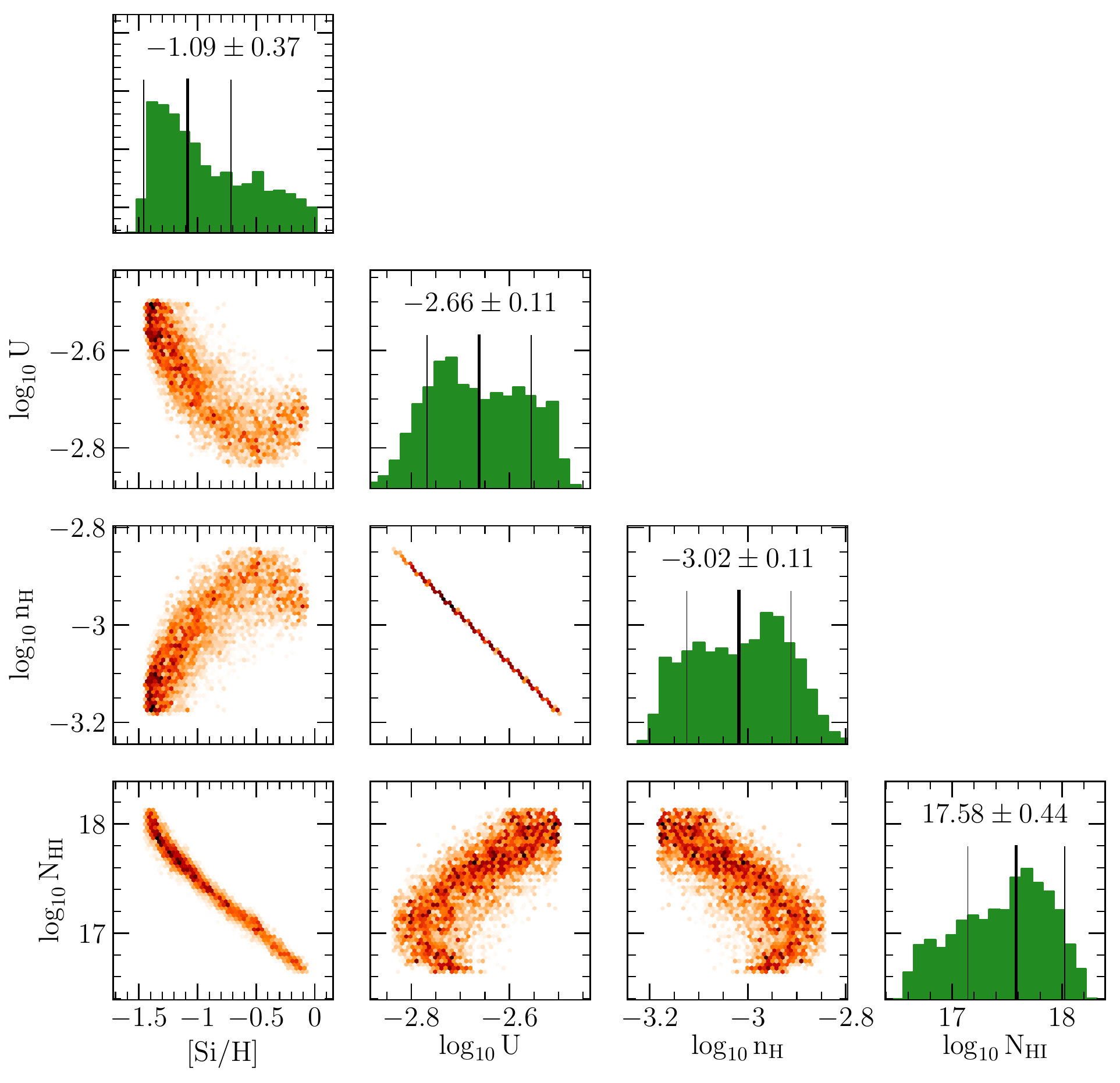}
	\caption{The posterior distribution profiles from the MCMC analysis of the Cloudy grids for J$1555$, $z_{gal} = 0.189201$, as for figure \ref{fig:Q0122_0.2119_par}.}
	\label{fig:J1555_1894_par}
\end{figure}
\begin{figure*}[hp]
	\centering
	\includegraphics[width=\linewidth]{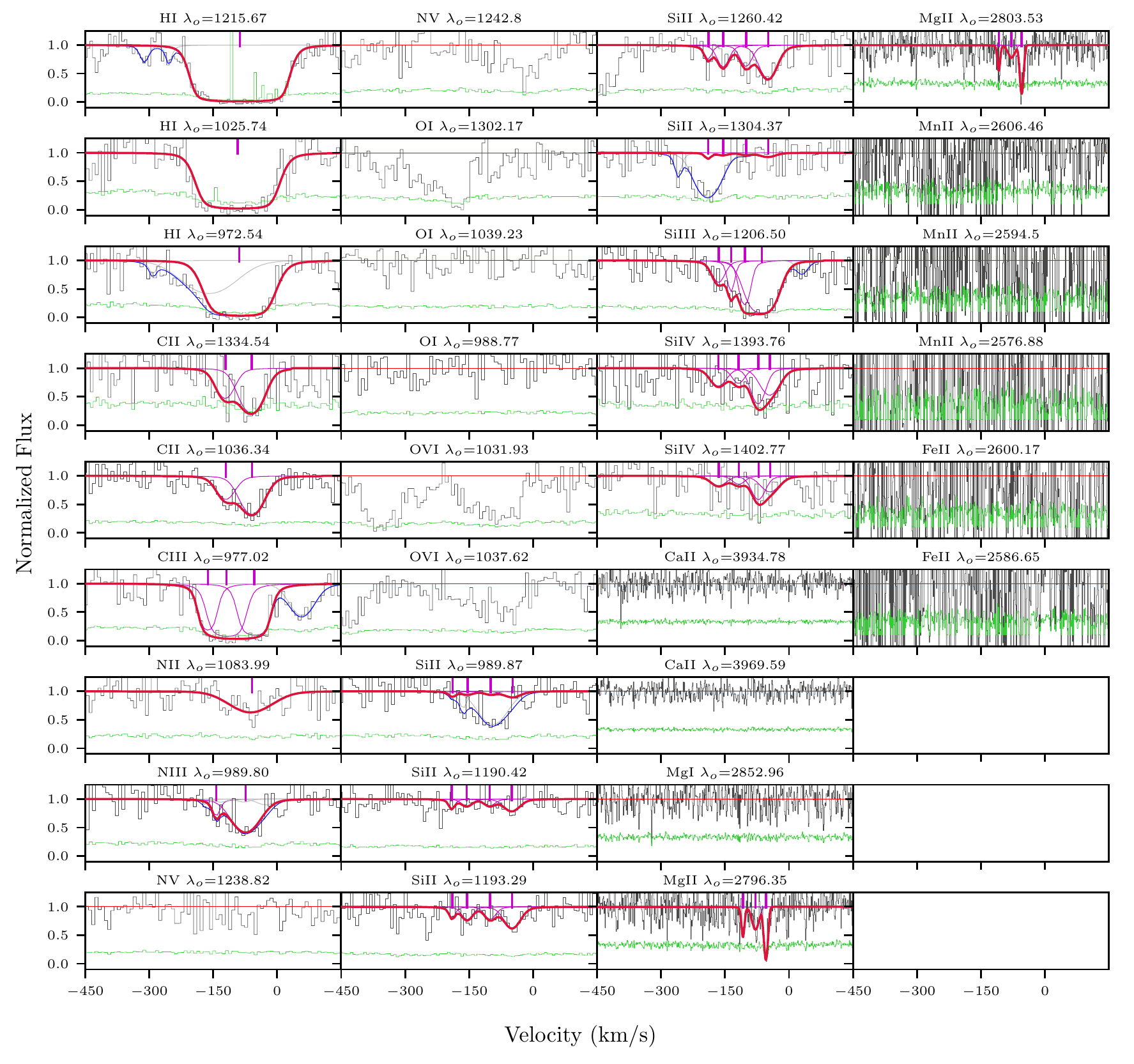}
	\caption{The fits for the system J$1555$, $z_{gal} = 0.189201$, as for figure \ref{fig:Q0122_0.2119}. Despite the presence of small, unknown blends in the {\HI} 1215~{\AA} and 972~{\AA} transitions, the column density was constrained using the 1025~{\AA} transition. The blend in the {\CIII} transition is sufficiently distinct from the absorption such that the column density could be calculated. The {\SiII} and {\NIII} 989~{\AA} lines are blended together. Additionally, an unknown blend was present in the {\SiII} 1304~{\AA} transition. The presence of the other {\SiII} transitions constrains the column densities. }
	\label{fig:J1555_1894}
\end{figure*}

\begin{figure*}[hp]
	\centering
	\includegraphics[width=\linewidth]{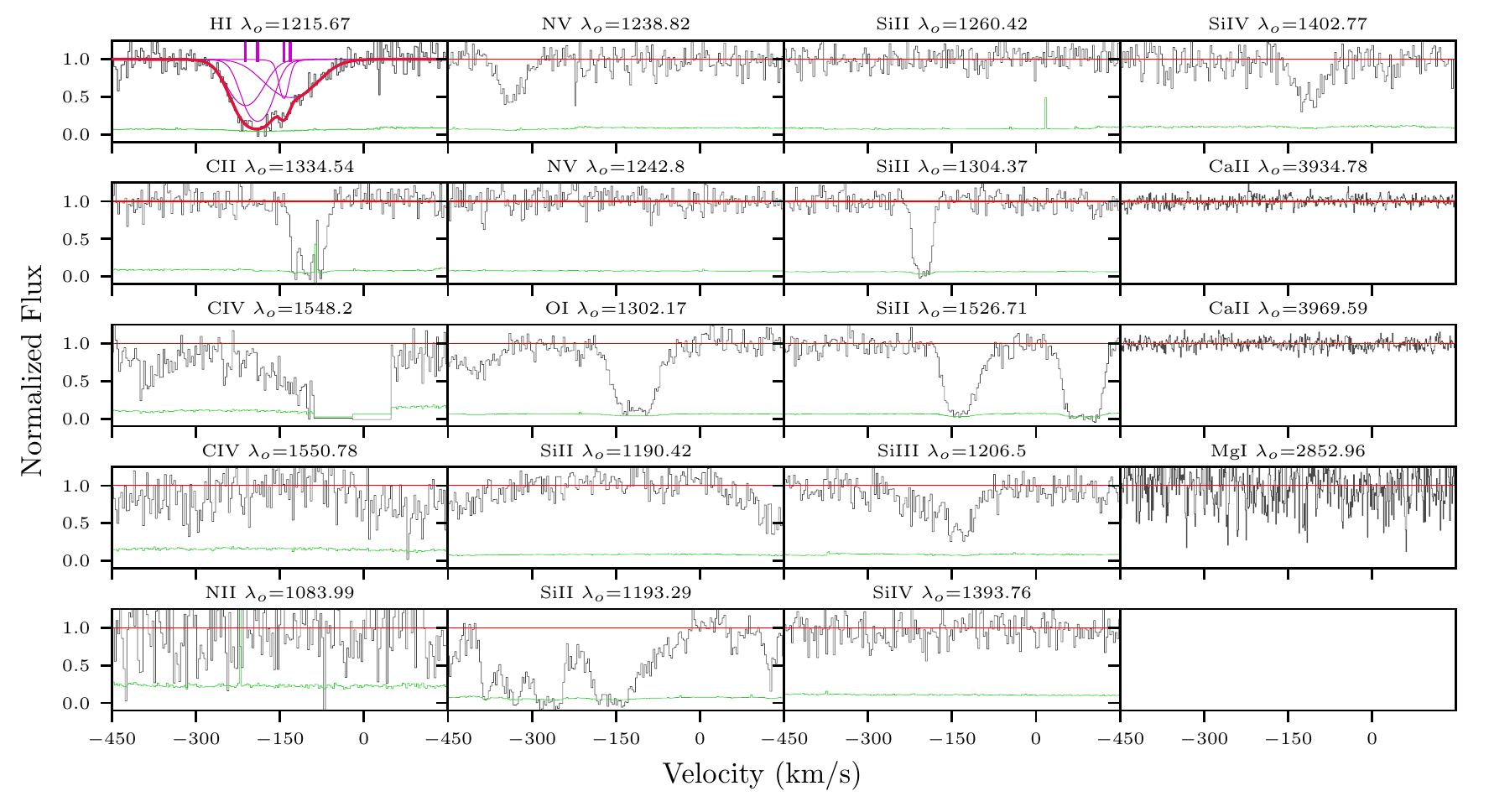}
	\caption{The fits for the system J$1704$, $z_{gal} = 0.0921$, as for figure \ref{fig:Q0122_0.2119}. No other metals are detected here.}
	\label{fig:J1704_0921}
\end{figure*}
\clearpage
\begin{deluxetable}{ccc}[hp]
	\tablecolumns{8}
	\tablewidth{\linewidth}
	\setlength{\tabcolsep}{0.06in}
	\tablecaption{J$1704$, $z_{gal} = 0.0921$ Column Densities\label{tab:J1704_0921}}
	\tablehead{
		\colhead{Ion}           	&
        \colhead{$\log N~({\cms})$}    &
		\colhead{$\log N$ Error~({\cms})}}
	\startdata
	{\HI}   & $14.27$   &$0.02$\\
{\CII}  & $<12.82$   &$\cdots$\\
{\NII}  & $<13.62$   &$\cdots$\\
{\NV}   & $<12.84$   &$\cdots$\\
{\OI}   & $<13.16$   &$\cdots$\\
{\SiII} & $<12.06$   &$\cdots$\\
{\SiIII}& $<11.85$   &$\cdots$\\
{\SiIV} & $<12.34$   &$\cdots$\\[-5pt]

	\enddata
\end{deluxetable}
\begin{figure}[hp]
	\centering
	\includegraphics[width=\linewidth]{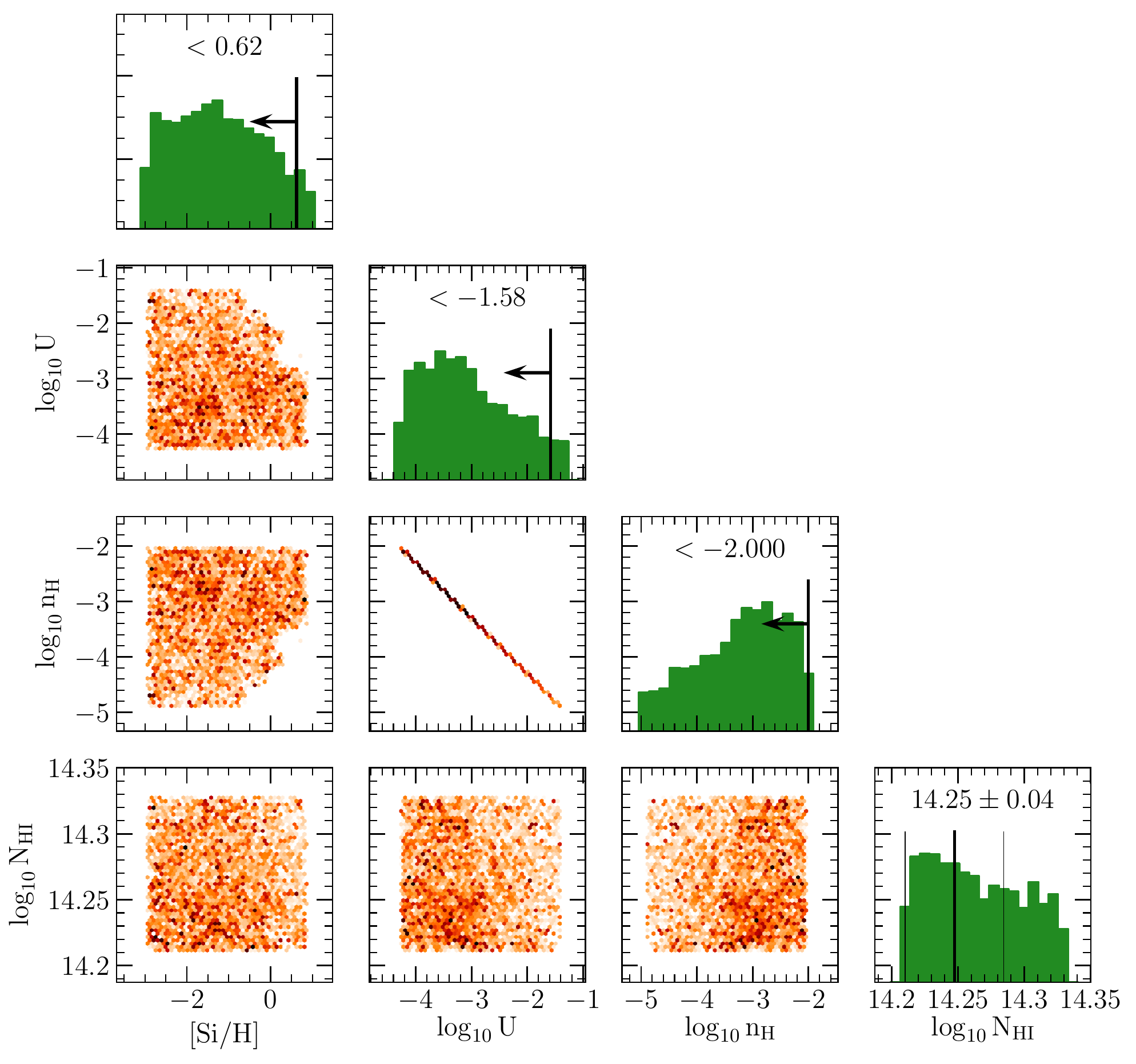}
	\caption{The posterior distribution profiles from the MCMC analysis of the Cloudy grids for J$1704$, $z_{gal} = 0.0921$, as for figure \ref{fig:Q0122_0.2119_par}.}
	\label{fig:J1704_0921_par}
\end{figure}
\newpage
\begin{deluxetable}{ccc}[hp]
	\tablecolumns{8}
	\tablewidth{\linewidth}
	\setlength{\tabcolsep}{0.06in}
	\tablecaption{J$2131$, $z_{gal} = 0.430200$ Column Densities\label{tab:J2131_4302}}
	\tablehead{
		\colhead{Ion}           	&
        \colhead{$\log N~({\cms})$}    &
		\colhead{$\log N$ Error~({\cms})}}
	\startdata
	{\HI}   & $19.88$&$0.10$\\
{\CII}  & $14.71$&$0.12$\\
{\CIII} & $14.48$&$0.03$\\
{\NII}  & $14.31$&$0.03$\\
{\NIII} & $14.62$&$0.03$\\
{\NV}   & $<13.18$&$\cdots$\\
{\OI}   & $<13.74$&$\cdots$\\
{\SiII} & $14.05$&$2.83$\\
{\SiIII}& $13.65$&$0.03$\\
{\CaII} & $12.00$&$0.01$\\
{\MgI}  &$12.19$&$0.01$\\
{\MgII} & $13.53$&$0.01$\\
{\FeII} & $13.83$&$0.04$\\

	\enddata
\end{deluxetable}
\begin{figure}[hp]
	\centering
	\includegraphics[width=\linewidth]{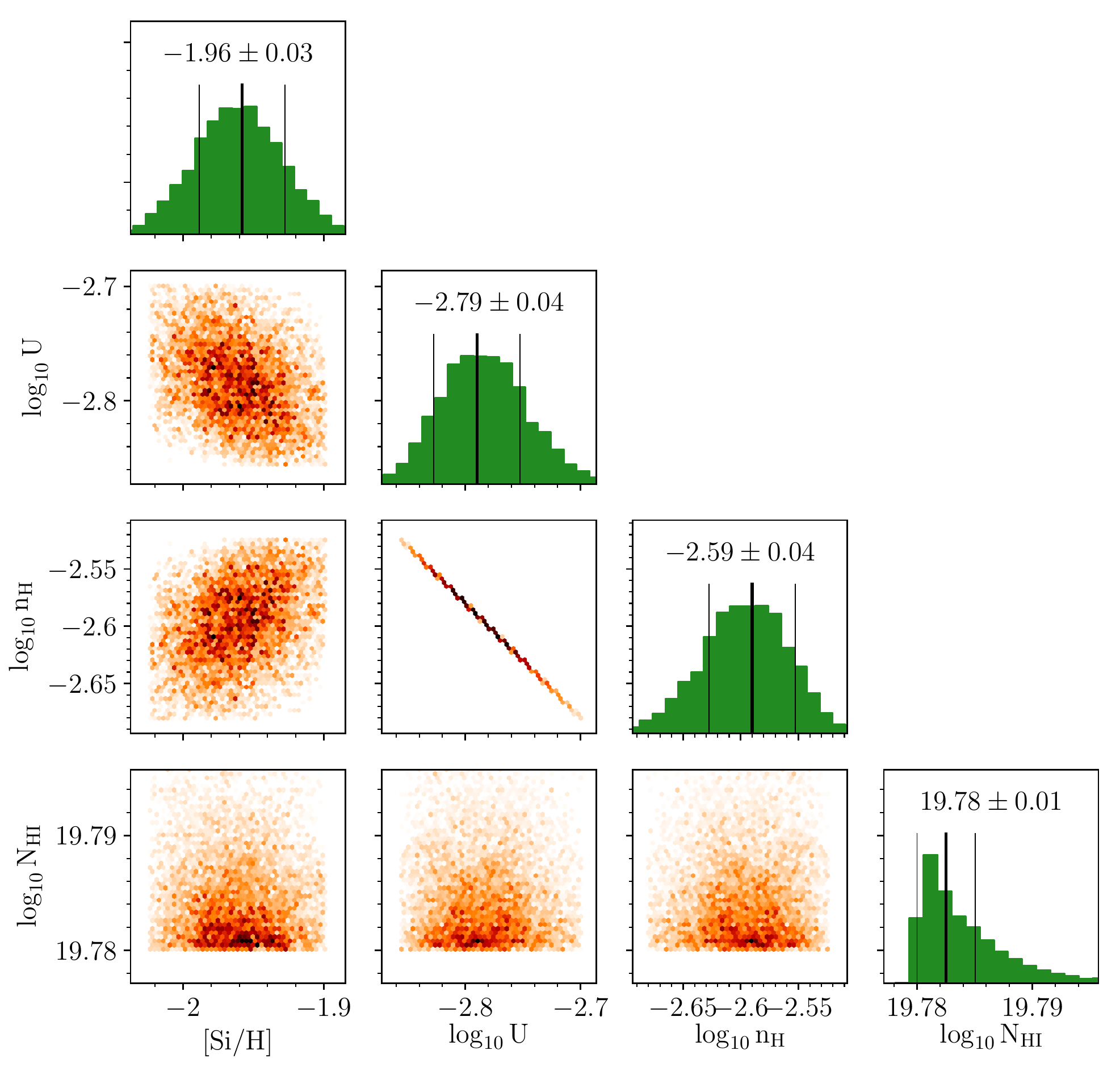}
	\caption{The posterior distribution profiles from the MCMC analysis of the Cloudy grids for J$2131$, $z_{gal} = 0.430200$, as for figure \ref{fig:Q0122_0.2119_par}.}
	\label{fig:J2131_4302_par}
\end{figure}
\begin{figure*}[hp]
	\centering
	\includegraphics[width=\linewidth]{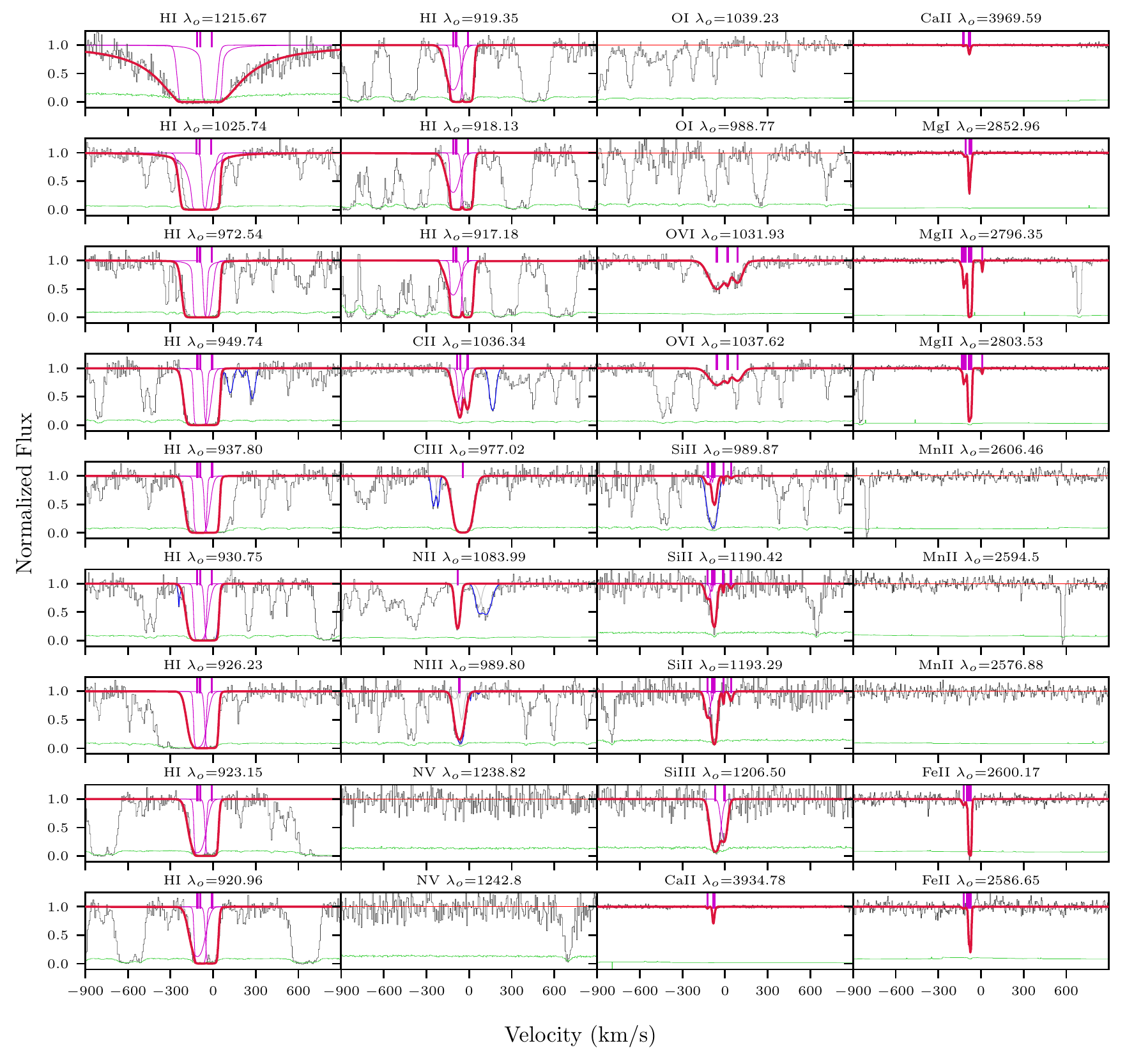}
	\caption{The fits for the system J$2131$, $z_{gal} = 0.430200$, as for figure \ref{fig:Q0122_0.2119}. The {\HI} column density is well constrained by the presence of the DLA, despite unknown blends in the {\HI} 949~{\AA} and 930~{\AA} transitions. The blends in the {\CII}, {\CIII} and {\NII} transitions are sufficiently offset such that the column densities of the ions can be constrained.  The {\SiII} and {\NIII} 989~{\AA} lines are blended together. The presence of the other {\SiII} transitions constrains the column densities. The total {\OVI} fits from \citet{nielsenovi} are shown here for completeness, although they are not used in the ionization modelling.}
	\label{fig:J2131_4302}
\end{figure*}

\begin{figure*}[hp]
	\centering
	\includegraphics[width=\linewidth]{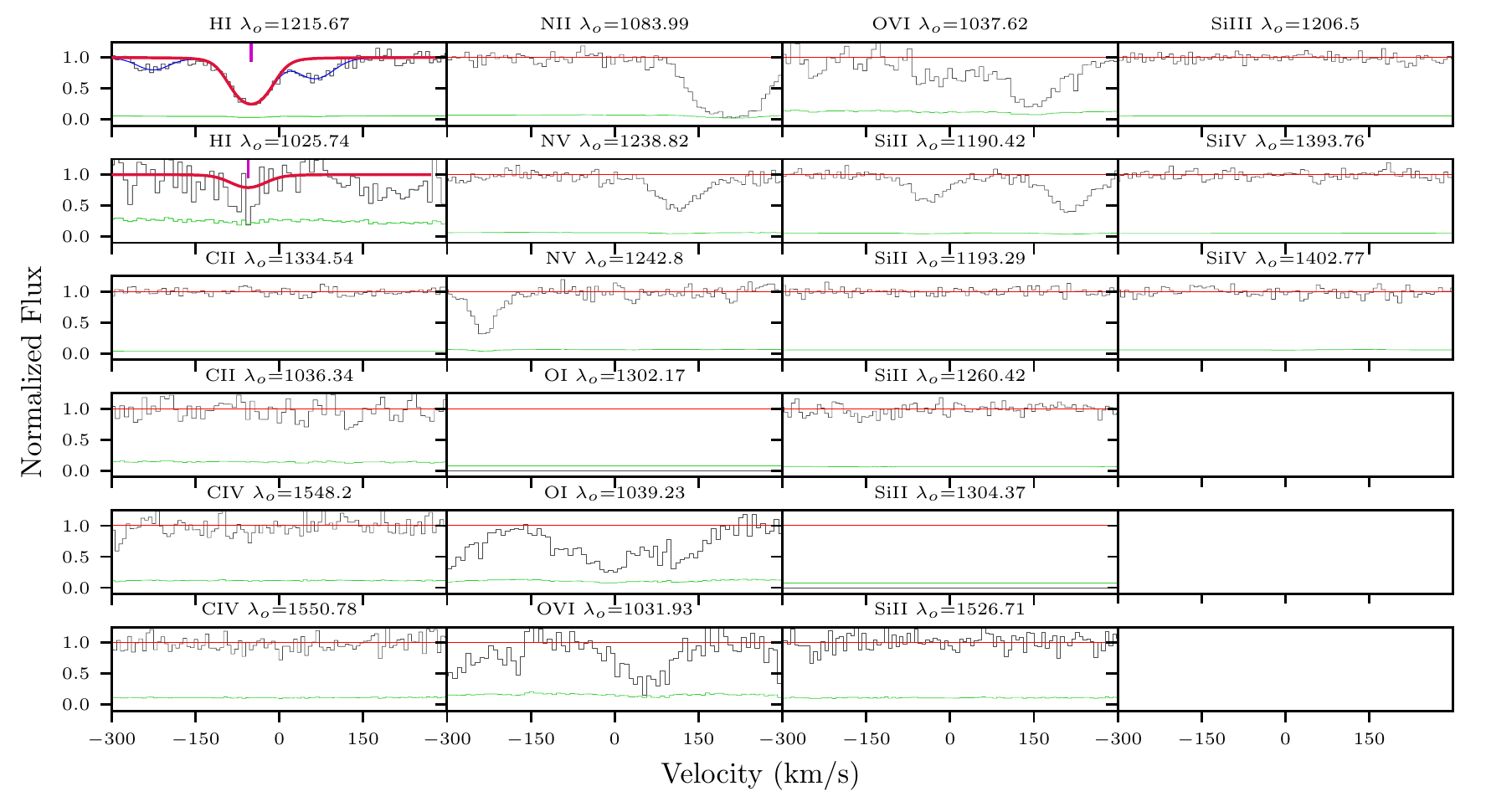}
	\caption{The fits for the system J$2137$, $z_{gal} = 0.0752$, as for figure \ref{fig:Q0122_0.2119}. The {\HI} column density was constrained using fit to the 1025~{\AA} line. There were no other metals detected for this absorber.}
	\label{fig:J2137_0755}
\end{figure*}
\clearpage
\begin{deluxetable}{ccc}[hp]
	\tablecolumns{8}
	\tablewidth{\linewidth}
	\setlength{\tabcolsep}{0.06in}
	\tablecaption{J$2137$, $z_{gal} = 0.0752$ Column Densities\label{tab:J2137_0755}}
	\tablehead{
		\colhead{Ion}           	&
        \colhead{$\log N~({\cms})$}    &
		\colhead{$\log N$ Error~({\cms})}}
	\startdata
	{\HI}   & $13.96$   &$0.02$\\
{\CII}  & $<12.68$   &$\cdots$\\
{\NII}  & $<13.14$   &$\cdots$\\
{\NV}   & $<13.84$   &$\cdots$\\
{\OI}   & $<14.63$   &$\cdots$\\
{\SiII} & $<12.02$   &$\cdots$\\
{\SiIII}& $<11.79$   &$\cdots$\\
{\SiIV} & $<12.19$   &$\cdots$\\[-5pt]

	\enddata
\end{deluxetable}
\begin{figure}[hp]
	\centering
	\includegraphics[width=\linewidth]{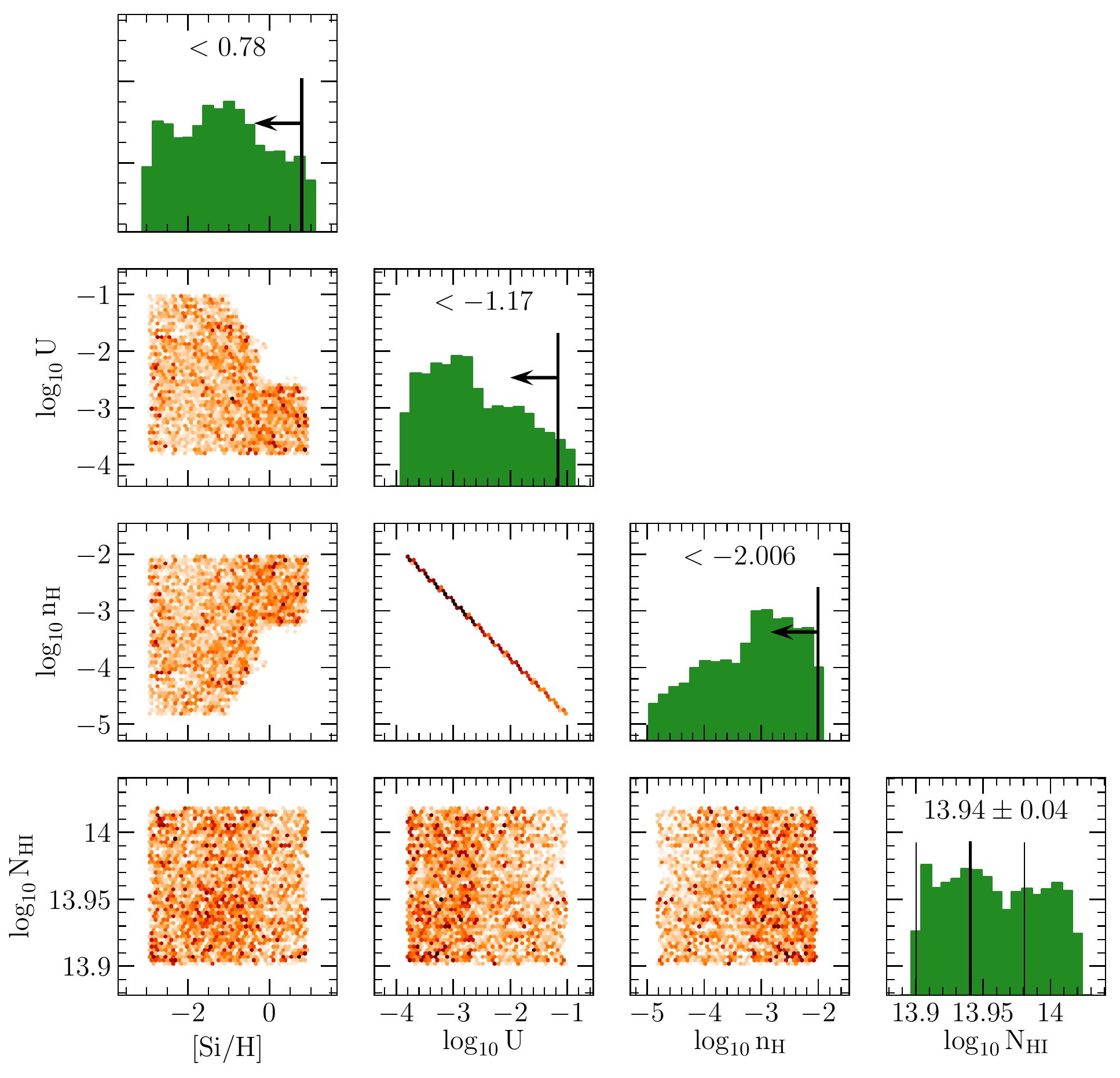}
	\caption{The posterior distribution profiles from the MCMC analysis of the Cloudy grids for J$2137$, $z_{gal} = 0.0752$, as for figure \ref{fig:Q0122_0.2119_par}.}
	\label{fig:J2137_0755_par}
\end{figure}
\newpage
\begin{deluxetable}{ccc}[hp]
	\tablecolumns{8}
	\tablewidth{\linewidth}
	\setlength{\tabcolsep}{0.06in}
	\tablecaption{J$2253$, $z_{gal} = 0.352787$ Column Densities\label{tab:J2253_3527}}
	\tablehead{
		\colhead{Ion}           	&
        \colhead{$\log N~({\cms})$}    &
		\colhead{$\log N$ Error~({\cms})}}
	\startdata
	{\HI}   & $14.53$   &$0.05$\\
{\CII}  & $<14.70$   &$\cdots$\\
{\CIII} & $<13.98$   &$\cdots$\\
{\NII}  & $<13.39$   &$\cdots$\\
{\NIII} & $<13.84$   &$\cdots$\\
{\NV}   & $<13.45$   &$\cdots$\\
{\OI}   & $<14.13$   &$\cdots$\\
{\SiII} & $<12.63$   &$\cdots$\\
{\SiIII}& $<12.40$   &$\cdots$\\
{\CaII} & $<10.92$   &$\cdots$\\
{\MgI}  & $<10.69$   &$\cdots$\\
{\MgII} & $<11.26$   &$\cdots$\\
{\MnII} & $<12.80$   &$\cdots$\\
{\FeII} & $<11.96$   &$\cdots$\\[-5pt]

	\enddata
\end{deluxetable}
\begin{figure}[hp]
	\centering
	\includegraphics[width=\linewidth]{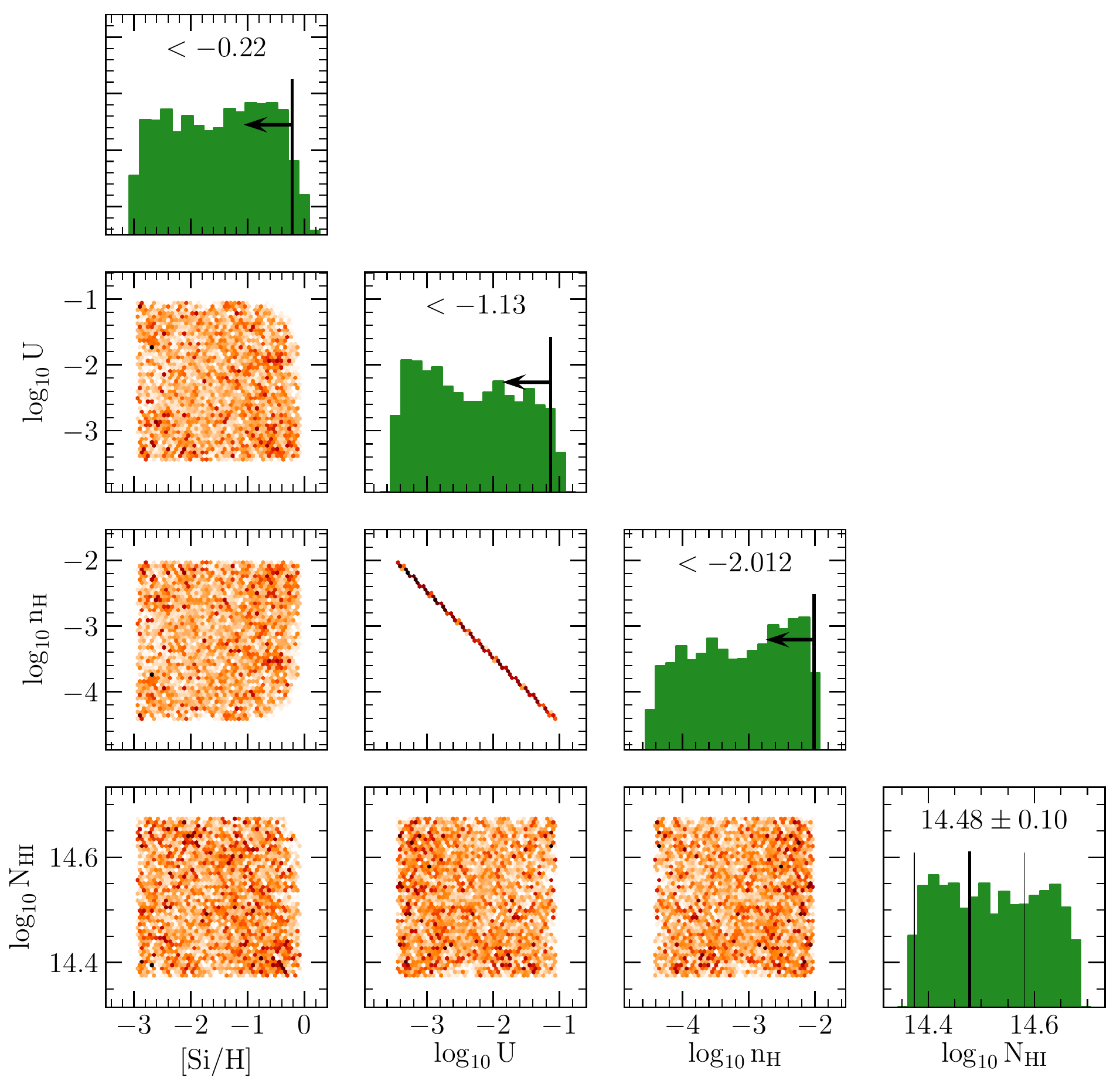}
	\caption{The posterior distribution profiles from the MCMC analysis of the Cloudy grids for J$2253$, $z_{gal} = 0.352787$, as for figure \ref{fig:Q0122_0.2119_par}.}
	\label{fig:J2253_3527_par}
\end{figure}
\begin{figure*}[hp]
	\centering
	\includegraphics[width=\linewidth]{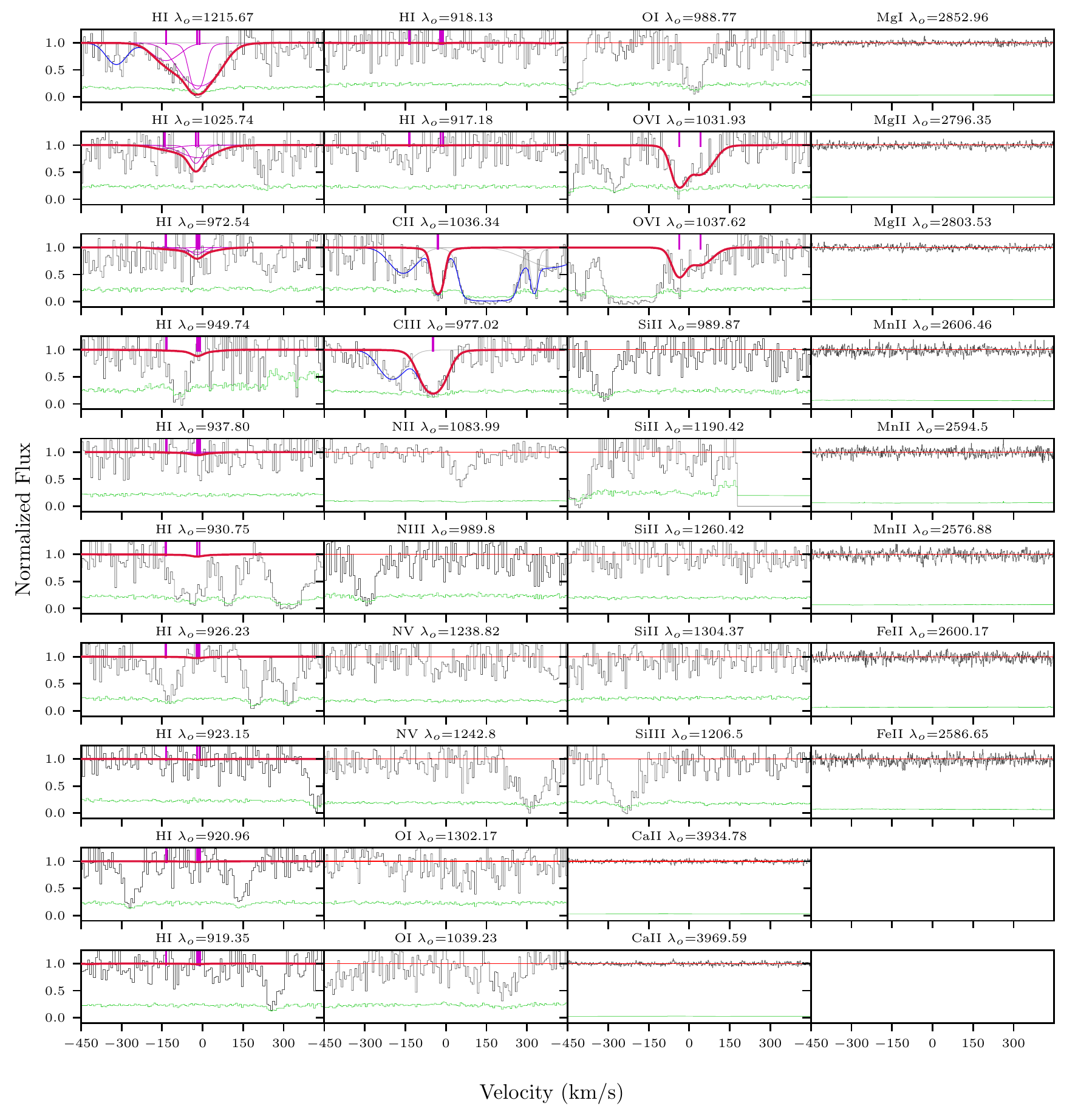}
	\caption{The fits for the system J$2253$, $z_{gal} = 0.352787$, as for figure \ref{fig:Q0122_0.2119}. The blends in the {\CII} and {\CIII} transitions were sufficiently offset that the absorption column density for each ion could be constrained. The total {\OVI} fits from \citet{nielsenovi} are shown here for completeness, although they are not used in the ionization modelling. }
	\label{fig:J2253_3527}
\end{figure*}
\end{appendices}
\end{document}